\documentclass[aps,prd,twocolumn,superscriptaddress,amssymb,eqsecnum,showpacs,showkeyes,secnumarabic,graphics,floatfix,nofootinbib,tightenlines,longbibliography]{revtex4-1}

\usepackage{relsize}
\usepackage{blindtext}
\usepackage{enumitem}
\usepackage{xcolor}
\usepackage{graphicx} 
\usepackage{epsfig}
\usepackage{bm}
\usepackage{cancel}
\usepackage{float}
\usepackage{dcolumn}
\usepackage{mathtools,amsmath} 
\usepackage{hhline} 
\usepackage[utf8]{inputenc}
\usepackage{longtable}
\usepackage[breaklinks=true,colorlinks=true,
linkcolor=blue,urlcolor=blue,citecolor=cyan,
bookmarks=true,bookmarksopenlevel=2]{hyperref} 
\usepackage{afterpage} 
\usepackage{natbib}

\def\apj{{Astroph.\@ J.\ }}

\def\aj{{Astron.\@ J.\ }}

\def\nat{{Nature\ }}

\def \beq  {\begin{equation}}
\def \eeq  {\end{equation}}
\def \ber  {\begin{eqnarray}}
\def \eer  {\end{eqnarray}}

\usepackage{titlesec}

\titleformat{\chapter}[display]
{\normalfont\huge\bfseries}{\chaptertitlename\ \thechapter}{20pt}{\Huge}
\titleformat{\section}
{\normalfont\Large\bfseries}{\thesection}{1em}{}
\titleformat{\subsection}
{\normalfont\large\bfseries}{\thesubsection}{1em}{}
\titleformat{\subsubsection}
{\normalfont\normalsize\bfseries}{\thesubsubsection}{1em}{}
\titleformat{\paragraph}[runin]
{\normalfont\normalsize\bfseries}{\theparagraph}{1em}{}
\titleformat{\subparagraph}[runin]
{\normalfont\normalsize\bfseries}{\thesubparagraph}{1em}{}

\titlespacing*{\chapter} {0pt}{50pt}{40pt}
\titlespacing*{\section} {0pt}{3.5ex plus 1ex minus .2ex}{2.3ex plus .2ex}
\titlespacing*{\subsection} {0pt}{3.25ex plus 1ex minus .2ex}{1.5ex plus .2ex}
\titlespacing*{\subsubsection}{0pt}{3.25ex plus 1ex minus .2ex}{1.5ex plus .2ex}
\titlespacing*{\paragraph} {0pt}{3.25ex plus 1ex minus .2ex}{1.5ex plus .2ex}
\titlespacing*{\subparagraph} {\parindent}{3.25ex plus 1ex minus .2ex}{1em}

\setcitestyle{authoryear,semicolon,open={(},close={)}}
\begin{document}
\newcommand{\newc}{\newcommand}

\newc{\be}{\begin{equation}}
\newc{\ee}{\end{equation}}
\newc{\ba}{\begin{eqnarray}}
\newc{\ea}{\end{eqnarray}}
\newc{\bea}{\begin{eqnarray*}}
\newc{\eea}{\end{eqnarray*}}
\newc{\D}{\partial}
\newc{\ie}{{\it i.e.} }
\newc{\eg}{{\it e.g.} }
\newc{\etc}{{\it etc.} } 
\newc{\etal}{{\it et al.}}
\newc{\lcdm}{$\Lambda$CDM }
\newc{\plcdm}{Planck/$\Lambda$CDM }
\newcommand{\nn}{\nonumber}
\newc{\ra}{\Rightarrow}

\title{Challenges for $\Lambda$CDM: An update} 

\author{L. Perivolaropoulos}\email{leandros@uoi.gr} 
\author{F. Skara}\email{f.skara@uoi.gr}
\affiliation{Department of Physics, University of Ioannina, 45110 Ioannina, Greece}

\date {\today} 

\begin{abstract}
A number of challenges to the standard $\Lambda$CDM model have been emerging during the past few years as the accuracy of cosmological observations improves. In this review we discuss in a unified manner many existing signals in cosmological and astrophysical data that appear to be in some tension ($2\sigma$ or larger) with the standard $\Lambda$CDM model as specified by the Cosmological Principle, General Relativity and the Planck18 parameter values. In addition to the well-studied $5\sigma$ challenge of $\Lambda$CDM (the Hubble $H_0$ tension) and other well known tensions (the growth tension, and the lensing amplitude $A_L$ anomaly), we discuss a wide range of other less discussed less-standard signals which appear at a lower statistical significance level than the $H_0$ tension some of them known as 'curiosities' in the data) which may also constitute hints towards new physics. For example such signals include cosmic dipoles (the fine structure constant $\alpha$, velocity and quasar dipoles), CMB asymmetries, BAO Ly$\alpha$ tension, age of the Universe issues, the Lithium problem, small scale curiosities like the core-cusp and missing satellite problems, quasars Hubble diagram, oscillating short range gravity signals etc.  The goal of this pedagogical review is to collectively present the current status (2022 update) of these signals and their level of significance, with emphasis on the Hubble tension and refer to recent resources where more details can be found for each signal. We also briefly discuss theoretical approaches that can potentially explain some of these signals.

\end{abstract}
\maketitle

\tableofcontents 

\section{Introduction}  
\label{sec:Introduction} 
The concordance or standard  $\Lambda$ Cold Dark Matter ($\Lambda$CDM) cosmological model \citep{Peebles:1984ge,Peebles:2002gy,Carroll:2000fy} is a well defined,  predictive and simple cosmological model \citep[see][for a review]{Bull:2015stt}.
It is defined by a set of simple assumptions:
\begin{itemize}
 \item
  The Universe consists of radiation (photons, neutrinos),  ordinary matter (baryons and leptons), cold  (non-relativistic) dark matter (CDM) \citep{Zwicky:1933gu,Zwicky:1937zza,Freeman:1970mx,Rubin:1970zza,Rubin:1980zd,Bosma:1981zz,Bertone:2004pz} being responsible for structure formation and cosmological constant $\Lambda$ \citep{Carroll:1991mt,Carroll:2000fy}, a homogeneous form of energy which is responsible for the late time observed accelerated expansion. The cosmological constant is currently associated with a dark energy or vacuum energy whose density remains constant even in an expanding background \citep[see][for a review]{Peebles:2002gy,Padmanabhan:2002ji,Padmanabhan:2004av,Weinberg:1988cp}.
    \item 
     General Relativity (GR) \citep{Einstein:1917ce} is the correct theory that describes gravity on  cosmological scales. Thus, the action currently relevant on cosmological scales reads
       \begin{align}
     S=\int d^4 x&\sqrt{-g}\left[\frac{1}{16\pi G}(R-2\Lambda)\right. \nonumber\\
+& \left. \frac{1}{4\alpha}F_{\mu\nu}F^{\mu\nu}+\mathcal{L}_m(\psi,A) \right]
\label{lcdmdef}
  \end{align}
 where $\alpha$ is the fine structure constant, $G$ is Newton's constant, $F_{\mu\nu}$ is the electromagnetic field-strength tensor and $\mathcal{L}_m$ is the Lagrangian density for all matter fields $\psi_m$. 
\item
The Cosmological Principle (CP) states that the Universe is statistically homogeneous and isotropic in space and  matter at sufficiently large  scales ($\gtrsim 100\,Mpc$).
\item
   There are six independent (free) parameters: the baryon $\omega_b=\Omega_{0b}h^2$ and cold dark matter $\omega_c=\Omega_{0c}h^2$ energy densities (where $h=H_0/100$ $km$ $s^{-1}$ $Mpc^{-1}$ is the dimensionless Hubble constant and $\Omega_X\equiv \rho_X/\rho_{crit}$ is the density of component $X$ relative to the critical density, $\rho_{crit}= 3H^2/8\pi G$), the angular diameter distance to the sound horizon at last scattering $\theta_s$, the amplitude $A_s$ and tilt $n_s$ of primordial scalar fluctuations and the reionization optical depth $\tau$.
\item
The spatial part of the cosmic metric is assumed to be  flat described by the Friedmann-Lema$\hat{ı}$tre-Roberson-Walker (FLRW) metric 
\be 
ds^2=dt^2-a(t)^2(dr^2+r^2d\theta^2+r^2sin^2\theta d\phi^2)
\ee
which emerges from the CP. 

Assuming this form of the metric and  Einstein’s field equations with a $\Lambda$-term we obtain the Friedmann equations  which may be written as
\be 
H^2\equiv \frac{\dot{a}^2}{a^2}=\frac{8\pi G\rho+\Lambda c^2}{3}
\ee
\be 
\frac{\ddot{a}}{a}=-\frac{4\pi G}{3}(\rho+\frac{3p}{c^2})+\frac{\Lambda c^2}{3}
\ee
where $a$ is the scale factor $a=\frac{1}{1+z}$ (with $z$ the redshift). 
The cosmological constant may also be viewed as a cosmic dark energy fluid  with equation of state parameter
\be 
w=\frac{p_{\Lambda}}{\rho_{\Lambda}}=-1
\ee
where $\rho_{\Lambda}$ and $p_{\Lambda}$ are the energy  density and the pressure of the dark energy respectively.
\item
A primordial phase of cosmic inflation (a period of rapid accelerated expansion) is also assumed in order to address the horizon and flatness problems \citep{Starobinsky:1980te,Guth:1980zm,Linde:1981mu,Albrecht:1982wi}.    
During this period, Gaussian scale invariant primordial fluctuations are produced from  quantum fluctuations in the  inflationary epoch.
\end{itemize}
Fundamental generalizations of the standard 
$\Lambda$CDM model may be produced by modifying the defining action (\ref{lcdmdef}) by generalizing the fundamental constants to dynamical variables in the existing action or adding new terms. Thus the following extensions of \lcdm emerge:
\begin{itemize}
\item
Promoting Newton’s constant to a dynamical degree of freedom by allowing it to depend on a scalar field $\Phi$ as $G\rightarrow G(\Phi(r,t))$ where the dynamics of $\Phi$ is determined by kinetic and potential terms added to the action. This class of theories is known as 'scalar-tensor theories' with its most general form with second order dynamical equations the Horndeski theories \citep{Horndeski:1974wa,Deffayet:2011gz}  \citep[see also][for a comprehensive review]{Kase:2018aps,Kobayashi:2019hrl}.
\item
Promoting the cosmological constant to a dynamical degree of freedom by the introduction of a scalar field (quintessence) with $\Lambda\rightarrow V(\Phi(r,t))$ and the introduction of a proper kinetic term. 
\item
Allowing for a dynamical Fine Structure Constant (Maxwell Dilaton theories)  with $\alpha\rightarrow \alpha(\Phi(r,t))$ \citep{Bekenstein:1982eu,Sandvik:2001rv,Barrow:2001iw,Barrow:2011kr,Barrow:2013uza} \citep[see also][for a review]{Martins:2017yxk}.
\item
Addition of new terms to the action which may be functions of the Ricci scalar, the torsion scalar or other invariants ($(f(R), f(T),…)$) \citep{Starobinsky:1980te,Nojiri:2006ri,Nojiri:2010wj,DeFelice:2010aj,Sotiriou:2008rp,Ferraro:2006jd,Nesseris:2013jea,Cai:2015emx,Capozziello:2002rd}.
\end{itemize}
The $\Lambda$CDM model has been remarkably successful in explaining most properties of a wide range of cosmological  observations including the accelerating expansion of the Universe \citep{Riess:1998cb,Perlmutter:1998np}, the power spectrum and statistical properties of the cosmic microwave background (CMB) anisotropies \citep{Page:2003fa}, the spectrum and statistical properties of large  scale  structures of the Universe \citep{Bernardeau:2001qr,Bull:2015stt} and the observed abundances of different types of light nuclei hydrogen, deuterium, helium, and lithium \citep{Schramm:1997vs,Steigman:2007xt,Iocco:2008va,Cyburt:2015mya}.
 
Despite of its remarkable successes and  simplicity, the validity of the cosmological standard model $\Lambda$CDM is currently under intense investigation \citep[see][for a review]{Abdalla:2022yfr,Buchert:2015wwr,DiValentino:2021izs,Schoneberg:2021qvd, Anchordoqui:2021gji,Schmitz:2022hsz}. This is motivated by a range of profound theoretical and observational difficulties of the  model. 

The most important theoretical difficulties that plague $\Lambda$CDM are the fine tuning \citep{Weinberg:1988cp,Martin:2012bt,Burgess:2013ara} and  coincidence problems \citep{1997cpp..conf..123S,Velten:2014nra}. The  first fundamental problem is associated with the fact that there is a large discrepancy between observations and theoretical expectations on the value of the cosmological constant $\Lambda$ (at least $60$ orders of magnitude) \citep{Weinberg:1988cp,Copeland:2006wr,Martin:2012bt,Sola:2013gha} and the second is connected to the coincidence between the observed vacuum energy density $\Omega_{\Lambda}$ and the matter density $\Omega_m$ which are approximately equal nowadays despite their dramatically different evolution properties. The anthropic principle has been considered as a possible solution to these problems. It states that these 'coincidences' result from a selection bias towards the existence of human life in the context of a multiverse \citep{Susskind:2003kw,Weinberg:1987dv}.

In addition to the above theoretical challenges, there are signals in cosmological and astrophysical data that appear to be in some tension ($2\sigma$ or larger) with the standard $\Lambda$CDM model as specified by the Planck18 parameter values \citep{Planck:2018vyg,Planck:2018nkj}. The most intriguing large scale tensions are the following\footnote{We use the term 'curiosity' as a term describing a discrepancy between datasets in \lcdm best fit parameter values at a level  with a statistical significance $\lesssim 3\sigma$.} \citep{Abdalla:2022yfr} \citep[see also][for a recent overview of the main tensions]{DiValentino:2020zio,DiValentino:2020vvd}:
\begin{itemize}
    \item 
    {\bf The Hubble tension ($>4\sigma$):} (see Section \ref{sec:Hubble tension}) Using a distance ladder approach, the local (late or low redshift)  measurements of the Hubble constant $H_0$  are measured to values that are significantly higher than those inferred using the angular scale of fluctuations of the CMB in the context of the $\Lambda$CDM model. Combined local direct measurements of $H_0$ are in  $5\sigma$ tension (or more if combinations of local measurements are used) with CMB indirect measurements of $H_0$ \citep{Wong:2019kwg,DiValentino:2020vnx,Riess:2019qba}. The \plcdm best fit value is  $H_0=67.4\pm0.5$ $km$ $ s^{-1} Mpc^{-1}$ \citep{Planck:2018vyg} while the local measurements using Cepheid calibrators by the SH0ES Team indicate $H_0=73.04\pm1.04$ $km$  $ s^{-1} Mpc^{-1}$ ($\sim 5\sigma$) \citep{Riess:2021jrx} \citep[see ][for a review]{DiValentino:2021izs,Shah:2021onj,CANTATA:2021ktz}. In the previous analysis by the SH0ES Team \citep{Riess:2020fzl} using the Gaia Early Data Release $3$ (EDR$3$) parallaxes \citep{gaiacollaboration2020gaia} a value of $H_0= 73.2\pm 1.3$ is obtained, at a $4.2\sigma$ tension with the prediction from Planck18 CMB  observations. A wide range of local observations appear to be consistently larger than the \plcdm measurement of $H_0$ at various levels of statistical significance \citep{Wong:2019kwg,DiValentino:2020vnx,Riess:2019qba}. Theoretical models addressing the Hubble tension utilize either a recalibration of the \plcdm standard ruler (the sound horizon) assuming new physics before the time of recombination \citep{Karwal:2016vyq,Poulin:2018cxd,Agrawal:2019lmo} or a deformation of the Hubble expansion rate $H(z)$ at late times \citep{Alestas:2020mvb,DiValentino:2016hlg} or a transition/recalibration of the SnIa absolute luminosity due to late time new physics \citep{Marra:2021fvf} \citep[see in][for a relevant talk]{pres1}. Also, for more detailed discussions of the proposed new-physics models see in \citep{Verde:2019ivm,DiValentino:2021izs,Schoneberg:2021qvd,Anchordoqui:2021gji}.  
    \item
    {\bf The growth tension ($2-3\sigma$):} (see Subsection \ref{Growth Tension}) Direct measurements of the growth rate of cosmological perturbations (Weak Lensing, Redshift Space Distortions (peculiar velocities), Cluster Counts) indicate a lower growth rate than that indicated by the Planck/$\Lambda$CDM parameter values at a level of about $2-3\sigma$ \citep{Joudaki:2017zdt,Abbott:2017wau,Basilakos:2017rgc}. In the context of General Relativity such lower growth rate would imply a lower matter density and/or a lower amplitude of primordial fluctuation spectrum than that indicated by \plcdm \citep{Macaulay:2013swa,Nesseris:2017vor,Kazantzidis:2018rnb,Skara:2019usd}.
    \item
    {\bf CMB anisotropy anomalies ($2-3\sigma$):} (see Subsection \ref{CMB Anomalies}) These anomalies include lack of power on large angular scales, small vs large scales tension (different best fit values of cosmological parameters), cold spot anomaly, hints for a closed Universe (CMB vs BAO), anomaly on super-horizon scales, quadrupole-octopole alignment, anomalously strong ISW effect, cosmic hemispherical power asymmetry, lensing anomaly, preference for odd parity correlations, parity violating rotation of CMB linear polarization (cosmic birefringence) etc. \citep[see][for a review]{Schwarz:2015cma,Akrami:2019bkn}.
    \item
   {\bf Cosmic dipoles ($2-5\sigma$):} (see Subsection \ref{Cosmic Dipoles})  The large scale velocity flow dipole \citep{Watkins:2008hf,Kashlinsky:2008ut}, the Hubble flow variance in the cosmic rest frame \citep{Wiltshire:2012uh}, the dipole anisotropy in radio source count  \citep{Bengaly:2017slg}, the quasar density dipole \citep{Secrest:2020has} and the fine structure constant dipole (quasar spectra) \citep{King:2012id,Webb:2010hc} indicate that the validity of the cosmological principle may have to be reevaluated.
   \item
    {\bf Baryon Acoustic Oscillations (BAO) curiosities ($2.5 - 3\sigma$):} (see Subsection \ref{BAO curiosities})
    There is a discrepancy between galaxy and Lyman-$\alpha$ (Ly$\alpha$) BAO  at an effective redshift of $z \sim 2.34$  \citep{Cuceu:2019for,Evslin:2016gre,Addison:2017fdm}.
     \item
    {\bf Parity violating rotation of CMB linear polarization (Cosmic Birefringence):} (see Subsection \ref{Parity violating rotation of CMB linear polarization}) The recent evidence of the non zero value of birefringence poses a problem for standard $\Lambda$CDM cosmology and indicates a hint of a new ingredient beyond this standard model. In particular using a novel method developed in  \citet{Minami:2019ruj,Minami:2020xfg,Minami:2020fin}, a non-zero value of the isotropic cosmic birefringence $\beta_a=0.35\pm 0.14$ deg ($68\%$ C.L) was recently detected in the Planck18 polarization data at a $2.4\sigma$ statistical significance level by  \citet{Minami:2020odp}. 

    \item
     {\bf Small-scale curiosities:} (see Subsection \ref{Small-scale curiosities})
     Observations on galaxy scales indicate that the $\Lambda$CDM model faces several problems (core-cusp problem, missing satellite problem, too big to fail problem, angular momentum catastrophe, satellite planes problem, baryonic Tully-Fisher relation problem, void phenomenon etc.) in describing structures at small scales  \citep[see][for a review]{DelPopolo:2016emo,Bullock:2017xww}.
      \item
    {\bf Age of the Universe:} (see Subsection \ref{Age of the Universe}) 
    The age of the Universe as obtained from local measurements using the ages of oldest stars in the Milky Way (MW) appears to be marginally larger and in some tension with the corresponding age obtained using the CMB Planck18 data in the context of $\Lambda$CDM cosmology \citep{Verde:2013wza}.
     \item
    {\bf The Lithium problem ($2-4\sigma$): } (see Subsection \ref{The Lithium problem}) Measurements of old, metal-poor stars in the Milky Way’s halo find $5$ times less lithium than that BBN predicts \citep{Fields:2011zzb}.
    \item
     {\bf Quasars Hubble diagram ($\sim 4\sigma$):} (see Subsection \ref{Quasars Hubble diagram})  
     The distance modulus-redshift relation for the sample of 1598 quasars at higher redshift ($0.5< z <5.5$) is in some tension with the concordance \lcdm model indicating some hints for phantom late time expansion \citep{Risaliti:2018reu,Banerjee:2020bjq,Lusso:2019akb}.
    \item
     {\bf Oscillating signals in short range gravity experiments:} (see Subsection \ref{Oscillating signals in short range gravity experiments})
     A reanalysis of short range gravity experiments has indicated the presence of an oscillating force signal with sub-millimeter wavelength  \citep{Perivolaropoulos:2016ucs,Antoniou:2017mhs}.
     \item
    {\bf Anomalously low baryon temperature ($\sim 3.8\sigma$):} (see Subsection \ref{Anomalously low baryon temperature})
    The Experiment to Detect the Global Epoch of Reionization Signature (EDGES) collaboration \citep{Bowman:2018yin} using global (sky-averaged) $21$-$cm$ absorption signal, reports anomalously low baryon temperature $T_b\approx 4K$ at $z\approx 17$ (half  of its expected value).
    \item
    {\bf Colliding clusters with high velocity ($\sim 6\sigma$):} (see Subsection \ref{Too-rapid formation})
    The El Gordo (ACT-CL J0102-4915) galaxy cluster at $z=0.87$ is in its formation process which occurs by a collision  of two subclusters with mass ratio $3.6$ merging at a very high velocity $V_{infall}\simeq 2500 km/s$. Such cluster velocities at such a redshift are extremely rare in the context of \lcdm as demonstrated by  \citet{Asencio:2020mqh} using the estimation of \citet{Kraljic:2014soa} for the expected number of merging clusters from interrogation of the DarkSky simulations. 
     \end{itemize}
The well known Hubble tension and the other less discussed curiosities of $\Lambda$CDM at a lower statistical significance level may hint towards new physics \citep[see][for a review]{Huterer:2017buf}.

In  the context of the above observational puzzles the  following strategic questions emerge
\begin{itemize}
    \item What are the current cosmological and astrophysical datasets that include the above non-standard signals?
    \item What is the statistical significance of each signal?
    \item 	Is there a common theoretical framework that may explain simultaneously many non-standard signals?
\end{itemize}
These questions will be discussed in what follows. There have been previous works \citep{Perivolaropoulos:2008ud,Perivolaropoulos:2011hp} collecting and discussing signals in data that are at some statistical level in tension with the standard $\Lambda$CDM model but these are by now outdated and the more detailed and extended update provided by the present review may be a useful resource. 

The plan of this review is the following: In  the next section (\ref{sec:Hubble tension}) we focus on the Hubble tension. We provide a list of observational probes that can lead to measurements of the Hubble constant, point out the current tension level among different probes and discuss some of the possible generic extensions of $\Lambda$CDM model that can address this tension. In section \ref{sec:Other Tensions - Curiosities} we present the current status of other less significant tensions, their level of significance and refer to recent resources where more details can be found for each signal. We also discuss possible theoretical approaches that can  explain the non-standard nature of these signals. Finally, in section \ref{sec:Discussion} we conclude and discuss potential future directions of the reviewed research. \\

In Table \ref{acronym} of of the Appendix \ref{List of used acronyms} we list the acronyms used in this review.

\section{Hubble tension}  
\label{sec:Hubble tension}

\subsection{Methods for measuring $H_0$ and data}
\label{Methods for measuring $H_0$}

The measurement of the Hubble constant $H_0$ which is the local expansion rate of the Universe, is of fundamental importance to cosmology. This measurement has improved in accuracy through number of probes \citep[see][for a review of most well established probes]{Weinberg:2012es}.

Distances to cosmological objects constitute the most common way to probe the cosmic metric and the expansion history of the Universe. In this subsection we review the use of the two main cosmological distances used to probe the cosmic expansion history.
\begin{itemize}
\item
{\bf Luminosity distance }

Consider a luminous cosmological source of absolute luminosity $L$ (emitted power) and an observer (Fig. \ref{fig1}) at a distance $d_L$ from the luminous source. In a static cosmological setup, the power radiated by the luminous source is conserved and distributed in the spherical shell with area $4\pi d_L^2$ and therefore the apparent luminosity $l$ (energy flux) detected by the observer is
\be 
l=\frac{L}{4\pi d_L^2}
\label{flux}
\ee
\begin{figure}
\begin{centering}
\includegraphics[width=0.45\textwidth]{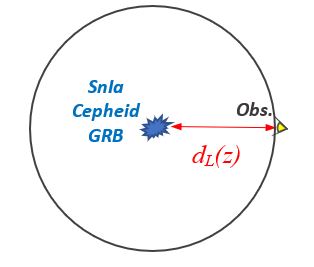}
\par\end{centering}
\caption{The luminosity distance is obtained from the apparent and absolute luminosities.} 
\label{fig1}
\end{figure}
Eq. (\ref{flux}) defines the  quantity $d_L$ known as {\it luminosity distance}.
It is straightforward to show that in an expanding flat Universe, where the energy is not conserved due to the increase of the photon wavelength and period with time, the luminosity distance can be expressed as \citep{Dodelson:2003ft,Perivolaropoulos:2006ce}
\be 
d_L(z)_{th}=c(1+z)\int_0^z \frac{dz'}{H(z')}
\label{dlz}
\ee

The luminosity distance is  an  important  cosmological  observable that is measured using standard candles (see Subsection \ref{Standard candles: late time calibrators})

\item
{\bf Angular diameter distance }
\begin{figure}
\begin{centering}
\includegraphics[width=0.45\textwidth]{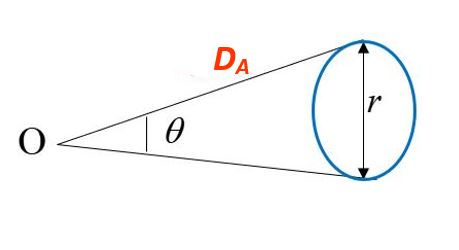}
\par\end{centering}
\caption{The angular diameter distance is obtained from the angular and physical scales.} 
\label{fig2}
\end{figure}

Consider a source (standard ruler) with a physical scale $r$ that subtends an angle $\theta$ in the sky (Fig. \ref{fig2}). In Euclidean space, assuming that  $\theta$ is small, the physical angular diameter distance $D_A$ is defined as  \citep[e.g.][]{Dodelson:2003ft,Hobson:2006se}
\be
D_A(z)=\frac{r}{\theta}
\label{angdiamdist}
\ee
A particularly useful standard ruler is the sound horizon at recombination calibrated by the peaks of the CMB anisotropy spectrum and observed either directly through the CMB anisotropies or through its signatures in the  large scale structure (Baryon Acoustic Oscillations (BAO)) (see Subsection \ref{Standard Rulers: early time calibrators}).

It is straightforward to show that in an expanding flat Universe  the physical angular diameter distance can be expressed as \citep[e.g.][]{Dodelson:2003ft} 
\be 
D_A(z)_{th}=\frac{c}{(1+z)}\int_0^z \frac{dz'}{H(z')}
\label{dAthdef}
\ee
\end{itemize}
The luminosity and angular diameter distances can be measured using standard candles and standard rulers thus probing the cosmic expansion rate at both the present time ($H(z=0)\equiv H_0$) and at higher redshifts ($H(z)$).

\subsubsection{Standard candles as probes of luminosity distance }  
\label{Standard candles: late time calibrators} 

The luminosity distance to a source may be probed using  standardizable candles  like Type Ia supernovae (SnIa) ($z<2.3$) \citep{Riess:1998cb,Perlmutter:1998np,Betoule:2014frx,Scolnic:2017caz} and gamma-ray bursts (GRBs) ($0.1<z\lesssim 9$) \citep{Amati:2008hq,Amati:2018tso,Salvaterra:2009ey,Tanvir:2009zz,Samushia:2009ib,Schaefer:2006pa,2011ApJ...736....7C,Wang:2015cya,Demianski:2016zxi,Tang:2019wlb,Cardone:2010rr,Khadka:2020hvb,Khadka:2021vqa,Dainotti:2013cta,Dainotti:2017iqb,Dirirsa:2019fcs,Demianski:2019vzl,Cao:2021irf,Luongo:2021nqh,Cao:2022wlg,Liu:2022srx,Hu:2021ycz,Dai:2021pob,Luongo:2021pjs}.

Surveys can indicate the distance-redshift relation of SnIa by measuring their peak luminosity that is tightly correlated with the shape of their characteristic light curves (luminosity as a function of time after the explosion) and the redshifts of host galaxies. The latest and largest SnIa dataset available that incorporates data from six different surveys is the Pantheon sample consisting of a total of $1048$ SnIa in the redshift range $0.01<z <2.26$ (the number of SnIa with $z>1.4$ is only six) \citep{Scolnic:2017caz}. More recently, the Pantheon+ sample which comprises 18 different samples has been released  \citep{Brout:2022vxf,Scolnic:2021amr} \citep[see also][]{Peterson:2021hel,Brownsberger:2021uue}. \citet{Brout:2022vxf,Scolnic:2021amr} present 1701 light curves of $1550$ distinct SnIa in the redshift range $0.001< z <2.26$ including SnIa which are in very nearby galaxies ($z\lesssim 0.01$) with measured Cepheid distances. For determination of $H_0$ the SH0ES team \citep{Riess:2021jrx} use as calibrator sample $42$ SnIa in the $37$ Cepheid hosts and $277$ SnIa in the Hubble flow ($0.0233<z< 0.15$) from the Pantheon+ sample. 

The apparent magnitude\footnote{The apparent magnitude $m$ of an astrophysical source detected with flux $l$ is defined as 
\be
m=-2.5 \; log_{10}\left(\frac{l}{l_0}\right)
\label{apmagdef}
\ee
where $l_0$ is a reference flux (zero point). The absolute magnitude $M$ of an astrophysical source is the apparent magnitude the source would have if it was placed at a distance of $10\, pc$ from the observer.} $m_{th}$  of SnIa in the context of a specified form of $H(z)$, is related to their luminosity distance $d_L(z)$ of Eq. (\ref{dlz}) in Mpc as
\be 
m(z)_{th}=M+5\log_{10}\left[\frac{d_L(z)}{Mpc}\right]+25
\label{apmagd}
\ee
Using now the dimensionless Hubble free luminosity distance
\be 
D_L(z)=\frac{H_0d_L(z)}{c}
\label{Dlz}
\ee
the apparent magnitude can be written as
\be 
m(z)_{th}=M+5\log_{10}\left[D_L(z)\right]+5\log_{10}\left[\frac{c/H_0}{Mpc}\right]+25
\label{apmag}
\ee
The use of Eq. (\ref{apmag}) to measure $H_0$ using the measured apparent magnitudes of SnIa requires knowledge of the value of the SnIa absolute magnitude $M$ which can be obtained using calibrators of local SnIa at $z<0.01$ (closer than the start of the Hubble flow) in the context of a distance ladder  \citep[e.g.][]{Sandage:2006cv} using calibrators like Cepheid stars.

In the cosmic distance ladder approach each step of the distance ladder uses parallax methods and/or the known intrinsic luminosity of a standard candle source to determine the absolute (intrinsic) luminosity of a more luminous standard candle residing in the same galaxy. Thus highly luminous standard candles are calibrated for the next step in order to reach out to high redshift luminosity distances. 

\paragraph{SnIa standard candles and their calibration}

\begin{itemize}
\item
{\bf SnIa-Cepheid:}
Geometric anchors may be used to calibrate the Cepheid variable star standard candles at the local Universe (primary distance indicators) whose luminosities are correlated  with their periods of variability\footnote{The period–luminosity (PL) relation  is also  called  the Leavitt law \citep{Leavitt:1908vb,Leavitt:1912zz}.}. The MW,  the Large Magellanic Cloud (LMC) and NGC 4258 are used as distance geometric anchor galaxies. For Cepheids in the anchor galaxies there are three different ways of geometric distance calibration of their luminosities: trigonometric parallaxes in the MW \citep{Benedict:2006cp,vanLeeuwen:2007xw,Casertano:2015dso,Riess:2014uga,2016A&A...595A...4L,Riess:2018uxu,Riess:2018byc,Riess:2020fzl}, Detached Eclipsing Binary Stars (DEBs) in the LMC \citep{2019Natur.567..200P} and water masers (see  Subsection \ref{Megamaser technique}) in NGC 4258 \citep{Yuan:2022kxa,Reid:2019tiq}. The DEBs technique relies on surface-brightness relations and is a one-step distance determination to nearby galaxies independent from Cepheids \citep{Pietrzynski:2013gia}.

Using the measured distances of the calibrated Cepheid stars  the intrinsic luminosity of nearby SnIa residing in the same galaxies as the Cepheids is obtained. This SnIa calibration  which fixes $M$ is then used for SnIa at distant galaxies to measure $H_0$ ($z\in [0.01,0.1]$) and $H(z)$ ($z\in [0.01,2.3]$). 
\item
{\bf SnIa-TRGB:}
Instead of Cepheid variable stars, the Tip of the Red Giant Branch (TRGB) stars in the  Hertzsprung-Russell diagram \citep{Beaton:2016nsw,Freedman:2020dne} and Miras  \citep{Huang:2018dbn,Huang:2019yhh}  \citep[see also][for a review]{Czerny:2018asb} can be used as calibrators of SnIa. The Red Giant stars have nearly exhausted the hydrogen in their cores and have just began helium burning (helium flash phase). Their brightness can be standardized using parallax methods and they can serve as bright standard candles visible in the local Universe for the subsequent calibration of SnIa. 
\item
{\bf SnIa-Miras:}
Miras (named for the prototype star Mira) are highly evolved low mass variable stars at the tip of  asymptotic giant branch (AGB) stars \citep[e.g.][]{Iben:1983ts}. The water megamaser as distance indicator (see Subsection \ref{Megamaser technique}) can be used to calibrate the Mira period–luminosity (PL) relation \citep{Huang:2018dbn}. Miras with short period ($<400$ days) have low mass progenitors and are present in all galaxy types or in the halos of galaxies, eliminating the necessity for low inclination SnIa host galaxies.

\item
{\bf SBF:} 
Another method to determine the Hubble constant based on calibration of the peak absolute magnitude of SnIa is the Surface Brightness Fluctuations (SBF) method \citep{Jensen:2000cg,Cantiello:2018ffy,Khetan:2020hmh}.  SBF is a secondary\footnote {Nearby Cepheids or stellar population models are used for the empirical or theoretical calibration of the SBF distances respectively.} luminosity distance indicator that uses stars in the old stellar populations (II) and can reach larger distances than Cepheids even inside the Hubble flow region where the recession velocity is larger than local peculiar velocities ($z>0.01$) \citep{1988AJ.....96..807T,Tonry:1996ug,Blakeslee:1998pb,Mei:2005ns,Biscardi:2008jk,Blakeslee:2009tc,Blakeslee:2012fi}. For SBF calibration \citet{Blakeslee:2021rqi} use both Cepheids and TRGB demonstrating that these calibrators  are consistent with each other.

Assume that a galaxy includes a finite number of stars covering a range of luminosity. Using SBF in the galaxy image  for the determination of its distance, the ratio $\bar{L}$ of the second and first moments of the stellar luminosity function in the galaxy is used along with the mean flux per star $\bar{l}$ as follows \citep{1988AJ.....96..807T,Blakeslee:1998pb} 
\be
d^2=\frac{\bar{L}}{4\pi \bar{l}}
\ee
where 
\be 
\bar{L}=\frac{\int n(L) L^2 dL}{\int n(L) L dL}=\frac{\sigma_L^2}{\langle L \rangle}
\ee
where $n(L)$ is the expectation number of stars with luminosity $L$. Thus SBF can be viewed  as providing an average brightness. A galaxy with double distance appears with double smoothness due to the effect of averaging.
\end{itemize}
\paragraph{Alternative cosmological standard candles}

\begin{itemize}
\item
{\bf SneII:}
An independent method to determine the Hubble constant utilizes Type II supernovae (SneII) as cosmic distance indicators \citep{deJaeger:2020zpb}. SneII are characterised by the presence of hydrogen lines in their spectra \citep{Filippenko:1997ub,Filippenko:2000yf}. This feature distinguishes SneII from other types of supernovae. Their light curve shapes include a plateau of varying steepness and length differ significantly from those of SnIa. The use of SneII as standard candles is motivated by the fact that they are more abundant than SnIa \citep{Li:2010kc,Graur:2016lca} \citep[although $1$-$2$ mag fainter][]{Richardson:2014gqa} and are produced by different stellar populations than SnIa which are more difficult to standardize. The SneII progenitors (red super giant stars) however are better understood than those of SnIa.

Different SneII distance-measurement techniques have been proposed and tested. These include, the expanding photosphere method \citep{1974ApJ...193...27K,1996ApJ...466..911E,Dessart:2005gg}, the spectral-fitting expanding atmosphere method \citep{Baron:2004wb,Dessart:2007rt}, the standardized candle method \citep{Hamuy:2002tj},  the photospheric magnitude method \citep{Rodriguez:2014pla} and the photometric color method \citep{T.:2015ald}. For example, the standardized candle method is based on the relation between the luminosity and the expansion velocity of the photosphere \citep{Hamuy:2002tj,E.:2010ug,deJaeger:2017qxx,deJaeger:2020ixu}. 

\item

{\bf GRBs:} In addition to SnIa and SneII, GRBs are widely proposed  as standard candles to trace the Hubble diagram at high redshifts \citep{Lamb:1999qv,Basilakos:2008tp,Wang:2015ira,Khadka:2020hvb,Khadka:2021vqa}. However GRBs distance calibration is not easy and various cosmology independent methods  \citep[e.g.][]{Liu:2014vda} or phenomenological relations  \citep[e.g.][]{Amati:2002ny,Ghirlanda:2004me} have been proposed for their calibration.

Furthermore GRBs can be combined with other probes to study the redshift evolution of Hubble constant \citep{Dainotti:2022bzg} \citep[see][for a review]{Moresco:2022phi}.\\

\end{itemize}

\paragraph{Using SnIa to measure $H_0$ and $H(z)$:}

The best fit values of the  parameter $H_0$ and the deceleration parameter $q_0$ may be obtained\footnote{ $q_0$ is the deceleration parameter $q_0\equiv -\frac{1}{H_0^2}\frac{d^2a(t)}{dt^2}\Big|_{t=t_0}$} \citep{Camarena:2019moy}  using local distance ladder measurements (e.g. Cepheid calibration up to  $z\simeq 0.01$) to measure directly $M$, low $z$ measurements of the SnIa apparent magnitude $m(z)$  and a kinematic local expansion of $D_L(z)$  ($z<0.1$) as  \citep[e.g.][]{Weinberg:2008zzc}
\be 
D_L(z,q_0)= z\left[1+\frac{1}{2}(1-q_0)z\right]
\label{dlexp}
\ee

Alternatively, $q_0$ may be fixed to its $\Lambda$CDM value $q_0=-0.55$ and $H_0$ may be fit as a single parameter \citep{Riess:2011yx,Riess:2016jrr,Riess:2019cxk}.

Using higher $z$ SnIa the best fit parameters of  $\Lambda$CDM may be obtained by fitting the $\Lambda$CDM expansion rate $H(z)$
\be 
H^2(z) =H_0^2\left[\Omega_{0m}(1+z)^3 +(1-\Omega_{0m})\right]
\label{Hubblez}
\ee 
where $\Omega_{0m}$ is the matter density parameter today. 
\begin{figure*}
\begin{centering}
\includegraphics[width=1\textwidth]{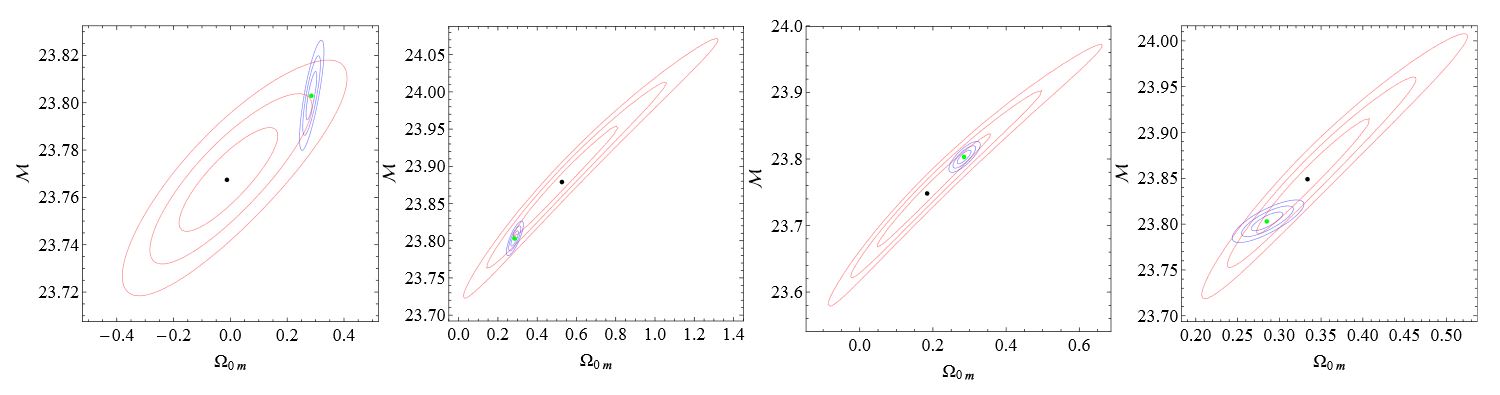}
\par\end{centering}
\caption{The $1\sigma - 3\sigma$ confidence contours in the parametric space ($\Omega_{0m}$, $\mathcal{M}$).  The blue contours correspond to the $1\sigma - 3\sigma$ full Pantheon dataset ($1048$ SnIa datapoints) best fit, while the red contours describe the $1\sigma - 3\sigma$ confidence contours of the four bins (from left to right).  The black points represent the best fit of each bin, while the green dot represents the best fit value indicated by the full Pantheon dataset ($\Omega_{0m}=0.285$ and $\mathcal{M}=23.803$)  \citep[from][]{Kazantzidis:2020tko}.}
\label{figPantheon}
\end{figure*}
\begin{figure*}
\begin{centering}
\includegraphics[width=1\textwidth]{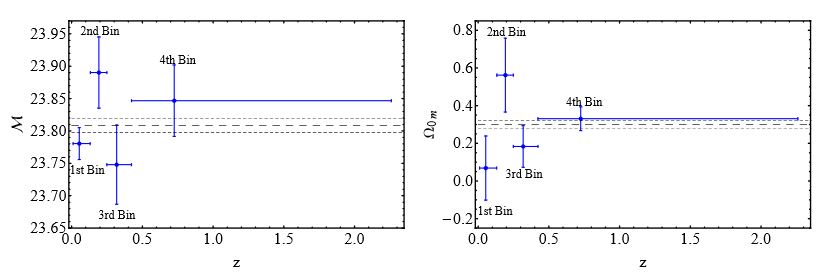}
\par\end{centering}
\caption{The best fit values of $\mathcal{M}$ (left panel) and $\Omega_{0m}$ (right panel) as well as the $1\sigma$ errors for the four bins, including the systematic uncertainties.  This oscillating behaviour relatively improbable in the context of constant underlying $\mathcal{M}$ and $\Omega_{0m}$  \citep[from][]{Kazantzidis:2020tko}.}
\label{figosc}
\end{figure*}
Using  Eqs. (\ref{dlz}), (\ref{Dlz}) and (\ref{Hubblez}), the Hubble free luminosity distance  can be written as
\be 
D_L(z,\Omega_{0m})=(1+z)\int_0^z \frac{dz'}{\left[\Omega_{0m}(1+z')^3 +(1-\Omega_{0m})\right]^{1/2}}
\ee
A key assumption in the use of SnIa in the measurement of $H_0$ and $H(z)$ is that they are standardizable and after proper calibration they have a fixed absolute magnitude independent of redshift\footnote{The possibility for intrinsic luminosity evolution of SnIa with redshift was first pointed out by  \citet{1968ApJ...151..547T}. Also, the assumption that the luminosity of SnIa is independent of host galaxy properties (e.g. host age, host morphology, host mass) and local star formation rate has been discussed in  \citet{Kang:2019azh,Rose:2019ncv,Jones:2018vbn,Rigault:2018ffm,2018ApJ...854...24K}.}. This assumption has been tested in \citet{Colgain:2019pck,Kazantzidis:2019nuh,Kazantzidis:2020tko,Sapone:2020wwz,Koo:2020ssl,Kazantzidis:2020xta,Lukovic:2019ryg,Tutusaus:2018ulu,Tutusaus:2017ibk,Drell:1999dx}. 

Using the degenerate combination 
\be 
\mathcal{M}=M+5\log_{10}\left[\frac{c/H_0}{Mpc}\right]+25
\label{combM}
\ee 
into Eq. (\ref{apmag}) we obtain
\be 
m(z,M,H_0,\Omega_{0m})_{th}=\mathcal{M}(M,H_0)+5\log_{10}\left[D_L(z,\Omega_{0m})\right]
\label{aparmagz}
\ee
The theoretical prediction (\ref{aparmagz}) may now be used to compare with the observed $m_{obs}$ data and to obtain the best fits for the parameters $\mathcal{M}$ and $\Omega_{0m}$. Using the maximum likelihood analysis the best fit values for these parameters may be found by minimizing the quantity 
\be 
\chi^2(\mathcal{M},\Omega_{0m})=\sum_i\frac{\left[m_{obs,i}-m_{th}(z_i;\mathcal{M},\Omega_{0m})\right]^2}{\sigma_i^2}
\ee

The results from the recent analysis by \citet{Kazantzidis:2020tko} using the latest  SnIa  (Pantheon) data \citep{Scolnic:2017caz} (consisting  of 1048  datapoints in the redshift range $0.01< z <2.3$ sorting them from lowest to highest redshift and dividing them in four equal uncorrelated bins) in  the  context  of  a $\Lambda$CDM model are shown in Figs. \ref{figPantheon} and \ref{figosc}\footnote{For $M=-19.24$ as indicated by Cepheid  calibrators \citep{Camarena:2019moy} of SnIa at $z<0.01$ and the SnIa local determination $H_0=74\, km\, s^{-1}Mpc^{-1}$ \citep{Riess:2019cxk}   \citet{Kazantzidis:2020tko} find ${\cal M}=23.80$ which is consistent with the full Pantheon SnIa best fit shown in Fig. \ref{figPantheon}.}. An oscillating signal for $\mathcal{M}$ and $\Omega_{0m}$ ($2\sigma$) is apparent in Fig. \ref{figosc} and its statistical significance may be quantified using simulated data \citep{Kazantzidis:2020xta,Koo:2020ssl}. 

The presence of large scale inhomogeneities at low $z$ including voids or a supercluster \citep{Grande:2011hm} can be a plausible physical explanation for this curious behavior. In the context of  a local void model the analysis by \citet{Kazantzidis:2020tko} indicated that the value of $H_0$ increases by  $2-3\%$ which is less than the $9\%$ required to address the Hubble tension. The bias and systematics induced by such inhomogeneities on the  Hubble diagram within a well-posed fully relativistic framework  \citep[light cone averaging formalism][]{Gasperini:2011us} has been discussed in \citet{Fanizza:2019pfp}. 

\citet{Marra:2021fvf} have pointed out that this $H_0$ tension is related to the mismatch between the SnIa absolute magnitude calibrated by Cepheids at $z<0.01$  \citep{Camarena:2019moy,Camarena:2021jlr}
\be
M^< = -19.2334 \pm 0.0404
\label{maglate}
\ee
 and the SnIa absolute magnitude   using the parametric free inverse distance ladder calibrating SnIa absolute magnitude using the sound horizon scale \citep{Camarena:2019rmj}
 \be
 M^>=-19.401\pm 0.027
 \label{magearly}
 \ee
 
 Thus a transition in the absolute magnitude with amplitude $\Delta M\simeq0.2$ may provide a solution to this tension \citep[see Subsection \ref{Late time modifications2} and in][]{Alestas:2020zol,Marra:2021fvf}. Alternatively if this discrepancy is not due to systematics \citep{Follin:2017ljs,Verde:2019ivm}, it could be an indication of incorrect estimate of the sound horizon scale due e.g. to early dark energy \citep{Chudaykin:2020acu} or to late phantom dark energy \citep{Alestas:2020mvb}. 

Note also that \citet{Rameez:2019wdt} find discrepancies between 'Joint Light-curve Analysis' (JLA) SnIa and Pantheon SnIa datasets which imply an uncertainty in the calibration of the absolute magnitude or equivalently of the Hubble constant which is large enough to undermine the claim for Hubble tension.

\begin{figure*}
\begin{centering}
\includegraphics[width=1\textwidth]{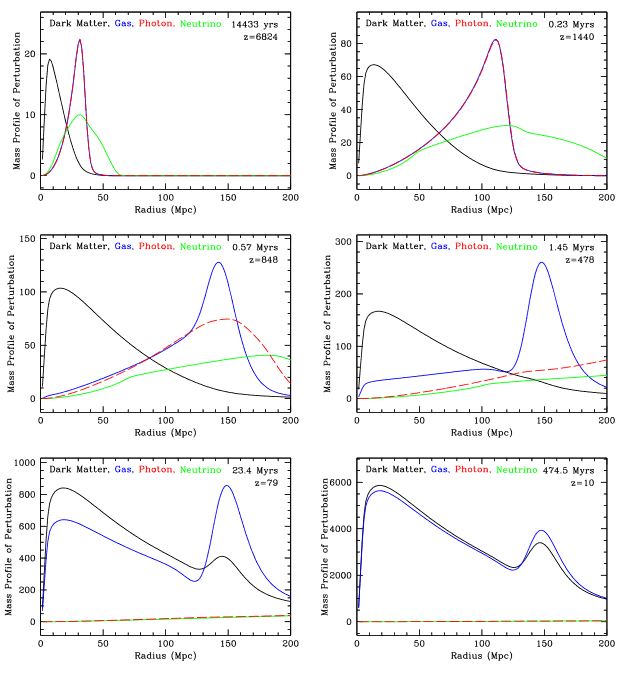}
\par\end{centering}
\caption{The snapshots show the radial mass profile of perturbation as a function of the comoving radius of an initially point-like overdensity located at the origin for redshifts $z=6824,1440,848,478,79,10$. The time after the Big Bang are given in each snapshots. The black, blue, red, and green lines correspond to the dark matter, baryons, photons, and neutrinos (all perturbations are fractional for that species), respectively. The top snapshots are for the early time before recombination where the overdensities in photons and baryons evolve together, the middle snapshots for soon after but close to recombination where the baryons freeze at the location reached with the photons forming a thick spherical shell, and the bottom snapshots are for long after recombination where the baryon overdensities start to gravitationally grow like dark matter overdensities  \citep[from][]{Eisenstein:2006nj}.}
\label{figmasspert}
\end{figure*}
\begin{figure}
\begin{centering}
\includegraphics[width=0.53\textwidth]{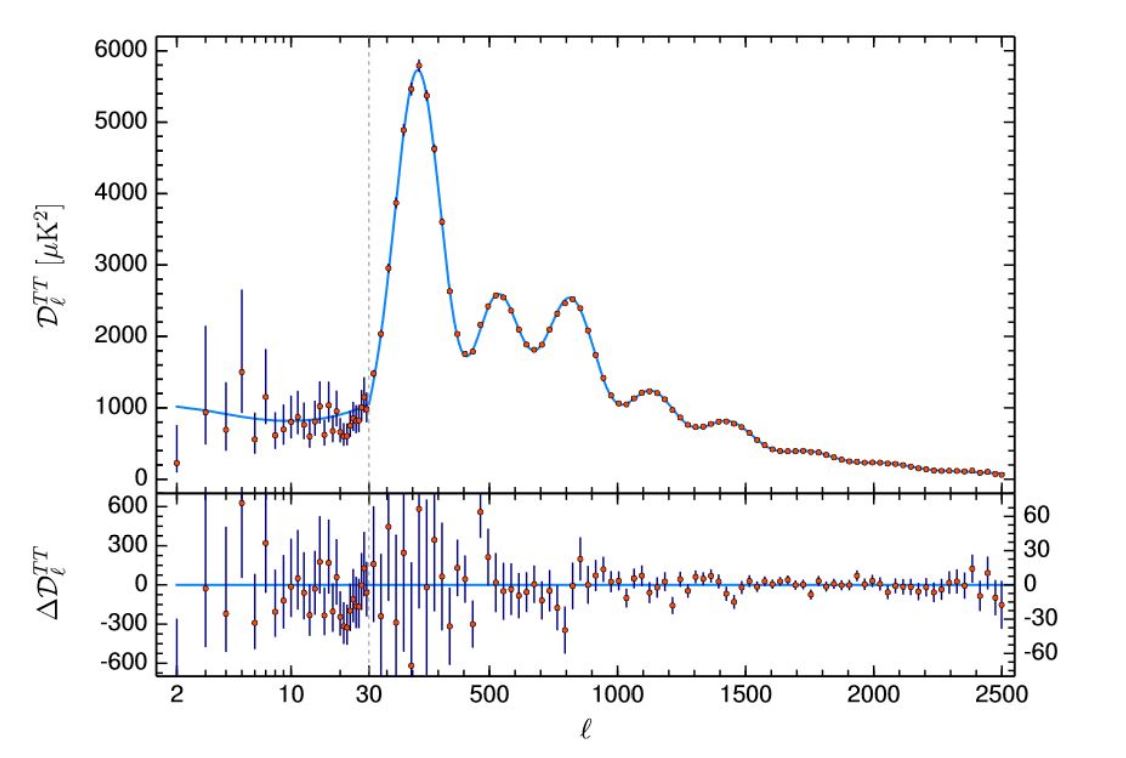}
\par\end{centering}
\caption{The Planck$18$ CMB angular power spectrum $\mathcal{D}_l^{TT}\equiv l(l+1)/(2\pi) C_l^{TT}$ (top) and residual angular power spectrum (bottom) of temperature fluctuations as a function of  multipole moment $l$. The light blue line in the upper panel is the best-fitting to  the Planck TT, TE, EE+lowE+lensing likelihoods assuming the  base-$\Lambda$CDM cosmology. The red points correspond to the binned Planck data. The lowest multipole range ($l\leq 30$) is dominated by cosmic variance (approximated as Gaussian), while positions and amplitudes of the acoustic peaks are accurately constrained  \citep[from][]{Planck:2018vyg}.}
\label{figcmb}
\end{figure}

\paragraph{Observational data - Constraints:} 
\label{SnIa} 

\begin{itemize}
\item
{\bf SnIa-Cepheid:}
Using the analysis of the Hubble Space Telescope (HST) observations \citep{Sandage:2006cv} the Hubble constant $H_0$ value has been measured from Cepheid-calibrated supernovae (using 70 long-period Cepheids in  the LMC) by the Supernovae $H_0$ for the Equation of State (SH0ES) of dark energy collaboration \citep{Riess:2011yx,Riess:2016jrr,Riess:2019cxk}. The analysis by the SH0ES Team using this local model-independent measurement refers $H_0=73.04\pm1.04$ $km$  $ s^{-1} Mpc^{-1}$ \citep{Riess:2021jrx}, which results in $5 \sigma$ tension with the value estimated by CMB Planck18 \citep{Planck:2018vyg} assuming the $\Lambda$CDM model while in previous analysis by SH0ES Team \citep{Riess:2020fzl} using the Gaia Early Data Release $3$ (EDR$3$) parallaxes \citep{gaiacollaboration2020gaia} and reaching $1.8\%$ precision by improving the calibration a value of $H_0= 73.2\pm 1.3$ is obtained, a $4.2\sigma$ tension with the prediction from Planck18 CMB  observations. \citet{Riess:2018uxu}  analysing the HST data,  using Cepheids as distance calibrators reports $H_0=73.48\pm1.66$ $km$  $ s^{-1} Mpc^{-1}$. A reanalysis of the SH0ES collaboration results using a cosmographic method allowing also  the deceleration parameter $q_0$ to be a free parameter by \cite{Camarena:2021jlr} leads to $H_0=74.30\pm 1.45$ $km$  $ s^{-1} Mpc^{-1}$.

\citet{Breuval:2020trd} considered  companion and average cluster parallaxes instead of direct Cepheid parallaxes and obtained $H_0=72.8\pm 1.9\, (statistical+systematics) \pm 1.9$ (ZP) $km$  $ s^{-1} Mpc^{-1}$ when all Cepheids are considered and $H_0=73.0\pm 1.9\, (statistical+systematics)\pm 1.9 $ (ZP) $km$  $ s^{-1} Mpc^{-1}$ for fundamental mode pulsators only (where ZP is the second Gaia data release (GDR2) \citep{Brown:2018dum} parallax zero point).

Various other previous estimates of $H_0$ have been obtained by treatments of the distance ladder \citep{Dhawan:2017ywl,CSP:2018rag,Feeney:2017sgx}. In particular, \citet{Dhawan:2017ywl} find $H_0=72.8\pm 1.6$ (statistical)$\pm 2.7 $ (systematic) $km$  $ s^{-1} Mpc^{-1}$ using SnIa as standard candles in the near-infrared (NIR), \citet{CSP:2018rag} find $H_0=73.2\pm 2.3$ $km$  $ s^{-1} Mpc^{-1}$ analysing the final data release of the Carnegie Supernova Project\footnote{\url{https://csp.obs.carnegiescience.edu}}  (CSP) I \citep{Krisciunas:2017yoe} and \citet{Feeney:2017sgx} find $H_0=73.15\pm1.78$ $km$  $ s^{-1} Mpc^{-1}$ using a Bayesian hierarchical model of the local distance ladder.

\item
{\bf SnIa-TRGB:}
The Carnegie–Chicago Hubble Program\footnote{\url{https://carnegiescience.edu/projects/carnegie-hubble-program}}  (CCHP) \citep{Beaton:2016nsw} using calibration of SnIa with the TRGB method estimates $H_0=69.8\pm 0.8 \,(\pm 1.1\% stat)\pm 1.7\, (\pm 2.4\% sys)$ $km$  $ s^{-1} Mpc^{-1}$ \citep{Freedman:2019jwv} and a revision of their measurements has lead to  $H_0=69.6\pm 0.8 \,(\pm 1.1\% stat)\pm 1.7\, (\pm 2.4\% sys)$ $km$  $ s^{-1} Mpc^{-1}$ \citep{Freedman:2020dne}. Recently, the updated TRGB calibration applied to a distant sample of SnIa from the CCHP lead to a value of the Hubble constant of $H_0=69.8 \pm 0.6\,({\rm stat}) \pm 1.6\,({\rm sys}){\rm \,km\,s^{-1}\,Mpc^{-1}}$ \citep{Freedman:2021ahq}. Using the LMC and the NGC 4258 as TRGB calibration of the SnIa distance ladder, the SH0ES team finds $H_0=72.4\pm 2$ $km$  $ s^{-1} Mpc^{-1}$ \citep{Yuan:2019npk} and $H_0=71.1 \pm 1.9{\rm \,km\,s^{-1}\,Mpc^{-1}}$ \citep{Reid:2019tiq} respectively. \citet{Freedman:2021ahq,Freedman:2020dne,Hoyt:2021irv} argue that the difference in the derived value of $H_0$ by SH0ES team compared to CCHP was due to incorrect assumptions regarding calibration of the TRGB in the LMC made by \citet{Yuan:2019npk}. A value of $H_0=65.8 \pm 3.5\,({\rm stat}) \pm 2.4\,({\rm sys}){\rm \,km\,s^{-1}\,Mpc^{-1}}$ is obtained by \citet{Kim:2020gai} using peculiar velocities and TRGB distances of 33 galaxies located between the Local Group and the Virgo cluster ($\sim 16.5\,Mpc$) \citep[mainly the sample of Virgo infall galaxies from][]{2018ApJ...858...62K}.

More recently, \citet{Soltis:2020gpl} have reported a measurement of  $H_0=72.1\pm 2.0$ $km$  $ s^{-1} Mpc^{-1}$  using the TRGB distance indicator calibrated from the European Space Agency (ESA) Gaia mission Early Data Release 3 (EDR3) trigonometric  parallax  of  Omega Centauri \citep{gaiacollaboration2020gaia}. \citet{Anand:2021sum} find $H_0=71.5 \pm 1.8 \,km\,s^{-1}\,Mpc^{-1}$ combining TRGB measurements with either the Pantheon or CSP samples of supernova. Finally, \citet{Jones:2022mvo} using NIR only cosmological analysis and TRGB distances to calibrate the SnIa luminosity of the CSP and RAISIN (an anagram for “SnIa in the IR”) samples \citep{Jones:2016cnm,DES:2018zzt} and \citet{Dhawan:2022yws} using  TRGB calibration of SnIa observed by the Zwicky Transient Facility (ZTF) \citep{Bellm_2018,Graham:2019qsw} report $H_0=72.4 \pm 3.3\rm \,km\,s^{-1}\,Mpc^{-1}$ and  $H_0=76.94 \pm 6.4\rm \,km\,s^{-1}\,Mpc^{-1}$ respectively.
\item
{\bf SnIa-Miras:}
Calibration of SnIa in the host NGC 1559 galaxy with the Miras method using a sample of 115 oxygen-rich Miras\footnote{Miras can be divided into oxygen- and carbon-rich Miras.} discovered in maser host NGC 4258 galaxy, has lead to a measurement of the Hubble constant as $H_0=73.3\pm 4$ $km$ $ s^{-1} Mpc^{-1}$ \citep{Huang:2019yhh}.
\item
{\bf SBF:}
Calibrating the SnIa luminosity with SBF method and extending it into the Hubble flow by using a sample of $96$ SnIa in the redshift range $0.02<z<0.075$, extracted from the Combined Pantheon Sample has lead to the measurement $H_0=70.50\pm 2.37$ (stat)$\pm 3.38$ (sys) $km\, s^{-1} Mpc^{-1}$ by  \citet{Khetan:2020hmh}. Previously \citet{Cantiello:2018ffy} combining distance measurement with the corrected recession velocity of NGC $4993$  reported a Hubble  constant $H_0=71.9\pm 7.1\,km\, s^{-1} Mpc^{-1}$. A new measurement of the Hubble constant $H_0=73.3\pm0.7\pm2.4 km\, s^{-1} Mpc^{-1}$ has recently been obtained based on a set of $63$ SBF \citep{Blakeslee:2021rqi} distances extending out to $100\, Mpc$.
 
\item
{\bf SneII:}
SneII have also been used for the determination of $H_0$. Using $7$ SneII as cosmological standardisable candles with host-galaxy distances measured from Cepheid variables or the TRGB  the Hubble constant was measured to be $H_0=75.8_{-4.9}^{+5.2}\,km\, s^{-1} Mpc^{-1}$  \citep{deJaeger:2020zpb}. More recently,  \citet{deJaeger:2022lit} find $H_0=75.4^{+3.8}_{-3.7} \,km\,s^{-1}\,Mpc^{-1}$ using 13 SneII.
 \end{itemize}

\subsubsection{Sound horizon as standard ruler: early time calibrators }  
\label{Standard Rulers: early time calibrators} 
\begin{figure}
\begin{centering}
\includegraphics[width=0.48\textwidth]{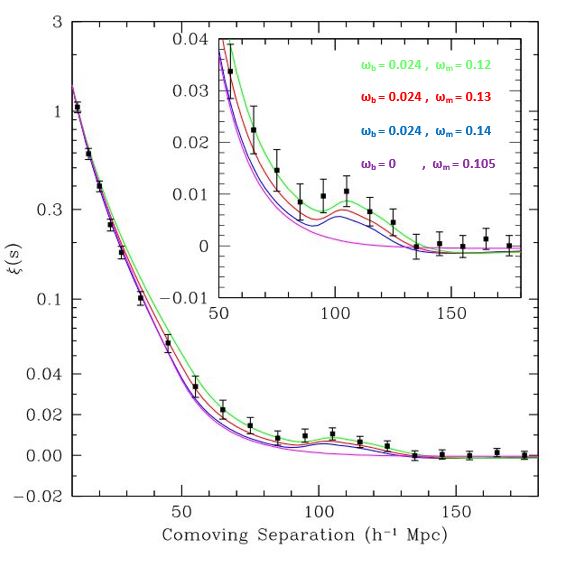}
\par\end{centering}
\caption{The signature of baryonic acoustic oscillations in galaxy two-point correlation function $\xi(s)$ as measured by  \citet{Eisenstein:2005su} using the luminous red galaxy samples of the Sloan Digital Sky Survey. The data show the existence of a baryonic acoustic peak in the galaxy correlation function $\xi(s)$ around the comoving separation scale $100\, h^{-1} Mpc$. The solid green, red, and blue lines correspond to model predictions with $\Omega_{0m} h^2=0.12, 0.13$ and $0.14$, respectively.  All models are taken to have the same $\Omega_{0b}h^2=0.024$ and $n=0.98$. The magenta line corresponds to a model with no baryons and $\Omega_{0m}h^2=0.105$, which has no acoustic peaks \citep[from][]{Eisenstein:2005su}.}
\label{figbaols}
\end{figure}

Before recombination ($z> 1100$), the  primeval plasma of coupled baryons to photons (baryon-photon fluid) oscillates as spherical sound waves emanating from baryon gas perturbations are driven by photon pressure. At recombination when the Universe has cooled enough the electrons and protons combine to form hydrogen  \citep[e.g.][]{Peebles:1968ja}, photons decouple from baryons and  propagate freely since the pressure becomes negligible. Thus the spherical sound wave shells of baryons become frozen. This process which was first detected in the galaxy power spectrum by \citet{Cole:2005sx,Eisenstein:2005su} is illustrated in Fig. \ref{figmasspert}. 
It inflicts a unique Baryon Acoustic Oscillations (BAO) scale on the CMB anisotropy spectrum peaks shown in Fig. \ref{figcmb} and on the  matter large scale structure (LSS) power spectrum on large scales at the radius of the sound horizon (the distance that the sound waves have traveled before recombination). This scale emerges as a peak in the correlation function $\xi(s)$\footnote{The correlation function is defined as the excess probability of one galaxy to be found within a given distance of another. Using the Landy-Szalay estimator \citep{Landy:1993yu} this function can be computed \citep{Eisenstein:2005su}
\be
\xi(s)\equiv\langle\delta(x)\delta(x+s)\rangle=\frac{DD(s)-2DR(s)+RR(s)}{RR(s)}
\label{corfun}
\ee
where $s$ is the comoving galaxy separation distance and  $DD(s)$, $RR(s)$ and $DR(s)$ correspond to the number of galaxy pairs with separations $s$ in real-real, random-random and real-random catalogs, respectively.} as illustrated in Fig. \ref{figbaols}  or  equivalently as damped  oscillations in the LSS power spectrum  \citep{Eisenstein:1997ik,Eisenstein:2005su,Meiksin:1998ra,Matsubara:2004fr}. 
The characteristic BAO scale is also imprinted in the Lyman-$\alpha$ (Ly$\alpha$) forest absorption lines of neutral hydrogen in the intergalactic medium (IGM) detected in quasar (QSO) spectra. 

The measured angular scale of the sound horizon $\theta_s$ at the drag epoch when photon pressure vanishes can be used to probe the Hubble expansion rate using the standard ruler relation  \citep[e.g.][]{1980lssu.book.....P,Amendola:2015ksp}
\be 
\theta_s=\frac{r_s}{d_A}
\label{thetadef}
\ee
where $d_A\equiv \frac{D_A}{a}\equiv (1+z)\,D_A=c\int_0^z \frac{dz'}{H(z')}$ is the comoving angular diameter distance to last scattering (at redshift $z\approx 1100$) and $r_s$ is the radius of sound horizon at last scattering. 

The radius $r_s$ of the sound horizon at last scattering can be calculated by the distance that sound can travel from the Big Bang, $t = 0$, to time $t_d$ at the drag epoch when the photon pressure can no longer prevent gravitational instability in baryons. This happens shortly after the time $t_s$ of the last scattering when the optical depth due to Thomson scattering reaches unity \citep{Eisenstein:1997ik}. Thus \citep{Aubourg:2014yra}
\begin{align}
 r_s =&\int_0^{t_d} \frac{c_s(a)}{a(t)}dt=\int_{z_{d}}^\infty \frac{c_s(z)}{H(z;\rho_b,\rho_{\gamma},\rho_c)}dz= \nonumber\\
&=\int_0^{a_d}\frac{c_s(a)}{a^2H(a;\rho_b,\rho_{\gamma},\rho_c)}da
\label{rsdef}
\end{align}
where the drag redshift $z_d$ corresponds to time $t_d$, $\rho_b$, $\rho_c$ and $\rho_\gamma$ denote the densities for baryon, cold dark matter and radiation (photons) respectively and $c_s$ is the  sound  speed  in  the  photon-baryon  fluid given by \citep{Efstathiou:1998xx,Komatsu:2008hk} 
\be 
c_s=\frac{c}{\sqrt{3 \left(1+\frac{3\rho_b}{4\rho_{\gamma}} \right)}} =\frac{c}{\sqrt{3 \left(1+\frac{3\omega_b}{4\omega_{\gamma}}a \right)}}
\label{csdef}   
\ee

The expansion rate $H(z)$ depends on the ratio of the matter density to radiation density and the sound speed determined by the  baryon-to-photon ratio. Both  the  matter-to-radiation ratio and  the  baryon-to-photon ratio can be estimated from the details of the acoustic peaks in CMB anisotropy power spectrum  \citep[e.g.][]{Hu:2001bc}. Thus the CMB is possible to lead to an independent determination of the radius of sound horizon. Alternatively an independent determination of the radius of sound horizon can obtained using primordial deuterium measurements \citep{Addison:2013haa,Addison:2017fdm}.  
Now using the Eqs. (\ref{dAthdef}) and (\ref{thetadef}) we can write the angular size of the sound horizon as
\be 
\theta_s=\frac{H_0 r_s}{c\int_0^{z_{d}} \frac{dz'}{E(z')}}
\label{thetas}
\ee 
where $E(z)$ is the dimensionless normalized Hubble parameter and for a flat $\Lambda$CDM model is given by
\be
E(z)\equiv \frac{H(z)}{H_0}=\left[\Omega_{0m}(1+z)^3 +(1-\Omega_{0m})\right]^{1/2}
\ee
Eq. (\ref{thetas}) indicates that there is a degeneracy between $r_s$, $H_0$ and $E(z)$. Thus $H_0$ can not be derived using the BAO data alone which constrain $E(z)$ and the degeneracy is broken when $r_s$ is fixed  using either CMB power spectra \citep{Zarrouk:2018vwy} or deuterium abundance \citep{Addison:2013haa,Addison:2017fdm}. 

For example  $r_s=147.05\pm 0.30 Mpc$ is inferred from Planck18  TT,TE,EE+lowE CMB data \citep{Planck:2018vyg}. Using the independent determination of $r_s$,  measuring the angular acoustic scale $\theta_s$ from the location of the first acoustic peak in the CMB spectrum and fitting the integral in Eq. (\ref{thetas}) using low z BAO or SnIa data, the Hubble constant $H_0$ can be derived. This is the 'inverse  distance ladder' approach \citep{Aubourg:2014yra,Cuesta:2014asa,Cai:2022dkh} which uses the sound horizon scale calibrated by the CMB peaks or by Big Bang Nucleosynthesis (BBN)  \citep{Schoneberg:2019wmt} instead of the SnIa absolute magnitude $M$ calibrated by Cepheid stars to obtain $H_0$. 

The deformation of the expansion rate $H(a)$ before recombination using additional components like early dark energy that increase $H(a)$ in Eq. (\ref{rsdef}) and thus decrease $r_s$ and increase the predicted value of $H_0$ for fixed measured $\theta_s$ in Eq. (\ref{thetas}), has been used as a possible approach to the solution of the Hubble tension. A challenge for this class of models is the required fine-tuning so that the evolution of $H(z)$ returns quickly to its standard form after recombination for consistency with lower $z$ cosmological probes and growth measurements \citep{Pogosian:2010tj}. The assumed increase of $H(z)$ at early times has been claimed to lead to a worsened growth tension \citep{Jedamzik:2020zmd} as discussed below even though the issue is under debate \citep{Smith:2020rxx,Chudaykin:2020igl}.

\paragraph{Observational data - Constraints}
\label{Early CMB}
\begin{itemize}
\item
{\bf CMB:}
The measurement of the Hubble constant $H_0$ using the sound horizon at recombination as standard ruler calibrated by the CMB anisotropy spectrum is model dependent and is based on assumptions about the nature of dark matter and dark energy as well as on an uncertain list  of relativistic  particles \citep[see][for a review]{Chang:2022tzj}. The best fit value obtained by the Planck18/$\Lambda$CDM  CMB temperature, polarization, and lensing power spectra is $H_0=67.36\pm 0.54$ $km$ $ s^{-1} Mpc^{-1}$ \citep{Planck:2018vyg}. The measurements of the CMB from the combination Atacama Cosmology Telescope (ACT)\footnote{\url{https://act.princeton.edu}} and Wilkinson Microwave Anisotropy Probe (WMAP) estimated the Hubble constant to be $H_0=67.6\pm 1.1\, km\, s^{-1} Mpc^{-1}$ and from ACT alone to be $H_0=67.9\pm 1.5\, km\, s^{-1} Mpc^{-1}$ \citep{ACT:2020gnv}. Note that the analysis of the nine-year data release of WMAP \citep{2013ApJS..208...19H} alone prefers a value for the Hubble constant $H_0=70.0\pm 2.2\, km\, s^{-1} Mpc^{-1}$. More recently, \citet{SPT-3G:2021wgf} obtain CMB-based constraints on Hubble parameter $H_0=67.49\pm 0.53\, km\, s^{-1} Mpc^{-1}$ using combined South Pole Telescope\footnote{\url{https://pole.uchicago.edu}} (SPT), Planck, and ACT DR4 datasets.  \citet{SPT-3G:2021eoc} find $H_0=68.8\pm 1.5\, km\, s^{-1} Mpc^{-1}$ using SPT-3G data alone,  while a previous analysis of SPT data by \citet{SPT:2017jdf} results in $H_0=71.3\pm 2.1\, km\, s^{-1} Mpc^{-1}$.
\item
{\bf BAO:}
The analysis of the wiggle patterns of BAO is an independent way of measuring cosmic distance using the CMB sound horizon as a standard ruler. This measurement has  improved in accuracy through a number of galaxy surveys which detect this cosmic distance scale:  the Sloan Digital Sky Survey (SDSS) supernova survey \citep{York:2000gk,Tegmark:2006az} encompassing the Baryon  Oscillation  Spectroscopic Survey (BOSS) which has  completed three different phases \citep{Dawson:2012va}. Its fourth phase (SDSS-IV) \citep{Blanton:2017qot} encompasses the Extended Baryon Oscillation Spectroscopic Survey (eBOSS) \citep{Dawson:2015wdb} \citep[see also][]{Alam:2020sor,Gil-Marin:2020bct,Neveux:2020voa,Raichoor:2020vio,Hou:2020rse}, the WiggleZ Dark Energy Survey \citep{Drinkwater:2009sd,Blake:2011en,Blake:2012pj},  the $2$-degree Field Galaxy Redshift Survey ($2$dFGRS) \citep{Colless:2001gk,Cole:2005sx}, the $6$-degree Field Galaxy Survey ($6$dFGS) \citep{Jones:2009yz,Beutler:2011hx,Beutler:2012px}. 

More recently, BAO measurements have been extended in the context of quasar redshift surveys and Ly$\alpha$ absorption lines of neutral hydrogen in the IGM detected in QSO spectra using the complete eBOSS survey. The measurement of BAO scale using first the auto-correlation of Ly$\alpha$ function \citep{Delubac:2014aqe,Agathe:2019vsu,Bautista:2017zgn} and then the Ly$\alpha$-quasar cross-correlation function \citep{Font-Ribera:2013wce,Blomqvist:2019rah} or both the auto- and cross-correlation functions \citep{duMasdesBourboux:2020pck} pushed BAO measurements to higher redshifts ($z\sim 2.4$). Recent studies present BAO measurements from the Ly$\alpha$ using the eBOSS sixteenth data release (DR$16$) \citep{Ahumada:2019vht} of the SDSS IV  \citep[e.g.][]{duMasdesBourboux:2020pck}. 

As discussed in subsection \ref{Standard Rulers: early time calibrators} BAO data alone cannot constrain $H_0$ because BAO  observations  measure the combination $H_0r_s$ rather than $H_0$ and $r_s$ individually (where $r_s$ is the radius of sound horizon). Using the CMB calibrated physical scale of the sound horizon and the combination of BAO with SnIa data (i.e inverse  distance ladder) the value $H_0=67.3\pm 1.1$ $km$ $ s^{-1} Mpc^{-1}$ was reported  which is in agreement with the value obtained by CMB data alone \citep{Aubourg:2014yra}. The analysis by \citet{Wang:2017yfu} using a combination of BAO measurements from  6dFGS \citep{Beutler:2011hx}, Main Galaxy Sample (MGS) \citep{Ross:2014qpa}, BOSS DR12 and eBOSS DR14 quasar sample in a flat $\Lambda$CDM cosmology reports $H_0=69.13\pm 2.34$ $km$ $ s^{-1} Mpc^{-1}$. Using BAO measurements and CMB data from WMAP,   \citet{Zhang:2018air} reported the constraints of $H_0=68.36_{-0.52}^{+0.53}$. The analysis by \citet{Addison:2017fdm} combining galaxy and Ly$\alpha$ forest BAO with a precise estimate of the primordial deuterium abundance (BBN) results in $H_0= 66.98\pm 1.18 \,km\, s^{-1}Mpc^{-1}$ for the flat $\Lambda$CDM model. \citet{eBOSS:2020yzd} find $H_0=67.35 \pm 0.97\rm\,km\,s^{-1}\,Mpc^{-1}$ using BOSS galaxy and eBOSS, with the BBN prior independent from the CMB anisotropies. \citet{DAmico:2019fhj} obtain $H_0= 68.5\pm 2.2 \,km\, s^{-1}Mpc^{-1}$ performing a analysis for the cosmological parameters of the DR12 BOSS data using the Effective Field Theory of Large-Scale Structure (EFTofLSS) formalism\footnote{The  EFTofLSS formalism can provide a  prediction of the LSS clustering in the mildly non-linear regime \citep{DAmico:2020kxu,Baumann:2010tm,Carrasco:2012cv,Porto:2013qua,Perko:2016puo}.} and \citet{Colas:2019ret} obtain $H_0= 68.7\pm 1.5 \,km\, s^{-1}Mpc^{-1}$ assuming a BBN prior on the baryon fraction of the energy density instead of the baryon/dark-matter ratio.

Recently, \citet{Pogosian:2020ded} report the constraints of $H_0=69.6\pm 1.8\,km\, s^{-1}Mpc^{-1}$ using BAO data, including the released eBOSS DR16, and CMB data from Plank. \citet{Zhang:2021yna} infer $H_0=68.19\pm 0.99$ imposing BBN priors on the baryon density and combining the BOSS Full Shape with the BAO measurements from BOSS and eBOSS. Also, a new analysis of galaxy 2-point functions in the BOSS survey, including full-shape information and post-reconstruction BAO by \citet{Chen:2021wdi} results in $H_0=69.23\pm 0.77\,km\, s^{-1}Mpc^{-1}$ and a full-shape analysis of BOSS DR12 by \citet{Philcox:2021kcw} results in $H_0=68.31_{-0.86}^{+0.83}\,km\, s^{-1}Mpc^{-1}$. A previous analysis of BOSS DR12 on anisotropic galaxy clustering in Fourier space by \citet{Ivanov:2019pdj} gives $H_0=67.9 \pm 1.1 \,km\,s^{-1}\,Mpc^{-1}$. Finally, analyzing the BOSS DR12 galaxy power spectra using a new approach based on the horizon scale at matter-radiation equality \citet{Farren:2021grl} find $H_0=69.5_{-3.5}^{+3.0}\,km\, s^{-1}Mpc^{-1}$ and adding Planck lensing \citet{Philcox:2020xbv} find $H_0=70.6_{-5.4}^{+3.0}\,km\, s^{-1}Mpc^{-1}$.
\end{itemize}

\subsubsection{Time delays: gravitational lensing }  
\label{Time delays: gravitational lensing}
Gravitational lensing time-delay cosmography can be used to measure $H_0$. This approach was first proposed by \citet{Refsdal:1964nw} and recently implemented by  \citet{Shajib:2019toy,Wong:2019kwg,Birrer:2020tax} \citep[see also][for clear reviews]{Treu:2016ljm,Suyu:2018vqs}.
\begin{figure}
\begin{centering}
\includegraphics[width=0.51\textwidth]{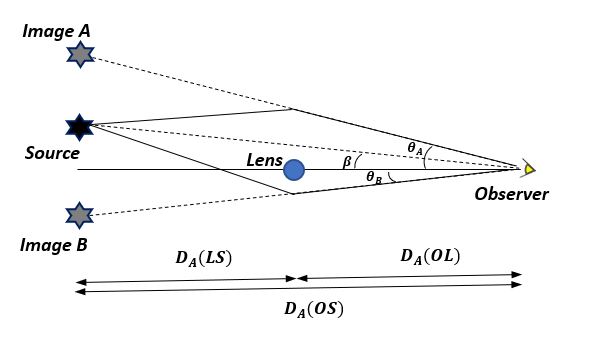}
\par\end{centering}
\caption{Schematic illustration of a typical gravitational lens system.}
\label{figslnew}
\end{figure}
Strong gravitational lensing \citep{Refsdal:1964nw} arises from the gravitational deflection of light rays of a background source when an intervening lensing mass distribution (e.g. a massive galaxy or cluster of galaxies) exists along the line of sight. The light rays go through different paths such that multiple images of the background source appear around the intervening lens \citep{Oguri:2019fix}.

The time delay $\Delta t_{AB}$ between two images ${\bf \theta_A }$ and ${\bf \theta_B }$ by a single deflector  originating from the same source at angle ${\bf \beta}$ shown in  Fig. \ref{figslnew} is given as \citep{Suyu:2009by}
\be 
\Delta t_{AB}=\frac{1+z_L}{c}\frac{D_A(OL) D_A(OS)}{D_A(LS)}\left[\phi({\bf \theta_A},{\bf \beta})-\phi({\bf \theta_B},{\bf \beta})\right]
\label{dtdef}
\ee
where $z_L$ is the lens redshift, $D_A(OL)$ is  the  angular  diameter distance to the lens, $D_A(OS)$ is the angular diameter distance to the source, $D_A(LS)$ is the angular diameter distance between the lens and the source and $\phi({\bf \theta},{\bf \beta})$ is the Fermat potential \citep[e.g.][]{Suyu:2009by}
\be
\phi({\bf \theta},{\bf \beta})=\frac{({\bf \theta}-{\bf \beta})^2}{2}-\psi({\bf \theta})
\ee
with $\psi({\bf \theta})$ the lensing potential at the image direction.
The  time  delay $\Delta t_{AB}$ in Eq. (\ref{dtdef}) is thus connected to the time delay distance defined as  \citep[e.g.][]{Suyu:2009by,Suyu:2016qxx}
\be 
D_{\Delta t}=\frac{1+z_L}{c}\frac{D_A(OL) D_A(OS)}{D_A(LS)}
\label{ddtdef}
\ee
This distance is inversely proportional to $H_0$ 
\be
D_{\Delta t}\propto \frac{1}{H_0}
\label{ddth0rel}
\ee
and thus its measurement constrains $H_0$. Strongly lensed quasars (bright and time variable sources) lensed by a foreground lensing mass  are used to measure the above observable time delay on cosmologically interesting scales \citep{Keeton:1996kf,Kochanek:2002zm,Oguri:2006qp,Bonvin:2016crt,Wong:2019kwg}. Active galactic nuclei (AGN) constitute another background source which may be used to measure the time delay \citep{Kochanek:2005ge,Fassnacht:1999re,Eigenbrod:2005yy}. Recently, \citet{Qi:2022sxm} proposed the strongly lensed SnIa as a precise late-universe probe to improve the measurements on the Hubble constant and cosmic curvature. The inference of $H_0$ from $D_{\Delta t}$ is relatively insensitive to the assumed background cosmology.

Note that a source of systematic effects in time delay cosmography is the uncertainty of the mass along the line of sight modeling with respect to the mass sheet transformation (MST). This is a mathematical degeneracy \citep[e.g.][]{1985ApJ...289L...1F,1988ApJ...327..693G,Saha:2006hq,Schneider:2013wga,Kochanek:2002rk,Kochanek:2019ruu} and can bias the strong lensing determination of Hubble constant \citep{Kochanek:2020crs}.

\paragraph{Observational data - Constraints:}
\label{Strong gravitational lensing }
Strong gravitational lensing time delay measurements of $H_0$ are consistent with the local measurements using late time calibrators and in mild tension with Planck  \citep[e.g.][]{Bonvin:2016crt}. The method of the measurement of $H_0$ Lenses in COSMOGRAIL’s  Wellspring (H$0$LiCOW) collaboration \citep{Wong:2019kwg} is independent of the cosmic distance ladder and is based on time delays between multiple images of the same source, as occurs in strong gravitational lensing. 

Using joint analysis of six gravitationally lensed quasars with measured time delays from the COSmological MOnitoring of GRAvItational Lenses (COSMOGRAIL) project,  the value $H_0=73.3_{-1.8}^{+1.7}$ $km$  $ s^{-1} Mpc^{-1}$ was obtained which is in  $3.1\sigma$ tension  with Planck CMB. Assuming the Universe is flat and using lensing systems from the lensing program H$0$LiCOW and the Pantheon supernova compilation a value of $H_0=72.2\pm 2.1$ $km$  $ s^{-1} Mpc^{-1}$ was reported by the analysis of Ref. \citep{Liao:2019qoc}. A similar value of $H_0=72.8_{-1.7}^{+1.6}$ $km$ $ s^{-1} Mpc^{-1}$ was found using updated H$0$LiCOW dataset consisting of six lenses \citep{Liao:2020zko}. The reanalysis of the four publicly released lenses distance posteriors from the H0LiCOW by \citet{Yang:2020eoh} leads to $H_0= 73.65_{-2.26}^{+1.95}$  $km$ $ s^{-1}Mpc^{-1}$. The analysis of the strong lens system DES $J0408-5354$ by \citet{Shajib:2019toy} for strong lensing insights into dark energy survey collaboration (STRIDES),  infers $H_0=74.2_{-3.0}^{+2.7}$ $km\,s^{-1} Mpc^{-1}$ in the $\Lambda$CDM cosmology. The analysis by  \citet{Birrer:2020tax} based on the strong lensing and using Time-Delay COSMOgraphy (TDCOSMO\footnote{TDCOSMO collaboration \citep{Millon:2019slk} was formed by members of H0LiCOW, STRIDES, COSMOGRAIL and SHARP.}\textsuperscript{,} \footnote{\url{http://www.tdcosmo.org/}}) data set  alone infers $H_0=74.5_{-6.1}^{+5.6}$ $km$  $ s^{-1} Mpc^{-1}$ and using a joint hierarchical analysis of the TDCOSMO and Sloan Lens ACS (SLACS) \citep{Bolton:2005nf} sample reports $H_0=67.4_{-3.2}^{+4.1}$ $km$  $ s^{-1} Mpc^{-1}$. \citet{Chen:2019ejq} based on a joint analysis of 3 strong lensing system, using ground-based adaptive optics (AO) from SHARP AO effort and the HST find $H_0=76.8\pm 2.6$ $km$  $ s^{-1} Mpc^{-1}$. A reanalysis of six of the TDCOSMO lenses using a power-law mass profile model results in $H_0=74.2\pm 1.6$ $km$  $ s^{-1} Mpc^{-1}$ \citep{Millon:2019slk}.  Analysing  $8$ strongly, quadruply lensing systems \citet{Denzel:2020zuq} present a determination of the Hubble constant $H_0=71.8_{-3.3}^{+3.9}$ $km$  $ s^{-1} Mpc^{-1}$ which is consistent with both early and late Universe observations. The value $H_0 = 73.6^{+1.8}_{-1.6} \,km\,s^{-1}\,Mpc^{-1}$ was reported by \citet{Qi:2020rmm} by combining the observations of ultra-compact structure in radio quasars and strong gravitational lensing with quasars acting as background source.

\subsubsection{Standard sirens: gravitational waves}  
\label{Standard sirens: gravitational waves}

An independent and potentially highly effective approach for the measurement of $H(z)$ and the  Hubble constant is the use of gravitational wave (GW) observations and in particular those GW bursts that have an electromagnetic counterpart (standard sirens) \citep{Schutz:1986gp,Holz:2005df,Dalal:2006qt,Nissanke:2009kt,Nissanke:2013fka}. In analogy with the traditional standard candles, it is possible to use standard sirens to directly measure the luminosity distance $d_L$ of the GW source. 

Standard sirens involve the combination of a GW signal and its independently observed electromagnetic (EM) counterpart. Such counterpart may involve short gamma-ray bursts (SGRBs) signal  from binary neutron star mergers \citep{Eichler:1989ve} or associated isotropic kilonova emission \citep{Coulter:2017wya,Soares-Santos:2017lru} and  enables the immediate identification of the host galaxy. In contrast to traditional standard  candles such  as  SnIa calibrated by Cepheid  variables, standard sirens do not require any form of cosmological distance ladder. Instead they are calibrated in the context of general relativity through the observed GW waveform. 

The  simultaneous observations of the GW signal and its EM counterpart (multi-messenger observations) of nearby compact-object merger leads to a measurement of the luminosity distance which depends on the inclination angle of the binary orbit with respect to the line of sight and the redshift (measured using photons) of the host galaxy respectively. An EM counterpart detected with a GW observation can further constrain the inclination angle and may also indicate the source’s sky position and the GW merger’s time and phase \citep[e.g.][]{Nissanke:2013fka}. 

In the case of GW events with small enough localization volumes without an observed EM counterpart (dark sirens) \citep{Chen:2017rfc}  a statistical analysis over a set of potential host galaxies within the event localization region may provide redshift information. A candidate for such statistical method is a merger of stellar-mass binary black holes\footnote{The stability analysis of the structures around black holes have been widely employed in the literature \citep[see e.g.][]{Kiuchi:2011re,Belczynski:2016jno,Cunha:2017eoe,Aretakis:2011ha,Alestas:2020wwa}.} (BBH) which is usually not expected to result in bright EM counterparts unless it takes place in significantly gaseous environment \citep{Graham:2020gwr}. For example,  GW190521  \citep{Abbott:2020mjq} is a  possible  candidate  with EM counterpart corresponding to a stellar-origin BBH merger in active galactic nucleus (AGN) disks \citep{McKernan:2019hqs} detected by ZTF \citep{Bellm_2018,Graham:2019qsw}. 

Alternatively, in the absence of an EM counterpart the redshift can be determined by exploiting information on the properties of the source (e.g. the knowledge of neutron star equation of state) to derive frequency-dependent features in the waveform \citep{Messenger:2011gi} or using the gravitational waveform to determine the redshift of  the mass distribution of the sources \citep{Taylor:2012db,Farr:2019twy}. Also, \citet{Leandro:2021qlc} use an alternative method,  presented in \citet{Ding:2018zrk}, for redshift determination by the statistical knowledge of the redshift distribution of sources. \citet{Trott:2021fnx} argue that any absolute determination of $H_0$ may be biased due to the fundamental degeneracy between redshift and $H_0$ and therefore can not lead to reliable determination of $H_0$. According to \citet{Trott:2021fnx} the reliable determination of $H_0$ with GW can only be achieved using standard sirens.

The luminosity distance-redshift relation Eq. (\ref{dlz}) determines the Universe’s expansion history and the associated cosmological parameters including the  Hubble constant $H_0$ \citep{Abbott:2017xzu,Fishbach:2018gjp}.
In particular using the mergers of binary neutron stars (BNS), or a binary of a neutron star with a stellar-mass black hole (NS-BH), which are excellent standard sirens, both the luminosity distance (from the gravitational wave waveform) and redshift of the host galaxy (from the electromagnetic counterpart) can be measured.

Using a BNS or a NS-BH merger, the distance to the source can be estimated from the detected amplitude $\langle h \rangle$ (r.m.s. - averaged over detector and source orientations) of the GW signal by the expression \citep{Schutz:1986gp,Andersson:1996pn,Jaranowski:1996hs,Kokkotas:2005vr,Kokkotas:2008hn}
\be 
d=Cf^{-2}\langle h \rangle ^{-1}\tau ^{-1}
\label{gwd}
\ee
where $f$ is the gravitational wave frequency, $\tau\equiv f/\dot{f}$ is the timescale of frequency change, $C$ is a known numerical constant. Assuming a flat\footnote{In an open (closed) Universe the distance in Hubble’s law is given $d(z)=\frac{1}{1+z}\frac{\chi}{\sinh{\chi}}d_L(z)$ ($d(z)=\frac{1}{1+z}\frac{\chi}{\sin{\chi}}d_L(z)$)} Universe the luminosity distance can then be obtained from the relation
\be 
d(z)=\frac{1}{1+z}d_L(z)
\label{gwdz}
\ee
For nearby sources, the recession velocity using the Hubble's law is determined  by the expression
\be
\upsilon(z)=H_0d(z)
\label{hlaw}
\ee
and using Eqs. (\ref{dlz}),  (\ref{Dlz}) and (\ref{gwdz}) is given by
\be 
\upsilon(z)=\frac{H_0d_L(z)}{1+z}=\frac{cD_L(z)}{1+z}=cH_0\int_0^z \frac{dz'}{H(z')}
\label{gwv}
\ee
At low redshifts using the local expansion Eq. (\ref{dlexp}) we obtain
\be
\upsilon(z)=\frac{cz}{1+z}\left[1+\frac{1}{2}(1-q_0)z\right]
\label{gwvz}
\ee
which is approximated for $d\leq 100$ $Mpc$ (or $z\leq 0.03$ ) as
\be 
\upsilon(z)=cz=H_0d
\label{gwv}
\ee
Using Eqs. (\ref{hlaw}) and (\ref{gwvz}), the equation for the determination of $H_0$ as a function of observables, $z$ and $d$ is  \citep[e.g.][]{Zhang:2017aqn}
\be
H_0=\frac{cz}{d(1+z)}\left[1+\frac{1}{2}(1-q_0)z\right]
\ee
where the deceleration parameter may be set by a fit to the GW data or may be fixed to its \plcdm best fit form ($q_0=-0.55$).
\begin{figure*}
\begin{centering}
\includegraphics[width=0.90\textwidth]{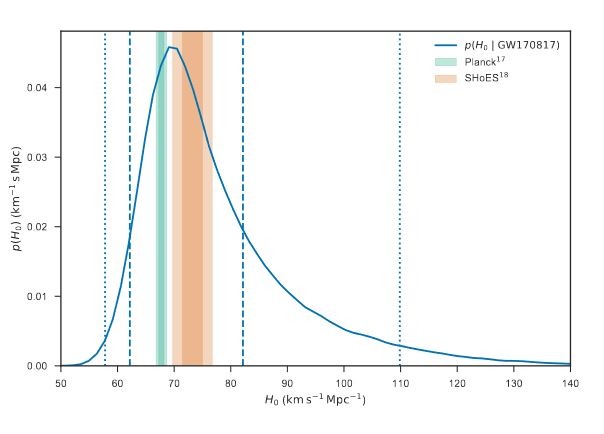}
\par\end{centering}
\caption{The probability of different values of $H_0$ with the maximum at $H_0=70.0_{-8.0}^{+12.0}$ $km$ $ s^{-1} Mpc^{-1}$ (solid blue curve) derived by BNS event GW170817. The dashed and dotted lines show minimal $68.3\%$ ($1\sigma$) and $95.4\%$ ($2\sigma$) credible  intervals. The shaded green and orange bands show the $1\sigma$ and $2\sigma$ constraints from the analysis of the CMB data obtained by the Planck \citep{Ade:2015xua} and from the analysis of the SnIa data obtained by SH0ES \citep{Riess:2016jrr} respectively \citep[from][]{Abbott:2017xzu}.}
\label{figgwprobh}
\end{figure*}

\paragraph{Observational data - Constraints:}
\label{Local Gravitational waves}

The first multi-messenger detection of a BNS merger, GW170817, by LIGO \citep{TheLIGOScientific:2014jea} and  Virgo \citep{TheVirgo:2014hva} interferometers enabled the first standard siren measurement of the Hubble constant $H_0$. Using the BNS merger GW170817, the distance to the source was estimated to be $d=43.8_{-6.9}^{+2.9}$ $Mpc$ (i.e. at redshift $z\sim 0.01$) from the detected amplitude $\langle h \rangle$ (r.m.s. - averaged over detector and source orientations) of the GW signal by the Eq. (\ref{gwd}) \citep{Abbott:2017xzu}. Also using the Hubble flow velocity $\upsilon_H=3017\pm166$ $km$ $s^{-1}$ inferred from measurement of the redshift of the host galaxy, NGC 4993 (NGC 4993 was identified as the unique host galaxy), the Hubble constant was determined to be $H_0=70.0_{-8.0}^{+12.0}$ $km$ $ s^{-1} Mpc^{-1}$ \citep{Abbott:2017xzu}  (see Fig. \ref{figgwprobh}) by using Eq. (\ref{gwv}).

Using  continued  monitoring of the the radio counterpart of GW17081 combining with earlier GW and EM data \citet{Hotokezaka:2018dfi} obtain a improved  measurement of $H_0=68.9_{-4.6}^{+4.7}$ $km$ $ s^{-1} Mpc^{-1}$. Note that using the BNS merger GW170817 in \citet{Fishbach:2018gjp} and a statistical analysis (as first proposed in  \citet{Schutz:1986gp}) over a catalog of potential host galaxies, the Hubble constant was determined  to be $H_0=77.0_{-18.0}^{+37.0}$ $km$ $ s^{-1} Mpc^{-1}$. Using  density-estimation Likelihood-Free Inference (LFI) \citet{Gerardi:2021gvk} focused on the inference of the cosmological expansion $H_0$ from GW-selected catalogues of BNS mergers with EM counterparts. 

Also using the BBH merger GW170814 as a standard (dark) siren in the absence of an electromagnetic counterpart, combined with a photometric redshift catalog from the Dark Energy Survey (DES) \citep{Abbott:2018jhe} the analysis by \citet{Soares-Santos:2019irc} results in $H_0=75_{-32}^{+40}$ $km$  $ s^{-1} Mpc^{-1}$. Using multiple GW observations (the BNS event GW170817 and the BBH  events  observed  by  advanced LIGO and Virgo in their first and second observing runs) in \citet{Abbott:2019yzh}  the Hubble constant was constrained as  $H_0=69.0_{-8.0}^{+16.0}$ $km$ $ s^{-1} Mpc^{-1}$. Using the event GW190814 from merger of a black hole with a lighter compact object the Hubble constant was measured to be $H_0=75_{-13}^{+59}$  $km$ $ s^{-1} Mpc^{-1}$ \citep{Abbott:2020khf}. In \citet{Mukherjee:2020kki} the BBH merger GW190521 was analysed choosing the NRSur7dq4 waveform\footnote{NRSur7dq4 is a numerical relativity surrogate 7-dimensional approximate waveform model of binary black hole merger with mass ratios $q\equiv \frac{m_1}{m2}\leq4$  \citep{Varma:2019csw}. This model is made publicly available through the gwsurrogate (see \url{https://pypi.org/project/gwsurrogate}) and surfinBH (see \url{https://pypi.org/project/surfinBH}) Python packages.}  for the estimation of luminosity distance, after marginalizing over matter density $\Omega_{0m}$  when the $\Lambda$CDM model is considered and using its EM counterpart ZTF19abanrhr\footnote{The  ZTF19abanrhr event was reported by ZTF \citep{Bellm_2018}. This  candidate EM counterpart is flare after a kicked BBH merger in the accretion disk of an AGN \citep{McKernan:2019hqs} with peak luminosity occurred $50$ days after the BBH event GW19052. The ZTF19abanrhr was first observed after $34$ days from the GW detection at the sky direction ($RA=192.42625^0$, $Dec=34.82472^0$) and was associated with an AGN J124942.3 + 344929 at redshift $z = 0.438$ \citep{Graham:2020gwr}.} as identified in  \citet{Graham:2020gwr} the Hubble constant was measured to be  $H_0= 50.4_{-19.5}^{+28.1}$  $km$ $ s^{-1} Mpc^{-1}$. The same study  \citep{Mukherjee:2020kki} choosing different types of waveform finds $H_0= 43.1_{-11.4}^{+24.6}$  $km$ $ s^{-1} Mpc^{-1}$ and $H_0= 62.2_{-19.7}^{+29.5}$  $km$ $ s^{-1} Mpc^{-1}$. Combining their results with the binary neutron star event GW170817 leads to $H_0= 67.6_{-4.2}^{+4.3}$  $km$ $ s^{-1}Mpc^{-1}$.
In \citet{Chen:2020gek} for the same GW-EM event, assuming a flat wCDM model has obtained $H_0=48_{-10}^{+23}$ $km$ $ s^{-1} Mpc^{-1}$.

The analysis by \citet{LIGOScientific:2021aug} using 47 gravitational-wave sources from the Third LIGO–Virgo–KAGRA Gravitational-Wave Transient Catalog (GWTC–3), infers $H_0=68^{+12}_{-7}$ $km$ $ s^{-1} Mpc^{-1}$. \citet{Palmese:2021mjm} find $H_0= 72.77^{+11.0}_{-7.55}{\rm \,km\,s^{-1}\,Mpc^{-1}}$ using the best available gravitational wave
events, uniform galaxy catalog from the Dark  Energy  Spectroscopic  Instrument (DESI) \citep{Aghamousa:2016zmz,Aghamousa:2016sne}  Legacy Survey and combining with the GW170817. The value $H_0=88.6^{+17.1}_{-34.3}$ $km$ $ s^{-1} Mpc^{-1}$ for GW190521 event was reported, and $H_0=73.4^{+6.9}_{-10.7}$ $km$ $ s^{-1} Mpc^{-1}$ was obtained when combing the GW190521 with the results of the neutron star merger GW170817 \citep{Gayathri:2020mra}. More recently, \citet{Mukherjee:2022afz} report $H_0= 67^{+6.3}_{-3.8}\rm \,km\,s^{-1}\,Mpc^{-1}$ combining the bright standard siren measurement from GW170817 with a better measurement of peculiar velocity.

\subsubsection{Megamaser technique}
\label{Megamaser technique}
Observations of water megamasers which are found in the accretion  disks around  supermassive  black holes (SMBHs) in active galactic nuclei (AGN) have been demonstrated to be powerful  one-step geometric probes for measuring extragalactic distances \citep{Herrnstein:1999kd,Humphreys:2013eja,Reid:2012hm}. 

Assuming a Keplerian circular orbit around the SMBH, the centripetal acceleration and the velocity of a masing cloud are given as \citep{Reid:2012hm}
\be
A=\frac{V^2}{r}
\ee
\be 
V=\sqrt{\frac{GM}{r}}
\ee 
where $G$ is the Newton's constant, $M$ is the mass of the central supermassive black hole, and $r$ is the distance of a masing cloud from the supermassive black hole.

The angular scale $\theta$ subtended by $r$ is given by
\be 
\theta=\frac{r}{d}
\ee
where $d$ is the distance to the galaxy.

Thus, from the velocity and acceleration measurements obtained from the maser spectrum, the distance to the maser may be determined
\be
d=\frac{V^2}{A\theta}
\ee
where $A$ is measured from the change in Doppler
velocity with time by monitoring the maser spectrum on month timescales. Using Hubble's law the Hubble constant may be approximated as \citep{Reid:2012hm}
\be
H_0\approx\frac{\upsilon}{d}
\ee
where $\upsilon$ is the measured recessional velocity.

In order to constrain  the  Hubble constant the Megamaser Cosmology Project (MCP) \citep{Reid:2008nm} uses angular diameter distance measurements to disk megamaser-hosting  galaxies well into the Hubble flow ($50-200\, Mpc$). These distances are independent of standard candle distances and their measurements do not rely on distance ladders, gravitational lenses or the CMB \citep{Pesce:2020xfe}. Early measurements of $H_0$ using masers tended to favor lower values of $H_0\simeq 67\;km\,s^{-1}Mpc^{-1}$ while more recent measurements favor higher values $H_0\simeq 73\;km\,s^{-1}Mpc^{-1}$ as shown e.g. in Table \ref{hubblevalue}.

\paragraph{Observational data - Constraints:}
Recently, the Megamaser Cosmology Project (MCP) \citep{Reid:2008nm} using  geometric distance  measurements to megamaser-hosting galaxies and assuming a global velocity uncertainty of $250$ $km$ $s^{-1}$ associated with peculiar motions of the maser galaxies constrains the  Hubble constant to be $H_0=73,9\pm 3$ $km$ $ s^{-1} Mpc^{-1}$ \citep{Pesce:2020xfe}. Previously the MCP reported results on galaxies, UGC $3789$ with $H_0= 68.9\pm 7.1\,km\,s^{-1}Mpc^{-1}$ \citep{Reid:2012hm}, NGC $6264$ with $H_0= 68.0\pm 9.0\, km\,s^{-1}Mpc^{-1}$ \citep{Kuo:2012hg}, NGC $6323$ with $H_0=73_{-22}^{+26}\,km\, s^{-1}Mpc^{-1}$ \citep{Kuo:2014bqa} and NGC $5765b$ with $H_0=66.0\pm 6.0\,km\, s^{-1}Mpc^{-1}$ \citep{Gao:2015tqd}. \citet{Reid:2019tiq} use a improved distance estimation of the maser galaxy NGC $4258$ (also known as Messier $106$) to calibrate the Cepheid-SN Ia distance ladder combined with geometric distances from MW parallaxes and DEBs in the LMC. The measured value of the Hubble constant is 
$H_0=73.5\pm 1.4\,km\, s^{-1}Mpc^{-1}$.

\subsubsection{Tully-Fisher relation (TFR) as  distance indicator} 
\label{Tully-Fisher relation (TFR) as  distance indicator.} 
The Tully-Fisher (TF) method is a historically useful distance indicator based on the empirical relation between the intrinsic total luminosity (or the stellar mass) of a spiral galaxy\footnote{Similarly, in the case of a elliptical galaxy the Faber–Jackson (FJ) empirical power-law relation  $L\propto \sigma^{\gamma_{FJ}}$ (where $L$ is the luminosity of galaxy, $\sigma$ the velocity dispersion of its stars and $\gamma_{FJ}$ is a index close to $4$) \citep{Faber:1976sn} can be used as a distance indicator. The FJ relation is the projection of the fundamental plane (FP) of elliptical galaxies which defined as $R_{eff}\propto  \sigma^{s_1} I_{eff}^{s_2}$ (where $R_{eff}$ is the effective radius and $I_{eff}$ is the mean surface brightness within $R_{eff}$) \citep{Djorgovski:1987vx}.} and its rotation velocity (or neutral hydrogen (HI) $21\,cm$ emission line width) \citep{Tully:1977fu}. This method has been used widely in measuring extragalactic distances   \citep[e.g.][]{Sakai:1999aw}.

The Baryonic Tully Fisher  relation (BTFR) \citep{McGaugh:2000sr,Verheijen:2001an,Gurovich:2004vd,McGaugh:2005qe} connects the  rotation speed $V_c$ and total baryonic mass $M_b$ (stars plus gas) of a spiral galaxy as
\be
M_b= A_c V_c^s 
\label{btfr}
\ee
where $s$ \citep[with $s\approx 3-4$][]{McGaugh:2000sr,McGaugh:2005qe,Lelli:2015wst} is a parameter and $\log A_c$ is the zero point in a log-log BTFR plot. This relation has been measured for hundreds of galaxies. The rotation speed $V_c$ can be measured independently of distance while the total baryonic mass $M_b$ may be used as distance indicator since it is connected to the intrinsic luminosity.  Thus, the BTFR is a useful cosmic distance indicator approximately independent of redshift and thus can be used to obtain $H_0$.

The  BTFR has a smaller amount of scatter with a corresponding better accuracy as a distance indicator than the classic TF relation \citep{Lelli:2015wst}. In addition the BTFR recovers two decades in velocity and six decades in mass \citep{McGaugh:2000sr,McGaugh:2005qe,McGaugh:2011ac,Iorio_2016,Lelli:2019igz,Schombert:2020pxm}.

A simple heuristic analytical derivation for the BTFR is obtained \citep{1979ApJ...229....1A} by considering a star rotating with velocity $v$ in a circular orbit of radius $R$ around a central mass $M$. Then the star velocity is connected with the central mass as
\be
v^2=G\; M_b /R\implies v^4=(G\;M_b/R)^2\sim M_b \; S\; G^2
\label{v2m}
\ee
where $G$ is Newton's constant and $S$ the surface density $S\equiv M/R^2$ which may be shown to be approximately constant \citep{1970ApJ...160..811F}. From Eqs. (\ref{btfr}) and (\ref{v2m}) we have
\be
A_c\sim G^{-2} S^{-1}
\label{abg}
\ee
which indicates that the zero point intercept of the BTFR can   probe both galaxy formation dynamics (through e.g. $S$) and possible fundamental constant dynamics (through $G$) \citep{Alestas:2021nmi}.

\paragraph{Observational data - Constraints:}
The analysis by \citet{Kourkchi:2020iyz} using infrared
(IR) data of sample galaxies and the Tully Fisher relation determined the value of Hubble constant to be $H_0= 76.0\pm 1.1(stat.)\pm 2.3(sys.) \,km\, s^{-1}Mpc^{-1}$. In \citet{Schombert:2020pxm} a value of $H_0=75.1\pm 2.3 (stat.) \pm 1.5 (sys.)\,km\, s^{-1}Mpc^{-1}$ was found using Baryonic Tully Fisher relation for $95$ independent Spitzer photometry and  accurate  rotation curves (SPARC) galaxies\footnote{The SPARC catalogue contains $175$ nearby  (up to distances of $\sim130 Mpc$) late-type galaxies (spirals and irregulars) \citep{Lelli:2016zqa,Lelli:2015wst}. The SPARC data are publicly available at \url{http://astroweb.cwru.edu/SPARC}.} (up to distances of $\sim130 Mpc$).

\subsubsection{Extragalactic background light $\gamma$-ray attenuation}
This method is based on the fact that the extragalactic background light (EBL) which is a diffuse radiation field that fills the Universe from ultraviolet (UV) through infrared wavelength induces opacity for very high energy (VHE) photons ($\geq30 \, GeV$) induced by photon-photon interaction \citep{Hauser:2001xs}. In this process a $\gamma$-ray and an EBL photon in the intergalactic medium may annihilate and produce an electron-positron pair \citep{Gould:1966pza}. The induced attenuation in the spectra of $\gamma$-ray sources is characterized by an optical depth $\tau_{\gamma\gamma}$ that scales  as $n\sigma_Tl$ (where $n$ is the photon density of the EBL, $\sigma_T$ is the Thomson cross section, and $l$ is the  distance from the $\gamma$-ray source to  Earth). The cosmic evolution and the matter content of the Universe determine the $\gamma$-ray optical depth and the amount of $\gamma$-ray attenuation along the line of sight \citep{Dominguez:2019jqc,Zeng:2019mae}. Thus a derivation of $H_0$ can be obtained by measuring the $\gamma$-ray optical depth with the $\gamma$-ray telescopes \citep{Dominguez:2013mfa}. This derivation is independent and complementary to that based on the distance ladder and cosmic microwave background (CMB) and seems to favor lower values of $H_0$ as shown in Table \ref{hubblevalue}.

\paragraph{Observational data - Constraints:}
The analysis by \citet{Dominguez:2019jqc} using extragalactic background light $\gamma$ - ray attenuation data from Fermi Large Area Telescope (Fermi-LAT) derives $H_0=67.4_{-6.2}^{+6.0} \, km \, s^{-1}Mpc^{-1}$ and  $\Omega_{0m}=0.14_{-0.07}^{+0.06}$. The analysis by \citet{Zeng:2019mae} fitting the $>10\,GeV$ extragalactic background data with modeled extragalactic background spectrum results in $H_0=  64.9_{-4.3}^{+4.6} \, km \, s^{-1}Mpc^{-1}$ and $\Omega_{0m} = 0.31_{-0.14}^{+0.13}$.

\subsubsection{Cosmic chronometers}
Cosmic chronometers are objects whose evolution history is known.  For instance such objects are some types of galaxies. The observation of these objects at different redshifts and the corresponding differences in their evolutionary state has been used to obtain the value of $H(z)$ at each redshift $z$.

The cosmic chronometer technique for the determination of $H_0$ was originally suggested in  \citet{Jimenez:2001gg} and is based on the quasi-local ($0.07 \lesssim z \lesssim 2.36$) measurements along the Hubble flow of the Hubble parameter expressed as 
\be
H(z)=-\frac{1}{1+z}\frac{dz}{dt}
\ee
Thus, the expansion rate may be obtained by measuring the age difference $\Delta t$ between two old and passively evolving galaxies\footnote{These galaxies form only a few new stars and become fainter and redder with time. The time that has elapsed since they stopped star formation can be deduced.} which are separated by a small redshif interval $\Delta z$, to infer the $dz/dt$  \citep{Moresco:2012jh,Moresco:2016mzx}. 

This  approach determines the $H_0=H(z=0)$ independent of the early-Universe physics and is not based on the distance ladder  \citep[e.g.][]{Jimenez:2001gg,Chen:2016uno,Farooq:2016zwm,Yu:2017iju,Gomez-Valent:2018hwc}. The estimated $H_0$ values are more consistent with the values estimated from recent CMB and BAO data than those values estimated from SnIa. The value of $H_0$ can not be derived using the cosmic chronometers observations alone because there is a background degeneracy between $H_0$ and $\Omega_{0m}$ and this degeneracy is broken when these observations are combined. 

\paragraph{Observational data - Constraints:}
In \citet{Chen:2016uno} the value of Hubble constant was found to be $H_0=68.3_{-2.6}^{+2.7}$ in the flat $\Lambda$CDM model relying on $28\,H(z)$ measurements and their extrapolation to redshift zero. Analysing $31\,H(z)$ data determined by the cosmic chronometric (CCH) method, and $5\, H(z)$ data by BAO observations and using the Gaussian Process (GP) method \citep{Seikel:2012uu,Shafieloo:2012ht,Yahya:2013xma,Joudaki:2017zhq} to determine a continuous $H(z)$ function the Hubble constant is estimated to be $H_0\sim67\pm 4\, km \, s^{-1}Mpc^{-1}$ by \citet{Yu:2017iju}.  Also using the GP an extension of this analysis by  \citet{Gomez-Valent:2018hwc}, including the $H(z)$ measurements obtained from Pantheon compilation and HST CANDELS and CLASH Multi-Cycle Treasury (MCT) programs, finds $H_0= 67.06 \pm 1.68 \, km \, s^{-1}Mpc^{-1}$ which is more consistent again with the lower range of values for $H_0$. The GP method \citep{10.5555/1162254} is used as a 'non-parametric' technique which does not assume any parametrization or any cosmological model \citep[see][for a discussion about GP as model independent method]{OColgain:2021pyh}. The GP modeling approach has been performed by several authors to reconstruct cosmological parameters and thus to extract cosmological information directly from data  \citep[see e.g][]{Benisty:2022psx,Escamilla-Rivera:2021rbe,Ren:2022aeo,Holsclaw:2010nb,Keeley:2019hmw,Holsclaw:2010sk,Bengaly:2021wgc,Zhang:2016tto,Busti:2014dua,Seikel:2013fda,Sahni:2014ooa,LHuillier:2016mtc,LHuillier:2017ani,Marques:2018ctl,Renzi:2020fnx,Benisty:2020kdt,Bonilla:2021dql,Bonilla:2020wbn,Sun:2021pbu,Escamilla-Rivera:2021rbe,Dhawan:2021mel,Mukherjee:2021ggf,Keeley:2020aym,Avila:2021mbj,Ruiz-Zapatero:2022zpx,Sharma:2020unh,Renzi:2021xii,Nunes:2020hzy,LHuillier:2019imn,Belgacem:2019zzu,Pinho:2018unz,Cai:2017yww,Haridasu:2018gqm,Zhang:2018gjb,Wang:2017jdm,Shafieloo:2018gin,Bengaly:2019oxx,Bengaly:2019ibu,Briffa:2020qli,Li:2019nux}.

Recently, a analysis by \citet{Moresco:2022phi} reports $H_0=67.8^{+8.7}_{-7.2}\,km\,s^{-1}\,Mpc^{-1}$ and $H_0=66.5\pm 5.4\,km\,s^{-1}\,Mpc^{-1}$ for a generic open wCDM and for a flat $\Lambda$CDM respectively. The analysis by \citet{Moresco:2022phi} examine the possible effects that can systematically bias the measurement and can affect the CC method. It should be pointed out however that the quality and reliability of cosmic chronometer data has been challenged by some authors. This is partly due to the fact that these datapoints are not model independent and are obtained by combining several datasets \citep{Gomez-Valent:2018hwc}. This has improved significantly in the context of the aforementioned analysis by \citet{Moresco:2022phi} where a detailed study of the covariance matrix and the effects of systematics has been implemented.

\subsubsection{HII galaxy measurements}

The ionized hydrogen gas (HII) galaxies (HIIG) emit massive and compact bursts generated by the violent star formation (VSF) in dwarf irregular galaxies. The HIIG measurements can be used to probe the background evolution of the Universe. This method of $H_0$ determination is based on the standard candle calibration provided by a  $L-\sigma$ (luminosity-velocity dispersion)  relation. This relation exists in HIIGs and Giant extragalactic  HII regions (GEHR) in nearby spiral and irregular galaxies. The turbulent emission line ionized gas velocity dispersion $\sigma$ of the prominent Balmer lines\footnote{The Balmer series, or Balmer lines is one of a set of six named series describing the spectral line emissions of the hydrogen atom. This is characterized by the electron transitioning from $n\geq 3$ to $n = 2$ (where n is the principal quantum number of the electron. The transitions $n = 3$ to $n = 2$ and  $n=4$ to $n=2$ are called  H-alpha and H-beta respectively.} H-alpha ($H\alpha$) and H-beta ($H\beta$) relates with its integrated emission line luminosity $L$  \citep{Melnick:1999qb,Siegel:2004xs,Plionis:2011jj,Chavez:2012km,Chavez:2014ria,Terlevich:2015toa,Wei:2016jqa,Chavez:2016epc,Yennapureddy:2017vvb,Gonzalez-Moran:2019uij}. The relationship between $L(H\beta)$ and $\sigma(H\beta)$ has a small enough scatter to define a cosmic distance indicator (that can be utilized out to $z\sim4$) independently of redshift and can be approximated as \citep{Chavez:2012km,Chavez:2014ria,Terlevich:2015toa,Wei:2016jqa,Leaf:2017dcx,Chavez:2016epc,Yennapureddy:2017vvb,Cheng-Zong:2019iau,Gonzalez-Moran:2019uij}
\be
\log L(H\beta)=\nu \log \sigma(H\beta)+\kappa
\label{lumhii}
\ee
where $\nu$ and $\kappa$ are constants representing the slope and the logarithmic luminosity at $\log\sigma(H\beta)=0$.

From Eq. (\ref{flux}) the  luminosity $L(H\beta)$ is given by
\be
L(H\beta)=4\pi d_L^2 l(H\beta)
\ee
Thus using Eq. (\ref{lumhii}), the distance modulus $\mu\equiv m-M$ of an HIIG can be obtained \citep{Wei:2016jqa,Leaf:2017dcx,Chavez:2016epc,Yennapureddy:2017vvb,Cheng-Zong:2019iau,Gonzalez-Moran:2019uij}

\be
\mu_{obs}=2.5\left[\nu \log \sigma(H\beta)+\kappa-\log l(H\beta)\right]-100.2
\ee

This observational distance modulus can be compared with the theoretical distance modulus. From the Eq. (\ref{apmagd}) this is given

\be 
\mu(z)_{th}=5\log_{10}\left[\frac{d_L(z)}{Mpc}\right]+25
\label{muz11}
\ee
\vspace{3mm}
Using now the dimensionless Hubble free luminosity distance Eq. (\ref{Dlz}) this can be written as
\be 
\mu(z)_{th}=5\log_{10}\left[D_L(z)\right]+5\log_{10}\left[\frac{c/H_0}{Mpc}\right]+25
\ee
In order to obtain the best fit values for the parameters $\Omega_{0m}$ and $H_0$ this theoretical prediction may now be used to compared with the observed $\mu_{obs}$ data. Using the maximum likelihood analysis the best fit values for these parameters may be found in the usual manner by minimizing the quantity 
\be 
\chi^2(H_0,\Omega_{0m})=\sum_i\frac{\left[\mu_{obs,i}-\mu_{th}(z_i;H_0,\Omega_{0m})\right]^2}{\varepsilon_i^2}
\ee
where $\varepsilon_i$ is the uncertainty of the $ith$ measurement.

\paragraph{Observational data - Constraints:}
Using $156$ HII galaxy measurements as a new distance indicator and implementing the model-independent GP, the Hubble constant was found to be $H_0 = 76.12_{-3.44}^{+3.47} \, km \, s^{-1}Mpc^{-1}$ which is more consistent with the recent local measurements \citep{Wang:2016pag}. Using data of $130$ giant  HII regions in $73$ galaxies with Cepheid  determined distances the best estimate of the Hubble parameter is $H_0=71.0\pm 2.8\, (random) \pm 2.1\, (systematic) \, km \, s^{-1}Mpc^{-1}$ \citep{Fernandez-Arenas:2017isq}.

\subsubsection{Combinations of data}
\label{Combinations of data-Historic evolution}

The Hubble constant $H_0$ values at  $68\%$  CL through  direct  and  indirect  measurements obtained by the different methods described in  subsection \ref{Methods for measuring $H_0$} are shown in Table \ref{hubblevalue} and described in more detail below in Fig. \ref{figh0}. Also the relative probability density value of $H_0$ was derived  by recently published studies in the literature are shown in Fig. \ref{figprobh0}.

Cosmological parameter degeneracies from each individual probe can be broken using combination of probes. The multi-probe analysis are crucial for independent $H_0$ determination and are required in order to reduce systematic uncertainties \citep{Suyu:2012ax,Chen:2011ab} \citep[see][for a review]{Moresco:2022phi}.

The analysis by \citet{Wong:2019kwg} using a combination of SH$0$ES and H$0$LiCOW results reports $H_0=73.8\pm 1.1$ $km$  $s^{-1} Mpc^{-1}$ which raises the Hubble tension to $5.3\sigma$ between late Universe determinations of $H_0$ and Planck. This has been discredited by \citet{Kochanek:2020crs} who points out that an artificial reduction of the allowed degrees of freedom can lead  to very precise but inaccurate estimates of $H_0$ based on gravitational lens time delays.

The  analysis by \citet{Abbott:2017smn} using a combination of the Dark Energy Survey (DES) \citep{Troxel:2017xyo,Abbott:2017wau,Krause:2017ekm} clustering and weak lensing measurements with BAO and BBN experiments assuming a flat $\Lambda$CDM model with minimal neutrino mass ($\Sigma m_{\nu} = 0.06$ $eV$) finds $H_0=67.2_{-1.0}^{+1.2}$ $km$  $s^{-1} Mpc^{-1}$ which is consistent with the value obtained with CMB data.

Using an extension of the standard GP formalism, and a combination of low-redshift expansion rate data (SnIa+BAO+CC) the Hubble constant was estimated to be $H_0 = 68.52 _{-0.94}^{+0.94+2.51(sys)}\,km\,s^{-1} Mpc^{-1}$ by \citet{Haridasu:2018gqm}. Using an alternative method \citet{Baxter:2020qlr} analysing the current CMB lensing data from Planck combined with Pantheon supernovae and using conservative  priors, find an $r_s$ independent constraint of $H_0=73.5\pm 5.3$ $km$  $s^{-1} Mpc^{-1}$.  Analysing low-redshift  cosmological  data  from SnIa, BAO, strong gravitanional lensing, $H(z)$ measurements  using  cosmic chronometers  and  growth  measurements from LSS observations for  $\Lambda$CDM model \citet{Dutta:2019pio} find $H_0 = 70.30 _{-1.35}^{+1.36}\,km\,s^{-1} Mpc^{-1}$ which is in $\sim 2\sigma$ tension with various low and high redshift observations.

\begin{center} 
\centering 
\begin{longtable*}[htb]{c ccc  ccc }
\caption{The Hubble constant $H_0$ values at  $68\%$  CL through  direct  and  indirect  measurements by different methods.}\vspace{2mm}
\label{hubblevalue} \\

\hhline{=======}
\\
\multicolumn{1}{c}{\textbf{Dataset}} &\multicolumn{1}{c}{} &\multicolumn{1}{c}{\textbf{$H_0$ [$km$  $ s^{-1} Mpc^{-1}$]}} &\multicolumn{1}{c}{}& \multicolumn{1}{c}{\textbf{Year}}&\multicolumn{1}{c}{} &\multicolumn{1}{c}{\textbf{Refs. }} \\ 
\\
\hhline{=======} 
\endfirsthead

\multicolumn{7}{c}%

{{\tablename\ \thetable{} -- continued from previous page}} \\
\hhline{=======} 
\\
\multicolumn{1}{c}{\textbf{Dataset}} &\multicolumn{1}{c}{} &\multicolumn{1}{c}{\textbf{$H_0$ [$km$  $ s^{-1} Mpc^{-1}$]}} &\multicolumn{1}{c}{}& \multicolumn{1}{c}{\textbf{Year}}&\multicolumn{1}{c}{} &\multicolumn{1}{c}{\textbf{Refs. }} \\ 
\\
\hhline{=======}
\endhead
\hline
\multicolumn{7}{r}{{Continued on next page}} \\ 
\hline
\endfoot 

\hhline{=======}
\endlastfoot
\\
{\bf Planck CMB} &&${\bf  67.27\pm 0.60}$&&{\bf 2020}&& \citep{Planck:2018vyg} \\  
 Planck CMB+lensing &&$ 67.36\pm 0.54$&&2020&& \citep{Planck:2018vyg} \\ 
Planck+SPT+ACT CMB &&$ 67.49\pm 0.53$&&2021&& \citep{SPT-3G:2021wgf} \\  
eBOSS+Planck CMB &&$ 69.6\pm 1.8$&&2020&& \citep{Pogosian:2020ded} \\
SPT-3G CMB &&$ 68.8\pm 1.5$&&2021&& \citep{SPT-3G:2021eoc} \\
ACT CMB  &&$67.9\pm 1.5 $&&2020&&\citep{ACT:2020gnv}\\ 
ACT+WMAP CMB  &&$ 67.6\pm 1.1 $&&2020&&\citep{ACT:2020gnv}\\
SPT CMB &&$ 71.3\pm 2.1$&&2018&& \citep{SPT:2017jdf} \\
WMAP9 CMB &&$ 70.0\pm 2.2$&&2013&& \citep{2013ApJS..208...19H} \\
BAO+WMAP CMB &&$68.36_{-0.52}^{+0.53}$&&2019&&\citep{Zhang:2018air}\\ BOSS correlation function+BAO+BBN&&$68.19\pm 0.99$&&2022&&\citep{Zhang:2021yna}\\
P+BAO+BBN&&$69.23\pm 0.77$&&2022&&\citep{Chen:2021wdi}\\
P+Bispectrum+BAO+BBN&&$68.31_{-0.86}^{+0.83}$&&2022&&\citep{Philcox:2021kcw}\\
BAO+BBN &&$66.98\pm 1.18$&&2018&&\citep{Addison:2017fdm}\\
BOSS DR12+BBN&&$68.5\pm 2.2$&&2020&&\citep{DAmico:2019fhj}\\
BOSS DR12+BBN&&$68.7\pm 1.5$&&2020&&\citep{Colas:2019ret}\\
BOSS DR12+BBN&&$67.9\pm 1.1$&&2020&&\citep{Ivanov:2019pdj}\\
BOSS+eBOSS+BBN&&$67.35\pm 0.97$&&2020&&\citep{eBOSS:2020yzd}\\
LSS $t_{eq}$ standard ruler&&$69.5_{-3.5}^{+3.0}$&&2022&&\citep{Farren:2021grl}\\
LSS $t_{eq}$ standard ruler+lensing &&$70.6_{-5.4}^{+3.0}$&&2020&&\citep{Philcox:2020xbv}\\
BAO+RSD  &&$ 69.13\pm 2.34 $&&2017&&\citep{Wang:2017yfu}\\
   & \\                                                    
  \hline  \\                                  
SnIa-Cepheid&&$73.04\pm 1.04$&&2022&&\citep{Riess:2021jrx}\\
SnIa-Cepheid&&$74.30\pm 1.45$&&2021&&\citep{Camarena:2021jlr}\\
   SnIa-Cepheid&&$73.20\pm 1.30$&&2021&&\citep{Riess:2020fzl}\\
  SnIa-Cepheid&&$74.03\pm 1.42$&&2019&&\citep{Riess:2019cxk}\\
   SnIa-Cepheid&&$73.48\pm 1.66$&&2018&&\citep{Riess:2018uxu}\\
  SnIa-Cepheid&&$72.80\pm 2.70$&&2020&&\citep{Breuval:2020trd}\\
  SnIa-Cepheid&&$73.00\pm 2.70$&&2020&&\citep{Breuval:2020trd}\\
SnIa-TRGB &&$76.94 \pm 6.4$&&2022&&\citep{Dhawan:2022yws}\\   
SnIa-TRGB &&$72.4 \pm 3.3$&&2022&&\citep{Jones:2022mvo}\\   
SnIa-TRGB &&$71.5\pm 1.8$&&2021&&\citep{Anand:2021sum}\\   
SnIa-TRGB &&$69.8\pm 1.7$&&2021&&\citep{Freedman:2021ahq}\\
SnIa-TRGB &&$65.8\pm 4.2$&&2021&&\citep{Kim:2020gai}\\ 
  SnIa-TRGB &&$72.10\pm 2.10$&&2020&&\citep{Soltis:2020gpl}\\
 SnIa-TRGB &&$69.60\pm 1.90$&&2020&&\citep{Freedman:2020dne}\\
  SnIa-TRGB&&$69.80\pm 1.90$&&2019&&\citep{Freedman:2019jwv}\\
SnIa-TRGB&&$71.1\pm 1.9$&&2019&&\citep{Reid:2019tiq}\\  
 SnIa-TRGB &&$72.40\pm 2.00$&&2019&&\citep{Yuan:2019npk}\\
SnIa-Miras&&$73.30\pm 4.00$&&2020&&\citep{Huang:2019yhh}\\
  SBF&&$73.30\pm 2.50$&&2021&&\citep{Blakeslee:2021rqi} \\
 SBF&&$70.50\pm 4.10$&&2020&&\citep{Khetan:2020hmh}\\
  SBF&&$71.90\pm 7.10$&&2018&&\citep{Cantiello:2018ffy} \\
SneII &&$75.4^{+3.8}_{-3.7}$&&2022&&\citep{deJaeger:2022lit}\\  
   SneII &&$75.8_{-4.9}^{+5.2}$&&2020&&\citep{deJaeger:2020zpb}\\
  Time-delay (TD) lensing &&$71.8_{-3.3}^{+3.9}$&&2021&&\citep{Denzel:2020zuq}\\ 
  TD lensing &&$73.3_{-1.8}^{+1.7}$&&2020&&\citep{Wong:2019kwg}\\ 
  TD lensing&&$72.8_{-1.7}^{+1.6}$&&2020&&\citep{Liao:2020zko}\\ 
 TD lensing&&$72.2\pm 2.1$&&2020&&\citep{Liao:2019qoc}\\ 
 TD lensing&&$73.65_{-2.26}^{+1.95}$&&2020&&\citep{Yang:2020eoh}\\ 
TD lensing&&$74.2\pm 1.6 $&&2020&&\citep{Millon:2019slk}\\ 
TD lensing&&$73.6_{-1.6}^{+1.8} $&&2021&&\citep{Qi:2020rmm}\\  
TD lensing&&$ 74.2_{-3.0}^{+2.7}$&&2020&&\citep{Shajib:2019toy}\\
TD lensing &&$74.5_{-6.1}^{+5.6}$&&2020&&\citep{Birrer:2020tax}\\ 
TD lensing+SLACS&&$67.4_{-3.2}^{+4.1}$&&2020&&\citep{Birrer:2020tax}\\ 
 TD  lensing+SLACS&&$76.8\pm 2.6$&&2019&&\citep{Chen:2019ejq}\\ 
GW Standard Sirens&&$67^{+6.3}_{-3.8}$&&2022&&\citep{Mukherjee:2022afz}\\  
GW Standard Sirens&&$68^{+12}_{-7}$&&2021&&\citep{LIGOScientific:2021aug}\\
GW Standard Sirens&&$72.77^{+11.0}_{-7.55}$&&2021&&\citep{Palmese:2021mjm}\\ 
GW Standard Sirens&&$73.4^{+6.9}_{-10.7}$&&2021&&\citep{Gayathri:2020mra}\\ 
 GW Standard Sirens&&$75_{-13}^{+59}$&&2020&&\citep{Abbott:2020khf}\\ 
GW Standard Sirens&&$50.4_{-19.5}^{+28.1}$&&2020&&\citep{Mukherjee:2020kki}\\ 
  GW Standard Sirens&&$67.6_{-4.2}^{+4.3}$&&2020&&\citep{Mukherjee:2020kki}\\ 
  GW Standard Sirens&&$ 48_{-10}^{+23}$&&2020&&\citep{Chen:2020gek}\\
   GW Standard Sirens&&$69.0_{-8.0}^{+16.0}$&&2019&&\citep{Abbott:2019yzh}\\ 
GW Standard Sirens&&$75_{-32}^{+40}$&&2019&&\citep{Soares-Santos:2019irc}\\ 
  GW Standard Sirens&&$68.9_{-4.6}^{+4.7}$&&2019&&\citep{Hotokezaka:2018dfi}\\
 GW Standard Sirens  &&$77.00_{-18.00}^{+37.00}$&&2019&&\citep{Fishbach:2018gjp}\\ 
 GW Standard Sirens &&$ 70.0_{-8.0}^{+12.0}$&&2017&&\citep{Abbott:2017xzu}\\ 
    Masers  &&$ 73.90\pm 3.00  $&&2020&& \citep{Pesce:2020xfe} \\ 
   Masers  &&$ 73.50\pm 1.40  $&&2019&& \citep{Reid:2019tiq} \\ 
  Masers &&$66.0\pm 6.0$&&2016&&\citep{Gao:2015tqd}\\ 
  Masers &&$ 73.0_{-22.0}^{+26.0}$&&2015&&\citep{Kuo:2014bqa} \\ 
   Masers &&$ 68.0\pm 9.0 $&&2013&&\citep{Kuo:2012hg}\\ 
   Masers &&$ 68.9\pm 7.1$&&2013&&\citep{Reid:2012hm}\\ 
   Tully Fisher&&$76.00\pm 2.60$&&2020&& \citep{Kourkchi:2020iyz} \\
   Tully Fisher&&$75.1\pm 2.80$&&2020&&\citep{Schombert:2020pxm}\\ 
    $\gamma$-ray attenuation&&$67.4_{-6.2}^{+6.0}$&&2019&&\citep{Dominguez:2019jqc}\\ 
   $\gamma$-ray attenuation &&$64.9_{-4.3}^{+4.6}$&&2019&&\citep{Zeng:2019mae}\\ 
      HII galaxy &&$71.00\pm 2.8$&&2018&&\citep{Fernandez-Arenas:2017isq}\\ 
   HII galaxy  &&$76.12_{-3.44}^{+3.47}$&&2017&&\citep{Wang:2016pag}\\ 
Cosmic chronometers, flat $\Lambda$CDM with systematics &&$66.5\pm5.4$&&2022&&\citep{Moresco:2022phi}\\   
Cosmic chronometers, open $w$CDM with systematics&&$67.8^{+8.7}_{-7.2} $&&2022&&\citep{Moresco:2022phi}\\   
 Cosmic chronometers, without systematics &&$67.06 \pm 1.68 $&&2018&&\citep{Gomez-Valent:2018hwc}\\ 
  Cosmic chronometers, without systematics &&$ 67.00\pm 4.00 $&&2018&&\citep{Yu:2017iju} \\ 
   Cosmic chronometers, without systematics &&$68.3_{-2.6}^{+2.7}$&&2017&&\citep{Chen:2016uno}\\
   &\\
 H(z)+BAO+SN-Pantheon+SN-DES+QSO+HIIG+ GRB&&$69.7\pm 1.2$&&2022&&\citep{Cao:2022ugh} \\
   CMB ($r_s$-independent)+lensing+Pantheon&&$73.5\pm 5.3$&&2021&&\citep{Baxter:2020qlr} \\
SnIa-Cepheid and TD lensing &&$73.8\pm 1.1$&&2020&&\citep{Wong:2019kwg}\\ 
 SnIa+BAO+TD lensing+cosmic chronometers+ LSS&&$70.30 _{-1.35}^{+1.36}$&&2019&&\citep{Dutta:2019pio}\\ 
 BAO+BBN+WL-CC &&$67.20_{-1.0}^{+1.2}$&&2018&&\citep{Abbott:2017smn} \\ SnIa+BAO+CC &&$68.52 _{-0.94}^{+0.94+2.51(sys)}$&&2018&&\citep{Haridasu:2018gqm} \\
  &\\
\end{longtable*}  
\end{center} 

\begin{figure*}
\begin{centering}
\includegraphics[width=0.92\textwidth]{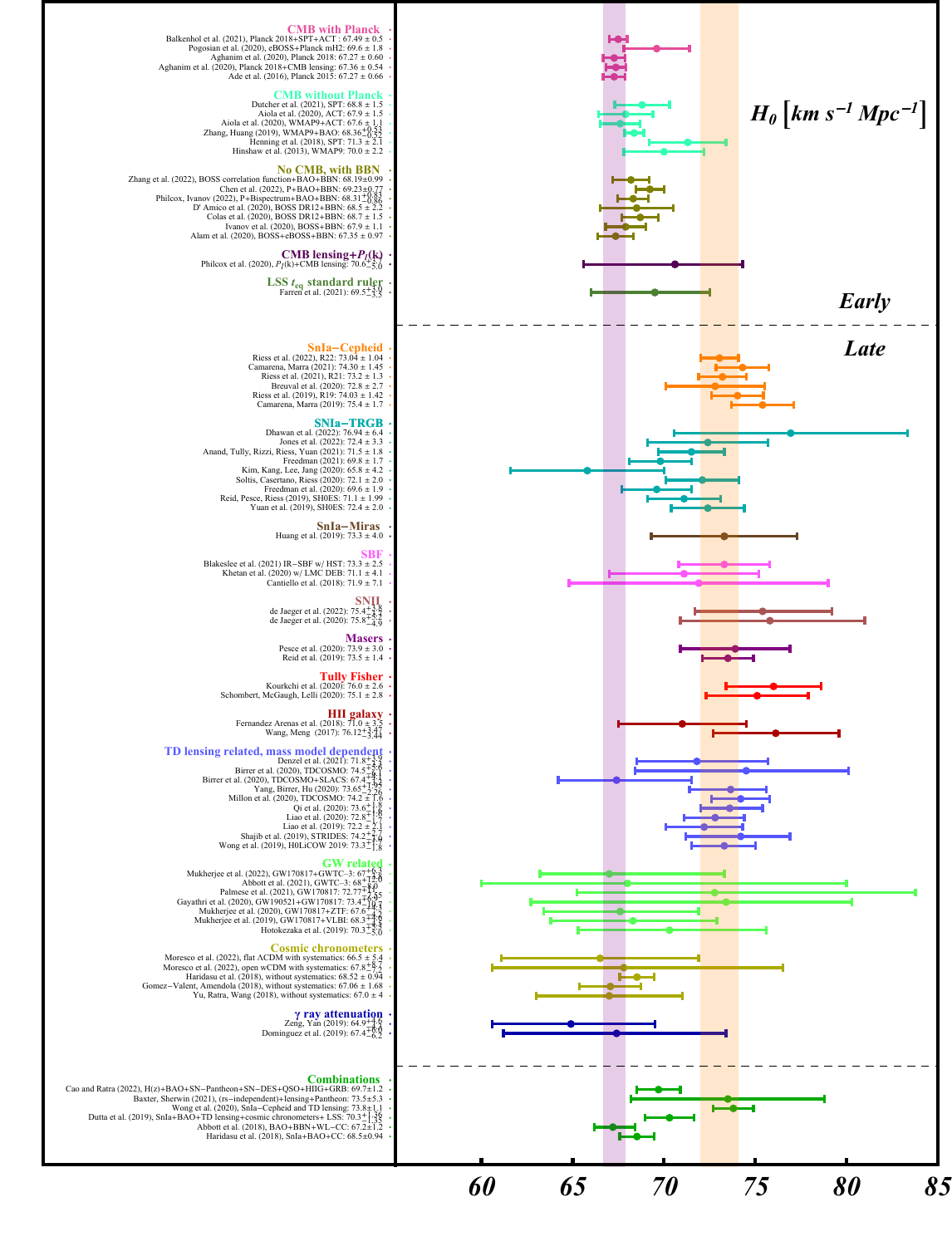}
\par\end{centering}
\caption{The Hubble constant $H_0$ values with the $68\%$ CL constraints derived by recent measurements. The value of the Hubble constant $H_0$ is derived by early time approaches based on sound horizon, under the assumption of a $\Lambda$CDM background.}
\label{figh0}
\end{figure*}

More recently,  the joint analysis of lower-redshift, non-CMB, data such as BAO, $H(z)$, SnIa, QSO, HII and GRBs by \citet{Cao:2022ugh} gives a model-independent determinations of the Hubble constant, $H_0=69.7\pm 1.2$ $km$  $s^{-1} Mpc^{-1}$ \citep[see also][for previous joint analyses]{Cao:2021ldv,Cao:2021cix,Cao:2020evz,Cao:2020jgu}. 
 
Many other estimates of $H_0$ have been obtained in the literature  within the standard $\Lambda$CDM model or in alternative scenarios by using joint analysis  \citep{Bonilla:2020wbn,Renzi:2020fnx}. 
In addition, many analyses using various combinations of data assuming a $\Lambda$CDM model or a extended model beyond $\Lambda$CDM cosmology investigate  whether the $H_0$ tension persists (or not). For example \citet{Okamatsu:2021jil} use  non-CMB data and specifically adopt the data from BAO, BBN, and SnIa to study the $H_0$ tension. They show that this tension exists in a broad framework beyond the standard $\Lambda$CDM model.

\begin{figure*}
\begin{centering}
\includegraphics[width=0.8\textwidth]{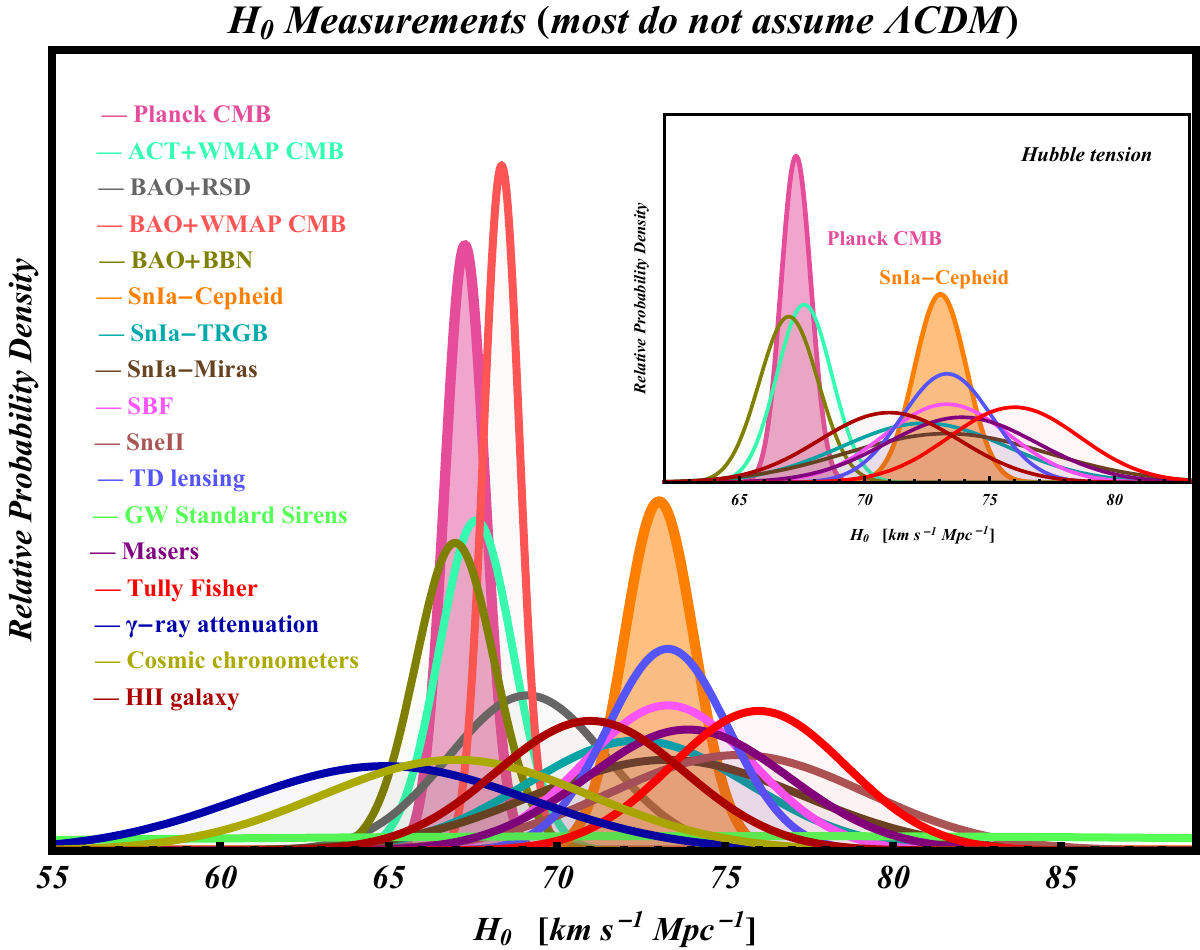}
\par\end{centering}
\caption{The one dimensional relative probability density value of $H_0$ derived  by recent measurements (Planck CMB \citep{Planck:2018vyg}, ACT+WMAP CMB \citep{ACT:2020gnv}, BAO+RSD \citep{Wang:2017yfu},  BAO+WMAP CMB \citep{Zhang:2018air},  BAO+BBN \citep{Addison:2017fdm},  SnIa-Cepheid \citep{Riess:2021jrx},  SnIa-TRGB \citep{Jones:2022mvo}, SnIa-Miras \citep{Huang:2019yhh},  SBF \citep{Blakeslee:2021rqi},   SneII \citep{deJaeger:2022lit}, TD lensing \citep{Wong:2019kwg}, GW Standard Sirens \citep{Abbott:2020khf}, Masers \citep{Pesce:2020xfe}, Tully Fisher \citep{Kourkchi:2020iyz},  $\gamma$-ray attenuation \citep{Zeng:2019mae}, cosmic chronometers \citep{Yu:2017iju},  HII galaxy \citep{Fernandez-Arenas:2017isq}). All measurements are shown as normalized Gaussian distributions. Notice that the tension is not so much between early and late time approaches but more between approaches that calibrate based on low $z$ ($z\lesssim 0.01$) gravitational physics and those that are independent of this assumption. For example cosmic chronometers and $\gamma$-ray  attenuation which are late time but independent of late gravitational physics are more consistent with the CMB-BAO than with late time calibrators. }
\label{figprobh0}
\end{figure*}

\begin{figure*}
\begin{centering}
\includegraphics[width=0.88\textwidth]{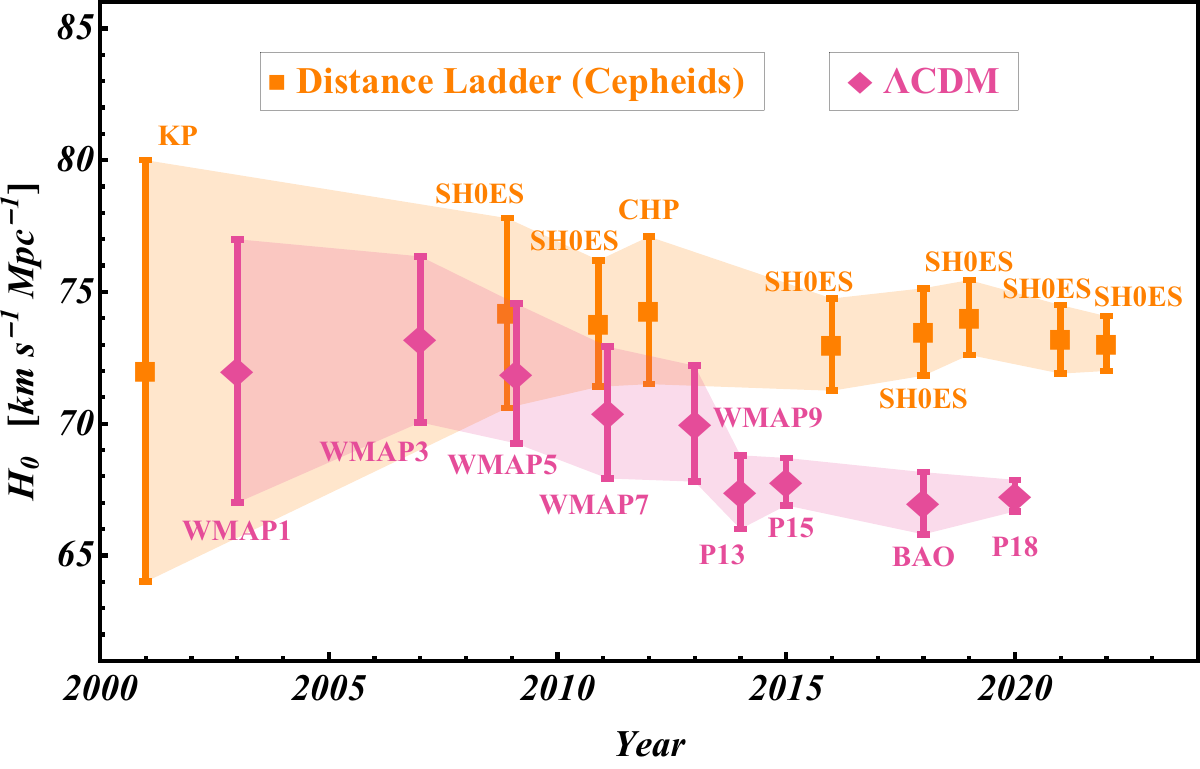}
\par\end{centering}
\caption{The Hubble constant as a function of publication date, using a set of different tools.  Symbols in orange denote values of $H_0$ determined in the late Universe with a calibration based on the Cepheid distance scale (Key Project (KP) \citep{HST:2000azd}, SH0ES \citep{Riess:2009pu,Riess:2011yx,Riess:2016jrr,Riess:2018uxu,Riess:2019cxk,Riess:2020fzl,Riess:2021jrx}, Carnegie Hubble Program (CHP) \citep{Freedman:2012ny}). Symbols in purple denote derived values of $H_0$  from analysis of the CMB data based on the sound horizon standard ruler (First Year WMAP (WMAP1) \citep{Spergel:2003cb}, Three Year WMAP (WMAP3) \citep{Spergel:2006hy}, Five Year WMAP (WMAP5) \citep{Dunkley:2008ie}, Seven Year WMAP (WMAP7) \citep{Komatsu:2010fb}, Nine Year WMAP (WMAP9) \citep{Bennett:2012zja}, Planck13 (P13) \citep{Ade:2013zuv}, Planck15 (P15) \citep{Ade:2015xua}, Planck18 (P18) \citep{Planck:2018vyg}, BAO \citep{Addison:2017fdm}). The orange and purple shaded regions demonstrate the evolution of the uncertainties in these values which have been decreasing for both methods. The most recent measurements disagree at greater than $4\sigma$.}
\label{figh0year}
\end{figure*}

\begin{figure*}
\begin{centering}
\includegraphics[width=0.96\textwidth]{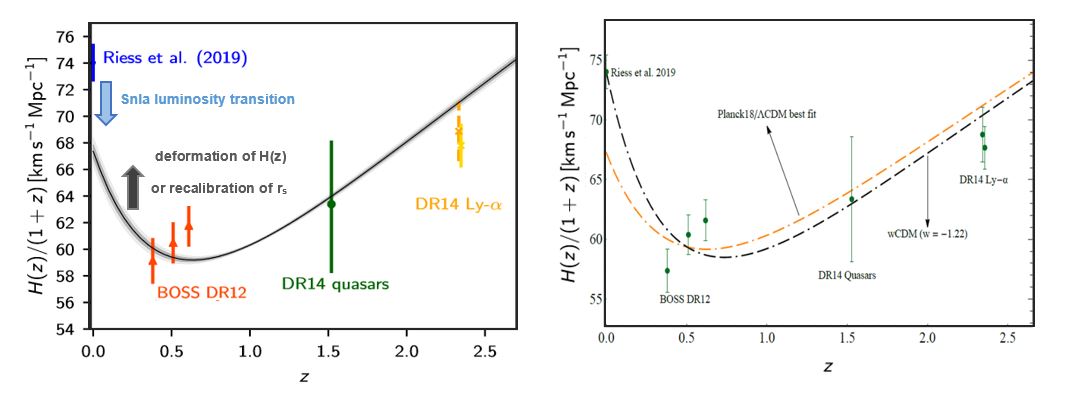}
\par\end{centering}
\caption{Left panel: The comoving Hubble parameter as a function of redshift. The black line corresponds to the best fit obtained from the Planck$18$ CMB when the $\Lambda$CDM model is considered, while the grey areas are the $1\sigma$ regions. The blue point at redshift zero denotes the inferred Hubble measurement by HST survey \citep{Riess:2019cxk}. The orange points, green point, and yellow points correspond to BAO data from BOSS DR12 survey \citep{Alam:2016hwk}, BOSS DR14 quasar sample \citep{Zarrouk:2018vwy}, and SDSS DR12 Ly$\alpha$ sample \citep{Riess:2018uxu} respectively. The arrows indicate approaches for the resolution of the Hubble tension: Down arrow (blue) corresponds to decrease of the Riess et. al. datapoint due to systematics or transition of the absolute magnitude $M$ (light blue arrow). Up arrow (black) corresponds to recalibration of $r_s$ which shifts the whole curve up or  and late time deformation of $H(z)$  \citep[adapted from][]{Planck:2018vyg}. Right panel: The comoving Hubble parameter as a function of redshift for a wCDM phantom modification of $\Lambda$CDM model which drives upward the low $z$ part of the $H(z)$ curve shown in left panel. Thus it brings the $z=0$ prediction of the CMB closer to the $H_0$ result of the local measurements (late time $H(z)$ deformation). }
\label{fighzm}
\end{figure*}

\subsubsection{The current status - Historic evolution}
Hubble's initial value in 1929 for the expansion rate, now called the Hubble constant, was approximately $500 km\,s^{-1} Mpc^{-1}$. From the 1970s, through the 80s and into the 90s the value of $H_0$ was estimated to be between 50 and 100 $km\,s^{-1} Mpc^{-1}$ \citep{1988Natur.334..209T}. Of interest is the historical Hubble constant debate  between, for example, long series of papers by Gérard de Vaucouleurs, who claimed that the value of $H_0$ is  $90<H_0<100\,km\,s^{-1} Mpc^{-1}$ \citep[e.g.][]{1985ApJ...297...27D,1986ASIC..180....1D}, and Allan Sandage , who claimed the value is $50<H_0<55\,km\,s^{-1} Mpc^{-1}$ \citep{1975ApJ...196..313S,1984Natur.307..326S} \citep[see][for a historical review]{Turner:2022gvw}. 

During the last decades there has been remarkable progress in measuring the Hubble constant.  The  available  technology  and measurement methods determine the  accuracy of this quantity. 
The Hubble constant as a function of publication date, using a set of different methods is shown in Fig. \ref{figh0year}. The values of $H_0$ determined in the late Universe with a calibration based on the Cepheid distance scale and the derived values of $H_0$ from analysis of the CMB anisotropy spectrum data are shown. The uncertainties in these values have been decreasing for both methods and the recent measurements disagree beyond $4\sigma$.

Furthermore the comoving Hubble expansion rate as a function of redshift obtained from the Planck$18$ CMB is shown in Fig. \ref{fighzm} along with a few relevant data-points demonstrating the Hubble tension. 

The basic strategic questions emerge 
\begin{itemize}
    \item 
    How can $H(z)$ derived from Cepheid late time calibrators (blue point in Fig. \ref{fighzm})  become consistent with $H(z)$ derived from the sound horizon early time calibrator (black line in Fig. \ref{fighzm})?
     \item
    What type of systematics could move the blue point down or shift black line up in Fig. \ref{fighzm} in early and late time calibrators? 
    \item
    To what extend can dynamical dark energy address the Hubble tension by distorting the black line in Fig. \ref{fighzm}?
\end{itemize}

These important Hubble tension questions will be discussed in the next subsection.

\begin{figure*}
\begin{centering}
\includegraphics[width=0.9\textwidth]{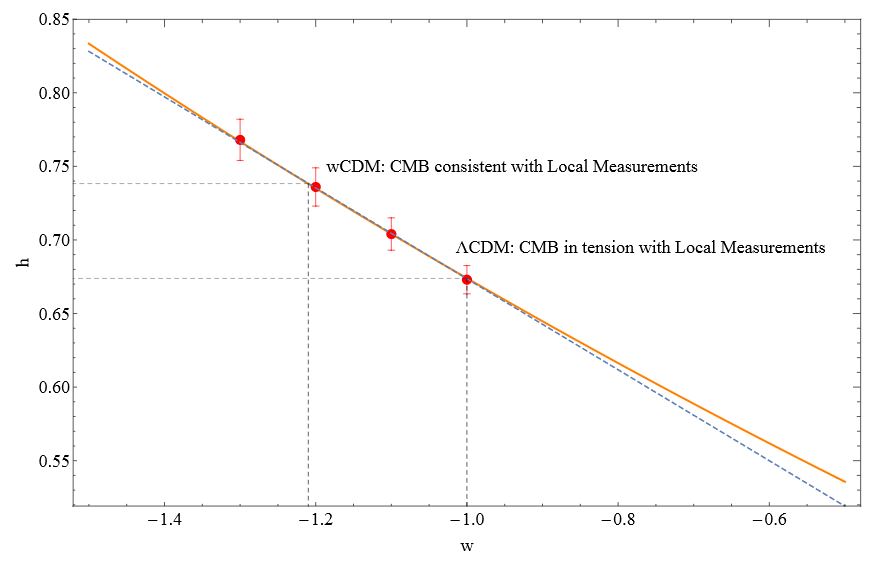}
\par\end{centering}
\caption{The predicted value of $h$ as a function of the fixed $w$ assuming one parameter dark energy (wCDM) model. The theoretically predicted best fit values of $h$ for different values of $w$ in the case of the wCDM model (orange line), whereas the linear fitting that has been made (dashed blue line). The redpoints correspond to the actual best fit values, including the errorbars, of $h$ for specific values of $w$ obtained by fitting these models to the CMB TT anisotropy  \citep[from][]{Alestas:2020mvb}.}
\label{figphantom}
\end{figure*}

\subsection{Theoretical models}
\label{Theoretical Models}
A wide range of models have been used to address the $H_0$ tension by introducing additional degrees of freedom to $\Lambda$CDM model where additional parameters are allowed to vary such as quintessence \citep{Fujii:1982ms,Ford:1987de,Wetterich:1994bg,Wetterich:1987fm,Ratra:1987rm,Chiba:1997ej,Ferreira:1997au,Ferreira:1997hj,Copeland:1997et,Caldwell:1997ii,Zlatev:1998tr,Tsujikawa:2013fta,Scherrer:2022umm,Steinhardt:1999nw,Liddle:1998xm}, in  which a scalar field plays the role of dark energy or modified gravity \citep{EspositoFarese:2000ij,Amendola:2006we,Fay:2007uy,Nojiri:2006su,Perivolaropoulos:2005yv,Boisseau:2000pr}, in which General  Relativity is modified on cosmological scales \citep[see][for a review]{Copeland:2006wr,Bamba:2012cp}. 

The extensions of $\Lambda$CDM model which can be used to resolve the Hubble constant $H_0$ tension fall into two categories: models with late time and models with early time modification (in the epoch  before the recombination)  \citep[see][for a review]{Verde:2019ivm,DiValentino:2021izs,Schoneberg:2021qvd}. 

The models with late time modification can be divided in four broad classes: deformations of the Hubble expansion rate $H(z)$ at late times  \citep[e.g. late time phantom dark energy ][]{Alestas:2020mvb,DiValentino:2016hlg}, deformations of the Hubble expansion rate $H(z)$  with additional interactions/degrees of freedom (e.g. interacting dark energy \citealt{DiValentino:2017iww,Yang:2019uog} and decaying dark matter \citealt{Berezhiani:2015yta}),  deformations of the Hubble expansion rate $H(z)$  due to inhomogeneous/anisotropic modifications \citep[e.g. inhomogeneous causal horizons][]
{Fosalba:2020gls} and  transition/recalibration of the SnIa absolute luminosity \citep{Kazantzidis:2020tko}  \citep[or combination of the previous classes e.g. late  $w-M$ phantom transition][]{Alestas:2020zol}.

Model selection statistical tools and approaches include the Akaike Information Criterion (AIC) \citep{Akaike:1974}, the Bayesian Information Criterion (BIC) \citep{Schwarz:1978tpv} and the Deviance Information Criterion (DIC) \citep{Spiegelhalter:2002yvw} and Bayesian model comparison \citep[e.g.][]{Nesseris:2012cq,Saini:2003wq,Mehrabi:2022mnr,Keeley:2021dmx,Koo:2021suo}. These tools have been developed and used to test, discriminate and compare the proposed models \citep{Liddle:2004nh,Liddle:2007fy,Arevalo:2016epc,Kerscher:2019pzk} \citep[see also][for a list of statistical tools]{Abdalla:2022yfr}.

\subsubsection{Late time deformations of the Hubble expansion rate $H(z)$ }
\label{Late time modifications1}
These late time models for the solution of the Hubble tension use a late time smooth deformation of the Hubble expansion rate Planck18/$\Lambda$CDM $H(z)$ so that it can match the locally measured value of $H_0$ while keeping the radius $r_s$ of the sound horizon at the last scattering  surface (see Subsection \ref{Standard Rulers: early time calibrators}). Many of these models effectively fix the comoving distance to the last scattering surface and the matter energy density  $\omega_m=\Omega_{0m} h^2$ to values consistent with \plcdm to maintain consistency with the CMB anisotropy spectrum while introducing late time phantom dark energy to deform $H(z)$ so that it matches the local measurements of $H(z)$. The required phantom behavior of such $H(z)$ deformations can not be provided by minimally coupled quintessence models and therefore such models have been shown to be unable to resolve the Hubble tension \citep{Banerjee:2020xcn,Lee:2022cyh}. 
These models have three problems 
\begin{itemize}
\item
They tend to worsen the fit to low z distance probes such as BAO and SnIa  \citep[e.g.][]{Alestas:2020mvb}
\item
They tend to worsen level of the growth tension \citep{Alestas:2021xes}.
\item
They tend to predict a lower value of SnIa absolute magnitude  than the one determined by local Cepheid calibrators shown in Eq. (\ref{apmagd}) \citep{Marra:2021fvf,Camarena:2019moy,Camarena:2021jlr}.
\end{itemize}
Thus, these models can not fully resolve the Hubble tension \citep[see  also][]{Alestas:2021luu,Alestas:2021nmi,Yang:2021flj,Benevento:2020fev,Efstathiou:2021ocp,Arendse:2019hev,Theodoropoulos:2021hkk,Krishnan:2021dyb, Bernal:2021yli,Vagnozzi:2021tjv,Cai:2021weh,Cai:2022dkh,Escamilla-Rivera:2021rbe,Benisty:2022psx}. 

Physical models where the deformation of $H(z)$ may be achieved include the following: phantom dark energy \citep[e.g.][]{Alestas:2020mvb} (see in Paragraph \ref{phantom dark energy}), running vacuum model \citep[e.g.][]{Sola:2005nh} (see in Paragraph \ref{running vacuum model}), phenomenologically emergent dark energy \citep{Li:2019yem} (see in Paragraph \ref{phenomenologically emergent dark energy}), vacuum phase transition \citep[e.g.][]{DiValentino:2017rcr} (see in Paragraph \ref{vacuum phase transition}), phase transition in dark energy \citep[e.g.][]{Khosravi:2017hfi} (see in Paragraph \ref{phase transition in dark energy}).  Plethora of late dark energy models with an equation of state $w\neq -1$ ($w<-1$ or $w>-1$) both constant or dynamical with redshift \citep[e.g.][]{Martinelli:2019krf} were proposed to address the Hubble tension. Recently, using a model-independent approach and a fully analytical analysis  \citet{Heisenberg:2022gqk,Heisenberg:2022lob} derive a set of necessary conditions that any late dark energy model must satisfy in order to potentially address both the Hubble and the growth tensions. In particular, solving the $H_0$ tension requires $w(z)<-1$ at some $z$ and solving both the $H_0$ and $\sigma_8$ tensions demands time-varying dark energy equation of state which cross the phantom divide. However \citet{Alestas:2021xes} have shown that $H(z)$ deformation approaches to the Hubble tension tend to worsen the $\sigma_8$ growth tension.

The following models may be classified in this class of theories: the holographic dark energy \citep{Guo:2018ans,Dai:2020rfo,Colgain:2021beg,vanPutten:2017bqf,10.1093/mnrasl/slz158,daSilva:2020bdc,Hernandez-Almada:2021aiw,Adhikary:2021xym}, the considering Chevallier - Polarski - Linder (CPL) \citep{Chevallier:2000qy,Linder:2002et,Kitazawa:2020qdx} parameterization \citep{Yang:2018prh},  the considering $w$ dependence on non-vanishing spatial curvature \citep{Miao:2018zpw}, the phantom brane dark energy \citep{Alam:2016wpf,Bag:2021cqm}, the  negative cosmological constant \citep{Visinelli:2019qqu,Calderon:2020hoc,Sen:2021wld}, the negative dark energy \citep{Dutta:2018vmq}, the graduated dark energy \citep{Akarsu:2019hmw}, the simple-graduated dark energy \citep{Acquaviva:2021jov}, the $\Lambda_{\rm s}$CDM model (sign-switching) \citep{Akarsu:2021fol}, the transitional dark energy \citep{Keeley:2019esp},  the frame dependent dark energy \citep{Adler:2019fnp}, the running $H_0$ with redshift \citep{Dainotti:2021pqg,Krishnan:2020vaf}, the varying gravitational constant \citep{Sakr:2021nja}, the deviation from the cold dark matter \citep{Elizalde:2021kmo} and  the phantom crossing \citep{DiValentino:2020naf}. For example in the case of the holographic dark energy  model \citep{Dai:2020rfo} and phantom crossing \citep{DiValentino:2020naf} models the tension on $H_0$ appears to be significantly alleviated within $1\sigma$ even though the three problems mentioned above do remain.

\paragraph{Phantom dark energy:}
\label{phantom dark energy}

The deformation of $H(z)$ through the implementation of late time phantom dark energy  \citep{Alestas:2020mvb,DiValentino:2016hlg,DiValentino:2017zyq,DiValentino:2019dzu,Vagnozzi:2019ezj,Huang:2016fxc} can address the Hubble tension as shown in Fig. \ref{fighzm}.

The analysis by \citet{Alestas:2020mvb} indicates that mildly phantom models with mean equation of state  parameter $w=-1.2$ have the potential to alleviate this tension. It was shown that the best fit value of $H_0$ in the context of the CMB power spectrum is degenerate with a constant equation of state parameter $w$. The CMB anisotropy spectrum was shown to be unaffected when changing $H(z)$ provided that specific parameter combinations remain unchanged. These cosmological parameters fix to high accuracy the form of the CMB anisotropy spectrum. The values of these parameters as determined by the Planck/$\Lambda$CDM CMB temperature power spectrum are the following \citep{Planck:2018vyg}.

\begin{figure*}
\begin{centering}
\includegraphics[width=0.92\textwidth]{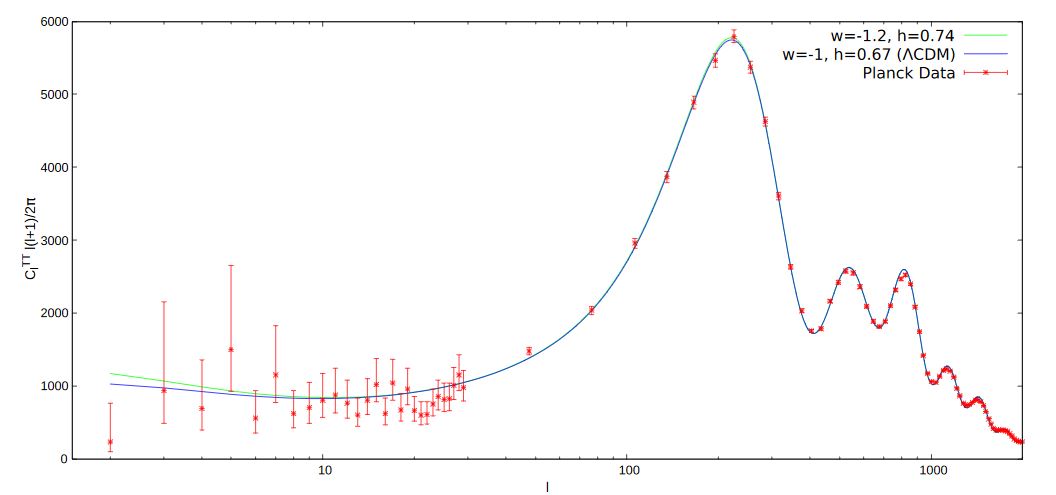}
\par\end{centering}
\caption{The predicted form of the CMB TT anisotropy spectrum with $w=-1$, $h= 0.67$, $\Omega_{0m}= 0.314$  for $\Lambda$CDM (blue line)  and with $w=-1.2$, $h=0.74$,  $\Omega_{0m}=0.263$ (green line). Red points correspond to the binned high-$l$ and low-$l$ Planck data  \citep[from][]{Alestas:2020mvb}.}
\label{figcmbphantom}
\end{figure*}
\begin{align}
&\omega_{m,Planck} = 0.1430\pm 0.0011 \label{ohm} \\
&\omega_{b,Planck} = 0.02237\pm 0.00015 \label{ohb}\\
&\omega_{r,Planck} = (4.64\pm 0.3)\: 10^{-5} \label{ohr}\\
&\omega_{k,Planck} = -0.0047\pm 0.0029 \label{ohk}\\
&d_{A,Planck}=(4.62\pm 0.08)\, (\:100\:km\:s^{-1}Mpc^{-1})^{-1}\label{dap}
\end{align}
where $\omega_{i,Planck}=\Omega_{0i,Planck}h^2$ is the energy density of component $i$ and $d_{A,Planck}$ is the comoving angular diameter distance.

Using the Eq. (\ref{dAthdef}) the comoving angular diameter distance $d_A$ to the recombination surface is ($c=1$)
\be
d_A=\int_0^{z_r} \frac{dz}{H(z)}=\int_0^{z_r} \frac{dz}{h(z)\:100\:km\:s^{-1}Mpc^{-1}}
\ee
where $z_r\simeq1100$  is  the  redshift  of  recombination and $h(z)=H(z)(\:100\:km\:s^{-1}Mpc^{-1})^{-1}$ is the dimensionless Hubble parameter which in general takes the form
\begin{align}
h(z)=&\left[\omega_r(1+z)^4 +\omega_m(1+z)^3+\right.\nonumber\\
&\left.+(h^2-\omega_r-\omega_m)f_{DE}(z)\right]^{1/2}
\end{align}
where $h=h(z=0)$ and $f_{DE}(z)$ determines the evolution of dark energy.

In the context of a simple one parameter parametrization where the equation of state $w$ remains constant in time and redshift (wCDM model), $f_{DE}(z)$ takes the simple form
\be
f_{DE}(z)=(1 +z)^{3(1+w)}
\ee

If the four energy densities Eqs.  (\ref{ohm}), (\ref{ohb}), (\ref{ohr}) and  (\ref{ohk}) and the observed value of the comoving angular diameter Eq. (\ref{dap}) are fixed then they provide the analytically predicted best fit value of the Hubble parameter $H_0$ (or $h$) given the dark energy equation of state parameter $w(w_0, w_1,...,z)$  where $w_0, w_1$,... are  the  parameters  entering the $w(z)$ parametrization. Thus assuming a flat Universe ($\omega_k= 0$) and solving the following equation with respect to $h$

\begin{align}
&d_A(\omega_{m,Planck},\omega_{r,Planck},\omega_{b,Planck},h=0.674,w=-1)=\nonumber\\
&d_A(\omega_{m,Planck},\omega_{r,Planck},\omega_{b,Planck},h,w)
\end{align}
it is straightforward to derive the degeneracy function $h(z=0,w)\equiv h$ shown in Fig. \ref{figphantom} (continuous orange line). In the range $w\in [-1.5,-1]$,\:  $h(w)$ is approximated as a straight line (dashed blue line in Fig. \ref{figphantom})
\be
h(w)\approx - 0.3093 w + 0.3647
\ee

For $w=-1$, this linear degeneracy equation leads to the best fit dimensionless Hubble constant $h=0.674$ as expected while for $w=-1.217$ the corresponding predicted CMB best fit is $h=0.74$ which is consistent with the value obtained by local distance ladder measurements. The  invariance of the CMB power spectrum when the cosmological parameters are varied along the above described degeneracy directions is shown in Fig. \ref{figcmbphantom}. This method of  \citet{Alestas:2020mvb} can be used to find general degeneracy relations between $f_{DE}(z)$ and $H_0$
and fixing $h=0.74$ gives infinite $f_{DE}(z)$, $w(z)$ forms that can potentially resolve the $H_0$ problem if they can also properly fit the low $z$ date (e.g. BAO, SnIa, Cepheid value of absolute luminosity $M$). Low $z$ distance data (BAO and SnIa) will determine which one of these forms is observationally favored. However, none of these forms can provide a quality of fit to low z data equally good or better than $\Lambda$CDM despite the introduced additional parameters. In addition, these models suffer from the other two problems mentioned above (worse growth tension and lower value of SnIa absolute magnitude).

\paragraph{Running  vacuum  model:}
\label{running vacuum model}
The running vacuum models (RVM) \citep{Sola:2005nh,Sola:2005et,Shapiro:2004ch,Shapiro:2008yu,Shapiro:2009dh,Basilakos:2012ra,Perico:2013mna,Basilakos:2015yoa,Perico:2013mna,Tsiapi:2018she,Banerjee:2019kgu,Farrugia:2018mex,Basilakos:2019acj,Lima:2013dmf,Moreno-Pulido:2020anb,Basilakos:2019wxu,Mavromatos:2021pxc,Rezaei:2021qwd,Mavromatos:2021urx,Mavromatos:2020kzj,Shapiro:2003kv,Basilakos:2020qmu} \citep[see][for a review]{Peracaula:2022vpx,Sola:2011qr,Sola:2013gha,Sola:2015rra,Sola:2014tta,Sola:2016zeg,Mavromatos:2021sew,Mavromatos:2021hai} attempts to address both the Hubble constant $H_0$ tension \citep{Sola:2017znb} and the $\sigma_8$ growth tension using a mechanism that has common features with the IDE models  \citep[e.g.][]{Sola:2015wwa,Sola:2016jky,Gomez-Valent:2017idt,Gomez-Valent:2018nib,SolaPeracaula:2021yid,SolaPeracaula:2021gxi} (for relaxing the growth tension, see Subsection \ref{Theoretical Models s8}). 

The RVM of the cosmic evolution is well motivated by the generic idea of renormalization group formalism which is used in Quantum Field Theory (QFT) \citep{Sola:2007sv,Shapiro:1999zt,Shapiro:2000dz} \citep[see also][for a approach using adiabatic regularization and renormalization techniques]{Moreno-Pulido:2020anb,Moreno-Pulido:2022phq}. In the RVM the cosmological constant, the corresponding vacuum energy density and pressure are assumed to be functions of the Hubble rate e.g. a power series of the Hubble rate and its cosmic time derivative with even time derivatives of the scale factor \citep{Gomez-Valent:2014rxa}
\be
\Lambda=a_0+\sum_{k=1}a_kH^{2k}+\sum_{k=1}b_k\dot{H}^k,
\ee
$\rho_{\Lambda}=\rho_{\Lambda}(H)=\frac{\Lambda(H)}{8\pi G}$ and $p_{\Lambda}=p_{\Lambda}(H)=-\rho_{\Lambda}(H)$  respectively. 

For the current Universe the vacuum energy density can be written in the relatively simple form  \citep[e.g.][]{Sola:2015rra,Gomez-Valent:2014rxa,Sola:2017jbl,Sola:2017znb}  
\be
\rho_{\Lambda}(H)=\frac{\Lambda(H)}{8\pi G}=\frac{3}{8\pi G}(c_0+\nu H^2)
\ee
where $c_0=H_0^2(\Omega_{0\Lambda}-\nu)\simeq\frac{\Lambda_0}{3}$ (with $\Lambda_0$ the current value) is an integration constant which is fixed  by the boundary  condition $\rho_{\Lambda}(H_0)=\rho_{\Lambda,0}$ (with $\rho_{\Lambda,0}$ the current value) and $\nu$ is a dimensionless running  parameter which characterizes the dynamics of the vacuum at low energy. For $\nu=0$ the vacuum energy remains constant at all times and for $\nu>0$ the vacuum energy density decreases with the  time. In QFT the running  parameter is $|\nu|\simeq 10^{-6}-10^{-3}$ \citep{Sola:2007sv} but in RVM it has been treated as a free parameter by fitting to the observational data  \citep[e.g.][]{Gomez-Valent:2014rxa,Sola:2017jbl}.

\paragraph{Phenomenologically emergent dark energy:}
\label{phenomenologically emergent dark energy}
Phenomenologically emergent dark energy (PEDE) is a zero freedom dark energy scenario proposed by  \citet{Li:2019yem}.  In this model the dark energy density has the following form
\be
\tilde{\Omega}_{DE}(z)=\Omega_{0DE}\left[1-\tanh\left(\log_{10}(1+z)\right)\right]
\ee
where $\Omega_{0DE}=1-\Omega_{0m}-\Omega_{0r}$.

The  dark energy in this model has no effective presence in the past and emerges at the later times and with the same number (six) of parameters compared to the spatially flat $\Lambda$CDM scenario. It has the potential for alleviating the $H_0$ tension  \citep{Li:2019yem,Pan:2019hac,Yang:2021egn,Liu:2020vgn,Rezaei:2020mrj,Yang:2020ope,DiValentino:2021rjj}. The generalised emergent dark energy (GEDE) model has one extra dimensionless free parameter $\Delta$ including both  $\Lambda$CDM model as well as the PEDE model as two of its special limits introduced by \citet{Li:2020ybr}. In the GEDE model the dark energy density has the following form \citep{Yang:2021eud}
\be
\tilde{\Omega}_{DE}(z)=\Omega_{0DE}\frac{1-\tanh\left(\Delta\log_{10}(\frac{1+z}{1+z_t})\right)}{1+\tanh\left(\Delta\log_{10}(1+z_t)\right)}
\ee
where $z_t$ is the transition redshift where dark energy density equals to matter density. For $\Delta=0$ and  $\Delta=1$  this model recovers $\Lambda$CDM and  PEDE model respectively. 
Using the latest observational Hubble dataset \citet{Hernandez-Almada:2020uyr}  revisited  and  constrained the free parameters of the PEDE and GEDE models. 

Other versions of the PEDE model are the Modified Emergent Dark Energy (MEDE) \citep{Benaoum:2020qsi}  and the Transitional Dark Energy (TDE) \citep{Zhou:2021xov} models. The MEDE model with one extra degree of freedom  reduces the Hubble tension to $2.4\sigma$ \citep{Benaoum:2020qsi} even though it also suffers from the three problems of the late time $H(z)$ deformation models.

\paragraph{Vacuum phase transition:}
\label{vacuum phase transition}
Vacuum phase transition \citep{DiValentino:2017rcr,DiValentino:2020kha,DiValentino:2021zxy,DiValentino:2021rjj} based on vacuum metamorphosis (VM) or vacuum cold dark matter model (VCDM) \citep{Parker:2000pr,Parker:2003as,Caldwell:2005xb} has the potential to address the $H_0$ tension.
This mechanism with six free parameters as the spatially flat $\Lambda$CDM. It also assumes a phase transition in the nature of the vacuum similar to Sakharov’s induced gravity \citep{Sakharov:1967pk}.
The phase transition occurs when the evolving Ricci scalar curvature $R$ becomes equal to the value of scalar field mass squared $m^2$ \citep{DiValentino:2017rcr}
\be 
R=6(\dot{H}+H^2)=m^2
\ee
where the dot corresponds to the derivative  with respect to cosmic time $t$. After the transition the Ricci scalar curvature remains constant with $R=m^2$ and this changes the expansion rate below ($z<z_t$) due to the phase transition 
\begin{align}
&\,\,\, \frac{H^2}{H_0^2}=\Omega_{0m}(1+z)^3+\Omega_{0m}(1+z)^3+ \nonumber\\  \,\,+M &\left\{ 1-\left[3\left(\frac{4}{3\Omega_{om}}\right)^4M(1-M)^3\right]^{-1}\right\},\, z>z_t
\end{align} 
\be
\frac{H^2}{H_0^2}=(1-M)(1+z)^4+M,\,\, z\leq z_t
\ee
where $M=\frac{m^2}{12H_0^2}$  and   $z_t=-1+\frac{3\Omega_{0m}}{4(1-M)}$ is the transition redshift.\\

\paragraph{Phase transition in dark energy: }
\label{phase transition in dark energy}
\begin{itemize}
\item
Phase transition in dark energy explored by
\citet{Khosravi:2017hfi,Banihashemi:2018has,Banihashemi:2018oxo,Moshafi:2020rkq} can address the Hubble tension.  Generalizing this model by assigning a more realistic time evolution of  dark energy \citet{Banihashemi:2020wtb} propose the critically emergent dark energy (CEDE) model.

In \citet{Farhang:2020sij} the form of phase transition parametrized phenomenologically by a hyperbolic  tangent function. This scenario for dark energy is similar used independently as PEDE and GEDE.

\item
Late dark energy (LDE) transition \citep{Benevento:2020fev}  at redshifts  $z\ll 0.1$  can reduce the Hubble tension.  This class of $H(z)$ deformation models has a more intense form of the third problem of the deformation class as they predict a significantly lower value of the SnIa absolute magnitude than the other $H(z)$ deformation models \citep{Alestas:2020zol,Camarena:2021jlr}.

In this scenario the true Hubble constant is given by \citep{Mortonson:2009qq,Benevento:2020fev}
\be
H_0^2=\tilde{H}_0^2(1+2\delta)
\ee
where $\tilde{H}_0$ is the prediction for a flat $\Lambda$CDM model in the context of a CMB sound horizon calibration.

In \citet{Benevento:2020fev,Dhawan:2020xmp} it was shown  that this model can not fully resolve the Hubble problem as it would imply a transition in the SnIa apparent magnitude which is not observed. These models however become viable in the context of a SnIa absolute magnitude transition \citep{Alestas:2020zol,Marra:2021fvf}.
\end{itemize}

\subsubsection{Deformations of the Hubble expansion rate $H(z)$  with additional interactions/degrees of freedom}
\label{IDE}
There exist several varieties of the models for the solution of the Hubble tension which use deformations of the Hubble expansion rate $H(z)$  with additional interactions/degrees of freedom. For example  the interacting dark energy models \citep[e.g.][]{DiValentino:2017iww,Yang:2019uog} (see in Paragraph \ref{interacting dark energy}) with an extra non-gravitational interaction between the components of the Universe and  the decaying dark matter models \citep[e.g.][]{Berezhiani:2015yta} (see in Paragraph \ref{decaying dark matter}) with additional degrees of freedom are able to alleviate the Hubble constant $H_0$ tension. 

The following models may be classified in this class of theories: multi-interacting dark energy \citep{Lucca:2021eqy}, new interacting dark energy \citep{Gao:2021xnk}, interacting vacuum energy \citep{Kumar:2021eev}, metastable dark energy  \citep{Li:2019san,Yang:2020zuk}, Quintom dark energy \citep{Panpanich:2019fxq}, cannibal dark matter \citep{Buen-Abad:2018mas},  baryons-dark energy interacting \citep{Jimenez:2020ysu} \citep[see also][]{Vagnozzi:2019kvw,Ferlito:2022mok}, swampland conjectures \citep{OColgain:2018czj,Colgain:2019joh,Agrawal:2019dlm}, nonlocal gravity \citep{Belgacem:2017cqo,Belgacem:2020pdz}, late time transitions in the quintessence field \citep{DiValentino:2019exe}, Galileon gravity \citep{Renk:2017rzu,Zumalacarregui:2020cjh,Peirone:2019aua,Frusciante:2019puu,Heisenberg:2020xak},  $f(R)$ gravity \citep{Odintsov:2020qzd,DAgostino:2020dhv,Wang:2020dsc,Cruz:2020cje},  $f(T)$ gravity \citep{Nunes:2018xbm,Wang:2020zfv,Ren:2021tfi,Cai:2019bdh,Hashim:2020sez,Briffa:2020qli,Benisty:2021sul,Bahamonde:2022ohm,Ren:2022aeo}, $f(T,B)$ gravity \citep{Escamilla-Rivera:2019ulu}, $f(Q)$ gravity \citep{Mandal:2020buf}, Brans-Dicke gravity \citep{SolaPeracaula:2019zsl,SolaPeracaula:2020vpg}, minimal theory of massive gravity \citep{deAraujo:2021cnd}, scale–dependent gravity \citep{Alvarez:2020xmk},  unimodular gravity \citep{Perez:2020cwa,LinaresCedeno:2020uxx}, the screened fifth forces \citep{Desmond:2019ygn,Desmond:2020wep}, the minimally modified gravity \citep{DeFelice:2020cpt}, the Lifshitz cosmology \citep{Berechya:2020vcy}, the Milne cosmology \citep{Vishwakarma:2020paa}, 4D Gauss-Bonnet gravity \citep{Wang:2021kuw}, the generalized Chaplygin gas \citep{Yang:2019nhz}, the unified cosmologies \citep{Yang:2019jwn}, the $\Lambda$-gravity \citep{Gurzadyan:2019yir,Gurzadyan:2021jrw},  the $\Lambda(t)$-model \citep{Benetti:2019lxu,Benetti:2021div}, the bulk viscous cosmology \citep{Elizalde:2020mfs,daSilva:2020mvk,Yang:2019qza,Normann:2021bjy} and the surface tension
hypothesis \citep{Ortiz:2020noa}. For instance in the case of the  metastable dark energy  \citep{Yang:2020zuk}, generalized Chaplygin gas \citep{Yang:2019nhz}  and Galileon gravity \citep{Renk:2017rzu} models the tension on $H_0$ appears to be significantly alleviated to within about $1\sigma$ even though the there problems of the $H(z)$ deformation models remain to be addressed.

\paragraph{Interacting dark energy:}
\label{interacting dark energy}

In the cosmological interacting dark energy (IDE) models \citep{Nunes:2022bhn,Linton:2021cgd,Hogg:2021yiz,Mancini:2021lec,Sharma:2021ivo,Carrilho:2021hly,Carrilho:2021rqo,Amendola:1999er,Wang:2004cp,Salvatelli:2013wra,Costa:2013sva,Pourtsidou:2013nha,Pettorino:2013oxa,Salvatelli:2014zta,Kumar:2016zpg,Xia:2016vnp,Kumar:2017dnp,vandeBruck:2017idm,DiValentino:2017iww,Yang:2017ccc,Yang:2018ubt,Yang:2018xlt,Yang:2018uae,Yang:2018euj,Li:2018ydj,An:2018vzw,Gonzalez:2018rop,Kumar:2019wfs,Pan:2019jqh,Yang:2019uzo,Yang:2019uog,Gomez-Valent:2020mqn,Aljaf:2020eqh,DiValentino:2019ffd,Martinelli:2019dau,DiValentino:2020kpf,Yao:2020hkw,Yao:2020pji,Lucca:2020zjb,Yang:2020uga,Yang:2020tax,Amirhashchi:2020qep,Johnson:2021wou,Pan:2019gop} \citep[see][for a review]{Wang:2016lxa,Bolotin:2013jpa} the dark components of the Universe i.e dark matter (DM) and dark energy (DE) have  an  extra  non-gravitational interaction. The IDE model was proposed to address the coincidence problem  \citep[e.g.][]{Comelli:2003cv,Huey:2004qv,Cai:2004dk,Pavon:2005yx,Zhang:2005rg,delCampo:2006vv,Berger:2006db,delCampo:2008jx,He:2008tn}.
In addition the interaction between the dark fluids has been shown to be effective in substantially alleviating the Hubble constant $H_0$ tension \citep{Yang:2018uae,Yang:2018euj,DiValentino:2017iww,DiValentino:2019jae,Pan:2019gop,Kumar:2016zpg,Yang:2019uog,Lucca:2020zjb,Wang:2021kxc,Gariazzo:2021qtg,Nunes:2021zzi,Guo:2021rrz,Pan:2020bur} or in addressing the structure growth $\sigma_8$ tension between the values inferred from the CMB and the WL measurements \citep{Pourtsidou:2016ico,An:2017crg,Barros:2018efl,Camera:2017tws} (see Subsection \ref{Theoretical Models s8}) or in solving  the two tensions simultaneously \citep{Kumar:2017dnp,Kumar:2019wfs,DiValentino:2019ffd}.

In IDE cosmology assuming spatially flat Friedmann-Lema$\hat{ı}$tre-Roberson-Walker background and pressureless dark matter ($w_c= 0$)  the equations of evolution of the dark matter and dark energy densities $\rho_c$ and $\rho_{DE}$ respectively are given by \citep{Gavela:2009cy}
\be 
\dot{\rho}_c+3H\rho_c=Q(t)
\ee
\be 
\dot{\rho}_{DE}+3H(1+w_{DE})\rho_{DE}=-Q(t)
\ee
where the dot corresponds to the derivative with respect to cosmic time $t$, $w_{DE}=\frac{p_{DE}}{\rho_{DE}}$ is the equation of state of dark energy and $Q$ represents the interaction rate between the dark sectors (i.e. the rate of energy transfer between the dark fluids). For $Q <0$ energy flows from dark matter to dark energy, whereas for $Q>0$ the energy flow is opposite.

These models combine the deformation of $H(z)$ with an extra modification of the growth rate of perturbations due to the tuned evolution of $\Omega_m(z)$ induced by the interaction term $Q$. This additional tuning allows for a simultaneous improvement of the growth tension in contrast to models that involve a simple $H(z)$ deformation. 

Various phenomenological IDE models were proposed in the literature where the rate of the interaction $Q$ has a variety of possible functional forms \citep{Pan:2020zza}. For example in some classes of IDE models the rate of the interaction $Q$ is proportional to the energy density of dark energy $Q= \delta\, H \rho_{DE}$ \citep{Kumar:2017dnp,Kumar:2019wfs,Yang:2019uog,DiValentino:2019ffd} or cold dark matter  $Q= \delta\, H \rho_c$ \citep{Kumar:2016zpg} (where $\delta$ is a constant and $\delta=0$ in the $\Lambda$CDM cosmology), or some combination of the two. Note that in the case of functional form $Q= \delta\, H \rho_c$ instabilities develop in the dark sector perturbations at early times \citep{He:2008si}.

\paragraph{Decaying dark matter:}
\label{decaying dark matter}
Decaying dark matter into dark radiation (i.e. an  unknown  relativistic species that is not directly detectable), which has been first analysed by \citet{Ichiki:2004vi} and studied by \citet{Chen:2020iwm,Chudaykin:2017ptd,Bjaelde:2012wi,Wang:2014ina,Anchordoqui:2015lqa,Xiao:2019ccl,Choi:2019jck,Choi:2020tqp,Nygaard:2020sow}, provides a promising scenario to relieve the Hubble constant $H_0$ tension  \citep[e.g.][]{Berezhiani:2015yta}. Also, it has been shown that this scenario can resolve the  $\sigma_8$ growth tension \citep{Enqvist:2015ara,Abellan:2020pmw} or the two tensions simultaneously \citep{Pandey:2019plg} by a similar mechanism as in the IDE models. However, using the Planck data the analysis of the model by \citet{Poulin:2016nat,Chudaykin:2016yfk} has shown that the cosmological tensions are only slightly alleviated \citep[see][for a different result]{Bringmann:2018jpr}.

In these models assuming spatially flat Friedmann-Lema$\hat{ı}$tre-Roberson-Walker Universe, pressureless dark matter, $w_c= 0$ and equation of state of dark radiation $w_{DR}=1/3$, the equations of evolution of the dark matter and dark radiation densities $\rho_c$ and $\rho_{DR}$ respectively are given by \citep{Gavela:2009cy}
\be 
\dot{\rho}_c+3H\rho_c=-\Gamma \rho_c
\ee
\be 
\dot{\rho}_{DR}+4H\rho_{DR}=\Gamma \rho_c
\ee
where $\Gamma=\frac{1}{\tau}$ is the decay rate of dark matter particles (with $\tau$ the particle’s lifetime). In the literature a variety of possible functional forms of the decay rate has been explored \citep{Enqvist:2015ara,Lesgourgues:2015wza,Poulin:2016nat,Bringmann:2018jpr}. For example in some cases the decay rate is proportional to the Hubble rate, $\Gamma\propto H$ \citep{Pandey:2019plg}. Constraints on the decay rate of dark matter have been obtained by the analysis of  \citet{Ando:2015qda,Enqvist:2015ara}.

A model with decaying dark matter into dark radiation in early/late Universe ($\tau\ll t_s$ / $\tau\gg t_s$, where $ t_s$ is the time of last scattering) increases/decreases the expansion rate $H(a;\rho_b,\rho_{\gamma},\rho_c,\rho_{DR},\rho_{DE})$ at high/low redshifts as it predicts a smaller matter content and a larger radiation content  as time evolves (the early/late Universe is dominated by the radiation/matter and the dark radiation density decreases more rapidly than the matter density, $\rho_{DR}\propto a^{-4}$ and $\rho_c\propto a^{-3}$). In the case of $\tau\ll t_s$, the faster cosmological expansion $H(z)$ decreases the scale of the sound horizon $r_s$ in Eq. (\ref{rsdef}) because the baryon-to-photon ratio, and thus $c_s$ in Eq. (\ref{csdef}), is tightly constrained by CMB fluctuations and BBN \citep{Ade:2015rim}. In the context of the degeneracy $H_0 r_s$ shown in Eq. (\ref{thetas}) the lower scale of the sound horizon $r_s$ yields a larger value of $H_0$. In the case of $\tau\gg t_s$, the lower dimensionless normalized Hubble rate $E(z)$ in the late-time leads to a larger value of $H_0$ since $\theta_s$ and $r_s$ must be kept fixed in Eq. (\ref{thetas}). Accordingly, both early and late decaying dark matter model are able to alleviate the Hubble constant $H_0$ tension \citep[see][for a detailed discussion]{Anchordoqui:2020djl,Anchordoqui:2022gmw}.

There are alternative decaying dark matter models such as the light dark matter \citep{Alcaniz:2019kah}, the dynamical dark matter \citep{Desai:2019pvs}, the many-body or $2$-body decaying cold dark matter scenarios \citep{Blackadder:2014wpa} and the decaying warm dark matter scenario \citep{Blinov:2020uvz}. In the $2$-body decaying cold dark matter scenario the decaying dark matter produces two particles, one  massive warm dark matter particle and one massless relativistic particle (dark radiation). This scenario can address the Hubble constant $H_0$ tension \citep{Vattis:2019efj,Clark:2020miy,Haridasu:2020xaa} and the $\sigma_8$ growth tension \citep{Abellan:2021bpx}.

A self-interacting dark matter model with a light force mediator coupled to dark radiation studied by \citet{Binder:2017lkj,Hryczuk:2020jhi}. This model can simultaneously reduce the tension between CMB and low-redshift astronomical observations of $H_0$ and $\sigma_8$.

\citet{Jaeckel:2020oet} pointed out that a  dark particle from reheating \citep{Jaeckel:2021gah}  can alleviate the $H_0$ tension through its decay to relativistic component which contributes to the dark radiation.

Recently, Ly-$\alpha$ constraints on possible models of dark-matter physics have been evaluated by \citet{Dienes:2021cxp}. In particular the Ly-$\alpha$ bounds on different classes of dark-matter velocity distributions have been obtained.

\subsubsection{Deformations of the Hubble expansion rate $H(z)$  with inhomogeneous/anisotropic modifications}

Models where the cosmological principle and the FLRW metric are relaxed by considering inhomogeneous/anisotropic modifications have the potential to resolve the Hubble problem \citep{Kasai:2019yqn}. Physical models where the deformation of $H(z)$ may be achieved with inhomogeneous/anisotropic modifications,  include the following: Chameleon  dark energy \citep[e.g.][]{Cai:2021wgv} (see in Paragraph \ref{Chameleon dark energy}), cosmic voids \citep{Wu:2017fpr} (see also Paragraph \ref{cosmic voids}) and inhomogeneous causal horizons
\citep{Fosalba:2020gls} (see also Paragraph \ref{inhomogeneous causal horizons}), charged dark matter \citep{Jimenez:2020bgw,BeltranJimenez:2020csl,BeltranJimenez:2021imo}, Bianchi type I spacetime  \citep{Akarsu:2019pwn} and emerging  spatial curvature \citep{Heinesen:2020sre,Bolejko:2017fos}.

\paragraph{Chameleon dark energy:}
\label{Chameleon dark energy}

Chameleon  dark energy \citep{Khoury:2003aq,Khoury:2003rn}  \citep[see also][]{Brax:2007vm,Banerjee:2008rs,Das:2008iq,Brax:2010kv,Wang:2012kj,Upadhye:2012vh,Khoury:2013yya,Vagnozzi:2021quy,Benisty:2021cmq} attempts to address the Hubble constant $H_0$ tension  by  introducing  a cosmic inhomogeneity in the Hubble expansion  rate at late-time from the chameleon field coupled to the local matter overdensities \citep{Cai:2021wgv}. This field trapped at a higher potential energy density acts as an effective cosmological constant and results in a faster local expansion rate than that of the background with lower matter density.

\paragraph{Cosmic voids:}
\label{cosmic voids}
In cosmic void models the local $H_0$ departs significantly from the cosmic mean $H_0$ because of the presence of an under-dense region (local void) \citep{Lombriser:2019ahl}. However in  \citet{Wu:2017fpr,Kenworthy:2019qwq} it was shown that this alternative theory is inconsistent with current observations. The analysis was based on the assumption of the validity of standard $\Lambda$CDM and a study of the sample variance in the local measurements of the Hubble  constant this alternative theory  has been shown inconsistent with current  observations.  \citet{Wu:2017fpr} estimated that the required radius of void to resolve the tension in $H_0$ is about $150$ $Mpc$ and density contrast of $\delta\equiv\frac{\rho-\bar{\rho}}{\bar{\rho}}\simeq -0.8$ which is inconsistent at $\sim 20\sigma$ with the $\Lambda$CDM cosmology \citep{Kenworthy:2019qwq,Haslbauer:2020xaa}.

In the context of this inconsistency,  \citet{Haslbauer:2020xaa} considered a cosmological Milgromian dynamics or modified  Newtonian dynamics (MOND) model \citep{Milgrom:1983pn} with  the  presence of $11 eV/c^2$ sterile neutrinos\footnote{Sterile neutrinos are a special kind of neutrino  with right handed chirality that might interact only through gravity \citep{Dodelson:1993je,Boyarsky:2009ix}  \citep[see][for a review]{Kusenko:2009up,Abazajian:2012ys,Drewes:2013gca}. They have been proposed to resolve some anomalies in neutrino data.}  to show that the Keenan-Barger-Cowie (KBC) void\footnote{The KBC void \citep{Keenan:2013mfa} is a large local underdensity between $40$ and $300\, Mpc$ (i.e. $0.01\lesssim z\lesssim 0.07$) around the Local Group.} has the potential to resolve the Hubble tension.

\paragraph{Inhomogeneous Causal Horizons:}
\label{inhomogeneous causal horizons}
\citet{Fosalba:2020gls} proposed a simple solution to the $H_0$ tension based on causally disconnected regions  of the CMB temperature anisotropy maps from Planck \citep{Akrami:2018vks}. It was pointed out that CMB maps show 'causal horizons' where cosmological parameters have distinct values. This could be justified by the fact that these regions of the Universe have never been in causal contact. Thus it was shown that the Hubble constant $H_0$ takes values which differ up to $20\%$ among different causally disconnected regions. These cosmological parameter inhomogeneities are in agreement with the model of the Universe proposed in  \citet{Gaztanaga:2020ksy} (see also \citealt{Gaztanaga:2021bgb,gaztanaga:hal-03344159,Gaztanaga:2022ktb} for details) where the cosmological constant is simply formulated as a boundary term in the Einstein equations and where 'Causal Horizons' naturally arise. Thus if there are similar 'causal horizons' in the local universe (i.e, $z<1100$), then $20\%$ variations between the local and high-z measures of $H_0$ are indeed to be expected \citep{Fosalba:2020gls}. 

\subsubsection{Late time modifications - Transition/Recalibration of the SnIa absolute luminosity}
\label{Late time modifications2}

This class of models can address the problems of the $H(z)$ deformation models (especially the low $M$ problem) by assuming a rapid variation (transition) of the SnIa intrinsic luminosity and absolute magnitude due e.g. to a gravitational physics transition at a redshift $z_t\lesssim 0.01$ \citep{Marra:2021fvf,Alestas:2020zol,Alestas:2021nmi}.

\paragraph{Gravity and evolution of the SnIa intrinsic luminosity:}

As shown in the recent analysis by \citet{Kazantzidis:2020tko} there are abnormal features which may be interpreted as evolution of the measured parameter combination $\mathcal{M}$ (see Section \ref{Standard candles: late time calibrators}). This measured parameter combination $\mathcal{M}$ in Eq. (\ref{combM}) depends on the absolute magnitude $M$ and on the Hubble constant $H_0$ ($M$ and $H_0$ are degenerate parameters). Any variation of the parameter $\mathcal{M}$ is due to a variation of $M$ which could be induced by a varying $\mu_G(z)\equiv\frac{G(z)}{G_0}$ (where $G_0$ is the local value of the Newton’s constant $G(z)$). If the calibrated SnIa absolute magnitude $M$ were truly constant then the parameter $\mathcal{M}$ should also be constant (independent of redshift).

A possible variation of the absolute magnitude $M$ and equivalently of the absolute luminosity
\be
L\sim 10^{-2M/5}
\ee
could be due to a variation of the fine structure constant $\alpha$ or the Newton’s constant $G$. 

If the absolute luminosity is proportional  to  the Chandrasekhar mass $L\sim M_{Chandr}$ we have \citep{GarciaBerro:1999bq,Gaztanaga:2001fh}
\be 
L\sim G^{-3/2}
\label{lumg}
\ee
Thus $L$ will increase as $G$ decreases\footnote{ Adopting a semi-analytical model which takes into account the stretch of SnIa light curves but assumes fixed mass of Ni,  obtains SnIa light curves in  the  context  of  modified  gravity \citet{Wright:2017rsu} have  claimed   that $L$ will increase as $G$ increases.}.

Under these assumptions, we obtain
\be 
M(z)-M_0=\frac{15}{4}\log \mu_G(z)
\ee
where $M_0$ corresponds to a reference local value of the absolute magnitude and $\mu_G \equiv \frac{G}{G_0}$ is the relative effective gravitational constant (with $G$ the strength of the gravitational interaction and $G_0$ the locally measured Newton’s constant).

Then, the Eq. (\ref{combM}) takes the following form
\be 
\mathcal{M}(z)=M_0+\frac{15}{4}\log \mu_G(z)+5\log_{10}\left[\frac{c/H_0}{Mpc}\right]+25
\label{combMz}
\ee 
and the apparent magnitude Eq. (\ref{aparmagz}) can be written as 

\be
m(z,H_0,\Omega_{0m})_{th}=\mathcal{M}(z,H_0)+
5\log_{10}\left[D_L(z,\Omega_{0m})\right]
\label{aparmagzmu}
\ee

A mild tension at $2\sigma$ level in the best fit value of $\cal M$ was found in between the low-redshfit ($z\lesssim0.2$) data and the full  Pantheon dataset  in the context of a $\Lambda$CDM  model.
This tension can be interpreted as \citep{Kazantzidis:2020tko}
\begin{itemize}
\item
a locally higher value of $H_0$ by about $2\%$, corresponding to a local matter underdensity. 
\item
a time variation of Newton’s constant which implies an evolving Chandrasekhar mass and thus an evolving absolute luminosity L and absolute magnitude M of low z SnIa.
\end{itemize}
In addition, the oscillating features shown in Fig. \ref{figosc} hint also to the possibility of evolutionary effects of $M$. As discussed below such evolutionary effects if they exist in the form of a transition may provide a solution to the Hubble and growth tensions.

\begin{figure*}
\begin{centering}
\includegraphics[width=0.96\textwidth]{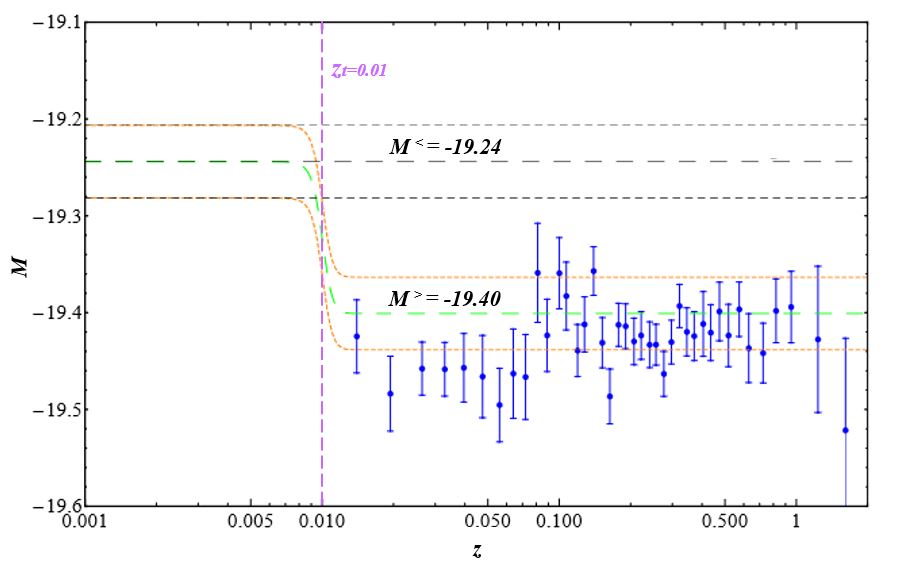}
\par\end{centering}
\caption{The Pantheon binned  SnIa absolute magnitudes Eq. (\ref{magpoint}) $M_i$ (blue  points) \citep{Scolnic:2017caz} for a Planck/$\Lambda$CDM luminosity distance. The data are inconsistent with the SnIa absolute magnitude $M^<=-19.24$ calibrated by Cepheids but the inconsistency disappears if there is a transition in the absolute magnitude with amplitude $\Delta M\simeq-0.2$ at redshift $z_t\simeq 0.01$  \citep[from][]{Marra:2021fvf}.}
\label{figtram}
\end{figure*}

\paragraph{Transition of the SnIa absolute magnitude $M$ at a redshift $z\simeq 0.01$:}

Recently, \citet{Marra:2021fvf} have  proposed that a rapid transition (abrupt deformation) at a  transition redshift $z_t\simeq 0.01$ in the value of the SnIa absolute magnitude $M$ of the form
\be
M^>(z) =M^<+ \Delta M\, \Theta (z-z_t) 
\ee
(where $\Theta$  is the  Heaviside  step  function) due to a rapid transition of the gravitational constant can address the Hubble tension. 

In particular the analysis by \citet{Marra:2021fvf} has shown that a $10\%$ rapid transition in the value of the relative effective gravitational constant $\mu_G$ at $z_t\simeq 0.01$ is sufficient to induce the required  reduction of $M$ 
\be
\Delta M\equiv M^>-M^<\simeq -0.2
\ee
where  $M^<$ is the SnIa  absolute magnitude of Eq. (\ref{maglate}) calibrated by  Cepheids  at $z<0.01$ \citep{Camarena:2019moy,Camarena:2021jlr} and  $M^>$ is the SnIa absolute magnitude of Eq. (\ref{magearly}) using the parametric-free inverse distance ladder of \citet{Camarena:2019rmj}. Fig. \ref{figtram} shows the Pantheon SnIa absolute magnitudes for a Planck/$\Lambda$CDM luminosity distance \citep{Scolnic:2017caz} obtained from
\be
M_i=m_i-5\log_{10}\left[\frac{d_L(z_i)}{Mpc}\right]-25
\label{magpoint}
\ee
where $m_i$ are the apparent magnitude datapoints.

The data are in disagreement with the SnIa absolute magnitude $M^<$ calibrated by Cepheids but they become consistent if there is a transition in the absolute magnitude with amplitude $\Delta M\simeq-0.2$  \citep{Marra:2021fvf}. Thus, this class of $M$-transition models avoids the $M$-problem of late time $H(z)$ deformation models.

Assuming the power law dependence Eq. (\ref{lumg}) and using RSD and WL data  \citep{Abbott:2017wcz,Skara:2019usd,Alestas:2020mvb} reported a best fit value $\Delta \mu_G\equiv \mu_G^>-\mu_G^<=-0.19\pm 0.09$ ($\mu_G^>$ corresponds to $z>0.01$ and  $\mu_G^<$ corresponds to $z<0.01$) in the context of a $\Lambda$CDM background $H(z)$. The analysis by \citet{Marra:2021fvf} showed that a rapid $\sim10\%$ increase of the effective gravitational constant roughly 150 million years ago can also solve $\Omega_m$-$\sigma_8$ growth tension.

Recently,  \citet{Alestas:2021xes} have demonstrated that this model has an advantage over both early time and late time deformations of $H(z)$ to fully resolve the Hubble tension while at the same time improving the level of the $\Omega_m$-$\sigma_8$  growth tension. In addition it has the potential to provide equally good fit to low $z$ distance probes such as BAO and SnIa as the Planck18/$\Lambda$CDM model. 

More recently, \citet{Perivolaropoulos:2022txg} generalized the symmetron screening mechanism\footnote{For reviews of modified gravity theories with screening mechanisms, such as the Vainshtein  \citep{Vainshtein:1972sx,Arkani-Hamed:2002bjr,Deffayet:2001uk} and the chameleon \citep{Khoury:2003aq,Khoury:2003rn,Gubser:2004uf,Brax:2004qh,Brax:2004px,Upadhye:2006vi,Mota:2006ed,Mota:2006fz,Brax:2008hh,Brax:2010kv} models see  \citep{Khoury:2010xi,Burrage:2017qrf,Sakstein:2013pda,Brax:2012gr,Jain:2010ka,Davis:2011qf,Hui:2009kc,Burrage:2016bwy,Joyce:2014kja,Brax:2021wcv,Baker:2019gxo,Sakstein:2018fwz} and for screening effects see \citep{Renevey:2021tcz}.} \citep{Hinterbichler:2010es,Hinterbichler:2011ca} by allowing for an explicit symmetry $Z_2$ breaking of the symmetron $\phi^4$ potential. The explicit symmetry breaking can create an asymmeron wall network pinned on matter overdensities separating regions with distinct gravitational properties which could constitute a physical mechanism for the realization of gravitational transitions in redshift space that could help in the resolution of the Hubble and growth tensions. Another theoretical model leading to a gravitational transition could include a pressure non-crushing cosmological singularity in the recent past \cite{Odintsov:2022eqm}.

\paragraph{Late (low-redshift) $w-M$ phantom transition:}

The late (low-redshift) $w-M$ phantom transition \citep{Alestas:2020zol} is a late time approach involving a combination of the previous two classes: the transition of the SnIa absolute luminosity and the deformation of the Hubble expansion rate $H(z)$.
A rapid phantom transition of the dark energy equation of state parameter $w$ at a  transition redshift $z_t<0.1$ of the form
\be 
w(z) =-1 + \Delta w\, \Theta (z_t-z)
\ee
with $\Delta w<0 $ and a similar transition in the value of the SnIa absolute magnitude $M$ of the form
\be
M(z) =M_C+ \Delta M\, \Theta (z-z_t) 
\ee
with $\Delta M<0 $ due to evolving fundamental constants  can address the Hubble tension \citep{Alestas:2020zol}. Where $\Theta$  is  the  Heaviside  step  function, $M_C$ is the  SnIa  absolute  magnitude Eq. (\ref{maglate}) calibrated by  Cepheids \citep{Camarena:2021jlr,Camarena:2019moy} at $z <0.01$ and $\Delta M$, $\Delta w$ are parameters to be fit by the data.  \citet{Alestas:2020zol}  find $\Delta M\simeq - 0.1$, $\Delta w\simeq - 4$  for $z_t= 0.02$ which imply a lower value of $\mu_G$ at $z>0.02$ (about $6\%$) compared to the pure $M$-transition model. 

The late (low-redshift) $w-M$ phantom transition (LwMPT) can lead to a resolution of the Hubble tension in a more consistent manner than smooth deformations of the Hubble tension and other types of late time transitions such as the Hubble expansion rate transition \citep{Benevento:2020fev,Dhawan:2020xmp}. Its main advantages include the consistency in the predicted value of the SnIa absolute magnitude M and the potential for simultaneous resolution of the growth tension.

\citet{Mortsell:2021nzg,Mortsell:2021tcx,Perivolaropoulos:2021bds} have analyzed the color-luminosity relation of Cepheids in anchor galaxies and SnIa host galaxies by identifying the color-luminosity relation for each individual galaxy instead of enforcing a universal color-luminosity relation to correct the NIR Cepheid magnitudes. A systematic brightening of Cepheids at distances larger than about $20\, Mpc$ which could be enough to resolve the Hubble tension was found.  In addition, \citet{Perivolaropoulos:2021bds} investigating the effects of variation of the Cepheid calibration empirical parameters (the color-luminosity parameter or the Cepheid absolute magnitude) find hints for the presence of a fundamental physics transition taking place at a time more recent than 100 Myrs ago. The magnitude of the transition lead to value of $H_0$ that is consistent with the CMB inferred value thus eliminating the Hubble tension. The distance range/timescale corresponding to this transition is consistent with solar system history data \citep{Perivolaropoulos:2022vql} indicating an increase of the rate of impactors on the Moon and Earth surfaces by about a factor of 2-3 during the past 100Myrs which correspond to $z<0.008$ \citep{1998JRASC..92..297S,1995hdca.book.....G,1997JGR...102.9231M,2001JGR...10632847G,Ward2007TerrestrialCC,2019Sci...365.9895M,Bottke} and low redshift galaxy surveys data \citep{Alestas:2022xxm}. Such a transition is also consistent with a recent analysis by \citet{Alestas:2021nmi} indicating a  transition in the context of the Tully-Fisher data.

In particular, using a robust dataset of 118 Tully-Fisher datapoints  \citet{Alestas:2021nmi} have demonstrated that evidence for a transition in the evolution of BTFR appears at a level of more than $3\sigma$. Such effect could be interpreted as a transition of the effective Newton's constant. The amplitude and sign of the gravitational transition are consistent with the mechanisms for the resolution of the Hubble and growth tension discussed above \citep{Marra:2021fvf,Alestas:2020zol}  \citep[see in][for a talk of the tensions of the $\Lambda$CDM and a gravitational transition ]{pres2}.

\subsubsection{Early time modifications of sound horizon}
\label{Early time modifications}

Modifying the scale of sound horizon $r_s$ (i.e. the scale of the standard ruler) by introducing new physics before recombination that deform  $H(z)$ at prerecombination redshifts $z\gtrsim1100$ can increase the CMB inferred value of $H_0$  \citep{Bernal:2016gxb,Poulin:2018zxs,Aylor:2018drw,Knox:2019rjx} and thus resolve the Hubble tension. Such deformation may be achieved by introducing various types of additional to the standard model components \citep[see][for a review]{Asadi:2022njl}. These models have the problem of predicting stronger growth of perturbations than implied by dynamical probes like redshift space distortion (RSD) and weak lensing (WL) data and thus may  worsen the $\Omega_m$-$\sigma_8$ growth tension \citep{Jedamzik:2020zmd,Alestas:2021xes}. 

A wide range of mechanisms has been proposed for the decrease of the the sound horizon scale at recombination. These mechanisms include the introduction of early dark energy, extra neutrinos or some other dark sector at recombination, features in the primordial power spectrum,  modified scenarios of recombination etc. The following models and theories may be classified in this class of mechanisms: early dark energy \citep[e.g.][]{Poulin:2018cxd} (see Paragraph \ref{early dark energy}),  dark radiation \citep[e.g.][]{Green:2019glg} (see Paragraph \ref{dark radiation}), neutrino self-interactions \citep[e.g][]{Kreisch:2019yzn} (see Paragraph \ref{neutrino self-interactions}), large primordial non-Gaussianities \citep{Adhikari:2019fvb} (see Paragraph \ref{large primordial non-Gaussianities}),  Heisenberg's uncertainty principle \citep{Capozziello:2020nyq} (see Paragraph \ref{Heisenberg's uncertainty principle}),  early modified gravity \citep{Braglia:2020auw} (see Paragraph \ref{early modified gravity}), cosmological inflation physics  \citep{Tanin:2020qjw,AresteSalo:2021lmp,Hazra:2018opk,Tram:2016rcw,DiValentino:2016ucb,DiValentino:2016ziq,Guo:2017qjt,Chiang:2018xpn,DiValentino:2018zjj,Liu:2019dxr,Keeley:2020rmo,Ye:2021nej,Takahashi:2021bti}, dark matter - photon coupling \citep{Kumar:2018yhh,Yadav:2019jio}, dark matter-neutrino interactions \citep{Paul:2021ewd}, interacting dark radiation \citep{Blinov:2020hmc}, ultralight dark photon \citep{Flambaum:2019cih}, primordial black holes \citep{Nesseris:2019fwr,Flores:2020drq}, primordial magnetic fields \citep{Jedamzik:2018itu,Jedamzik:2020krr,Thiele:2021okz}, non-standard recombination \citep{Liu:2019awo}, unparticles dark energy \citep{Artymowski:2020zwy}, varying fundamental constants \citep{Franchino-Vinas:2021nsf,Hart:2017ndk,Sekiguchi:2020teg,Hart:2019dxi,Hart:2021kad}, early-time thermalization of cosmic components \citep{Velten:2021cqj}, CMB monopole temperature shift \citep{Ivanov:2020mfr}, open and hotter universe \citep{Bose:2020cjb,Bengaly:2020vly}, Axi-Higgs cosmology \citep{Fung:2021wbz,Fung:2021fcj}, string Cosmology \citep{Anchordoqui:2020znj,Anchordoqui:2019amx} and dark massive vector fields \citep{Anchordoqui:2019yzc}. In this list of proposed cosmological models the tension on $H_0$ is alleviated with a significance ranging from the $1\sigma$ to the $3\sigma$ level.
\begin{figure*}
\begin{centering}
\includegraphics[width=0.97\textwidth]{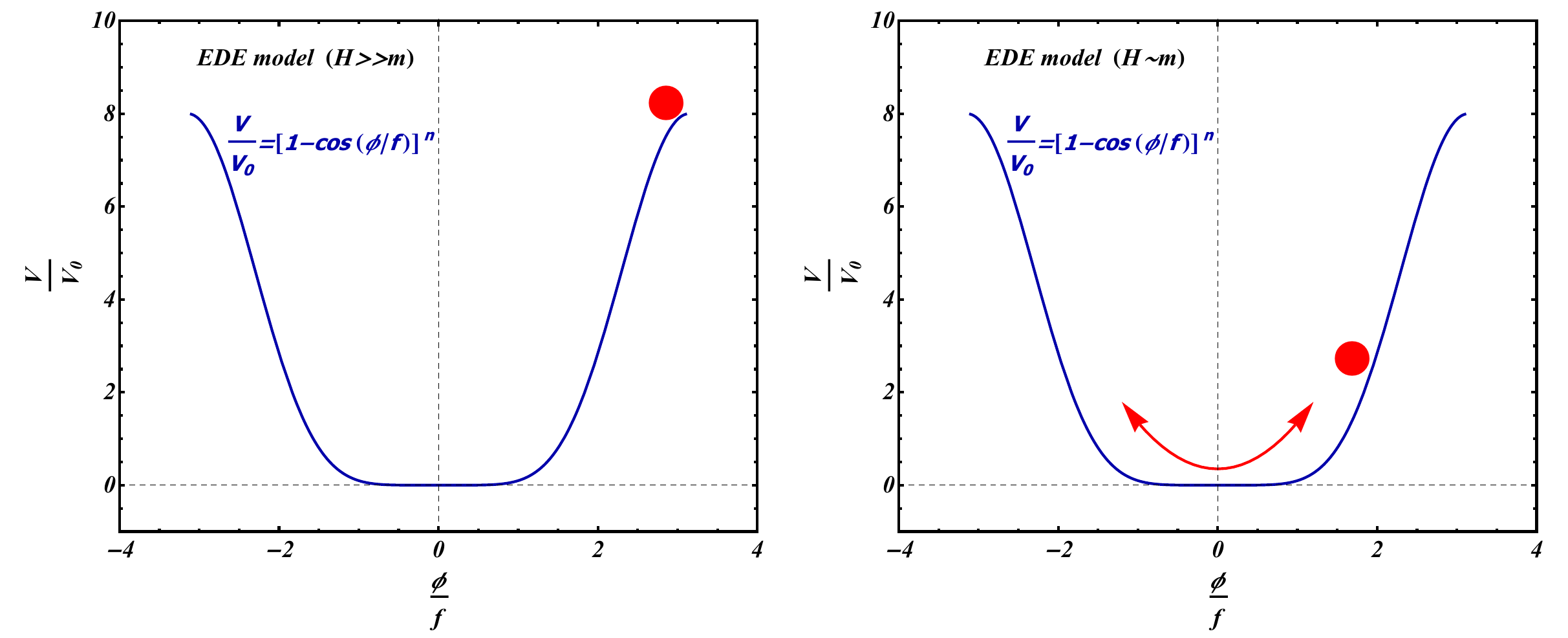}
\par\end{centering}
\caption{The potential $V/V_0$ (with $V_0=m^2f^2$, $n=3$ in Eq. (\ref{potede})) as a function of $\phi/f$  at early times ($H\gg m$)  (left panel) when the field $\phi$ is initially frozen in its potential due to Hubble friction and acts as a cosmological constant with equation of state $w_{\phi}=-1$, and  at a critical redshift $z_c$ when the Hubble parameter drops below some value ($H\sim m$) (right panel)  and the field becomes dynamical and begins to oscillate around its minimum which is locally $V\sim \phi^{2n}$.}
\label{figedepot}
\end{figure*}
\begin{figure}
\begin{centering}
\includegraphics[width=0.48\textwidth]{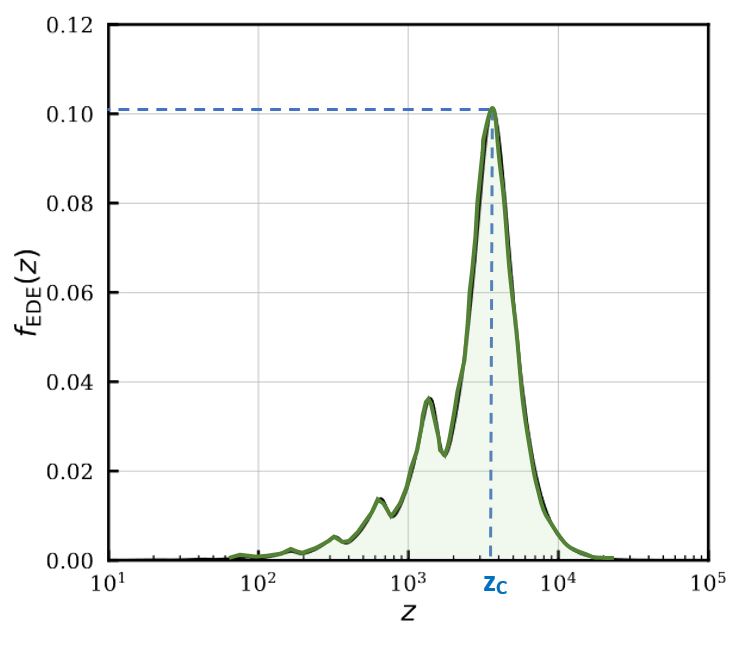}
\par\end{centering}
\caption{Fractional contribution of EDE to the cosmic energy  budget as a function of redshift \citep[adapted from][]{Ivanov:2020ril}.}
\label{figede}
\end{figure}
\begin{figure}
\begin{centering}
\includegraphics[width=0.48\textwidth]{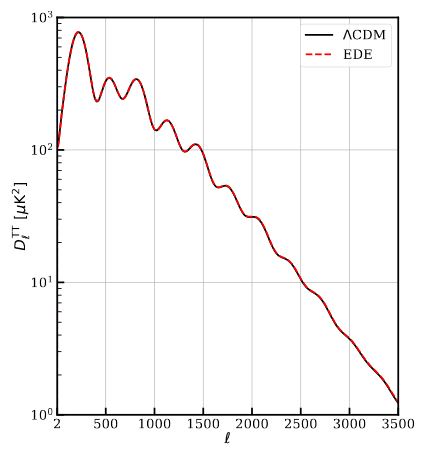}
\par\end{centering}
\caption{CMB TT  power spectrum. The black solid and the red dashed lines correspond to $\Lambda$CDM model with $H_0= 68.07$ $km$ $s^{-1}Mpc^{-1}$ and EDE model with $H_0= 71.15$ $km$ $s^{-1}Mpc^{-1}$ respectively  \citep[from][]{Ivanov:2020ril}.}
\label{figededl}
\end{figure}
\begin{figure*}
\begin{centering}
\includegraphics[width=0.96\textwidth]{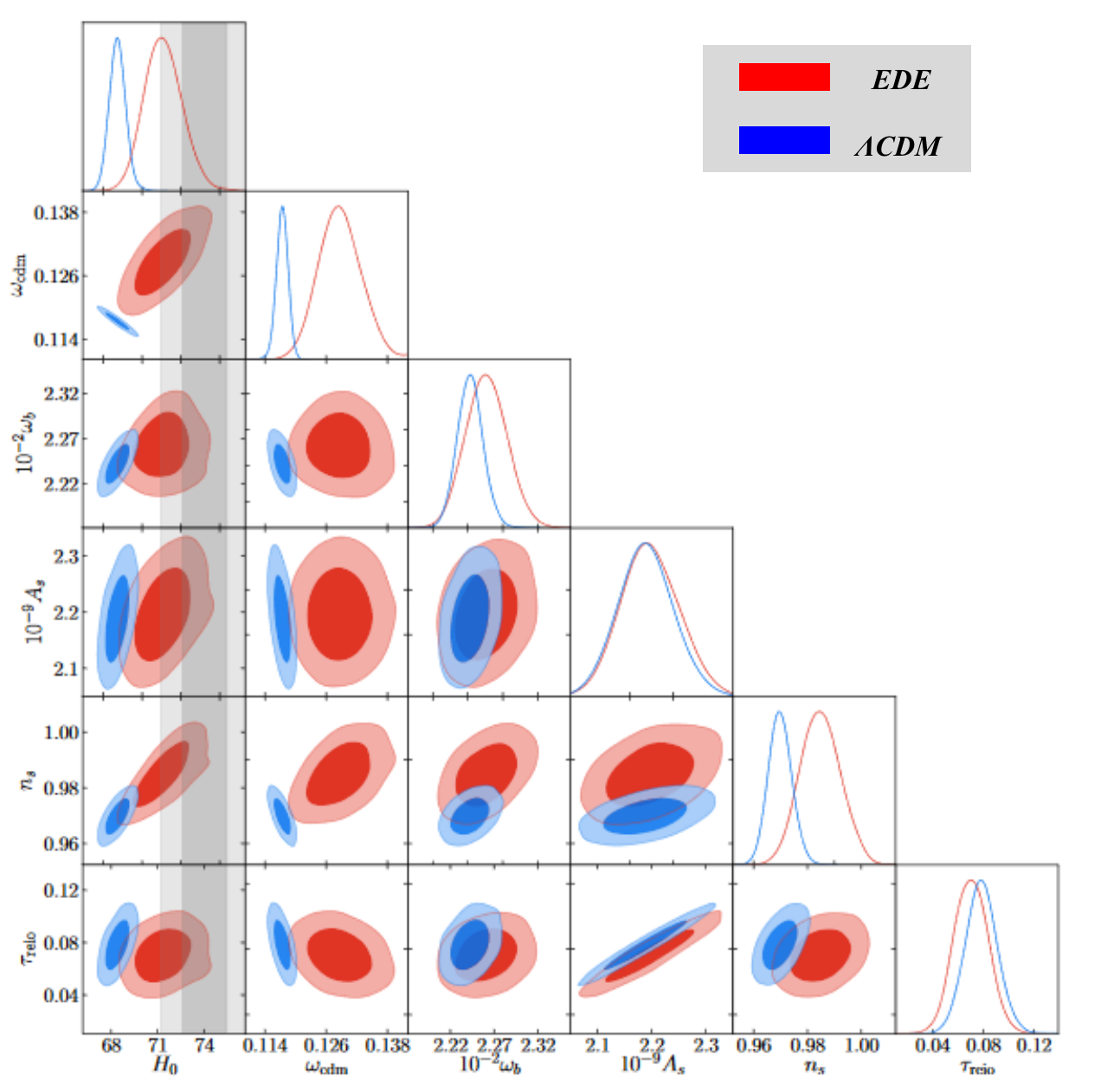}
\par\end{centering}
\caption{Posterior 1D and 2D distributions of the cosmological $\Lambda$CDM parameters reconstructed from a run to all data (including Planck high $l$ polarization) in EDE (red) and the  $\Lambda$CDM (blue) scenario. The gray bands correspond to the SH0ES determination of $H_0$  \citep[adapted from][]{Smith:2019ihp}.}
\label{figedecont}
\end{figure*}

\paragraph{Early dark energy: }
\label{early dark energy}
In the early dark energy (EDE) model \citep{Klypin:2020tud,Herold:2021ksg,LaPosta:2021pgm,Gomez-Valent:2021cbe,Fondi:2022tfp,Hill:2021yec,Poulin:2021bjr,Nojiri:2021dze,Moss:2021obd,Pettorino:2013ia,Poulin:2018cxd,Karwal:2016vyq,Sakstein:2019fmf,Agrawal:2019lmo,Lin:2019qug,Berghaus:2019cls,Smith:2019ihp,Lucca:2020fgp,Alexander:2019rsc,Mortsell:2018mfj,Kaloper:2019lpl,Murgia:2020ryi,Gogoi:2020qif,Haridasu:2020pms,Smith:2022hwi} an additional dynamical scalar field behaves like a cosmological constant at early times (near matter-radiation equality but before  recombination). This field decays rapidly after recombination thus leaving the rest of the expansion history practically unaffected up to a rescaling which modifies $H_0$. This rescaling allows for the resolution of the Hubble constant tension. Using the Eq. (\ref{rsdef}) in a EDE model the radius of sound horizon at last scattering can be calculated by
\begin{align}
 &r_s =\int_0^{t_d} \frac{c_s(a)}{a(t)}dt=\nonumber\\
 &=\int_{z_{d}}^\infty \frac{c_s(z)}{H(z;\rho_b,\rho_{\gamma},\rho_c,\rho_{DE})}dz= \nonumber\\
&=\int_0^{a_d}\frac{c_s(a)}{a^2H(a;\rho_b,\rho_{\gamma},\rho_c,\rho_{DE})}da 
\label{rsededef}
\end{align}
The baryon-to-photon ratio, and thus $c_s$ in Eq. (\ref{csdef}), is tightly constrained by CMB fluctuations and BBN \citep{Ade:2015rim}. As a consequence a EDE phase before and around the recombination epoch would increase $H(z)$ and thus decrease the scale of the sound horizon $r_s$ in Eq. (\ref{rsededef}). In the context of the degeneracy $H_0 r_s$  shown in Eq. (\ref{thetas}) this decrease of $r_s$ leads to an increased value of $H_0$ for a fixed measured value of $\theta_s$.

An EDE model can be implemented by several functional forms of scalar field which contribute to the cosmic energy shortly before matter-radiation equality. Possible functional forms of scalar field are the axion-like potential (higher-order periodic potential) inspired by string axiverse scenarios for dark energy \citep{Marsh:2011gr,Marsh:2015xka,Kamionkowski:2014zda,Karwal:2016vyq,Poulin:2018dzj}, the single axion-like particle potential consisting of two cosine functions which unifies the inflaton and DM while reheating the universe \citep{Daido:2017wwb,Daido:2017tbr}, the power-law potential \citep{Agrawal:2019lmo}, the acoustic dark energy \citep{Lin:2019qug,Lin:2020jcb,Yin:2020dwl}, the $\alpha$-attractor-like potential \citep{Braglia:2020bym} and others.

\citet{Poulin:2018cxd} consider two physical models. One that involves an oscillating scalar field and another with a slowly-rolling scalar field. 
In the case of the first model of the proposal of  \citet{Poulin:2018cxd}, the potential of the scalar field $\phi$ is a generalization of the axion potential of the form 
\be 
V(\phi)=m^2f^2\left(1-cos(\phi/f)\right)^n
\label{potede}
\ee
where $m$ is the field mass (for ultralight scalar field $m\sim 10^{-28}$ $eV$)  and $f$ is a decay constant.

Consider the time evolution of the EDE  scalar field which may be written as
\be
\ddot{\phi}+3H\dot{\phi}+V'(\phi)=0
\ee
where the dot and the prime denote the derivatives with respect to cosmic time $t$ and field $\phi$ respectively.

At early times, deep in the radiation era  the field $\phi$ is initially frozen in its potential due to Hubble friction ($H\gg m$) and acts as a cosmological constant with equation of state $w_{\phi}=-1$ (hence the name Early Dark Energy), but when the Hubble parameter drops below some value ($H\sim m$) at a critical redshift $z_c$ (for EDE this happens when $z_c\sim z_d$ for $m\sim 10^{-27}$ $eV$) the field becomes dynamical and begins to oscillate around its minimum which is locally $V\sim \phi^{2n}$ (Fig. \ref{figedepot}). It thus begins to behave like a fluid with an equation of state \citep{Turner:1983he}
\be 
w_{\phi}=\frac{n-1}{n+1}
\ee
The energy density of the field dilutes as $a^{-3(1+w_{\phi})}$ and thus when $n=1$, $n=2$ and $n\geq 3$ dilutes as cold dark matter ($a^{-3}$, $w_{\phi}=0$), as radiation ($a^{-4}$, $w_{\phi}=1/3$) and faster than radiation ($a^{-x}$ with $x>4$, $w_{\phi}>1/3$) respectively. Also when $n\rightarrow \infty$ the energy density dilutes as free scalar field (stiff matter \citep{Chavanis:2014lra}) ($a^{-6}$, $w_{\phi}= 1$) i.e. the scalar field is fully dominated by its kinetic energy. 

The EDE models are  parameterized  by the critical redshift $z_c$ , the dimensionless quantity $\theta_i=\phi_i/f$ (with $\phi_i$ the  initial  value  of  the  scalar  field and $0<\theta_i<\pi$) and the peak EDE energy density fraction of the Universe $f_{EDE}(z_c)$ which is given by
\be 
f_{EDE}(z_c)\equiv \frac{\Omega_{\phi}(z_c)}{\Omega_{tot}(z_c)}=\frac{\rho_{EDE}(z_c)}{3M_{pl}^2H(z_c)^2}
\ee
where $\Omega_{\phi}$ is the EDE energy density which evolves as \citep{Poulin:2018dzj,Poulin:2018cxd}
\be 
\Omega_{\phi}(z)=\frac{2\Omega_{\phi}(z_c)}{\left[(1+z_c)/(1+z)\right]^{3\left(w_n+1\right)}+1}
\ee
The fractional contribution of EDE to the cosmic energy budget as a function of redshift, i.e. $f_{EDE}(z)$, is shown in Fig. \ref{figede}  \citep[from the analysis by][]{Ivanov:2020ril}. Clearly, for $z\simeq z_c$  the EDE contributes the most to the  total  energy density ($\sim 10\%$), for  $z>z_c$ the EDE is not dynamically important while for  $z<z_c$ decays away  as radiation or faster than radiation leaving the later evolution of the Universe relatively unchanged. By construction, the EDE models can nicely match the CMB TT power spectrum of $\Lambda$CDM and therefore of Planck as illustrated in Fig. \ref{figededl}. The black solid and the red dashed lines (almost identical) correspond to $\Lambda$CDM model with $H_0= 68.07$ $km$ $s^{-1}Mpc^{-1}$ and EDE model with $H_0= 71.15$ $km$ $s^{-1}Mpc^{-1}$ respectively \citep{Ivanov:2020ril}.

EDE models face the fine-tuning issues \citep{Sakstein:2019fmf} and suffer from a coincidence problem \citep{Pettorino:2013ia}. \citet{Sakstein:2019fmf,CarrilloGonzalez:2020oac} proposed a natural explanation for this coincidence using the idea of neutrino-assisted early dark energy. 

EDE modifies growth and $H(z)$ at early times (around recombination) and higher matter density is required to compensate for this effect in the CMB. Higher matter density contradicts the required low value of matter density at late times from weak lensing and growth data as shown in Fig. \ref{figedecont}. In particular the analysis by \citet{Hill:2020osr,Clark:2021hlo} has shown that an EDE model can not practically resolve the Hubble tension because it results in higher value of the late-time density fluctuation amplitude $\sigma_8$ and thus the tension with LSS dynamical probes WL, RSD and CC data can get worse. In addition  \citet{Jedamzik:2020zmd} argued that any model which attempts to reconcile the CMB inferred value of $H_0$ by solely  reducing the sound horizon results into tension with either the BAO or the galaxy weak lensing data. Thus, a compelling and full resolution of the Hubble tension may require multiple modifications (more than just the size of the sound horizon) of the $\Lambda$CDM cosmology.

Recent studies by  \citet{Smith:2020rxx,Chudaykin:2020igl} reexamining the above issue and using combined data method show that the EDE scenario remains a potential candidate solution to the Hubble tension. Future observations will provide data with improved quality and thus will enable more detailed tests of the EDE model.

Many alternative models have been proposed to implement the basic EDE scenario such as Chain EDE \citep{Freese:2021rjq}, Axion EDE \citep{DAmico:2020ods,Jiang:2022uyg}, Anti-de Sitter EDE \citep{Wang:2022jpo,Ye:2020btb,Ye:2021iwa,Ye:2020oix,Jiang:2021bab,Jiang:2022uyg}, assisted quintessence EDE \citep{Sabla:2021nfy}, EDE with extra radiation \citep{Seto:2021xua}, EDE in the framework of the ultralight scalar decay to massless fields \citep{Gonzalez:2020fdy} and New EDE (NEDE) \citep{Niedermann:2021vgd,Niedermann:2021ijp,Niedermann:2019olb,Niedermann:2020dwg,Niedermann:2020qbw} which can potentially address the Hubble tension. In NEDE  a vacuum first-order phase transition of the NEDE scalar field is assumed to have taken place before recombination in the early Universe. The NEDE  sudden transition can be described by a scalar field whose potential at some critical point develops  two non-degenerate minima (true and false vacuum)\footnote{It has recently been pointed out \citep{Marra:2021fvf,Alestas:2020zol,Alestas:2022xxm,Perivolaropoulos:2022txg,Perivolaropoulos:2022vql,Perivolaropoulos:2021bds,Alestas:2021nmi} that a similar mechanism in the context of the ultra late transition taking place at a redshift $z\lesssim 0.01$ can lead to a resolution of the Hubble tension.}. \citet{Allali:2021azp} develop a phenomenological dark sector with decaying dark energy  and ultra-light axions which addresses the Hubble tension similarly to the EDE and NEDE scenarios and simultaneously can resolve the $S_8$ tension. \citet{Karwal:2021vpk,Sabla:2022xzj,McDonough:2021pdg} argue that a EDE model may require a more complicated dynamics in order to soften both the $H_0$ and $S_8$ tensions. In particular, \citet{McDonough:2021pdg} introduced the Early Dark Sector (EDS) model considering an EDE-dependence of the mass of dark matter. The considering form of the potential is given by Eq. (\ref{potede}) (with $n=3$) and the form of the field-dependent mass given by 
\be
m(\phi)=m_0 \exp(\frac{c \phi }{M_{pl}})
\ee
as motivated by the the Swampland Distance
Conjecture (SDC) \citep{Ooguri:2006in} and its extension to axions \citep{Klaewer:2016kiy,Scalisi:2018eaz,Blumenhagen:2017cxt,Baume:2016psm}.

\paragraph{Dark radiation:} 
\label{dark radiation}
Modifications in the light relic sector can relieve the tension by changing the early-time dynamics of the Universe \citep{Lancaster:2017ksf,Buen-Abad:2017gxg}.
The dark radiation model assumes an increased number of light relics \citep{Ghosh:2021axu,Wyman:2013lza,Archidiacono:2013fha,Ackerman:mha,Blennow:2012de,Conlon:2013isa,Vogel:2013raa,Dvorkin:2014lea,Leistedt:2014sia,Bernal:2016gxb,Aloni:2021eaq,Feng:2017nss,Vagnozzi:2019ezj,Seto:2021tad,Giare:2020vzo,Aboubrahim:2022gjb} which are weakly interacting components of radiation (i.e. relativistic species). For example the addition of hidden photons, sterile neutrinos \citep{Feng:2021ipq,Carneiro:2018xwq,Gelmini:2019deq,Gelmini:2020ekg}, Goldstone bosons, Majoron \citep{Fernandez-Martinez:2021ypo}, axions \citep{Gu:2021lni,DEramo:2018vss,Cuesta:2021kca} which are predicted in many extensions of the Standard Model (SM) increases the value in the effective number of relativistic particles $N_{eff}$ beyond its canonical expectation value   $N_{eff}^{SM}\simeq3.044$ \citep{Mangano:2005cc,deSalas:2016ztq,Akita:2020szl,EscuderoAbenza:2020cmq,Bennett:2020zkv,Froustey:2020mcq}. These extra particles modify the time of matter-radiation equality and would lead to a lower $r_s$ sound horizon. As a consequence a lower expansion rate of the Universe and a higher value of $H_0$ emerges from early-time physics \citep{Green:2019glg} (see Eq. (\ref{thetas})). 

Another interesting approach was presented by \citet{Lesgourgues:2015wza,Cyr-Racine:2013fsa,Schewtschenko:2015rno,Buen-Abad:2015ova,Chacko:2016kgg,Ko:2016uft,Ko:2016fcd,Tang:2016mot,Buen-Abad:2017gxg,Krall:2017xcw,Archidiacono:2017slj,Archidiacono:2019wdp,Ko:2017uyb,Becker:2020hzj,Choi:2020pyy}, in which dark matter (DM) interacts with a new form of dark radiation (DR) aimed at solving $H_0$ tension. Assuming  the  Effective  Theory of Structure Formation (ETHOS) paradigm \citep{Cyr-Racine:2015ihg,Vogelsberger:2015gpr} the interaction between the dark matter and dark radiation components is a 2-to-2 scattering
DM + DR $\leftrightarrow$ DM + DR.

\paragraph{Neutrino self-interactions: } 
\label{neutrino self-interactions}
The strong (massive) neutrino self-interactions cosmological model can provide a larger value of $H_0$ and smaller $\sigma_8$, hence can resolve the tensions between cosmological datasets \citep{Kreisch:2019yzn}.
The strong neutrino self-interactions were proposed in \citet{Cyr-Racine:2013jua} and further studied in  \citet{Lancaster:2017ksf,Oldengott:2017fhy}. The introduction of strong self-interacting neutrinos increases the value in the effective number of relativistic particles $N_{eff}= 4.02\pm0.29$ without extra neutrino species. This model modifies the standard neutrino free-streaming in the early Universe. The onset of neutrino free-streaming is delayed until close to the  matter radiation equality epoch. This late-decoupling of the neutrinos shifts the CMB power spectra peaks towards smaller scales as compared to $\Lambda$CDM  model. This shift modifies the scale of sound horizon $r_s$ that can resolve the Hubble constant $H_0$ tension \citep{Kreisch:2019yzn}.

Furthermore self-interactions between the neutrinos or between other additional light relics was studied by  \citet{Das:2020xke,Choudhury:2020tka,Mazumdar:2020ibx,Forastieri:2019cuf,He:2020zns,Berbig:2020wve,Lyu:2020lps,Hannestad:2013ana,Archidiacono:2014nda,Chu:2015ipa,Blinov:2019gcj,Ghosh:2019tab,Brinckmann:2020bcn,Archidiacono:2020yey,Corona:2021qxl,Archidiacono:2015oma,Archidiacono:2016kkh}. The strong neutrino self-interactions models are basically excluded by various existing data or experimental tests \citep{Blinov:2019gcj,Lyu:2020lps,Brune:2018sab,Deppisch:2020sqh,Brdar:2020nbj}. The analysis by \citet{Brinckmann:2020bcn} leads to conclusion that these models can not ease the Hubble tension more effectively than the $\Lambda$CDM$+N_{eff}$ approach alone.

Models with nonstandard neutrinos - dark matter interactions were studied by \citet{Huang:2021dba,Escudero:2021rfi,Boyarsky:2021yoh,Mangano:2006mp,Wilkinson:2014ksa,Mosbech:2020ahp,Stadler:2019dii,Choi:2019ixb,DiValentino:2017oaw,Arias-Aragon:2020qip,Escudero:2019gvw}. These models increase the value in the effective number of relativistic particle $N_{eff}$ and thus can provide a solution to the Hubble problem. However, in this class of models it is not possible to solve simultaneously the Hubble and growth tensions \citep{DiValentino:2017oaw}.

\paragraph{Large primordial non-Gaussianities:} 
\label{large primordial non-Gaussianities}
The presence of large primordial non-Gaussianity in the CMB can affect the higher-order $n$-point correlation functions statistics. A non-vanishing primordial trispectrum ($n=4$) which is the Fourier transform of the connected four-point correlation function leads to the non-Gaussian covariance of the angular power spectrum  estimators \citep{Hu:2001fa,Smith:2015uia,Adhikari:2018osh}. The trispectrum is nonzero when there is a strong coupling between long-wavelength (super-CMB) modes and short-wavelength modes. The non-Gaussian covariance scenario (Super-$\Lambda$CDM model) has two extra free parameters relative to those in $\Lambda$CDM and provides a larger value of $H_0$ reducing tension  with late Universe measurements of the Hubble constant \citep{Adhikari:2019fvb}.

\paragraph{\bf Heisenberg's uncertainty principle:}
\label{Heisenberg's uncertainty principle}
The Heisenberg's uncertainty principle \citep{aHeisenberg:1927zz,Robertson:1929zz} and the generalized uncertainty Principle \citep{Mead:1964zz,Maggiore:1993kv,Maggiore:1993rv,Maggiore:1993zu,Kempf:1994su,Hinrichsen:1995mf,Kempf:1996ss,Kempf:1996nm,Snyder:1946qz,Yang:1947ud,Karolyhazy:1966zz,Adler:2001vs,Ashoorioon:2004vm,Ashoorioon:2004rs,Ashoorioon:2004wd,Ashoorioon:2005ep,Nozari:2011gj,Faizal:2014mba,Ali:2015ola,Mohammadi:2015upa,Faizal:2016zlo,Lake:2017uzd,Zhao:2017xjj,Alasfar:2017loh,Lake:2020rwc}  \citep[see][for a review]{Tawfik:2015rva} can  provide constraints to the values for certain pairs of physical quantities of a particle and raise the possibility of the existence of observational signatures in cosmological data \citep[e.g.][]{Perivolaropoulos:2017rgq,Skara:2019uzz}. \citet{Capozziello:2020nyq} have argued that the Heisenberg's uncertainty principle can provide an explanation for the Hubble constant $H_0$ tension. In particular the authors equate the luminosity distance (expanded for low $z$ as in Eqs. (\ref{Dlz}) and (\ref{dlexp})) with the photon (assumed massive) Compton wavelength 
\be
\lambdabar_C=\frac{\hbar}{m\;c}
\ee
and express the corresponding effective “rest mass” of the photon as a function of the cosmological redshift
\be
m=\frac{\hbar H_0}{zc^2\left[ 1+\frac{z}{2}(1-q_0)\right]}
\label{mhup}
\ee
Thus, choosing $z=1$, fixing $q_0=-1/2$ and setting  $H_0=74$ $km$  $ s^{-1} Mpc^{-1}$ and $H_0=67$ $km$  $ s^{-1} Mpc^{-1}$ in Eq. (\ref{mhup}) find $m=1.61\times 10^{-69}kg$ and $m=1.46\times 10^{-69}kg$ respectively\footnote{The current upper limit on the photon mass is $m= 10^{-54}kg$ \citep{Tanabashi:2018oca}.}.  Thus using these results infer that the tension on the $H_0$ measurements can be the effect of the uncertainty on the photon mass i.e.
\be
\frac{\Delta m}{m}=\frac{\Delta H_0}{H_0}\simeq 0.1
\label{dmhup}
\ee
Note that the non-zero photon mass could emerge through the Heisenberg's uncertainty principle and through the recent analysis of the Standard-Model Extension\footnote{For studies of the massive photons in the Standard-Model Extension, see  \citet{Helayel-Neto:2019har,Spallicci:2020diu}.}  \citep{Bonetti:2016vrq,Bonetti:2017toa}.

\paragraph{Early modified gravity:}
\label{early modified gravity}
A scalar tensor modified  gravity model can be described by the following action
\begin{align}
S=\int& d^4 x\sqrt{-g}\left[\frac{F(\sigma)}{2}R - \frac{g^{\mu \nu}}{2}\partial_{\mu}\sigma\partial_{\nu}\sigma-\Lambda
-V(\sigma)\right]\nonumber\\
&+S_m    
\end{align}
where, $R$ is the Ricci scalar, $\Lambda$ is the cosmological  constant, $S_m$ is the action for matter fields, $\sigma$ is a  scalar field non-minimally coupled to the Ricci scalar, $F(\sigma)$ is the coupling to the Ricci scalar and $V(\sigma)$ is the potential for the scalar field. A variety of possible types of the non-minimal coupling of the scalar field to the Ricci $F(\sigma)$ and of the potential for the scalar field which can  alleviate the $H_0$ tension by reducing the sound horizon scale through modified early cosmic expansion, has been considered in the literature   \citep{Ballardini:2020iws,Braglia:2020iik,Rossi:2019lgt,Ballesteros:2020sik,Abadi:2020hbr}.

In particular \citet{Braglia:2020auw} introduce a model of early modified gravity\footnote{It should not be confused with the  previously introduced differed model with the same name 'Early Modified Gravity' \citep{Brax:2013fda,Pettorino:2014bka,Lima:2016npg}. In this model gravity is allowed to be modified after BBN,  before and during recombination.}. This model has a non-minimal coupling of the form \citep{Braglia:2020auw}
\be 
F(\sigma)=M_{pl}^2+\xi\sigma^2
\ee
and a  quartic potential
\be
V(\sigma)=\frac{\lambda \sigma^4}{4}
\ee
where $\lambda$ and $\xi$ are dimensionless parameters. For $\xi=0$ this model reduces to the EDE model of \citet{Agrawal:2019lmo}.
In the early modified gravity model, gravity changes with redshift in such a way that the $H_0$ estimate from CMB can have larger values.  \citet{Braglia:2020auw} have shown that this  model can resolve the Hubble tension and at the same time, in contrast to an EDE model, results in lower value of the late-time density fluctuation amplitude $\sigma_8$ and thus the tension with LSS dynamical probes WL, RSD and CC data can be at least partially resolved. In general early modified gravity model compared to the EDE can provide a better fit to LSS data and can imply better predictions on LSS  observables.

\section{Other Tensions - Curiosities}  
\label{sec:Other Tensions - Curiosities} 
In this section we provide a list of the non-standard signals in cosmological data and the curiosities of the $\Lambda$CDM cosmology beyond the Hubble tension which is currently the  most widely studied and among the most statistically significant tensions. In many cases the signals are controversial and there is currently debate in the literature on the possible physical or systematic origin of these signals. For completeness we refer to all signals we could identify in the literature referring also to references that dispute the physical origin of these signals.

\subsection{Growth tension}  
\label{Growth Tension}  
The \plcdm parameter values in the context of GR indicate stronger growth of cosmological perturbations than the one implied by observational data of dynamical probes. In this section we review the observational evidence for this tension also known as the $\Omega_{0m}-\sigma_8$ tension or simply 'growth tension'.

\subsubsection{Methods and data}
\label{Growth Tension Methods and data}

\begin{figure*}
\begin{centering}
\includegraphics[width=0.9\textwidth]{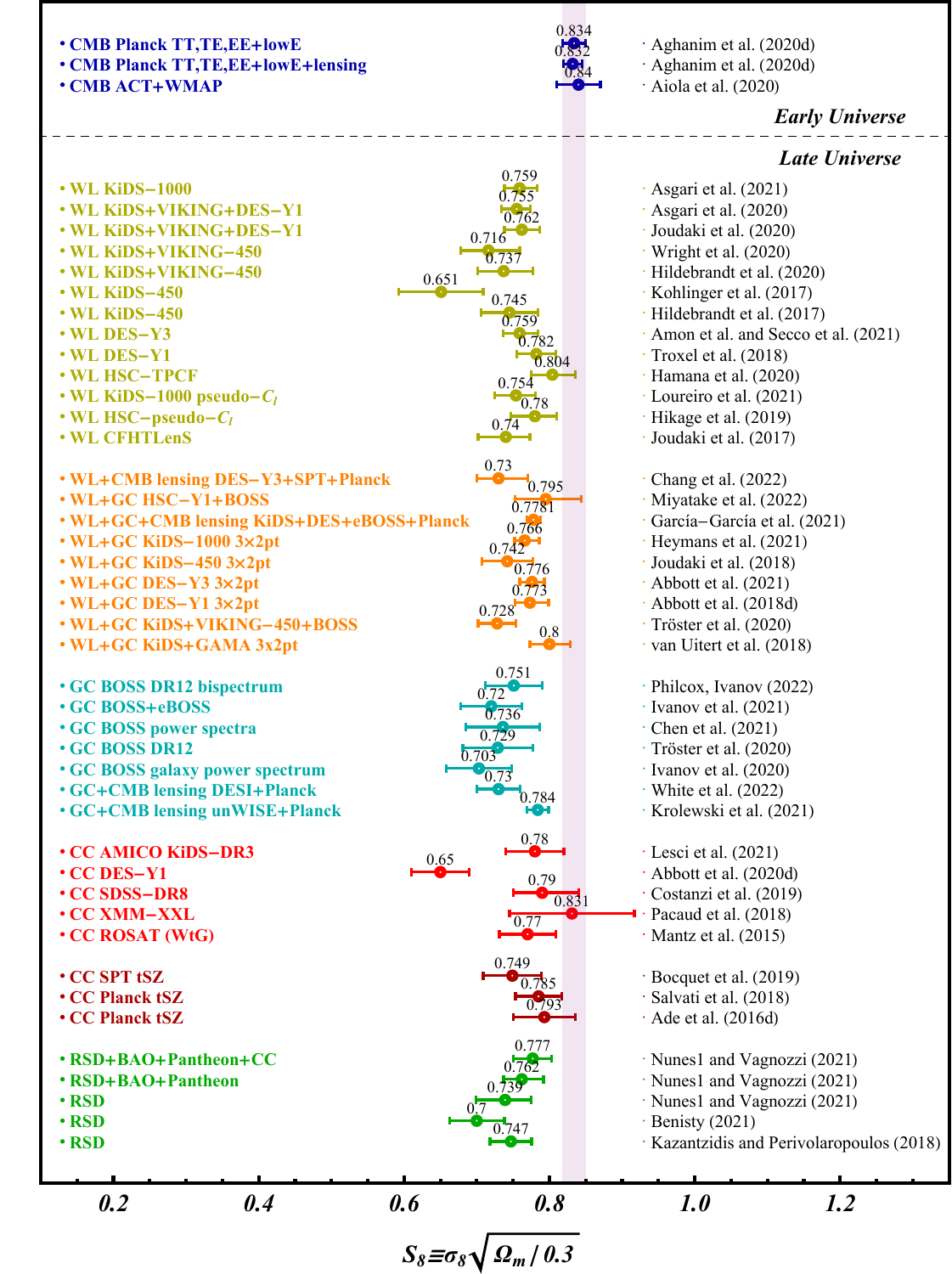}
\par\end{centering}
\caption{The  value of $S_8$ with the $68\%$  CL  constraints derived by recent measurements.}
\label{figs8}
\end{figure*}

The value of the growth parameter combination   $S_8\equiv\sigma_8(\Omega_{0m}/0.3)^{0.5}$ (where $\sigma_8$ is discussed in more detail in what follows) is found by weak lensing (WL) \citep{Hall:2021qjk,Asgari:2019fkq,Kohlinger:2017sxk,Joudaki:2017zdt,Hildebrandt:2016iqg,Abbott:2017wau,Abbott:2020knk}, cluster counts (CC)  \citep{Rapetti:2008rm,Rozo:2009jj,Mantz:2009fw,Ade:2015fva,Costanzi:2018xql,Costanzi:2020dgw} and redshift space distortion (RSD) data \citep{Macaulay:2013swa,Basilakos:2017rgc,Nesseris:2017vor,Sagredo:2018ahx,Kazantzidis:2018rnb,Kazantzidis:2019nuh,Perivolaropoulos:2019vkb,Skara:2019usd,Arjona:2020yum} to be lower compared to the  Planck CMB (TT,TE,EE+lowE) value $S_8=0.834\pm 0.016$ \citep{Planck:2018vyg} at a level of about $2-3\sigma$ as shown\footnote{The definition $S_8= \sigma_8 (\Omega_{0m}/0.3)^\alpha$ with $\alpha=1/2$ has been uniformly used for all points.  In those cases where $\alpha \neq 1/2$ has been used in some references, the value of $S_8$ with $\alpha =1/2$ was recalculated (along with the uncertainties) using the constraints on $\sigma_8$ and $\Omega_{0m}$ shown in those references, assuming their errors $\sigma_{\sigma_8}$ and $\sigma_{\Omega_{0m}}$ are Gaussian. The errors of the $S_8$ constraints are propagated according to $\sigma_{S_8}^2=(\Omega_{0m}/0.3)^{2\alpha}\sigma_{\sigma_8}^2+\sigma_8^2\alpha^2(\Omega_{0m}/0.3)^{2\alpha-2}\sigma_{\Omega_{0m}}^2$, with $\alpha=1/2$.} in Table \ref{tab:growthpar}  and in Fig. \ref{figs8}  \citep[see][for a  recent review of this tension]{DiValentino:2020vvd,Abdalla:2022yfr}. The tension is also confirmed by the latest ACT+WMAP  CMB analysis \citep{ACT:2020gnv} which finds  $S_8=0.840\pm0.030$.

This is also expressed by the fact that dynamical cosmological probes (WL, RSD, CC) favor lower value of the matter density parameter $\Omega_{0m}\approx 0.26\pm 0.04$ \citep{Alam:2020sor} than geometric probes (CMB, BAO, SnIa). This could be a signal of weaker gravity than the predictions of General Relativity in the context of a $\Lambda$CDM background \citep{Macaulay:2013swa,Tsujikawa:2015mga,Nesseris:2017vor,Kazantzidis:2018rnb,Skara:2019usd}  \citep[for a recent study on a weak gravity in the context of a $\Lambda$CDM background, see][]{Gannouji:2020ylf}.
 
\begin{table*}

\begin{center} 
\caption{The value of the structure growth parameter combination   $S_8\equiv\sigma_8(\Omega_{0m}/0.3)^{0.5}$, the matter density parameter $\Omega_{0m}$ and the the power spectrum amplitude $\sigma_8$ at  $68\%$  CL through  direct  and  indirect  measurements by different methods. }
\label{tab:growthpar} 
\vspace{2mm}

\begin{tabular}{c ccc  ccc c} 
\hhline{========}
   & \\
Dataset &&$S_8$&&$\Omega_{0m}$&&$\sigma_8$& Refs.\\
     & \\
   \hhline{========}
     & \\
CMB Planck  TT,TE,EE+lowE  &&$0.834\pm 0.016  $&& $0.3166\pm 0.0084$ &&$0.812\pm0.007$& \citep{Planck:2018vyg}   \\
CMB Planck  TT,TE,EE+lowE+lens. &&$0.832\pm0.013 $&& $0.3153\pm 0.0073$ &&$0.811\pm 0.006$& \citep{Planck:2018vyg}   \\
CMB ACT+WMAP &&$0.832\pm0.013 $&& $0.3153\pm 0.0073$ &&$0.840\pm0.030$& \citep{ACT:2020gnv}   \\
   & \\
\hline
   & \\
WL KiDS-1000&&$ 0.759_{-0.021}^{+0.024}  $&& - && - & \citep{Asgari:2020wuj}  \\  
WL KiDS + VIKING + DES-Y1&&$ 0.755_{-0.021}^{+0.019} $&& -  &&-&\citep{Asgari:2019fkq}   \\ 
WL KiDS + VIKING + DES-Y1&&$0.762_{-0.024}^{+0.025}   $&& - &&-&\citep{Joudaki:2019pmv}   \\
WL KiDS+VIKING-450 &&$0.716_{-0.038}^{+0.043} $&&-&&-& \citep{Wright:2020ppw}   \\
WL KiDS+VIKING-450 &&$0.737_{-0.036}^{+0.040}   $&& -&&-& \citep{Hildebrandt:2018yau}   \\
WL KiDS-450 &&$0.651\pm 0.058$&& - &&-& \citep{Kohlinger:2017sxk}   \\  
WL KiDS-450 &&$0.745\pm 0.039  $&& -&&-& \citep{Hildebrandt:2016iqg} \\
WL DES-Y3 &&$0.759_{-0.023}^{+0.025}$&& $0.290_{-0.063}^{+0.039} $ &&$0.783_{-0.092}^{+0.073} $& \citep{DES:2021bvc,DES:2021vln}   \\ 
WL DES-Y1&&$0.782_{-0.027}^{+0.027}   $&& - &&-& \citep{Troxel:2017xyo}   \\ 
WL HSC-TPCF &&$0.804_{-0.029}^{+0.032}  $&& $0.346_{-0.100}^{+0.052} $ &&$0.766_{-0.098}^{+0.110} $& \citep{Hamana:2019etx}  \\
WL KiDS-1000 pseudo-$C_l$ &&$0.754_{-0.029}^{+0.027} $&& - &&-& \citep{KiDS:2021opn} \\ 
WL HSC-pseudo-$C_l$ &&$0.780_{-0.033}^{+0.030} $&& - &&-& \citep{Hikage:2018qbn} \\ 
WL CFHTLenS &&$0.740_{-0.038}^{+0.033} $&& - &&-& \citep{Joudaki:2016mvz} \\ 
& \\
WL+CMB lensing DES-Y3+SPT+Planck &&$0.73_{-0.03}^{+0.04}$&& $0.25_{-0.04}^{+0.03}$ &&$0.82_{-0.07}^{+0.08}$& \citep{DES:2022ign} \\ 
WL+GC\footnote{ HSC-Y1+SDSSS-III/BOSS DR11} &&$0.795^{+0.049}_{-0.042} $&&$0.383^{+0.028}_{-0.053} $ &&$0.718^{+0.044}_{-0.031} $& \citep{Miyatake:2021sdd} \\
WL+GC+CMB lensing\footnote{KiDS+DES+eBOSS+Planck} &&$0.7781\pm 0.0094$&&$0.305^{+0.021}_{-0.025}$ &&$0.774\pm 0.033  $& \citep{Garcia-Garcia:2021unp}\\
WL+GC KiDS-1000  $3\times2$pt&&$0.766_{-0.014}^{+0.020}$&& $0.305_{-0.015}^{+0.010} $ &&$0.76_{-0.020}^{+0.025} $& \citep{Heymans:2020gsg}   \\ 
WL+GC KiDS-450  $3\times2$pt&&$0.742\pm 0.035$&& $0.243_{-0.045}^{+0.026} $ &&$0.832_{-0.079}^{+0.080} $& \citep{Joudaki:2017zdt}   \\ 
WL+GC KiDS+GAMA  $3\times2$pt&&$0.800_{-0.027}^{+0.029}$&& $0.33_{-0.06}^{+0.05} $ &&$0.78_{-0.08}^{+0.06} $& \citep{vanUitert:2017ieu}   \\ 
WL+GC DES-Y3  $3\times2$pt&&$0.776_{-0.017}^{+0.017}$&& $0.339_{-0.031}^{+0.032} $ &&$0.733_{-0.049}^{+0.039} $& \citep{DES:2021wwk}   \\ 
WL+GC DES-Y1  $3\times2$pt&&$0.773_{-0.020}^{+0.026}$&& $0.267_{-0.017}^{+0.030} $ &&$0.817_{-0.056}^{+0.045} $& \citep{Abbott:2017wau}   \\ 
WL+GC KiDS+VIKING-450+BOSS &&$0.728\pm0.026$&& $0.323_{-0.017}^{+0.014}$ &&$0.702\pm 0.029$&\citep{Troster:2019ean}\\
& \\
GC BOSS DR12 bispectrum&&$0.751\pm 0.039$&& $0.32_{-0.01}^{+0.01}$ &&$0.722_{-0.036}^{+0.032}$&\citep{Philcox:2021kcw}\\
GC  BOSS+eBOSS &&$0.72\pm 0.042$&& - &&-&\citep{Ivanov:2021zmi}\\
GC  BOSS galaxy power spectrum &&$0.703\pm 0.045$&& $0.293\pm 0.012$ &&$0.713\pm 0.045$&\citep{Ivanov:2019pdj}\\
GC  BOSS power spectra &&$0.736\pm 0.051$&& $0.303\pm0.0082$ &&$0.733\pm 0.047$&\citep{Chen:2021wdi}\\
GC  BOSS DR12 &&$0.729\pm0.048$&& $0.317_{-0.019}^{+0.015}$ &&$0.710\pm 0.049$&\citep{Troster:2019ean}\\
GC+CMB lensing  DESI+Plank &&$0.73 \pm 0.03$&& - &&-&\citep{White:2021yvw}\\
GC+CMB lensing  unWISE+Plank &&$0.784 \pm 0.015$&& $0.307 \pm 0.018$ &&$0.775 \pm 0.029$&\citep{Krolewski:2021yqy}\\
&\\
CC  AMICO KiDS-DR3 &&$0.78\pm 0.04$&& $0.24_{-0.04}^{+0.03}$ &&$0.86\pm 0.07$&\citep{Lesci:2020qpk}    \\   
CC  SDSS-DR8 &&$0.79_{-0.04}^{+0.05}$&& $0.22_{-0.04}^{+0.05}$ &&$0.91_{-0.10}^{+0.11}$&\citep{Costanzi:2018xql}    \\
CC  ROSAT (WtG)&&$0.77\pm 0.05  $&& $0.26\pm 0.03  $ &&$0.83\pm 0.04 $& \citep{Mantz:2014paa}   \\
CC  DES-Y1    &&$0.65_{-0.04}^{+0.04}  $&& $0.179_{-0.038}^{+0.031}$ &&$0.85_{-0.06}^{+0.04} $&\citep{Abbott:2020knk}  \\
CC  XMM-XXL   &&$0.83\pm 0.11  $&& $0.40\pm 0.09$ &&$0.72\pm 0.07    $& \citep{Pacaud:2018zsh}   \\
   & \\
CC SPT-tSZ   &&$0.749\pm0.055  $&& $0.276\pm 0.047   $ &&$0.781\pm 0.037   $& \citep{Bocquet:2018ukq}   \\ 
CC Planck tSZ &&$0.785\pm0.038  $&& $0.32\pm0.02  $ &&$0.76\pm0.03 $& \citep{Salvati:2017rsn}    \\
CC Planck tSZ  &&$0.792\pm0.056  $&& $0.31\pm0.04 $ &&$0.78\pm0.04  $& \citep{Ade:2015fva}    \\
    & \\ 
RSD+BAO+Pantheon+CC &&$0.777_{-0.027}^{+0.026} $&& $0.288\pm 0.008 $ &&$0.793_{-0.020}^{+0.018}$&\citep{Nunes:2021ipq}    \\    
RSD+BAO+Pantheon &&$0.762_{-0.025}^{+0.030} $&& $0.286\pm 0.008 $ &&$0.7808_{-0.019}^{+0.021}$&\citep{Nunes:2021ipq}    \\
RSD &&$0.739_{-0.040}^{+0.036} $&& $0.254_{-0.058}^{+0.038}$ &&$0.804_{-0.071}^{+0.048}$&\citep{Nunes:2021ipq}    \\  
RSD &&$0.700_{-0.037}^{+0.038} $&& $0.201_{-0.033}^{+0.036} $ &&$0.857_{-0.042}^{+0.044}$&\citep{Benisty:2020kdt}    \\
 RSD     &&$0.747\pm 0.029 $&& $0.279\pm 0.028$ &&$0.775\pm 0.018$& \citep{Kazantzidis:2018rnb}   \\
    &&&& && &   \\
 \hhline{========}   
\end{tabular} 
\end{center} 
\end{table*}

The observational evidence for weaker growth indicated by the dynamical probes of the cosmic expansion and the gravitational law on cosmological scales may be reviewed as follows:

\paragraph{Weak lensing:} 

The weak gravitational lensing from matter fluctuations along the line of sight slightly distorts the shapes (shear) and size (magnification) of distant galaxies  \citep[see][for a review]{Bartelmann:1999yn,Kilbinger:2014cea,Mandelbaum:2017jpr}. This distortion is a powerful and principal  cosmological probe of the mass distribution which can be predicted theoretically \citep{Bacon:2000sy,Kaiser:2000if,Wittman:2000tc,vanWaerbeke:2000rm}. Using various statistical methods shape distortions can be measured by analyzing the angular shear correlation function, or its Fourier transform, the shear power spectrum \citep{Hikage:2018qbn,Asgari:2020wuj}. A special type of WL is the galaxy-galaxy lensing (GGL) \citep{Brainerd:1995da,Hudson:1997bj} which is the slight distortion of shapes of source galaxies in the background of a lens galaxy arising  from  the  gravitational deflection of light due to the gravitational potential of the lens galaxy along the line of sight.

The WL surveys, the Kilo Degree Survey (KiDS) \citep{deJong:2012zb,deJong:2015wca,deJong:2017bkf,Kuijken:2015vca}, the Subaru Hyper Suprime-Cam lensing survey (HSC) \citep{Miyazaki:2015haa,Aihara:2017tri} and the Dark Energy Survey (DES) \citep{Abbott:2005bi,Jarvis:2015hvz} provide data useful for cosmic shear studies. In particular WL measurements of $S_8$ obtained from the shear catalogues by the lensing analysis of the Canada-France-Hawaii Telescope Lensing (CFHTLenS) \citep{Heymans:2012gg,Hildebrandt:2011hb,Erben:2012zw,Heymans:2013fya,Miller:2012am,Joudaki:2016mvz} and the KiDS \citep{Hildebrandt:2016iqg,Kohlinger:2017sxk} appear to be lower compared to the Planck value at a level of about $3\sigma$. The analysis by \citet{Hildebrandt:2016iqg} adopting a spatially flat $\Lambda$CDM model and using the KiDS-$450$ data reports $S_8=0.745\pm 0.039$ which results in $2.3\sigma$ tension with the value estimated by Planck$15$. This KiDS-Planck discordance has also been investigated in  \citet{Kohlinger:2017sxk} where applying the  quadratic estimator to KiDS-450 shear data reports $S_8=0.651\pm 0.058$ which is in tension with the Planck20$15$ results at the $3.2\sigma$ level. 

Using a combination of the measurements of KiDS-450 and VISTA Kilo-Degree infrared Galaxy Survey (VIKING) \citep{2007Msngr.127...28A},  \citet{Hildebrandt:2018yau} find $S_8=0.737_{-0.036}^{+0.040}$ which is discrepant with measurements from the Planck analysis at the  $2.3\sigma$ level. For the KiDS+VIKING-450 (or KV450) \citet{Wright:2020ppw} report an updated constraint of  $S_8= 0.716_{-0.038}^{+0.043}$. Meanwhile, using the DES first year (DES-Y1) data assuming a $\Lambda$CDM model \citet{Troxel:2017xyo} report $S_8=0.782_{-0.027}^{+0.027}$ which is in $\sim2.3\sigma$ tension\footnote{This tension was calculated by  \citet{Lemos:2020jry}. The authors have explored a number of different methods to quantify the tension relative to the best-fit Planck2018 cosmology.} with the Planck18 result. The constraint on $S_8$ from the combined tomographic weak lensing analysis of KiDS + VIKING + DES-Y1 adopting a flat $\Lambda$CDM model by \citet{Joudaki:2019pmv} is $S_8=  0.762_{-0.024}^{+0.025}$ which is in $2.5\sigma$ tension with Planck18 result and by  \citet{Asgari:2019fkq} is $S_8= 0.755_{-0.021}^{+0.019}$ which is in $3.2\sigma$ tension with Planck18 result. Analysing the most recent KiDS cosmic shear data release (KiDS-1000 \citealt{Kuijken:2019gsa}) alone and assuming a spatially flat $\Lambda$CDM model the value $S_8=0.759_{-0.021}^{+0.024}$ was estimated by \citet{Asgari:2020wuj}. Analysing the first-year data of HSC in the context of the flat $\Lambda$CDM model and  using the pseudo-spectrum (pseudo-$C_l$) method\footnote{For a realistic experiment the pseudo-$C_l$ statistics from cut-sky maps which provide incomplete data are applied in order to obtain unbiased estimates of the angular power and cross-power spectra by correcting for the convolution with the survey window \citep[see][for details of this method]{Hivon:2001jp,Brown:2004jn,Alonso:2018jzx,Wandelt:2000av}.},   \citet{Hikage:2018qbn} find $S_8=0.780_{-0.033}^{+0.030}$ and adopting the standard two-point correlation functions (TPCF) estimators, $\xi_{\pm}$,  \citet{Hamana:2019etx} find $S_8=0.804_{-0.029}^{+0.032}$. Recently, a analysis of the KiDS-1000 data using pseudo-$C_l$ method by \citet{KiDS:2021opn} has lead to $S_8=0.754_{-0.029}^{+0.027}$. The latest cosmic shear analysis of the DES third Year (DES-Y3) \citep{DES:2021bvc,DES:2021vln} in the context of the $\Lambda$CDM model constrains the clustering amplitude as $S_8=0.759_{-0.023}^{+0.025}$. Also, recently \citet{DES:2022ign} found $S_8 = 0.73_ {-0.03}^{+0.04}$ using the cross-correlations of galaxy positions and shears from DES-Y3 with CMB lensing maps from SPT and Planck.

The analysis of galaxy clustering and weak gravitational lensing of the DES-Y1 data combining three two-point functions (the so-called $3\times2$pt analysis) of gravitational lensing and galaxy positions (the cosmic shear correlation function, the galaxy clustering angular autocorrelation function, the galaxy-galaxy lensing cross-correlation function)  by  \citet{Abbott:2017wau} gives $S_8=0.773_{-0.020}^{+0.026}$  and $\Omega_{0m}= 0.267_{-0.017}^{+0.030}$ in flat $\Lambda$CDM model. This value is in $\sim2.3\sigma$ tension with Planck18 result. In the  latest analysis by \citet{DES:2021wwk} the constraints $S_8=0.776_{-0.017}^{+0.017}$ and $\Omega_{0m}= 0.339_{-0.031}^{+0.032}$ in flat $\Lambda$CDM model are obtained using an improvement in signal-to-noise of the DES-Y3 $3\times2$pt data relative to DES-Y1 by a factor of $2.1$. Also, \citet{Heymans:2020gsg} using $3\times2$pt analysis of KiDS-1000+BOSS+ 2-degree Field Lensing Survey\footnote{\url{https://2dflens.swin.edu.au}} ($2$dFLenS) \citep{Blake:2016fcm} data finds $S_8 = 0.766^{+0.020}_{-0.014}$. While previous analyses $3\times2$pt of KiDS+GAMA data and  KiDS-450+BOSS+$2$dFLenS data by \citet{vanUitert:2017ieu} and \citet{Joudaki:2017zdt} obtained $S_8 = 0.800^{+0.029}_{-0.027}$ and $S_8 = 0.742\pm 0.035$ respectively. A combined analysis  of KiDS+VIKING-450+BOSS data by \citet{Troster:2019ean} resulted in $S_8 = 0.728\pm0.026$. Performing a Joint analysis of galaxy-galaxy weak lensing and galaxy clustering from first-year data of HSC and SDSSS-III/BOSS DR11 \citet{Miyatake:2021sdd} found $S_8 = 0.795^{+0.049}_{-0.042}$. Also, from a combined analysis of KiDS-1000 and DES-Y1 cosmic shear and galaxy clustering, eBOSS quasars, DESI, Planck CMB lensing data \citet{Garcia-Garcia:2021unp} obtains a constraint $S_8 = 0.7781\pm 0.0094$.

Clearly, the tension between WL and CMB measurements is a level more than $2\sigma$ as seen in Table \ref{tab:growthpar} and in Fig. \ref{figs8}. In addition, the tension with more recent measurements persists at the level of $\sim 2\sigma$. Finally, combined analyses of WL with galaxy clustering does not change the tension level.

\paragraph{Cluster counts:}

Galaxy clusters which are related to peaks in the matter density field on large scales constitute a  probe of the growth history of structures \citep{1989ApJ...341L..71E,1989ApJ...347..563P}  \citep[see][for a review]{Allen:2011zs,Kravtsov:2012zs}. Current analyses from the number counts of galaxy clusters use catalogs from surveys at different wavelengths of the electromagnetic spectrum. Such surveys include Planck\footnote{\url{https://www.cosmos.esa.int}}, South Pole Telescope (SPT) and Atacama  Cosmology  Telescope (ACT) in the microwave (millimeter) via  the thermal Sunyaev-Zel’dovich (tSZ) effect\footnote{The inverse Compton scattering between CMB photons and hot electrons in the intracluster medium (ICM) \citep[see][for a review]{Birkinshaw:1998qp,Carlstrom:2002na,Mroczkowski:2018nrv}} \citep{Sunyaev:1970er,Sunyaev:1972eq,Sunyaev:1980vz}, extended Roentgen survey with an  imaging telescope array (eROSITA\footnote{\url{https://www.mpe.mpg.de/eROSITA}})  \citep{Predehl:2010vx,Merloni:2012uf,Pillepich:2011zz,Hofmann:2017luz} in the X-ray that finds extended sources and measures the X-ray luminosity and temperature, Sloan Digital Sky Survey\footnote{\url{https://www.sdss.org/}} (SDSS)  and Dark  Energy  Survey\footnote{\url{https://www.darkenergysurvey.org}} (DES) in the optical/NIR. These surveys find peaks in the galaxy distribution and measure the richness of the corresponding clusters. The microwave/tSZ and X-ray surveys detection techniques are based on the hot ICM \citep{Hasselfield:2013wf,Bleem:2014iim} and in some cases require auxiliary data to obtain useful constraints e.g. redshift estimates  \citep[see][for recent methods]{Klein:2017fwg,Bleem:2019fvp}. 

The CC method is based on the predicted halo abundance (number density) $n(M,z)$ of halos with mass less than $M$ at redshift $z$ which is also known as the halo mass function (HMF). This formalism was originally introduced  by Press and Schechter  \citep{Press:1973iz}. A general mathematical form for the comoving number density expression of haloes is  \citep[e.g.][]{Jenkins:2000bv,Sheth:2001dp,White:2002at,Warren:2005ey,Tinker:2008ff}
\be
\frac{dn}{dM}=f(\sigma)\frac{\bar{\rho}_m}{M}\frac{d\ln\sigma^{-1}}{dM}
\ee
where $\bar{\rho}_m=\rho_{crit}\Omega_m$ is the mean matter density of the Universe, $\sigma$ is the rms variance of the linear density field smoothed on a spherical volume containing a mass $M$, and $f(\sigma)$ is a model-dependent ‘universal’ halo multiplicity function\footnote{For a publicly available cluster toolkit Python package, see in \url{https://cluster-toolkit.readthedocs.io/en/latest/source/massfunction.html}.}.
There are numerous parametrizations of the multiplicity function $f(\sigma)$ based on numerical N-body simulations or theoretical models. A popular parametrization  provided by \citet{Tinker:2008ff} is
\be 
f(\sigma)=A\left[\left(\frac{\sigma}{b}\right)^{-\chi}+1\right]e^{-c/\sigma^2}
\ee
where $A,\chi,b,c$ are four free parameters that depend on the halo definition and need to be calibrated.

Measurements  of  the  abundance of galaxy clusters $n(M,z)$ provide consistent constraints on the density of matter $\Omega_{0m}$, the root mean square density fluctuation $\sigma_8$, the parameter combination $S_8(\alpha)=\sigma_8(\Omega_{0m}/0.3)^{\alpha}$  \citep[e.g.][]{Kilbinger:2012qz,Hikage:2018qbn,Asgari:2020wuj} (where $\alpha\sim 0.2-0.6$  and $S_8\equiv S_8(\alpha = 0.5)$), the dark energy equation-of-state $w$ and the sum of the neutrino masses  $\sum m_{\nu}$ (massive neutrinos can suppress the matter power spectrum on small scales and this directly affect the growth of cosmic structure) \citep{Weinberg:2012es}. More recently, \citet{Sabti:2021xvh,Sabti:2021unj} used a method of clustering measurements at higher redshift ($z=4-10$) based on UV galaxy luminosity function data from the Hubble Space Telescope \citep[see e.g.][]{Atek:2018nsc,Bouwens:2014fua}. They derive the large-scale matter clustering amplitude to be $\sigma_8=0.76_{-0.14}^{+0.12}$.

Using cluster abundance analysis in the SDSS DR8 for a flat $\Lambda$CDM cosmological model with massive neutrinos  \citet{Costanzi:2018xql} find $S_8=0.79_{-0.04}^{+0.05}$.  \citet{Mantz:2014paa} using  Weighting the Giant (WtG) \citep{vonderLinden:2012kh,Applegate:2012kr} lensing analysis of the X-ray ROentgen SATellite (ROSAT) cluster catalogs \citep{1993Sci...260.1769T} find
$S_8=0.77\pm 0.05$. The analysis of the counts and weak lensing signal of of the DES-Y1 dataset by  \citet{Abbott:2020knk} gives $S_8=0.65\pm0.04$  and $\Omega_{0m}= 0.267_{-0.017}^{+0.030}$ in flat $\Lambda$CDM. Also, assuming a flat $\Lambda$CDM model and performing a galaxy cluster abundance analysis in the AMICO KiDS-DR3 catalogue \citet{Lesci:2020qpk} obtains $S_8=0.78\pm 0.04$.

Using galaxy clusters observed in millimeter wavelengths through the tSZ effect  \citet{Salvati:2017rsn} report $S_8=0.785\pm0.038$ assuming $\Lambda$CDM model. The analysis of the Planck 2015 cluster counts via the tSZ signal by \citet{Ade:2015fva} finds $S_8=0.792\pm0.056$. Recently, assuming a flat $\Lambda$CDM model, in which the total neutrino mass is a free parameter, the analysis of SPT tSZ cluster counts by  \citet{Bocquet:2018ukq} results in $S_8=0.749\pm0.055$. Using  X-ray clusters detected from the XMM-XXL survey \citep{Pierre:2015cqe} for a flat $\Lambda$CDM cosmological model  \citet{Pacaud:2018zsh} report $S_8=0.83\pm0.10$. Also, constraints on structure growth parameter combination $S_8$ from cluster abundance data have been obtained by  \citet{deHaan:2016qvy,Abdullah:2020qmm,Kirby:2019mrb,Schellenberger:2017usb,Ntampaka:2019cww,Zubeldia:2019brr}. For example using GalWCat19 \citep{Abdullah:2019hsx}, a catalog of 1800 galaxy clusters was derived from the SDSS-DR13 \citep{SDSS:2016ear} and assuming a flat $\Lambda$CDM cosmology \citet{Abdullah:2020qmm} measured the matter density  and the amplitude of fluctuations to be 
$\Omega_m =0.310^{+0.023}_{-0.027}\pm 0.041$ (systematic) and $\sigma_8=0.810^{+0.031}_
{-0.036}\pm 0.035$ (systematic) respectively. 

The results of $S_8$ from all cluster count experiments as seen in Table \ref{tab:growthpar} and in Fig. \ref{figs8} are in agreement with WL measurements and similarly  prefer a lower value compared to the CMB measurements. 

\paragraph{Redshift space distortion-Galaxy clustering:}

Peculiar motions of galaxies falling towards overdense region generate large scale galaxy clustering, anisotropic in redshift space. Measuring this illusory  anisotropy that distorts the distribution of galaxies in redshift space (i.e. RSD) we can quantify the galaxy velocity field. This important probe of LSS can be used to constrain the growth rate of cosmic structures \citep{Kaiser:1987qv,Hamilton:1992zz,Hamilton:1997zq}. 

In particular the RSD is sensitive to the cosmological growth  rate of matter  density  perturbations $f$ which depends on the theory of gravity and is defined as \citep{Wang:1998gt,Linder:2005in,Polarski:2007rr}
\be
f(a)\equiv\frac{ d\ln \delta(a)}{ d\ln a}\simeq [\Omega_m(a)]^{\gamma(a)} 
\label{fa}
\ee
where $a=\frac{1}{1+z}$ is  the scale factor, $\delta\equiv \frac{\delta\rho_m}{\rho_m}$ is the matter overdensity field (with $\rho_m$ is the matter density of the background) and $\gamma$ is the growth index  \citep[e.g.][]{Lahav:1991wc}. The nearly constant and scale-independent value $\gamma\simeq \frac{6}{11}\simeq 0.545$ corresponds to General Relativity (GR) prediction in the context of $\Lambda$CDM  \citep[e.g.][]{Linder:2005in}.

The observable combination $f\sigma_8(a)\equiv f(a)\cdot \sigma(a)$ is measured at various  redshifts by different surveys as a probe of the growth of matter density perturbations. The theoretically predicted value of this product can be obtained from the solution $\delta(a)$ of the equation \citep{Nesseris:2015fqa}  
\be
\delta''(a)+\left(\frac{3}{a}+\frac{H'(a)}{H(a)}\right)\delta'(a)
-\frac{3}{2}
\frac{\Omega_{0m} G_{eff}(a)/G}
{a^5 H(a)^2/H_0^2} 
\delta(a)=0 
\label{eq:odedeltaa}
\ee
using the definition 
\be 
\sigma(a)\equiv\sigma_8 \frac{\delta(a)}{\delta(a=1)} 
\ee
where $G$ is Newton’s constant as measured by local experiments, $G_{eff}$ is the effective gravitational coupling which is related to the  growth of matter perturbation, $\sigma(a)$ is the redshift dependent rms  fluctuations of the linear density field within spheres of radius $R=8 h^{-1} Mpc$ and $\sigma_8$ is its value today.

Hence, the more robust bias free quantity $f\sigma_8$ is given by
\be
f\sigma_8(a)=\frac{\sigma_8}{\delta(a=1)}~a~\delta'(a) 
\label{fs8}
\ee

RSD growth data in the form of $f\sigma_8$\footnote{For an extensive compilation of RSD data points $f\sigma_8$, see in \citet{Skara:2019usd} and for other compilations, see in \citet{Basilakos:2016nyg,Nesseris:2017vor,Kazantzidis:2018rnb,Sagredo:2018ahx,Kazantzidis:2018jtb}. Also for a publicly available RSD likelihood for MontePython see in \citet{Arjona:2020yum,Cardona:2020ama}.} have been provided by wide variety of surveys including the $2$-degree Field Galaxy Redshift Survey ($2$dFGRS) \citep{Peacock:2001gs,Hawkins:2002sg}, VIMOS-VLT Deep  Survey  (VVDS) \citep{Guzzo:2008ac}, SDSS \citep{Tegmark:2006az,Samushia:2011cs,Reid:2012sw,Tojeiro:2012rp,Chuang:2013hya,Howlett:2014opa}, WiggleZ  \citep{Blake:2011rj}, $6$dFGS \citep{Beutler:2012px,Johnson:2014kaa}, Galaxy  and Mass Assembly (GAMA) \citep{Simpson:2015yfa}, BOSS \citep{White:2014naa,Alam:2016hwk,Li:2016bis,Gil-Marin:2016wya}, Subaru Fiber Multi-Object Spectrograph (FMOS) galaxy redshift survey (FastSound) \citep{Okumura:2015lvp}, VIMOS Public Extra-galactic  Redshift Survey (VIPERS) \citep{delaTorre:2013rpa,Pezzotta:2016gbo,Mohammad:2018mdy}, eBOSS  \citep{Zhao:2018gvb,Tamone:2020qrl,Zhao:2020bib,Bautista:2020ahg,Zhao:2020tis,deMattia:2020fkb,Alam:2020sor}, DESI \citep{Aghamousa:2016zmz,Aghamousa:2016sne}. Using such data the $\Omega_{0m}$ and $\sigma_8$ parameters in the context of a $\Lambda$CDM background can be constrained. Thus,  \citet{Kazantzidis:2018rnb} using a compilation of $63$ RSD datapoints find the $\Lambda$CDM best fit value $\Omega_{0m}=0.279\pm 0.028 $ and  $\sigma_8= 0.775\pm 0.018$. 
\citet{Benisty:2020kdt} using RSD selected data  and assuming $\Lambda$CDM model report $S_8= 0.700_{-0.037}^{+0.038}$, $\Omega_{0m}=0.201_{-0.033}^{+0.036} $ and  $\sigma_8= 0.857_{-0.042}^{+0.044}$ which are in $3\sigma$ tension with the Planck 2018 results. Recently, using RSD data and the RSD+BAO+Pantheon and RSD+BAO+Pantheon+CC dataset combinations \citet{Nunes:2021ipq} find $S_8= 0.739_{-0.040}^{+0.036}$, $S_8= 0.762_{-0.025}^{+0.030}$ and $S_8= 0.777_{-0.027}^{+0.026}$ respectively. 

Galaxy clustering methods, such as the galaxy power spectrum and bispectrum have also been used to constrain $S_8$. Constraints from the BOSS galaxy power spectrum \citep{Ivanov:2019pdj} gave $S_8=0.703\pm 0.045$ and  from BOSS DR12 bispectrum  \citep{Philcox:2021kcw} gave $S_8=0.751\pm 0.039$. Previous analysis of the BOSS DR12 data by \citet{Troster:2019ean} gave $S_8 = 0.729\pm0.048$. A analysis of the power spectrum of eBOSS  by \citet{Ivanov:2021zmi} resulted in $S_8=0.720\pm 0.042$. Recently, using the BOSS power spectra \citet{Chen:2021wdi} found $S_8 = 0.736\pm 0.051$. Also, the combination of the auto- and cross-correlation signal of unWISE \footnote{Wide-field Infrared  Survey  Explorer (WISE) \citep{Wright:2010qw} is a NASA infrared astronomy space telescope and is mapping the whole sky.} galaxies \citep{2019ApJS..240...30S} and Planck CMB lensing maps \citep{Planck:2018lbu} by \citet{Krolewski:2021yqy} gave $S_8 = 0.784 \pm 0.015$. Finally, using  the luminous red galaxies of the DESI in combination with Planck CMB lensing maps  \citet{White:2021yvw} found $S_8 = 0.73 \pm 0.03$. 

Clearly, as seen in Table \ref{tab:growthpar} and in Fig. \ref{figs8} the analyses of RSD data gives $S_8$ values in tension with CMB measurements at level more than $2\sigma$, in agreement with other dynamical cosmological probes (WL and CC). 

\subsubsection{Theoretical models}
\label{Theoretical Models s8}

Non-gravitational mechanisms can address the $S_8$ tension \citep[see][for a review]{Ishak:2018his}. Such mechanisms include the following:
\begin{itemize}
    \item
    Dynamical dark energy models \citep{Roy:2022fif,Linder:2002et,Cooray:1999da,Efstathiou:1999tm,Astier:2000as,Chevallier:2000qy,Weller:2001gf,Feng:2004ff,Nesseris:2005ur,Jassal:2005qc,Barboza:2008rh,Yang:2017amu,Melia:2016djn,Yang:2017alx,Pan:2017zoh,Rezaei:2017yyj,Joudaki:2016kym,Du:2018tia,Vagnozzi:2018jhn,Zhao:2017cud,Lambiase:2018ows,Ooba:2018dzf,Yang:2018qmz,Benisty:2021gde,DAmico:2020kxu} and running vacuum models \citep{Sola:2015wwa,Sola:2016jky,Gomez-Valent:2017idt,Gomez-Valent:2018nib,Wang:2018fng,Colgain:2021pmf,SolaPeracaula:2021gxi,BeltranJimenez:2021wbq}, which modify the cosmological background $H(z)$ to a form different from $\Lambda$CDM (see Subsection \ref{Late time modifications1}). This modification may involve the presence of dynamical dark energy dominant at late cosmological times or at times before recombination.
    \item 
    Interacting dark energy models, which modify the equation for the evolution of linear matter fluctuations as well as the $H(z)$ cosmological background \citep{Pourtsidou:2016ico,An:2017crg,Barros:2018efl,Camera:2017tws} as discussed in Subsection \ref{IDE}. This class of models can address the structure growth $\sigma_8$ tension between the values inferred from the CMB and the WL measurements.
    \item
    Effects of massive neutrinos \citep{Marulli:2011he,Battye:2013xqa,Costanzi:2014tna,Joudaki:2016kym,Zennaro:2017qnp,Poulin:2018zxs,DiazRivero:2019ukx,Biswas:2019uhy} which are relativistic at early times and contribute to radiation while at late times they become non-relativistic but with significant velocities (hot dark matter) \citep[see][for a review]{Lesgourgues:2006nd,Wong:2011ip,Lesgourgues:2012uu,Lesgourgues:2014zoa}. The change of radiation to hot dark matter affects the Hubble expansion. Simultaneously the residual streaming velocities are still large enough at late times to slow down the growth of structure \citep{Bond:1980ha}. This effect of massive neutrinos slows down the growth as required by the RSD data and relieves the $S_8$ tension coming from WL data \citep{DiazRivero:2019ukx}.
    \item
    Primordial magnetic fields \citep{Jedamzik:2018itu,Banerjee:2004df} \citep[see][for a review]{Durrer:2013pga,Subramanian:2015lua,Vachaspati:2020blt} induce additional mildly non-linear, small-scale baryon inhomogeneities present in the plasma before recombination. The required field results in a reduction of the sound horizon scale at recombination and has the potential to resolve both the $H_0$ and $S_8$ tension \citep{Jedamzik:2020krr,Rashkovetskyi:2021rwg,Thiele:2021okz}.
    \item
    Non-thermal dark radiation  \citep{Das:2021pof}  seems to help alleviate the $S_8$ tension to a great extent. However, the  inclusion of  BAO  data reduces significantly the quality of fit of this model.
\end{itemize}
In addition to these non-gravitational mechanisms discussed above that can slow down growth at low redshifts a possible interesting new fundamental physics approach can also reduce the $S_8$ tension. Such an approach  is most likely to affect three basic observable parameters: the Hubble parameter $H(z,w)$ (with $w$ the dark energy equation of state parameter), as well as the effective Newton constants for growth of perturbations 
\be
\mu_G(z,k)\equiv \frac{G_{eff}(z,k)}{G} 
\ee
and lensing 
\be
\Sigma (z,k)\equiv \frac{G_{L}(z,k)}{G}
\ee
where $G$ is the locally measured value of the Newton's constant.  According to $\Lambda$CDM $H(z)=H(z,w=-1)$, $\mu_G=1$, $\Sigma=1$.

The Bardeen potentials \citep{Bardeen:1980kt} (the Newtonian potential  $\Psi$ and the spatial curvature potential $\Phi$) appear in the scalar perturbed Friedmann-Lema\^{i}tre-Robertson-Walker (FLRW) metric in the conformal Newtonian gauge \citep{Mukhanov:1990me,Ma:1995ey,EspositoFarese:2000ij} 

\be  
ds^2=-(1+2\Psi)dt^2+a^2(1-2\Phi)d \vec{x}^2
\label{metric} 
\ee  

The  LSS probes are sensitive to the Bardeen potentials $\Psi$ and $\Phi$. In particular the WL probe is sensitive to $\nabla^2(\Psi+\Phi)$. The galaxy clustering arises from the gravitational attraction of matter and is sensitive only to the potential $\Psi$. The RSD probe is sensitive to the rate of growth of matter density perturbations $f$ (see Eq. (\ref{fa})) and  provides measurements of $f\sigma_8$ (see Eq. (\ref{fs8})) that depends on the potential $\Psi$.

At linear level, in modified gravity models, using the perturbed metric Eq. (\ref{metric}) and the gravitational field equations the following  phenomenological equations in Fourier space emerge for the scalar perturbation potentials defining the functions  $\mu_G(a,k)$ and $\Sigma(a,k)$ on subhorizon scales (i.e. $k/aH\gg1$) 
\be 
k^2(\Psi+\Phi)=-8\pi G\Sigma(a,k)a^2\rho\Delta 
\label{poissonsigma} 
\ee 
\be  
k^2\Psi=-4\pi G\mu_G(a,k)a^2\rho\Delta 
\label{poissonmu} 
\ee
where $\rho_m$ is the matter density of the background, $\Delta$ the comoving matter density contrast defined as $\Delta\equiv\delta+3Ha(1+w)\upsilon/k$ which is gauge-invariant \citep{Pogosian:2010tj}, $w =p/\rho$  is the equation-of-state parameter and $\upsilon^i=-\nabla^i u$  is the irrotational component of the peculiar velocity $u$ \citep{Boisseau:2000pr}.

Using the gravitational slip parameter $\eta$ (or anisotropic stress parameter) which describes the possible inequality \citep{Pogosian:2007sw,Jain:2010ka} of the two Bardeen potentials that may occur in modified gravity theories \be 
\eta(a,k)=\frac{\Phi(a,k)}{\Psi(a,k)} 
\label{slip}
\ee 
the two LSS parameters  $\mu_G$ and $\Sigma$ are related via  
\be 
\Sigma(a,k)=\frac{1}{2}\mu_G(a,k)\left[1+\eta(a,k)\right]
\ee
The the Hubble parameter $H(z)$ is usually parametrized as wCDM
\be
H(z)=H_0\left[\Omega_{0m}(1+z)^3 +(1-\Omega_{0m})(1+z)^{3(1+w)}\right]^{1/2}
\ee
while the two LSS parameters  $\mu_G$ and $\Sigma$ do not have a commonly accepted parametrization. A model and scale independent parametrization for $\mu_G$ and $\Sigma$ which reduce to the GR value at early times and at the present time as indicated by solar system (ignoring possible screening effects) and  BBN constraints ($\mu_G=1$ and $\mu_G'=0$ for $a=1$ and $\mu_G=1$ for $a\ll1$) \citep{Muller:2005sr,Pitjeva:2013xxa,Gannouji:2006jm} is  of the form \citep{Nesseris:2006hp,Nesseris:2017vor,Kazantzidis:2018rnb,Skara:2019usd}
\begin{widetext}
\be 
\mu_G =1+g_a(1-a)^n-g_a(1-a)^{2n}=1+g_a(\frac{z}{1+z})^n-g_a(\frac{z}{1+z})^{2n}
\label{muaz} 
\ee
\be 
\Sigma =1+g_b(1-a)^m-g_b(1-a)^{2m}=1+g_b(\frac{z}{1+z})^m-g_b(\frac{z}{1+z})^{2m}
\label{sigmaaz}  
\ee
\end{widetext}
where  $g_a$ and $g_b$ are parameters to  be  fit and $n$ and $m$ are integer parameters with  $n\geq2$ and $m\geq2$.

Alternatively, a rapid transition parametrization is of the form \citep{Alestas:2020zol,Marra:2021fvf}
\be
\mu_G^>(z) =\mu_G^<+ \Delta \mu_G\, \Theta (z-z_t) 
\ee
\be
\Sigma^>(z) =\Sigma^<+ \Delta \Sigma\, \Theta (z-z_t) 
\ee
where $\Theta$  is  the  Heaviside  step  function, $z_t$ is a transition redshift, $\mu_G^>$ and $\Sigma^>$ correspond to $z>z_t$ and  $\mu_G^<$ and $\Sigma^<$ correspond to $z<z_t$.

Various studies utilize modified gravity theories including Teleparallel theories of gravity\footnote{Many authors have studied the extensions of the Teleparallel gravity such as the scalar-torsion theories of gravity \citep{Geng:2011ka,Geng:2011aj,Gonzalez-Espinoza:2021qnv,Paliathanasis:2021nqa,Skugoreva:2014ena,Kofinas:2015hla,Leon:2022oyy} and the Teleparallel Horndeski theories \citep{Dialektopoulos:2021ryi,Bernardo:2021izq,Bahamonde:2021dqn,Bahamonde:2020cfv,Bahamonde:2019shr}. } \citep{DAgostino:2018ngy,Gonzalez-Espinoza:2018gyl} \citep[see][for a review]{Bahamonde:2021gfp}, Horndeski theories \citep{Kennedy:2018gtx,Linder:2018jil} or theories beyond Horndeski \citep{DAmico:2016ntq} to reduce the effective Newton’s constant $G_{eff}$ at low redshifts and slow down growth at low redshifts. The above parametrizations can be realized in the context of physical models based on the above theories. 

\subsection{CMB anisotropy anomalies}  
\label{CMB Anomalies}
There is a wide range of other less discussed no-standard signals and statistical anomalies of the large angle fluctuations in the CMB \citep{Rassat:2014yna} with a typical $2$ to $3\sigma$ significance. As mentioned a main assumption of the $\Lambda$CDM model is that the fluctuations are Gaussian and statistically homogeneous and isotropic. Diverse anomalies have been noticed in the CMB at large  angular scales by the space missions Cosmic Background Explorer (COBE) \citep{Mather:1991pc}, Wilkinson Microwave Anisotropy Probe (WMAP) \citep{Bennett:2003ba} and Planck satellite \citep{Adam:2015rua}, which appear to violate this assumption \citep[see][for a review]{Schwarz:2015cma,Akrami:2019bkn}.  \citet{Perivolaropoulos:2014lua} presents  possible explanations of the observed CMB anomalies and  \citet{Cayuso:2019hen} explore the kinetic and the polarized Sunyaev-Zel’dovich effects as potential probes of physical models of these anomalies. 

In what follows we discuss some of these signals. Note that some of these  may not be independent\footnote{The covariance of CMB anomalies in the standard $\Lambda$CDM model has been studied by \citet{Muir:2018hjv}. This study focusing on the correlation of observed anomalies (i.e. the relationship or connection between all of them) examines the independence of large-angle CMB feature quantities.}. Some of these signals have been attributed to the look-elsewhere effect. Based on this effect any large dataset will have a small number of peculiar features  when there is a careful search for such features. However, this argument may not be applicable when the considered statistics are simple and generic as  are most of the signals discussed below \citep[see][for a detailed discussion]{Bayer:2020pva,Peiris:2014lda,Gross:2010qma}.

\subsubsection{Hints for a closed Universe (CMB vs BAO)}
The Universe under the assumption of the cosmological principle is described by the Friedmann-Lema\^{i}tre-Roberson-Walker (FLRW) metric 
\be 
ds^2=dt^2-a(t)^2\left[\frac{dr^2}{1-Kr^2}+r^2(d\theta^2+sin^2\theta d\phi^2)\right]
\ee
where $K$ characterizes the constant spatial curvature of the spatial slices with $K=-1,0,+1$ corresponding to open hyperbolic space (negative spatial curvature), flat Euclidean space (zero spatial curvature), and closed hyperspherical space (positive spatial curvature) respectively. The curvature density parameter is defined as $\Omega_K \equiv -K/(Ha)^2$ so that a closed  Universe corresponds to $\Omega_K <0$ and an open Universe to $\Omega_K >0$. This parameter plays a crucial role in determining the evolution of the Universe, and is closely related with the early Universe physics.

The Planck$18$ temperature and polarization data \citep{Planck:2018vyg} show a preference ($\sim3.4\sigma$) for a closed Universe ($\Omega_K <0$) in the context of $\Lambda$CDM. In particular using these data from Planck$18$ the curvature density parameter was constrained to be $-0.095<\Omega_K<-0.007$ at $99\, \% $ C.L \citep{Planck:2018vyg,Planck:2019nip}. This anomaly may be connected with other asymmetries of the CMB anisotropy spectrum discussed below. The preference for closed universe however disappears when the CMB data are combined with the BAO data. \citet{DiValentino:2020srs,DiValentino:2019qzk,Handley:2019tkm} pointed out that Planck+BAO can give a biased result because Planck and BAO are in disagreement at more than $3\sigma$. Combining Planck18 data with recent BAO measurements the curvature density parameter was estimated to be  $\Omega_K=0.0008\pm 0.0019$ \citep{DiValentino:2020srs,DiValentino:2019qzk,Handley:2019tkm} in agreement with a spatially flat Universe.  Using  the full-shape galaxy power spectrum measurements $P(k)$, \citet{Vagnozzi:2020zrh} has also confirmed that the Planck data are in tension with both the full-shape power spectrum and BAO with respect to $\Omega_K$. The recent study by \citet{Efstathiou:2020wem} confirms the tension between Planck and BAO data in the context of cosmic curvature. \citet{Efstathiou:2020wem} used a new statistical analysis (the alternative Planck CamSpec likelihood TTTEEE instead of Plik as discussed in \citet{Efstathiou:2019mdh}) to show that Planck favors a closed Universe at more than $99\%$ CL. However, Planck+BAO was again found to be in  agreement with a spatially flat Universe with $\Omega_K=0.0004\pm 0.0018$ thus confirming previous studies by \citet{DiValentino:2019qzk,Handley:2019tkm} .

In an effort to further investigate this tension between Planck and BAO data, the analysis of  \citet{Vagnozzi:2020dfn} combined Planck18 CMB temperature and polarization data with cosmic chronometer measurements and was lead to confirm that the Universe is consistent with spatial flatness to $\mathcal{O}$($10^{-2}$) level. 

A positive curvature (closed Universe) may be a plausible source of the anomalous lensing amplitude \citep{DiValentino:2020srs,DiValentino:2019qzk,Handley:2019tkm} (see Subsection \ref{The lensing anomaly}).

\subsubsection{Anomalously strong ISW effect}
The decay of cosmological large-scale gravitational potential $\Psi$ causes the integrated Sachs-Wolfe (ISW) effect \citep{Sachs:1967er} which imprints tiny secondary anisotropies to the primary fluctuations of the CMB and is a complementary probe of dark energy  \citep{Kofman:1985fp,Fosalba:2003iy,Giannantonio:2008zi,Kable:2021yws}. Using a stacking technique  in the CMB data  \citep[see][for a detailed discussion]{Marcos-Caballero:2015lxp,Ade:2015hxq} anomalously strong integrated Sachs–Wolfe (ISW) signal ($>3\sigma$) has been detected for supervoids and superclusters on scales larger than $100h^{-1}Mpc$  \citep{Granett:2008ju,Granett:2008xb}. This stronger than expected within standard $\Lambda$CDM signal of the ISW effect first emphasised in \citet{Hunt:2008wp}  has  been studied by  \citet{Nadathur:2011iu,Flender:2012wu,Ilic:2013cn,Cai:2016rdk,Kovacs:2022dyg,Kovacs:2017hxj,Kovacs:2018irz,Dong:2020fqt}. 

In particular the analysis by \citet{Kovacs:2018irz} for DES data alone found an excess ISW imprinted profile with $A_{ISW}\equiv \Delta T^{data}/\Delta T^{theory}\approx 4.1\pm2.0$ amplitude (where $A_{ISW}=1$ corresponds to the $\Lambda$CDM prediction). Also a combination with independent BOSS data leads to  $A_{ISW}=5.2\pm 1.6$.  This is in  $2.6\sigma$ tension  with $\Lambda$CDM  cosmology.

The average expansion rate approximation (AvERA) inhomogeneous cosmological simulation \citep{Racz:2016rss} uses the separate Universe conjecture to calculates the spatial average of the expansion rate of local mini-Universes predicts. It indicates under the inhomogeneity assumption, about  $\sim 2-5$ times higher ISW effect than $\Lambda$CDM depending on the $l$ index of the spherical power spectrum \citep{Beck:2018owr}. Thus large scale spatial inhomogeneities could provide an explanation to this ISW excess signal. \citet{Giannantonio:2012aa} use angular cross-correlation techniques and combines  several tracer catalogues to report  $A_{ISW}\approx 1.38\pm 0.32$. 

\citet{Vagnozzi:2021gjh} investigated the early Integrated Sachs-Wolfe (eISW) effect \citep[see e.g.][]{Bowen:2001in,Galli:2010it} which is assumed to occur soon after recombination ($30<z<1100$), due to the presence of a non-negligible radiation. Constraints were thus imposed on the parameter $A_{eISW}$  introduced  by \citet{Hou:2011ec}. Using Planck CMB data, this parameter was constrained to $A_{eISW} = 0.988\pm 0.027$, in perfect agreement with $\Lambda$CDM. Note that in previous studies the parameter $A_{eISW}$ was constrained to $A_{eISW} = 0.979\pm0.055$ using data from WMAP7+SPT \citep{Hou:2011ec}, to $A_{eISW} = 1.06 \pm 0.04$ from the Planck 2015 data release \citep{Cabass:2015xfa}, and to $A_{eISW} = 1.064\pm0.042$ from the Planck 2018 temperature data alone \citep{Kable:2020hcw}.

In general the reported $A_{ISW}$ amplitude varies in the literature depending on the dataset and the assumptions of the analysis. Further investigation of this issue is needed.

\subsubsection{CMB cold spot}
The cold (blue) spot was first found in WMAP 1-year temperature data by \citet{Vielva:2003et} and was confirmed in Planck data \citep{Ade:2013nlj,Ade:2015hxq,Akrami:2019bkn} in the southern hemisphere at the galactic longitude and latitude $(l,b)=(209^0,-57^0)$. It is a statistical anomaly of the large-angle fluctuations in the CMB indicating non-Gaussian features. This inconsistency with Gaussian simulations has a p-value of $\sim1\%$. 

The cold spot is an unusually large region of low temperature with the mean temperature decrement $\Delta T\approx -100\,\mu K$  and is not consistent with the prediction of gaussianity of the standard $\Lambda$CDM model \citep{Cruz:2004ce,Cruz:2006sv,Cruz:2006fy}. 

\citet{Zhang:2009qg,Nadathur:2014tfa,Kovacs:2017hxj} pointed out that the anomalous nature of the cold spot corresponds to a rather cold area with an angular radius in the sky of about $5^0-10^0$ from the centre surrounded by a hot ring. 

Possible approaches for the explanation of the Cold Spot include: non-Gaussian feature due to a large statistical fluctuation \citep{Vielva:2003et}, an artifact of inflation \citep{Cruz:2004ce}, the foreground \citep{Cruz:2006sv,Hansen:2012xq}, multiple voids \citep{Naidoo:2015gab}, the imprint of a supervoid (about $ 140-200Mpc$ radius completely empty void at $z\leq 1$) through the ISW effect \citep{Inoue:2006rd,Inoue:2006fn,Rudnick:2007kw,Granett:2008ju}, the axis of rotation of the Universe \citep{Jaffe:2005pw}, cosmic texture \citep{Cruz:2004ce,Zhao:2012hz}, adiabatic perturbation on the last scattering surface \citep{Valkenburg:2011ty}  \citep[see][for a review]{Cruz:2009nd,Vielva:2010ng}.

\subsubsection{Cosmic hemispherical power asymmetry}
The cosmic hemispherical power asymmetry (or dipolar asymmetry) is a directional dependency of the CMB angular power spectrum \citep{Eriksen:2003db,Hansen:2004vq,Eriksen:2007pc,Paci:2010wp}. The continuous dipolar modulation of hemispherical power asymmetry corresponds to a hemispherical temperature variance asymmetry (signal in the CMB temperature field) \citep{Eriksen:2003db,Hansen:2004vq,Monteserin:2007fv,Hoftuft:2009rq,Akrami:2014eta,Bernui:2014gla,Ade:2015hxq,Akrami:2019bkn,ODwyer:2019rfg}. 

The dipolar modulated/observed  CMB temperature fluctuation $\frac{\Delta T}{T}\arrowvert_{mod}$ in the direction $\hat{n}$ which appears to extend to $l_{max} \simeq 64$ can be expressed as \citep{Gordon:2006ag,Bennett:2010jb,Akrami:2014eta}\footnote{Note that the hemispherical dipole is distinct from the usual CMB dipole. In the former case the {\it power spectrum} is assumed modulated  discontinuously across a circle on the sky and in the second the actual temperature map has a component modulated by a smooth cosine function across the sky \citep{Bennett:2010jb}.}
\be
\frac{\Delta T}{T}\arrowvert_{mod}(\hat{n})=\left[1+A_{dm} \hat{n}\cdot \hat{p}\right]\frac{\Delta T}{T}\arrowvert_{iso}(\hat{n})
\ee
where $\frac{\Delta T}{T}\arrowvert_{iso}$ is a  statistically unmodulated/isotropic  temperature  fluctuation, $A_{dm}$ denotes the amplitude of dipolar modulation and $\hat{n}\cdot \hat{p}$ corresponds to the dipolar modulation between the line-of-sight (LOS) of the observer (with unit vector $\hat{n}$) and the preferred dipolar direction (with unit vector $\hat{p}$). The amplitude of dipolar modulation $A_{dm}$ is large at large angular scales  $2<l\lesssim64$ ($k\lesssim 0.035 Mpc^{-1}$), small at small angular scales  $l\gtrsim64$ and vanishes by a  multipole moment of $\sim 500-600$  \citep{Ade:2013nlj,Ade:2015hxq,Akrami:2019bkn}. The scale dependence of the hemispherical power asymmetry was suggested by  \citet{Erickcek:2008sm,Liddle:2013czu,Lyth:2014mga,Cai:2015xba,Mukherjee:2015wra,Yang:2016wlz,Byrnes:2016zxb,Byrnes:2015dub,Byrnes:2016uqw,Ashoorioon:2015pia,Jazayeri:2017szw} and was investigated by  \citet{Shiraishi:2016omb,Li:2019bsg}.
 
According to the hemispherical asymmetry nearly aligned with the Ecliptic, the temperature fluctuations are larger on one side of the CMB sky than on the other, resulting in an unexpected dipole configuration in the CMB power spectrum with an anomalously lower value of the variance in the northern sky compared to the  southern sky \citep{Ade:2013nlj}. The preferred direction for the asymmetry from the Planck18 data is $(l,b) = (221^0,-20^0)$ in galactic coordinates and the amplitude is $A_{dm}\sim 0.07$ with statistically significant at the $\sim 3 \sigma$ level \citep{Akrami:2019bkn}. This amplitude is $\sim2$ times higher than expected asymmetry due to cosmic variance ($A_{dm}\sim 0.03$) and it is inconsistent with isotropy ($A_{dm}= 0$) at the $\sim3\sigma$ level.
The hemispherical power asymmetry in CMB can be explained by assuming a superhorizon perturbation \citep{Gordon:2005ai,Gordon:2006ag} or asymmetric initial states of  the quantum perturbations \citep{Ashoorioon:2015pia}.

\subsubsection{Quadrupole-octopole alignment}
The fluctuations in the standard $\Lambda$CDM model are Gaussian and statistically isotopic. Thus in harmonic space the quadrupole  ($l=2$) and octopole ($l=3$) harmonics are expected to have independent and random orientations and shapes. The  quadrupole and octopole have been observed to be planar and unexpectedly aligned with each other \citep{deOliveira-Costa:2003utu,Schwarz:2004gk,Copi:2005ff,Copi:2006tu,Copi:2010na,Copi:2013jna}. This implies a violation of statistical isotropy. 

In particular in this low multipole moment anomaly the quadrupole and octopole planes are found to be mutually aligned with the direction of the cosmic dipole or CMB dipole (see Subsection \ref {Cosmic Dipoles} and Table \ref{axes}) and perpendicular to the Ecliptic \citep{Schwarz:2015cma}. 

In order to study this large-angle anomaly one can use the maximum angular momentum dispersion \citep{deOliveira-Costa:2003utu} 
\be
\langle\psi|(\hat{{\bf n}}_l\cdot {\bf L})^2|\psi\rangle=\sum_{m=-l}^l m^2|a_{lm}(\hat{{\bf n}}_l)|^2
\ee
where the CMB map is represented by a wave function 
\be
\frac{\Delta T}{T}(\hat{{\bf n}}_l)\equiv\psi(\hat{{\bf n}}_l)
\ee
Here $a_{lm}(\hat{{\bf n}}_l)$ correspond to the spherical harmonic coefficients of the CMB map in a coordinate system with its $z$-axis in the the $\hat{{\bf n}}_l$-direction.

The preferred axis $\hat{{\bf n}}_l$ is the axis around which the angular momentum dispersion is maximized. The directions of the quadrupole $\hat{{\bf n}}_2$ and the octopole $\hat{{\bf n}}_3$ are  \citep{deOliveira-Costa:2003utu} 
\be
\hat{{\bf n}}_2=(-0.1145,-0.5265,0.8424)
\ee
\be
\hat{{\bf n}}_3=(-0.2578,-0.4207,0.8698)
\ee
with
\be
|\hat{{\bf n}}_2\cdot \hat{{\bf n}}_3|\simeq 0.9838
\ee
This unexpected alignment of the $\hat{{\bf n}}_2$ and $\hat{{\bf n}}_3$ directions has only a $1/62$  probability of happening.

An approach in the analysis of this large-angle anomaly may also involve the use the multipole vectors \citep{Copi:2003kt} (an alternative to the spherical harmonics) where each multipole order $l$ is represented by $l$ unit vectors i.e a dipole $l=1$ can be constructed by a vector, a quadrupole by the product of two vectors/dipoles, an octopole from three vectors/dipoles etc.

The alignment of low multipoles indicates the existence of a preferred direction in the CMB temperature anisotropy. Furthermore possible relation  between the quadrupole-octopole alignment and the dipolar asymmetry has been investigated by \citet{Hoftuft:2009rq,Gordon:2006ag}. A negligible relation between these anomalies was reported. However the analysis by \citet{Marcos-Caballero:2019jqj} has shown that a particular dipolar modulation including the scale dependence may be connected with the quadrupole-octopole alignment.

\subsubsection{Lack of large-angle CMB temperature correlations}
There is a lack of large-angle CMB temperature correlations as first was observed by COBE  satellite \citep{Hinshaw:1996ut} and was confirmed by the WMAP \citep{Bennett:2003bz,Spergel:2003cb} and Planck \citep{Ade:2015hxq,Akrami:2019bkn} temperature maps in the range $l=2$ to $32$. This is in tension with the $\Lambda$CDM  prediction. 

This anomaly is directly connected to the temperature $T$ two-point angular correlation function $C^{TT}(\theta)$ of the CMB at large angular scale ($\theta \gtrsim 60^0$) which is unexpectedly close to zero \citep{Copi:2008hw,Copi:2010na,Copi:2013cya}. In angular space the two-point angular correlation function is defined as
\be
C^{TT}(\theta)\equiv\langle T(\hat{n}_1)T(\hat{n}_2)\rangle=\frac{1}{4\pi}\sum_l (2l+1)C_l P_l(\cos\theta)
\ee
where the average is over all pairs of directions $\hat{n}$ with $\hat{n}_1\cdot \hat{n}_2=\cos \theta$, $P_l(\cos\theta)$ are the Legendre polynomials and $C_l$ is the angular power spectrum 
\be
C_l\equiv \frac{1}{2l+1}\sum_{m=-l}^l |a_{lm}|^2
\label{angps}
\ee
with $a_{lm}$ the spherical harmonic coefficients of the temperature fluctuations.

The simplest and most useful statistic is $S_{1/2}$ first introduced in the WMAP first-year release  \citep{Spergel:2003cb} in order to measure the  deviation of the angular correlation function from zero at angular scales $60^0< \theta <180^0$. It is defined as
\be
S_{1/2}=\int_{\mu_2}^{\mu_1}\left[C^{TT}(\theta)\right]^2 d(\cos \theta)
\ee
with $\mu_1\equiv \cos\theta_1=\cos 60^0=1/2$ and $\mu_2\equiv \cos\theta_2=\cos 180=-1$.

A number of  alternative statistics have been proposed in the literature  \citep{Hajian:2007pi,Efstathiou:2009di,Gruppuso:2013dba,Akrami:2019bkn}. For example a generalization of the $S_{1/2}$ statistic suggested by  \citet{Copi:2013zja}. This statistic known as $S^{TQ}$ uses the  two-point angular correlation function between fluctuations in the temperature $T$ and the Stokes parameter\footnote{The Stokes parameters Q and U \citep[for the Stokes parameters formalism see][]{Jackson:1998nia} are used to  describe the state of CMB polarization  \citep[e.g.][]{Zaldarriaga:1996xe,Dodelson:2003ft}. These parameters are directly related to the E and B modes \citep{Kamionkowski:1996ks,Lewis:2001hp,Bunn:2002df}. The polarization amplitude is given by $P=\sqrt{Q^2+U^2}$.} $Q$, $C^{TQ}(\theta)$, which can be expressed in terms of the two-point angular power spectrum, $C_l^{TE}$ (with $E$ the gradient mode of polarization). The  significance of a test statistic can be quantified by using the p-value\footnote{The probability value or p-value is the probability of measuring a test statistic equal to or more extreme as the observed one, considering that the null hypothesis is correct \citep{Ade:2013nlj}. It provides the lower value of significance at which the model would be ruled out. A low p-value means that there is strong indication of new physics beyond the null hypothesis.}, suggested by  \citet{Ade:2013nlj}.

No sufficient explanation has yet been suggested for this large-angle anomaly. \citet{Copi:2016hhq} study the ISW effect, \citet{Aurich:2021ofm,Bernui:2018wef} explore a non-trivial spatial topology of the Universe and \citet{Pranav:2018lox} study the topology of the Planck CMB temperature fluctuations in order to find a possible explanation to the suppression of large-angle CMB temperature correlations. Also the low observed power in the quadrupole is a potential explanation for the lack of correlation in the temperature maps.  \citet{Copi:2008hw} argue that there is a cancellation between the combined contributions of $C_l$ with multipoles $l\leq 5$ and the contributions of $C_l$ with multipoles $l\geq 6$.

\subsubsection{Anomaly on super-horizon scales}
\citet{Pranav:2021ted} analysed the topological characteristics of the CMB temperature fluctuation. Using mathematical investigations on persistent homology to describe the cosmic mass distribution and performing  experiments on  Planck 2020 data release 4 (DR4)  (based on the NPIPE data processing pipeline \citealt{Akrami:2020bpw}), \citet{Pranav:2021ted}  claimed a detection of an anomalous topological signature in the Planck CMB maps indicating non-Gaussian fluctuations. In particular \citet{Pranav:2021ted}  reports an anomaly in the behavior of the loops (a $4\sigma$ deviation in the number of loops) in the observed sky compared to the analysis of the redshift evolution of structure on simulations when the $\Lambda$CDM model is considered.

\subsubsection{The lensing anomaly}
\label{The lensing anomaly}
The recent Planck$18$ release by \citet{Planck:2018vyg} has confirmed the higher compared to that expected in the standard $\Lambda$CDM model, anomalous, lensing contribution in the CMB power spectra which is quantified by the phenomenological parameter, $A_L$ \citep{Calabrese:2008rt,Zaldarriaga:1998ar}. This weak lensing parameter $A_L$ rescales the lensing potential power spectrum as\footnote{Note that this is not the usual $C_l$ but it is the additional contribution due to lensing.}
\be
C_l^{\Psi} \rightarrow A_L C_l^{\Psi}
\ee
where $A_L=0$ corresponds to unlensed while $A_L= 1$ is the expected lensed result \citep{Calabrese:2008rt}  measuring the lensing effect in the CMB temperature power spectrum.

Since the main impacts of lensing on the CMB temperature power spectrum are to add power at small scales and to smooth the structure of the acoustic peaks and troughs (the peaks are reduced slightly, and the troughs between them filled in) \citep{Lewis:2006fu,Ade:2015zua} the adding of parameter $A_L$ changes the amount of smoothing of the CMB primary spectra peaks and troughs. A higher lensing amplitude ($A_L>1$) than predicted in the flat $\Lambda$CDM cosmology ($A_L= 1$) by roughly $10\%$ (at the level of $2.8\sigma$) has been found in the temperature power spectra by the Planck team \citep{Planck:2018vyg}. 

It should be noted that the oscillatory residuals between the Planck temperature power spectra and the best-fit $\Lambda$CDM model in the multipole range $l\in [900, 1700]$ are in opposite phase compared to the CMB and thus phenomenologically similar to the effects of gravitational lensing \citep{Aghanim:2016sns,Motloch:2019gux}. 

A plausible explanation of the anomalous lensing amplitude is a positive curvature (closed Universe) which was investigated by  \citet{DiValentino:2020srs,DiValentino:2019qzk,Handley:2019tkm}. Other possible sources which explain the lensing anomaly by mimicking a lensing effect are: a component of cold dark matter isocurvature (CDI) perturbation with a blue tilt  \citep[see][for a detailed discussion]{Akrami:2018odb} and oscillations in the primordial power spectrum which have the same frequency but opposite phase with the acoustic peaks \citep{Planck:2018vyg}. All these effects are degenerate with the smoothing effect of lensing.

Furthermore, the modified gravity models could be candidates for a solution of the lensing anomaly \citep{Ade:2015rim,DiValentino:2015bja,Moshafi:2020rkq,DiValentino:2020evt}. In particular the hints for $\Sigma_0>1$ (where $\Sigma_0$ the current value of parameter  $\Sigma$ which modifies the equation for the lensing potential i.e. Eq. (\ref{poissonsigma})) are directly connected to the lensing anomaly as characterized by $A_L> 1$ \citep{DiValentino:2015bja,DiValentino:2020evt}.

\subsubsection{High-low l consistency}

\citet{Addison:2015wyg} pointed out that there are internal inconsistencies in the Planck TT power spectrum. The $\Lambda$CDM parameter values derived by the high $l$ part of the CMB anisotropy spectrum ($l>1000$) are in $2-3\sigma$ tension  with the corresponding values of these parameters derived from the low $l$ part of the spectrum ($1<1000$). For example the low $l$ multipoles predict a lower value of the cold dark matter density parameter $\omega_c$ than the high $l$ multipoles, with discrepancy at $2.5\sigma$ \citep{Addison:2015wyg}. In addition it has been  shown that the value of $H_0$ predicted by Planck from $l>1000$, $H_0= 64.1\pm 1.7$ $km$ $s^{-1}Mpc^{-1}$, disagrees with the value predicted by Planck from $l<1000$, $H_0= 69.7\pm 1.7$ $km$ $s^{-1}Mpc^{-1}$ at the $2.3\sigma$ level. Thus it is found that the value of $H_0$ depends on the CMB $l$-range examined.

This anomaly is probably related to the lensing anomaly i.e. the fact that $\Lambda$CDM is more consistent with the low $l$ part of the spectrum that this not affected by the lensing anomaly \citep[see][for a discussion]{Riess:2016jrr,Aghanim:2016sns,Planck:2019nip}.

\subsubsection{The preference for odd parity correlations}

There is an anomalous power excess (deficit) of odd (even) $l$ multipoles in the CMB anisotropy spectrum on the largest angular scales ($2<l<30$), \citep{Land:2005jq,Kim:2010gd,Kim:2010gf,Kim_2011,Gruppuso:2010nd,Gruppuso:2017nap,Akrami:2019bkn}. A map consisting of odd (even) multipoles possesses odd (even) parity thus this effect may be considered as power (spectrum) asymmetry between even and odd parity map which is known as parity asymmetry.

In order to compare even and odd multipoles  \citet{Kim:2010gf} consider the parity asymmetry statistic defined as the ratio $P\equiv P^+/P^-$ of quantities $P^+$ and $P^-$ which represent the mean power in even and odd only multipoles respectively for the range $2\leq l \leq l_{max}$
\be
P^{\pm}=\sum_2^{l_{max}}\frac{\left[1\pm (-1)^l\right]l(l+1) C_l}{4\pi}
\ee
A different statistic to quantify the parity asymmetry has been proposed by  \citet{Aluri:2011wv}.

\begin{figure*}
\begin{centering}
\includegraphics[width=0.9\textwidth]{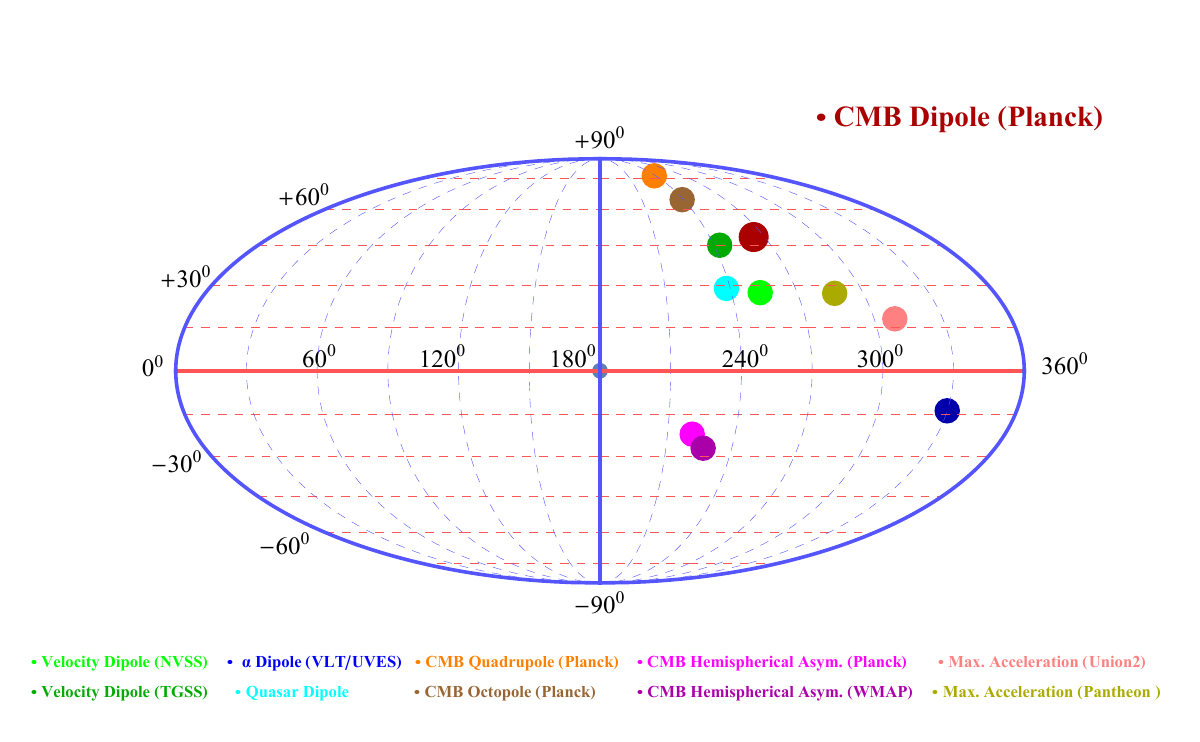}
\par\end{centering}
\caption{Mollweide-projection view of preferred directions in galactic coordinates for different cosmological observations (see Table \ref{axes}).}
\label{figsky}
\end{figure*}

\subsection{Cosmic dipoles}
\label{Cosmic Dipoles} 
There have been studies pointing out the presence of signals which indicate the violation of the cosmological principle. A physical mechanism producing such violation on Hubble scales is studied by  \citet{BuenoSanchez:2011wr}.  
Various other possible mechanisms have been suggested to explain the observed violations of statistical isotropy e.g.  superhorizon perturbations which introduce a preferred direction in our Universe \citep{Gordon:2005ai,Erickcek:2008jp}  \citep[see also][for a review]{Perivolaropoulos:2014lua}. The dipole amplitudes and the directions $(l, b)$ (galactic coordinates) from the different cosmological observations described below are shown in Fig. \ref{figsky} and along with the corresponding references in Table \ref{axes}.

\begin{center}      
\begin{table*}    
\centering 
\caption{The amplitudes and the directions $(l, b)$ (galactic coordinates) from different cosmological observations (Fig. \ref{figsky}) along with the corresponding references . The amplitude of CMB dipole has derived using the Eq. (\ref{ampldipo}) \citep[e.g.][]{Secrest:2020has}.}
\label{axes} 
\vspace{2mm}
\begin{tabular}{c ccc  ccc c} 
\hhline{========}
   & \\
Observations&& $l \,[deg]$ &&$b  \,[deg]$&& Amplitude & Refs.  \\
 
    & \\
   \hhline{========}
     & \\

CMB Dipole (Planck) &&$264.021\pm0.011$  && $48.253\pm0.005$  && $\sim0.007$  &   \citep{Akrami:2018vks,Aghanim:2018fcm}   \\

Velocity Radio Dipole (TGSS)  && $243.00\pm 12.00$  && $45.00\pm3.00$  && $0.070\pm0.004$&   \citep{Bengaly:2017slg}    \\ 
 Velocity Radio Dipole (NVSS) && $253.12\pm11.00$  && $27.28\pm3.00$  && $0.023\pm0.004 $&  \citep{Bengaly:2017slg}    \\ 
 Velocity Radio Dipole (NVSS) && $253.00\pm2.00$  && $28.71\pm12.00$  &&$0.019\pm0.002 $&  \citep{Colin:2017juj}    \\ 
 Velocity Radio Dipole (NVSS) && $253.00$  && $32.00\pm12.00$  &&$0.012\pm0.005$&   \citep{Tiwari:2015tba}    \\ 
 Quasar  Dipole  && $238.20$  &&$28.80$   && $0.01554$&   \citep{Secrest:2020has}   \\ 
 $\alpha$  Dipole (VLT/UVES) && $330\pm15$ &&$-13\pm10$   && $0.97_{-0.20}^{+0.22}\times10^{-5} $&\citep{King:2012id}   \\ 
 CMB Quadrupole (Planck SMICA\footnote{Spectral Matching  Independent Component Analysis (SMICA) is a method to extract the CMB component \citep{Akrami:2018mcd} described  in \citet{4703509}.})&&$238.5$   && $76.6$  &&    &    \citep{Ade:2013nlj}\\
 CMB Octopole (Planck SMICA) && $239.0$  && $64.3$  &&    &   \citep{Ade:2013nlj}  \\
CMB Hemispher. Asym. (Planck) && $221$  && $-22$  &&$ 0.07$&     \citep{Akrami:2019bkn}\\
 CMB Hemispher. Asym. (WMAP) && $227$  && $-27$  &&$0.07$  &    \citep{Axelsson:2013mva}\\
 Maximum Acceleration (Pantheon)  &&$286.93\pm 18.52$  && $27.02\pm 6.50$&& $0.0018\pm 0.0002$  &  \citep{Kazantzidis:2020tko}\\
 Maximum Acceleration (Union2) &&$309_{-3}^{+23}$  && $18_{-10}^{+11}$&&   & \citep{Antoniou:2010gw}\\
     & \\
 \hhline{========}   
\end{tabular} 
\end{table*} 
\end{center}
The physical origin of these dipoles is described in the following subsections.

\subsubsection{Velocity radio dipole}
A large scale velocity flow dipole\footnote{For peculiar velocities as variation of the Hubble expansion produced by nearby nonlinear structures see in \citet{Wiltshire:2012uh} and for dipole anisotropy in radio source count, see in \citet{Bengaly:2017slg}.} was pointed out in   \citet{Watkins:2008hf,Kashlinsky:2008ut}.  The dipole moment of the peculiar velocity field (dipole bulk flow) which is a sensitive probe of the amplitude and growth rate of fluctuations on large scales \citep{Koda:2013eya} was investigated  by \citet{Watkins:2008hf,Feldman:2009es,Kashlinsky:2008ut,Kashlinsky:2009dw,Nusser:2011tu,Ma:2012tt,Turnbull:2011ty,Kashlinsky:2012gy,Hong:2014jla}. In many cases the results are controversial and there is a debate in the literature on the consistency with the $\Lambda$CDM model. 

A recent detailed analysis has indicated that 'tilted observers' within the bulk flow can be misled into inferring acceleration \citep{Tsagas:2011wq,Tsagas:2015mua,Tsagas:2021dsl} \citep[see also][for observational constraints of the deceleration parameter in a tilted universe]{Asvesta:2022fts}.

\citet{Colin:2019opb} use SnIa JLA data to demonstrate that the indications for cosmic acceleration found in the SnIa data disappears if a bulk flow induced anisotropy is  allowed in the SnIa data.  Thus, a bulk flow dipole (at $3.9 \sigma$) aligned with the local bulk flow is identified while any monopole (which can be attributed to $\Lambda$) is consistent with zero (at $1.4 \sigma$).\footnote{A recent model independent analysis of SnIa data (Pantheon) data \citet{Arjona:2019fwb} implementing machine learning has confirmed a $\sim 4.5 \sigma$ detection of the accelerated expansion even though that analysis did not allow for anisotropic dipole effects.}

It is usually assumed that our local (solar system) peculiar motion with respect to the CMB rest frame produces the CMB dipole anisotropy ($l=1$) \citep{Kogut:1993ag,Fixsen:1996nj} \citep[also known as solar dipole][]{Aghanim:2016yuo,Aghanim:2018fcm}. In the standard model, this implies that the LSS distribution  should have a similar kinematic dipole known as the velocity dipole or radio dipole which arises from the Doppler boosting of the CMB monopole and from special relativistic aberration effects \citep{Itoh:2009vc}. 

In order to describe  the origin of this dipole, a population of sources with power-law spectra depending on frequency $\nu$ is usually assumed
\be
S_{\nu}\propto \nu^{-\alpha}
\ee
where $S_{\nu}$ is the flux  density and $\alpha$\footnote{The individual spectral index $\alpha$ should not be confused with the  fine-structure  constant $\alpha$.} is an individual spectral index with typically assumed value  $\alpha \sim 0.75$ \citep{Bacon:2018dui}.
The integral source counts per unit solid angle above some limiting flux density $S_{\nu}$ can be approximated by a power law
\be
\frac{dN}{d\Omega}(>S)\propto S_{\nu}^{-x}
\ee
where  $x\sim 1$ and can be different for each survey. An observer moving with velocity $\upsilon \ll c$ with respect to the frame in which these sources are isotropically distributed sees a dipole anisotropy $1+D\cos{\theta}$ over the sky with amplitude \citep{10.1093/mnras/206.2.377}
\be
D=[2+x(1+\alpha)]\frac{\upsilon}{c}
\label{ampldipo}
\ee

According  to the  most  recent  measurements the inferred velocity of the Sun relative to the CMB rest frame is \citep{Akrami:2018vks,Aghanim:2018fcm}
\be
\beta\equiv \frac{\upsilon}{c}=(1.23357\pm 0.00036)\times10^{-3}
\ee
\be
or\quad \upsilon= 369.82\pm 0.11\, km\, s^{-1}
\ee
along the direction with galactic longitude and latitude $(l,b)=(264.021^0 \pm 0.011^0, 48.253^0 \pm 0.005^0) $ or $RA\sim 168^0$, $Dec\sim-7^0$ \citep{Akrami:2018vks,Aghanim:2018fcm}.

The CMB rest frame is conventionally taken to correspond to the standard of cosmic rest frame and is assumed to be statistically homogeneous and isotropic in the context of the FLRW model. In this rest frame the Hubble flow should be most uniform (minimum Hubble variation frame) and the comoving observers should not see a kinematic dipole. However it has been observed \citep{Wiltshire:2012uh,McKay:2015nea} that the  dipole  structure of the velocity field is less in the reference frame of the Local Group of galaxies than in the CMB frame. This persistence of the dipole structure of the velocity flow in the CMB frame at large distances is not unexpected if we are located in an underdensity \citep{Kraljic:2016acj}.

According to the standard model if the Universe is isotropic our velocity with respect to the CMB rest frame and our velocity relative to the LSS should be identical. However, as was first noted by  \citet{Singal:2011dy}, while the direction of the radio dipole is consistent with that of the CMB, the velocity of our local motion obtained from the radio dipole exceeds that obtained from the CMB dipole. Radio continuum surveys which sample the Universe at intermediate redshifts ($z\sim1$) have been used as an excellent probe to large scale isotropy and a discrepancy between the predicted and measured amplitudes of the velocity have been revealed  \citep{Gibelyou:2012ri,Rubart:2013tx,Kothari:2013gya,Tiwari:2015tba,Colin:2017juj,Bengaly:2017slg}. In particular the analysis by  \citet{Bengaly:2017slg} has shown that the radio dipole using the sky distribution of radio sources from the NRAO VLA Sky Survey (NVSS) dataset \citep{Condon:1998iy} and TIFR GMRT Sky Survey (TGSS) dataset \citep{Bengaly:2017slg,Intema:2016jhx,Singal:2019pqq} is $\sim2$ and $\sim5$ times larger than predicted by the mock realisations within the context of $\Lambda$CDM cosmology respectively. The above observed discrepancy between the radio and CMB dipoles has been confirmed by independent groups and could imply the existence of an anisotropic Universe.

Possible explanations of the violation of statistical isotropy are: systematics due to the incomplete sky coverage of the radio continuum surveys \citep{Blake:2002gx,Gibelyou:2012ri,Tiwari:2015tba}, intrinsic dipole in the local LSS \citep{Fernandez-Cobos:2013fda}, nearby  nonlinear  structures  of  voids  and walls and  filaments \citep{Wiltshire:2012uh}, remnant of the  pre-inflationary Universe \citep{Turner:1991dn} and  superhorizon perturbation \citep{Ghosh:2013blq,Das:2021ssc}.

\subsubsection{Quasar dipole}
The distribution of quasars across the sky may provide independent probe of the cosmological principle \citep{Singal:2014wra}. As previously discussed there is an expected anisotropy related to the CMB dipole anisotropy (about 1 part in 1000) due to our motion with respect to the CMB rest frame.  

\citet{Secrest:2020has} used mid-infrared data from the Wide-field Infrared  Survey  Explorer (WISE) \citep{Wright:2010qw} to create reliable AGN/quasar catalogs and a custom quasar sample from the new CatWISE2020 data release \citep{Eisenhardt_2020}. It was shown that there is a statistically significant dipole in the density of distant quasars with direction $(l,b) =(238.2^0,28.8^0)$ which is $27.8^0$  away from  the direction of the CMB dipole. Its amplitude was found to be $0.01554$, $\sim2$  times larger than predicted,  with statistical significance at the $4.9\sigma$ level (or  with a p-value of $5\times10^{-7}$) for a normal distribution. This result is in conflict with the cosmological principle.

\subsubsection{Fine structure constant $\alpha$ dipole} 
In the past $20$ years there has been interest in the possibility of the variation of the fine structure constant $\alpha\equiv e^2/(4\pi\epsilon_0\hbar c)$ (where $e$, $\epsilon_0$, $\hbar$, and $c$ are the electron charge, the vacuum permittivity, the reduced Planck’s constant, and the  speed of light) \citep{Murphy:2003hw,Webb:2010hc,Molaro:2013saa,Berengut:2013dta,Webb:2014lpa,Evans:2014yva,Bainbridge:2017lsj,Hu:2018lwv,Milakovic:2020tvq,Lee:2021kjr}  \citep[for review of varying fine structure constant, see][]{Martins:2017yxk}. 

The analysis by \citet{King:2012id,Webb:2010hc} uses the “many multiplet” (MM) method \citep{Webb:1998cq,Dzuba:1999zz,Webb:2000mn,Murphy:2000pz} to analyze quasar absorption line spectra obtained using the Ultraviolet and Visual Echelle  Spectrograph (UVES) \citep{10.1117/12.395512} on  the Very Large Telescope (VLT). It indicates both the violation of the cosmological principle and the spatial variation of the fine structure constant $\alpha$ which is approximated as a spatial dipole with direction $(l,b) = (330^0\pm15^0,-13^0\pm10^0)$ and amplitude $0.97_{-0.20}^{+0.22}\times10^{-5} $, preferred over a simple monopole model with significance at the  $4.2\sigma$ \citep[for possible systematics in this analysis, see ][]{Cameron:2012ku,Kanekar:2012fy,Levshakov:2012kv}. 

The variation of $\alpha$ across the sky was shown to be well fit by an angular dipole model of the form \citep{King:2012id}
\be
\frac{\Delta \alpha}{\alpha}\equiv \dfrac{\alpha-\alpha_0}{\alpha}= C_A \cos{\theta}+C_B
\ee
where $\alpha_0$ is the present local value, $\theta$ is the angle with respect to the dipole direction, $C_A$ is the angular amplitude of the dipole term and $C_B$ is a monopole term.

It is worth to note that the analysis by \citet{Martins:2017qxd} suggests that there are no robust indications of time or space variations of $\alpha$. However recent measurements of quasar absorption-line spectra indicate a spatially dependent value of fine structure constant $\alpha$ at a $\sim 4\sigma$ significance level over a simple monopole (no-variation) model \citep{Dumont:2017exu,Wilczynska:2020rxx}. In addition, it is found that the fine structure constant $\alpha$ dipole is anomalously aligned with other dipoles and the preferred direction in $\Delta \alpha/\alpha$ is correlated with the one in the distribution of SNIa \citep{Mariano:2012wx,Mariano:2012ia}. 

\subsection{BAO curiosities}
\label{BAO curiosities} 
As mentioned above (see Subsection \ref{Standard Rulers: early time calibrators}) the BAO measurements can be classified in two classes: galaxy BAO and Ly$\alpha$ BAO (with Ly$\alpha$ auto-correlation function and  Ly$\alpha$-quasar cross-correlation function). A $2.5 - 3\sigma$ discrepancy between the BAO peak position in the Ly$\alpha$ at an effective redshift of $z \sim 2.34$ and the CMB predictions from Planck/$\Lambda$CDM cosmological model has been found \citep{Font-Ribera:2013wce,Aubourg:2014yra,Bourboux:2017cbm}. 

For example,  \citet{Aubourg:2014yra} use the Ly$\alpha$ auto-correlation function and the Ly$\alpha$-quasar cross-correlation function to report the measurements of the BAO scale in the line-of-sight direction 
\be
D_H(z=2.40)/r_s=8.94\pm 0.22
\ee
and in the transverse direction
\be 
D_M(z=2.40)/r_s=36.6\pm 1.2
\ee
where $D_H(z)\equiv\frac{c}{H(z)}$ is the Hubble distance and $D_M(z)\equiv(1+z)D_A(z)=d_A(z)$ is the comoving angular diameter distance. These values are in  $\sim 2.3\sigma$ tension with CMB predictions $D_H(z=2.40)/r_s=8.586\pm 0.021$ and $D_M(z=2.40)/r_s=39.77\pm 0.09$ by Planck 2015 flat $\Lambda$CDM cosmology \citep{Ade:2015xua}. 

The galaxy BAO peak position in the matter correlation function $\xi(s)$ (see Eq. (\ref{corfun}) and Fig. \ref{figbaols}) and the measurements $D_H(z=2.40)/r_s$ and $D_M(z=2.40)/r_s$ were found to be consistent with CMB predictions. This discrepancy between galaxy and Ly$\alpha$ BAO constitutes the BAO anomaly which has been investigated in  \citet{Cuceu:2019for,Evslin:2016gre,Addison:2017fdm}. 

Using new Ly$\alpha$ BAO measurements from the BOSS survey and  from its extended version eBOSS in the SDSS DR$14$ the tension with CMB predictions was reduced  to $\sim 1.7\sigma$ \citep{Blomqvist:2019rah,Agathe:2019vsu} and from
eBOSS in the SDSS DR$16$ to only $\sim 1.5\sigma$ \citep{duMasdesBourboux:2020pck}.
 
\citet{Evslin:2016gre} argues that this anomaly arises by cosmological effects at $z<2.34$ and the tension is caused  by evolution of dark energy equation of state $w(z)$ for redshift range $0.57< z <2.34$.

\subsection{Parity violating rotation of CMB linear polarization (Cosmic Birefringence)}
\label{Parity violating rotation of CMB linear polarization}

In the standard model of elementary particles and  fields, parity violation is observed only in the weak interaction sector \citep{Lee:1956qn,Wu:1957my}. A certain class of quintessence models should generically generate such parity asymmetric physics \citep{Carroll:1989vb,Carroll:1998zi}. In particular a parity violating (nearly) massless axionlike scalar field $\phi$ (dark matter or dark energy) would rotate CMB polarisation angles of CMB photons as they travel from the last scattering surface ($z\approx1000$) to the present by a non-zero angle $\beta_a$ (cosmic birefringence). 

A Chern–Simons coupling between a time-dependent axionlike field $\phi(t)$ and the electromagnetic tensor and its dual in the Lagrangian density  \citep[e.g.][]{Ni:1977zz,Turner:1987bw,Carroll:1989vb}
\be
\mathcal{L}=\frac{1}{4}g_{\phi\gamma}\phi F_{\mu\nu}\tilde{F}^{\mu\nu}
\ee
induces a cosmic isotropic birefringence angle \citep[e.g.][]{Harari:1992ea,Fujita:2020aqt}
\be
\beta_a=\frac{1}{2}g_{\phi\gamma}\int_{t_s}^{t_0}\dot{\phi}dt
\ee
and produces a non-zero observed $EB$ spectrum  \citep{Lue:1998mq}
\be
C_l^{EB}=\frac{1}{2}\sin{(4\beta_a)}(C_l^{EE}-C_l^{BB})
\ee
where $g_{\phi\gamma}$ is a Chern-Simons coupling constant which has mass-dimension $-1$, $\tilde{F}^{\mu\nu}$ is the dual of the electromagnetic tensor of $F_{\mu\nu}$, and $t_0$ and $t_s$ are the times at present and last scattering surface, respectively.

Using a novel method developed in  \citet{Minami:2019ruj,Minami:2020xfg,Minami:2020fin}, a non-zero value of the isotropic cosmic birefringence $\beta_a=0.35\pm 0.14\,\deg$  ($68\%$ C.L) was recently detected in the Planck18 polarization data at a $2.4\sigma$ statistical significance level by  \citet{Minami:2020odp}. More recently, using the latest Planck public data release 4 \citep{Planck:2020olo} a birefringence angle of $\beta_a=0.30 \pm 0.11\,\deg$ ($68\%$ C.L) was reported by \citet{Diego-Palazuelos:2022dsq}.

These recent evidences of the non zero value of birefringence poses a problem for standard $\Lambda$CDM cosmology and indicates a hint of a new ingredient beyond this model. 

An axion or an axion-like particle with a weak coupling to photon as a possible source  of the cosmic birefringence was investigated by  \citet{Fujita:2020ecn} and a two-axion alignment model with periodic potential was investigated by  \citet{Obata:2021nql}. \citet{Takahashi:2020tqv} showed that if an ultralight axion coupled to photons forms domain walls due to inflationary fluctuations, the domain-wall network can explain the hint for isotropic cosmic birefringence found by \citet{Minami:2020odp}. This model predicts a testable peculiar anisotropic cosmic birefringence as well. In contrast to the approach of  \citet{Fujita:2020ecn}, this scenario explains the birefringence with the photon anomalous coefficient of the axion-like particle  $\sim O(1)$. Furthermore, birefringence inducing axion-like particles could be candidates for an early dark energy resolution to the Hubble tension \citep{Fujita:2020ecn}.  \citet{Bianchini:2020osu,Namikawa:2020ffr} study the anisotropic birefringence and constraints are derived. The axion field fluctuations over space and time generate anisotropic birefringence.

\subsection{Small-scale curiosities}
\label{Small-scale curiosities}
On small scales (on scales of hundreds of $kpc$ and below) the predictions of $\Lambda$CDM model are in many cases inconsistent with observations \citep{Kroupa:2010hf,Weinberg:2013aya,Nakama:2017ohe}. In particular observations on galaxy scales indicate that the $\Lambda$CDM model faces several problems in describing structures at small scales  ($\lesssim 1Mpc$) \citep[see][for a review]{Kroupa:2012qj,Kroupa:2014ria,DelPopolo:2016emo,Bullock:2017xww,Salucci:2018hqu}. Alternative models that modify the nature of dark matter have been used to solve these problems e.g. warm \citep{Bode:2000gq,Abazajian:2005xn,Viel:2005qj,Schneider:2013wwa,Viel:2013fqw}, fuzzy \citep{Hu:2000ke,Hui:2016ltb,Schive:2014dra,Irsic:2017yje}, self-interacting \citep{Carlson:1992fn,Blennow:2016gde,Garcia-Cely:2017qpx,Tulin:2017ara} and meta-cold dark matter \citep{Strigari:2006jf} \citep[see also][for a review]{deMartino:2020gfi}. Other models which have the potential to provide a solution to these problems have been proposed by  \citet{Foot:2014uba,Foot:2016wvj,Sawala:2015cdf,Schewtschenko:2015rno,Vogelsberger:2015gpr,Archidiacono:2017slj}. In particular  \citet{Foot:2014uba,Foot:2016wvj} argued that the existence of a dissipative hidden dark matter sector  (dark matter coupled to a massless dark photon) can solve some of these problems (core-cusp, missing satellites, and plane of satellites problem). 

These small scale signals include the following:
\subsubsection{The core-cusp curiosity} 
The core-cusp curiosity \citep{Moore:1994yx,Flores:1994gz,deBlok:2009sp} (see also \citealt{Cooke:2022upv,Lelli:2022hvc}) refers to a discrepancy between the  density of a dark matter halo profile of low-mass galaxies $\rho(r)\propto r^{-x}$ in N-body simulations (an important tool for evaluating the predictions of the $\Lambda$CDM model) with $1\lesssim x\lesssim 1.5 $ (cusp profile)\footnote{The well know Navarro–Frenk–White profile \citep{Navarro:1995iw,Navarro:1996gj} is cusped with $\rho(r\rightarrow 0)\sim r^{-1}$.} \citep{Navarro:1996gj,Navarro:1996bv,Moore:1999gc,Ferrero:2011au} and the astronomical observed profile with $x\sim 0$ (core profile) \citep{Moore:1994yx,Davis:1985rj,Navarro:1996gj,Navarro:1996bv,Flores:1994gz,Battaglia:2008jz,Walker:2011zu,Amorisco:2011hb,Genina_2017}.  \citet{McGaugh:1998tq} probe this problem in low surface brightness galaxies.

\subsubsection{The missing satellites problem (or dwarf galaxy problem)} 
The missing satellites problem (or dwarf galaxy problem) \citep{Kauffmann:1993gv,Klypin:1999uc,Moore:1999nt,Bullock:2010uy}  refers to an over-abundance of the predicted number of halo substructures in detailed collisionless N-body simulations compared to the observed number of satellite galaxies in the Local Group. In particular the $\Lambda$CDM model predicts orders of magnitude larger number of satellites ($\sim 1000$) than the observed number of dwarf galaxies ($\sim 50$) \citep{Mateo:1998wg,Moore:1999nt}. 

\subsubsection{The Too Big To Fail (TBTF) problem} The Too Big To Fail (TBTF) problem \citep{BoylanKolchin:2011de,BoylanKolchin:2011dk,Garrison-Kimmel:2014vqa,Papastergis:2014aba,Tollerud:2014zha,Kaplinghat:2019svz} refers to an inconsistency between the predicted mass of dark matter subhaloes in $\Lambda$CDM theory and the observed central mass of brightest satellite galaxies in the Local Group \citep{Garrison-Kimmel:2014vqa,Papastergis:2014aba} (also in the MW \citealt{BoylanKolchin:2011de,BoylanKolchin:2011dk} or in the  Andromeda (M31) \citealt{Tollerud:2014zha}). 
       
In particular the $\Lambda$CDM predicted central densities of the most massive dark matter subhalos are systematically larger than the inferred from kinematics of the brightest Local Group satellites \citep{Read:2005zv,BoylanKolchin:2011dk,Garrison-Kimmel:2014vqa}. An observed bright satellite is more likely to reside in subhalos with lower mass than is expected in a $\Lambda$CDM model. The simulated massive dark matter subhalos {\it 'failed'} to form a comparatively bright satellite galaxy. 
       
This problem is possibly related to the missing satellites problem but it is a distinct problem which dependents on the internal structure of subhalos or the central shapes of density profiles of satellite halos \citep{Garrison-Kimmel:2014vqa}. 
       
Alternative models that modify the nature of dark matter have been investigated to solve this problem: non-trivial dark matter physics \citep{Lovell:2016nkp,Bozek:2018ekc}, interaction between the dark matter and dark radiation components \citep{Schewtschenko:2015rno,Vogelsberger:2015gpr}, self-interacting dark matter \citep{Zavala:2012us,Vogelsberger:2012ku} and fuzzy dark matter \citep{Schive:2014dra}
       
\subsubsection{The problem of satellite planes} 
In the problem of satellite planes \citep{Kroupa:2004pt,Ibata:2014csa,Conn:2013iu,Pawlowski:2014una,Pawlowski:2018sys}  several satellite galaxies of the MW, of neighboring Andromeda galaxy (M31), and  of Centaurus A (CenA) are part of thin plane that is approximately perpendicular to the Galactic disk.  Moreover measurement of the motions of satellite galaxies has shown that their orbits appear to be correlated \citep{Pawlowski:2012vz,Ibata:2013rh,Sohn_2020}. This flattened structure and coherent motions of satellite galaxy systems is in inconsistency with the prediction of the $\Lambda$CDM model as inferred from simulations \citep{Pawlowski:2018sys}. The simulations based on $\Lambda$CDM cosmology indicate uncorrelated and close to isotropic satellite structures \citep{Libeskind:2005hs,Zentner:2005wh}. In these simulations the observed structure formations with spatial and kinematic coherence distribution are very rare with a probability $\sim 10^{-3}$ \citep{Pawlowski:2014una,Pawlowski:2018sys}.

\subsubsection{The angular momentum catastrophe} 
The angular momentum catastrophe \citep{vandenBosch:2001bp}  concerns a catastrophic angular momentum loss of gas during disk galaxies formation in Smooth Particle Hydrodynamics (SPH) \citep{Monaghan:1992rr} simulations. The formed disks in simulations according to the predictions of $\Lambda$CDM  have smaller scale lengths by a factor of $2-3$ compared with observed ones \citep{Bullock:2000ry}. An axion dark matter model may resolve this discrepancy between the observed and predicted angular momentum distributions of baryons (ordinary cold dark matter) in the dwarf galaxies \citep{Banik:2013rxa}.

\subsubsection{Baryonic Tully-Fisher Relation (BTFR)} 
Baryonic Tully-Fisher Relation (BTFR) \citep{McGaugh:2011nv,Lelli:2015wst}.
As mentioned above, the well known Tully-Fisher (TF) \citep{Tully:1977fu} empirical relation connects the velocity of  rotation of a spiral galaxy with its intrinsic luminosity while the Baryonic Tully-Fisher Relation (BTFR) \citep{McGaugh:2000sr,Verheijen:2001an,Gurovich:2004vd} Eq. (\ref{btfr}) is a scaling relation between the observed total baryonic mass $M_b$ (stars plus gas) of a spiral galaxy and its rotation velocity $V_c$ (see Subsection \ref{Tully-Fisher relation (TFR) as  distance indicator.}). The problem for $\Lambda$CDM model as inferred from simulations  \citep[e.g.][]{Dutton:2012jh} is that the BTFR leads to existence of a higher intrinsic scatter ($\sim 0.15\, dex$) and a lower slope ($s=3$) compared to the observed ( $\sim 0.10\, dex$ and $s\sim 4$) \citep{Lelli:2015wst}. \citet{McGaugh:2011ac} suggests the Modified  Newtonian Dynamics (MOND) \citep{Milgrom:1983pn} as a possible solution to this problem. However some simulations or semi-analytic approaches of galaxy formation within a $\Lambda$CDM cosmological context can reproduce a realistic BTFR slope but not its small scatter  \citep[e.g.][]{Sales:2016dmm, DiCintio:2015eeq,Geha:2006mc,Santos-Santos:2015lna}. 

\subsubsection{The void phenomenon } 
The void phenomenon\citep{Peebles:2001nv}  refers to the emptiness of voids (the number of small galaxies in the void). Cosmological N-body simulations in the context of $\Lambda$CDM have established a clear prediction \citep{Gottloeber:2003zb} that many small dark matter haloes should reside in voids \citep{Peebles:2003pk,Peebles:2007qe}. This is consistent with observations on large scales but is inconsistent with observations on small scales. In particular the local void contains much fewer galaxies than expected from  $\Lambda$CDM theory \citep{Tikhonov:2008ss}.\\
     
\subsection{Age of the Universe}
\label{Age of the Universe} 
A lower limit can be set on the age of the Universe by the ages of the oldest stars (or oldest astrophysical objects)  because on cosmological timescales they form shortly after the Big Bang. In the context of $\Lambda$CDM cosmology, the standard theory \citep{Haiman:1995cr,Bromm:2001bi,Abel:2001pr,Bromm:2003vv,Bromm:2013iya} and cosmological numerical simulations  \citep{Yoshida:2003rw,Hirano:2015wxa,Xu:2016tkk} predict that the first stars, the so-called population III (Pop III), formed in dark matter minihaloes of typical mass $M\sim10^5 -10^6 M_{\odot} $ at redshifts $z\sim 20-30$ (about $100$ million years after the Big Bang i.e. about around the end of the cosmic dark ages) \citep[for models indicating late, $z\sim 2-7$, Pop III star formation, see][]{Tornatore:2007gb,Johnson:2009fh}. 

The age of the Universe $t_*$ as obtained from local measurements using the ages of oldest observed stars (the so-called population II (Pop II)) in the MW appears to be larger and in some tension with the corresponding age of the Universe $t_U$ obtained using the CMB Planck data in the context of $\Lambda$CDM \citep{Verde:2013wza}. 

The age of the Universe in the flat $\Lambda$CDM model is an observable  determined  by the integral 
\begin{widetext}
\be
t(z)=\int_0^{z_t}\frac{dz'}{(1+z')H(z')}=
\frac{1}{H_0}\int_0^{z_t}\frac{dz'}{(1+z')\left[\Omega_{0m}(1+z')^3 +\Omega_{0r}(1+z')^4+(1-\Omega_{0m})\right]^{1/2}}
\label{ageun}
\ee
\end{widetext}
where $t$ is the cosmic time corresponding to redshift $z_t$. Thus the age of the Universe is $t_U=t(z_t=\infty)$.

For example the age of the MW Population II halo, metal deficient, high velocity subgiant HD-$140283$ (also known as Methuselah star) is estimated to be $t_*=14.46\pm 0.31\,Gyr$ by  \citet{Bond:2013jka} and using new sets of stellar models is estimated to be $t_*=14.27\pm 0.80\,Gyr$ by \citet{VandenBerg_2014}. These estimates of the age of this star are slightly higher ($\sim 2\sigma$) than the age of Universe $t_U=13.800\pm 0.024\,Gyr$ inferred by CMB Planck18 data \citep{Planck:2018vyg} but within the errors it does not conflict with this age. 

Despite of the above indications the analysis by \citet{Jimenez:2019onw} using new parallaxes from the Gaia space mission \citep{Perryman:2001sp,Prusti:2016bjo} in place of the older HST, reports a revision of the age of HD-$140283$ to $t_*=13.5\pm 0.7\,Gyr$ which is more compatible with the age $t_U$ inferred by Planck data. Also the analysis by \citet{Valcin:2020vav} using populations of stars in globular clusters (very-low-metallicity stars) reports age of the Universe constrained to be larger than $t_*=13.5_{-0.14}^{+0.16}\,Gyr$.

Clearly, Eq. (\ref{ageun}) indicates that in a $\Lambda$CDM Universe the quantities  $H_0$, $t_U$ and $\Omega_{0m}$ are related. Therefore the determination of the age of older objects based on local Universe observations provides a test of the current cosmological model and plays an important role in the studies of Hubble and spatial curvature tensions \citep{Jimenez:2019onw,DiValentino:2019qzk,DiValentino:2020srs,Wei:2022plg}.

\subsection{The Lithium problem}
\label{The Lithium problem}
It has long been known (since the early 80’s) that absorption lines in the photospheres of old, metal-poor (Population II) halo stars in the Milky Way's halo indicate $\sim 3.5$ times less primordial abundance of lithium isotope $^7Li$ compared to  the prediction of the standard BBN theory \citep{Cyburt:2003fe,Asplund:2005yt,Cyburt:2008kw}. The observed value of the lithium abundance\footnote{Usually in the literature the abundance of lithium is expressed by  $A(^7Li)=12 +\log_{10} [n(^7Li)/n(H)]$ where $n$ is the number density of atoms and $12$ is the solar hydrogen abundance.}  $ ^7Li/H=(1.6\pm 0.3)\times 10^{-10}$ \citep{Zyla:2020zbs} is smaller than the theoretically expected value $^7Li/H=(5.62\pm 0.25)\times 10^{-10}$  \citep{Pitrou:2018cgg} at a level $\sim 5\sigma$. This constitutes the lithium problem \citep{Fields:2011zzb}. No such problem exists for the observed abundances of other light elements  $^2H$ (or $D$), $^3He$, and $^4He$ that are in broad quantitative agreement with BBN  predictions + WMAP/Planck cosmic baryon density $\Omega_b$ which is deduced by the CMB \citep{Cyburt:2015mya,Tanabashi:2018oca}. 

A number of theoretical or experimental studies in the literature have attempted to address the lithium problem  \citep[e.g.][]{Hammache:2013jdw,Pizzone:2014xrw,Yamazaki:2014fja,Kusakabe:2014moa,Poulin:2015woa,Sato:2016len,Goudelis:2015wpa,Salvati:2016jng,Hou:2017uap,Coc:2017pxv,Mori:2019cfo,Hayakawa:2020bjr,Ishikawa:2020fbm,Iliadis:2020jtc}.
For example the analysis by \citet{Clara:2020efx} shows that the variations in Nature’s fundamental constants on primordial nucleosynthesis provide a possible solution to the lithium problem. Specifically, they determined that if the value of the fine-structure constant $\alpha$ at the primordial nucleosynthesis epoch was larger than the present one by ten parts per million of relative variation, the lithium problem could be resolved. 

It was also proposed by \citet{DiBari:2013dna} that decaying dark matter into dark radiation in the early Universe can solve the long-standing lithium problem, leaving completely unaffected the abundance of other light elements. This mechanism was also proposed to alleviate the $H_0$ tension (see Subsection \ref{Late time modifications1}) but is severely constrained by the Planck data \citep{Anchordoqui:2020djl}. 

Measurements of lithium  \citep[e.g.][]{Sbordone:2010zi,Melendez:2010kw} may not be representative of the cosmological production mechanism \citep{Spite:2012us,Iocco:2012vg}. It is thus possible that the solution to the lithium problem lies in the effects of stars in the lithium abundance. Therefore a precise knowledge of the stellar formation process and physics of stellar atmosphere is necessary to provide a fully satisfactory solution.
Thus, possible solutions to this persistent problem can be classified into four categories  \citep[see][for a review]{Fields:2011zzb,Cyburt:2015mya,Mathews:2019hbi}: 
\begin{itemize}
    \item
    Cosmological solutions (e.g. new theory beyond the standard BBN including variations of fundamental constants) \citep{Franchino-Vinas:2021nsf,Clara:2020efx,Larena:2005tu,Kohri:2006cn,Berengut:2009js,Kawasaki:2010yh,Cheoun:2011yn,Coc:2006sx,Hou:2017uap,Luo:2018nth}
    \item
    Nuclear Physics solutions (e.g. reactions destroy lithium during or after BBN) \citep{Cyburt:2003ae,Boyd:2010kj,Chakraborty:2010zj,Coc:2011az,Broggini:2012rk,Mori:2019cfo,Hayakawa:2020bjr,Ishikawa:2020fbm}
    \item 
    Astrophysical solutions (e.g. stars destroy lithium after BBN) \citep{Pinsonneault:2001ub,Richard:2004pj,Korn:2006tv,10.1093/mnras/stv1384}
    \item
    Extensions of the standard model (e.g. simultaneous imposition of photon cooling after BBN, X-particle decay and a primordial magnetic field \citep{Yamazaki:2014fja,Yamazaki:2017uvc}, destruction of $^7$Be due to the decay of a sterile neutrino \citep{Salvati:2016jng} and including new particles or interactions \citep{Goudelis:2015wpa}.
 \end{itemize}
     
\subsection{Quasars Hubble diagram}
\label{Quasars Hubble diagram}   
The quasar distances can be estimated from their X-ray (coronal) emission generated by a plasma of hot relativistic electrons around the accretion disk. The emission is induced through inverse-Compton scattering processes and UV emission generated by the accretion disk where the gravitational energy of the infalling material is partially converted to radiation \citep{Risaliti:2018reu,Lusso:2019akb}. 

In recent years model independent derivation\footnote{\citet{Yang:2019vgk} argued that even though the data used in this approach are valid, their analysis involves significant uncertainties as it may lead to spurious artificial tensions.} of the distance modulus–redshift relation using high-$z$ quasars ($z\lesssim 7$) as distance indicators (quasars Hubble diagram) provides a new bright standard candle in the higher redshifts and earlier times beyond SnIa. The method used is based on a non-linear relation between the X-ray and the UV emissions at low redshift which is of the form \citep{1986ApJ...305...83A,Lusso:2009nq,Steffen:2006px,Risaliti:2015zla,Risaliti:2018reu}
\be
\log_{10}L_X=\gamma_q \log_{10} L_{UV}+\beta_q
\ee
where $L_X$ and $L_{UV}$ are the rest-frame monochromatic luminosities at $2$ $keV$ and at $2500$ $\AA$, respectively \citep{1986ApJ...305...83A}. Also $\gamma_q \sim 0.6$ \citep{Steffen:2006px,Lusso:2009nq,Salvestrini:2019thn,Vignali:2002ct,Vagnetti:2010rw,Bisogni:2021hue,Risaliti:2018reu} and $\beta_q$ are fitting parameters of the luminosities. 

Extending a Hubble diagram up to redshift $z=5.5$ shows hints for phantom dark energy \citep{Risaliti:2018reu,Banerjee:2020bjq,Lusso:2019akb}. In particular the distance modulus-redshift relation for a sample of $1598$ quasars at higher redshift ($0.5< z <5.5$)  is in disagreement with the concordance model at a $\sim 4\sigma$ significance level\footnote{The analyses of the high$-z$ quasar data has lead to a wide range of conclusions \citep{Melia:2019nev,Yang:2019vgk,Khadka:2019njj,Li:2021onq,Velten:2019vwo}. For example  \citet{Yang:2019vgk} concludes that the log polynomial expansion generically fails to recover flat $\Lambda$CDM beyond $z\sim2$, thus implying that the previously derived  $\sim 4 \sigma$ tension may be artificial.} \citep{Risaliti:2018reu}.  Moreover, the analysis by \citet{Lusso:2019akb} building a Hubble diagram by combining three samples of Pantheon, quasars, and gamma-ray bursts (GRBs) reported tension at more than the $\sim 4\sigma$  statistical level with the flat $\Lambda$CDM model. 

However \citet{Khadka:2020tlm} using an updated, larger QSO dataset \citep{Lusso:2020pdb} containing $2421$ QSO measurements with redshifts up to $z\sim 7.5$ have demonstrated that the $L_X$-$L_{UV}$ relation parameter values depend on the cosmological model thus cannot be used to constrain cosmological parameters. Recently \citet{Khadka:2021xcc} demonstrated  that the parameter values of the largest of seven sub-samples in this QSO dataset depend on the cosmological model while the second and third biggest sub-samples appear standardizable.

\subsection{Oscillating signals in short range gravity experiments}
\label{Oscillating signals in short range gravity experiments}
The most constraining test of gravity at very short distance (sub-millimeter) scales looking for departures from Newtonian gravity is implemented via torsion balance experiments. A reanalysis of short range gravity experiments has indicated the presence of an oscillating force signal with sub-millimeter wavelength  \citep{Perivolaropoulos:2016ucs,Antoniou:2017mhs}. In particular  \citet{Perivolaropoulos:2016ucs} has indicated the presence of a signal at $2\sigma$ level of spatially oscillating new force residuals in the torsion balance data of the Washington experiment \citep{Kapner:2006si}. As an extension of the previous analysis the study by  \citet{Antoniou:2017mhs} using  Monte  Carlo  simulation and analysing the data of the Stanford Optically Levitated Microsphere Experiment (SOLME) which involves force measurements an optically levitated  microsphere as a function of its distance $z$ from a gold  coated silicon cantilever \citep{Rider:2016xaq} reports a oscillating signal at about $2\sigma$ level. 

The sub-millimeter scale of the quantum nature of dark energy may be written as
\be
\lambda_{de}\equiv \sqrt[4]{\frac{\hbar c}{\rho_{de}}}\approx 0.085\,mm
\ee
where it is assumed that $\Omega_{0m}=0.3$ and $H_0=70$ $km$ $s^{-1}$ $Mpc^{-1}$.

Thus, if the accelerating expansion of the Universe is connected with effects of modified gravity due to quantum gravity \citep{Addazi:2021xuf} it would be natural to expect some modification of Newton's law at the submillimeter scale.

The deviations from Newton’s law of gravitation is usually described in the context of scalar-tensor \citep{EspositoFarese:2000ij,Gannouji:2006jm} and flat extra dimension theories \citep{ArkaniHamed:1998rs,ArkaniHamed:1998nn,Antoniadis:1998ig,Perivolaropoulos:2002pn,Perivolaropoulos:2003we,Floratos:1999bv,Kehagias:1999my} by a short range Yukawa type potential of the form
\be
V_{eff}=-G\frac{M}{r}\left(1+\alpha_Y e^{-m r}\right)
\ee
where $\alpha_Y$ and $m$ are parameters to be constrained by  the data.

Alternatively, a power law ansatz may also generalize the gravitational potential to the form 
\be
V_{eff}=-G\frac{M}{r}\left[1+\bar{\beta}^k\left(\frac{1}{mr}\right)^{k-1}\right]
\ee
This power law parametrization is motivated by some brane world models \citep{Donini:2016kgu,Benichou:2011dx,Bronnikov:2006jy,Nojiri:2002wn}.

For $m^2<0$ the Yukawa gravitational potential becomes oscillating and takes the form
\be 
V_{eff}=-G\frac{M}{r}\left[1+\alpha_Y\cos{(mr+\theta)}\right]
\ee
where $\theta$ is a parameter.

Recently a reanalysis of the data of the Washington experiment searching for modifications of Newton’s Law on sub-millimeter scales by  \citet{Perivolaropoulos:2019vkb} has indicated that a spatially oscillating signal is hidden in this dataset. In addition it is shown that even though this signal cannot be explained in the context of standard modified gravity theories\footnote{For a free massive scalar \citet{Perivolaropoulos:2020uqy} investigate the physical conditions that can eliminate the tachyonic instabilities or at least drastically change their lifetime.} (viable scalar tensor and $f(R)$ theories), it occurs naturally in nonlocal (infinite derivative) gravity theories \citep{Edholm:2016hbt,Kehagias:2014sda,Frolov:2015usa} that predict such spatial oscillations without the presence of ghosts (instabilities) and has a well-defined Newtonian limit. 

The origin of oscillating signals could be due to three possible effects:
\begin{itemize}
    \item
    A statistical fluctuation of the data.
    \item
    A  periodic  distance-dependent  systematic  feature in  the  data. 
    \item
    A signal for a short distance modification of GR (e.g. non-local modified theory of gravity). 
\end{itemize}
In the later case, it is important to identify modified theories that are consistent with such an oscillating signal and are not associated with instabilities  \citep[e.g.][]{Tomboulis:1997gg,Biswas:2013cha}. 

\subsection{Anomalously low baryon temperature}
\label{Anomalously low baryon temperature}
The Experiment to Detect the Global Epoch of Reionization Signature (EDGES) collaboration \citep{Bowman:2018yin} report anomalously low baryon temperature $T_b\approx 4K$ at $z\approx 17$ (half  of its expected value). This temperature was inferred from the detection of global (sky-averaged) $21$-$cm$ absorption signal which is centred at a frequency of $\sim78$ $MHz$. The absorption depth of cosmic CMB photons at redshifts range $15\lesssim z \lesssim 20$ estimated by EDGES is more than twice the maximal value expected in the $\Lambda$CDM model, at $\sim 3.8\sigma$ significance. 

Possible explanations of this discrepancy were investigated and various models were proposed  \citep[e.g.][]{Hill:2018lfx,Fraser:2018acy,Munoz:2018pzp,Boyarsky:2019fgp}. For example  \citet{Hill:2018lfx} argue that EDE can explain this anomaly.

The EDGES observation has been used to constrain various cosmological models of dark matter and dark energy  \citep[e.g.][]{Yang:2019nhz,Barkana:2018qrx,Kovetz:2018zan,Hill:2018lfx}.

\subsection{Colliding clusters with high velocity}
\label{Too-rapid formation}
Observed galaxy clusters like the massive ($\sim 10^{15} M_{\odot}$) high-redshift ($z=0.87$) interacting pair known as El Gordo (ACT-CL J0102-4915) \citep{Menanteau:2011xy} have a very high relative velocity. This implies that formation of large structures may have taken place earlier than expected in $\Lambda$CDM cosmology.  \citet{Asencio:2020mqh} based on light cone tomography estimated that the too-early formation of El Gordo rules out $\Lambda$CDM cosmology at $6.16\sigma$ confidence. The early and rapid formation of clusters which consist of two colliding massive galaxy clusters at a high redshift may constitute a problem of the $\Lambda$CDM model. \citet{Asencio:2020mqh} argue that MOND with light sterile neutrinos model as suggested by  \citet{Haslbauer:2020xaa} can resolve this issue.

\begin{center}      
\begin{table*}    
\centering 
\caption{Some existing and upcoming large-scale structure  missions/experiments.}
\label{explss} 
\vspace{2mm}
\begin{tabular}{c ccc  ccccccc ccccccc} 
\hhline{================}
   & \\
Experiments&& Type&&  Probes  && Redshift   && Wavelengths   && Operator &&   Duration &&Refs.  \\
  &&  &&  &&  &&     &&        &&   &&  &&    \\
    & \\
   \hhline{================}
     & \\
Euclid&& Space&& WL, BAO     && $z \lesssim 6$      && $550\, nm-2\,\mu m $ &&  ESA   && $>2022$  && \citep{Laureijs:2011gra}\\ 
Vera C. Rubin Obs.&& Ground&& WL, BAO  && $z \lesssim 7.5$       && $320-1060\, nm$   && LSST    && $>2022$    && \citep{Abell:2009aa} \\
Gaia && Space&&  Astrometry   && $z\simeq 0$    && $320-1000\,nm$   && ESA  && $>2013$   &&\citep{Prusti:2016bjo}  \\ 
 JWST&&  Space && WL  &&$z \lesssim 15$    && $0.6-28.3\,\mu m$   &&NASA-ESA-CSA   &&   $>2021$  && \citep{Gardner:2006ky} \\ 
GAUSS&& Space&&WL $3\times2$pt  &&$z \lesssim 5$       && $0.5-5\, nm$   &&    && $>2035$    &&\citep{Blanchard:2021ffq}   \\  
  & \\
 \hhline{================}   
\end{tabular} 

\end{table*} 
\end{center} 

\begin{center}      
\begin{table*}    
\centering 
\caption{Some existing and upcoming CMB missions/experiments.}
\label{expcmb} 
\vspace{2mm}
\begin{tabular}{c ccc  ccccccc cccccccc} 
\hhline{=================}
   & \\
Experiments&& Type&&Detectors  &&Frequencies\footnote{The main  CMB  channels have frequencies $70-217$ $GHz$ with spectrum peak at a frequency of $\sim 160\, GHz$.} &&  Resolution\footnote{ The angular resolution is a function of  frequency, with lower frequencies having a worse angular  resolution.} && Sensitivity\footnote{The sensitivity is a function of  frequency and its estimates scaled to $1^0$ assuming that the noise is white.}  &&Sky  &&Duration &&Refs.  \\
  &&  &&  &&  ($GHz$) && ($arcmin$) &&  ($\mu K\, arcmin$)   &&  Cover   &&  &&    \\
    & \\
   \hhline{=================}
     & \\
Planck&& Space&& $74$  && $25-1000$          &&$5-33$&& $\sim 30$    && All &&2009-2013    && \citep{Akrami:2018vks}    \\
CMB S4    && Ground&& $500\cdot10^3$   && $30-270$    &&$0.8-11$  && $\sim1$ && $70 \%$  && $>2027$  && \citep{Abazajian:2016yjj}   \\ 
SO LAT &&  Ground   && $30\cdot10^3$   && $27-280$  &&$0.1$ && $\sim 6$  && $40 \%$ &&$>2021$  && \citep{Ade:2018sbj}    \\ 
 SO SATs &&  Ground   && $30\cdot10^3$ && $90-280$  && 0.5 && $\sim2$  && $10 \%$ &&$>2021$  && \citep{Ade:2018sbj}    \\ 
SPT-3G: &&  Ground   && $16\cdot10^3$ && $90-280$  && 1 && $\sim 3.5, 6$  && $10 \%$ &&$>2017$  && \citep{SPT-3G:2014dbx}    \\ 
     & \\
 \hhline{=================}   
\end{tabular} 

\end{table*} 
\end{center} 

\begin{center}      
\begin{table*}    
\centering 
\caption{Some existing and upcoming GW experiments/observatories}
\label{expgw}
\vspace{2mm}
\begin{tabular}{c ccc  ccccccc ccccc} 
\hhline{==============}
   & \\
Experiments&& Type/Detectors  && Arms   &&Frequencies\footnote{GW spectrum could span a wide range of frequencies, thus there are numerous proposed detectors, including pulsar timing arrays (PTAs)  \citep{Detweiler:1979wn,1990ApJ...361..300F,Hellings:1983fr} ($10^{-9}$ to $10^{-6}\,Hz$) such as Square Kilometre Array (SKA) \citep{Janssen:2014dka} (with the three collaborations, North American Nanohertz Observatory for Gravitational-waves (NANOGrav) \citep{McLaughlin:2013ira}, European Pulsar Timing Array (EPTA) \citep{Desvignes:2016yex} and Parkes Pulsar Timing Array (PPTA) \citep{Hobbs:2013aka} members of the International Pulsar Timing Array (IPTA) \citep{Verbiest:2016vem}) and atom interferometery such as Atomic  Experiments for Dark Matter and Gravity Exploration (AEDGE) \citep{Bertoldi:2019tck}.} && Location &&   Duration &&Refs.  \\
  &&    &&  &&  ($Hz$)     &&        &&   &&  &&    \\
    & \\
   \hhline{==============}
     & \\
Adv. LIGO&& Ground/Laser interf.  &&  $2\times4\,km$     &&$10-10^3$    &&  Hanford, USA    && $>2015$    && \citep{TheLIGOScientific:2014jea}   \\ 
Adv. LIGO&& Ground/Laser interf. && $2\times4\,km$     &&$10-10^3$    && Livingston, USA&& $>2015$    && \citep{TheLIGOScientific:2014jea}   \\
Adv. Virgo&& Ground/Laser   interf.  &&  $2\times3\,km$   && $10-10^3$   &&   Pisa, Italy     &&$>2016$   &&\citep{TheVirgo:2014hva}     \\ 
KAGRA && Undergr./Laser interf.   && $2\times3\,km$  && $10-10^3$   &&     Kamioka, Japan  &&  $>2020$  && \citep{Somiya:2011np}   \\ 
CE && Ground/Laser interf.  && $2\times40\,km$  && $5-4\cdot10^3$   &&    USA &&  $>2030$  && \citep{Evans:2016mbw} \\ 
  LISA&& Space/Laser interf. && $3\times2.5 \,Gm$  &&  $10^{-4}-10^{-1}$   && Heliocentric  orbit    &&  $>2034$  &&\citep{Audley:2017drz}  \\
 Taiji&& Space/Laser interf.&& $3\times2 \,Gm$  &&  $10^{-4}-10^{-1}$   && Heliocentric  orbit    &&  $>2033$  &&\citep{Hu:2017mde} \\ 
TianQin&& Space/Laser interf.&& $3\times0.1 \,Gm$  &&  $10^{-4}-1$   && Geocentric  orbit    &&  $>2035$  &&\citep{Luo:2015ght} \\  
DECIGO&& Space/Laser interf. &&   $4\times3\times 1 \,Mm$  &&  $1-10 $   && Heliocentric  orbit    &&  $>2027$  &&\citep{Kawamura:2011zz,Kawamura:2006up} \\ 
ET&& Undergr./Laser interf.&& $3\times2\times10 \,km$  &&  $1-10^4 $   &&      &&  $>2035$  &&\citep{Punturo:2010zz} \\ 
  & \\
 \hhline{==============}   
\end{tabular} 
 
\end{table*} 
\end{center} 

\section{Conclusions-Discussion-Outlook} 
\label{sec:Discussion}  
In the present  review, we discussed  in a unified manner many existing  curiosities in cosmological and astrophysical data that appear to be in some tension ($2\sigma$ or larger) with the standard $\Lambda$CDM model as specified by the Planck18 parameter values. The Hubble tension is the most significant  observational indication that the current standard model $\Lambda$CDM may need to be modified after more than 20 years since its establishment.

In addition to the well known tensions ($H_0$ tension, $S_8$ tension and $A_L$ anomaly), we provided a list of the non-standard cosmological signals in cosmological data. We presented the current status of these signals and their level of significance and also referred  to recent resources where more details can be found for each signal. These signals have a lower statistical significance level than the $H_0$ tension but may also constitute hints towards new physics. We also briefly discussed possible theoretical approaches that have been considered in order to  explain the non-standard nature of these signals. 

We also discussed the possible generic extensions of $\Lambda$CDM model. Generic extensions of $\Lambda$CDM may allow for a redshift dependence of the parameters $w$, $\mu_G$, $\Sigma$ and $\alpha$ as well as a possible large scale spatial dependence of these parameters which could violate the cosmological principle. {\it Varying fundamental constants} can potentially address the Hubble tension, the fine structure constant $\alpha$ dipole, the lithium problem, the growth tension, the curious SnIa $M$ signals (variation of the SnIa absolute magnitude $\cal M$), quasar signals and the ISW CMB signal.

The strategic approach required for the identification of new physics may include the following three steps: 
\begin{itemize}
\item
Tune current missions towards verification or rejection of non-standard signals.
\item
Identify favored parametrizations of $H(z,w(z),r)$, $\mu_G(z,r)$, $\Sigma(z,r)$, $\alpha(z,r)$ assuming that at least some of the non-standard signals are physical.
\item
Identify theoretical models (field Lagrangians) that are consistent with these parametrizations that can address simultaneously more than one of these tensions. Interestingly, for example only a small subset of modified gravity models is consistent with the weak gravity + $\Lambda$CDM background \citep{Wittner:2020yfc,Pizzuti:2019wte,Gannouji:2020ylf,Gannouji:2018ncm} suggested in the context of the $S_8$ tension.  
\end{itemize}
In the next decades new observational data from existing and upcoming missions/experiments (see Tables \ref{explss}, \ref{expcmb} and \ref{expgw}) will improve measurements and open up a wide range of new directions in the explanation of the curiosities of $\Lambda$CDM cosmology and understanding of cosmological physics. Here we provide an incomplete list of these missions:
\begin{itemize}
\item
{\bf Euclid:} The European Space Agency (ESA) Euclid mission \citep{Laureijs:2011gra} is planned for launch in July-December 2022. The goals of Euclid are to investigate the nature of dark matter, dark energy and gravity and thus to provide a better knowledge  of the origin of the accelerated expansion of the Universe \citep{Laureijs:2011gra,Refregier:2008js,Cimatti:2009is,Sartoris:2015aga,Amendola:2012ys}. The optical and  NIR Euclid survey using the cosmological WL and BAO probes will detect a high number of galaxy clusters up to redshift $z \sim 2$ (and possibly higher) in a redshift range that is sensitive to dark energy \citep{Adam:2019ewx} and will provide consistent growth rate data in both the low-z and high-z regimes. Therefore, Euclid will improve significantly the constraints on cosmological parameters such as $\sigma_8$ and the mass density parameter $\Omega_{0m}$ with a precision of $\sim10^{-3}$ for $\Lambda$CDM  \citep{Sartoris:2015aga}. The Euclid survey will also measure the equation of state parameter of dark energy $w_{DE}$ with higher precision ($\sim 1\%$) than precursor surveys. Stochastic inhomogeneities are expected to lead to an intrinsic uncertainty in the values of cosmological parameters obtained with such high redshift surveys. The corresponding cosmic variance in the context of Euclid for the measurement of $H_0$ has been shown to be limited to about $0.1\%$ \citep{Fanizza:2021tuh}.  Thus Euclid and other deep surveys ($z\gtrsim 0.15$) will provide an estimation of the $H_0$ which will be more precise than the low redshift surveys ($ z\lesssim 0.15$). Such improved constraints from Euclid in combination with contemporary surveys \citep{EUCLID:2020syl,Euclid:2021cfn, Euclid:2021frk} will allow the verification or rejection of many of the non-standard signals discussed in this review and will also help distinguish among the favored theoretical models that have been proposed for the explanation of these signals.
\item
{\bf Vera C. Rubin Observatory Legacy Survey of Space and Time:} The Large Synoptic Survey Telescope (LSST), recently renamed the Vera C. Rubin Observatory LSST \citep{Ivezic:2008fe,Abell:2009aa} is a future survey of the southern sky planned for the beginning in 2022. The Vera C. Rubin Observatory based in Chile with an $8.4\,m$ ($6.5\, m$ effective) telescope in six bands, targeting at least $18,000\,deg^2$ of high galactic latitude sky, will provide databases including $25$ billion galaxies with  $\gtrsim 0.2$ arcsecond pixel sampling \citep{Marshall:2017wph}. The main cosmological goals of the Vera C. Rubin Observatory ground-based project are to investigate the nature of dark matter and the dynamical behavior of dark energy by measuring WL and BAO \citep{Marshall:2017wph}.  Vera C. Rubin observatory will detect enormous number of galaxies and in combination with the Euclid BAO survey will probe an unprecedented range of redshifts. These surveys can determine $w_{DE}(z)$ in bins of redshift and their dark energy constraining power could be orders of magnitude greater than that of precursor surveys \citep{Abate:2012za}. The provided improved constraints on cosmological parameters allow to address potential systematics and to ensure that any measured tension is robust. In addition, the Vera C. Rubin project would be a useful tool in testing the models which have been used to explain these tensions.

\item
{\bf CMB-S4:}
The fourth generation\footnote{Planck was the  third generation space mission  which mapped the anisotropies of CMB.} (Stage-$4$) ground-based CMB experiment (CMB-S4) \citep{Abazajian:2016yjj,Abazajian:2019eic,CMB-S4:2020lpa}, is planned to start observations in 2027. It is anticipated  to be the definitive  CMB polarization experiment. The goals of CMB-S4 are to detect the signature of primordial gravitational waves in order to shed light on models of inflation, to search  for previously undiscovered light relic particles in order to study the dark Universe, to map normal and dark matter in the cosmos separately and to explore the time-variable millimeter-wave sky \citep{Abazajian:2016yjj}. The CMB-S4 survey in combination with external cluster surveys which are sensitive to different redshift ranges such as the Vera C. Rubin will provide detailed cluster data \citep{Abazajian:2016yjj}. These data will be used to study the growth of cluster scale perturbations, to improve constraints on cosmological parameters and to test the alternative models or extensions of $\Lambda$CDM which can be used to clarify the origin of many of the tensions and non-standard signals referred in the present review. The CMB-S4  will also contribute to neutrino cosmology providing compelling sensitivity in the constraint of the effective number of relativistic species, $N_{eff}$ and of the sum of the neutrino masses  $\sum m_{\nu}$ (see \citealt{Planck:2018vyg,DiValentino:2021imh}, for cosmological constraints and  \citealt{Capozzi:2021fjo}, for three-neutrino  analysis). This project has also the potential to constrain  $\Delta N_{eff}\equiv N_{eff}-N_{eff}^{SM}\simeq 0.060$ at $95\%$ confidence level \citep{Abazajian:2019eic}. Planck has provided a constraint  $\Delta N_{eff}\simeq 0.126$ at $95\%$ confidence level using temperature and polarization TT, TE, EE + lowE data \citep{Planck:2018vyg}. The improved constraints on $N_{eff}$ will enable us to test the scenarios with modifications of $\Lambda$CDM model in the light relic sector.
\item
 {\bf Gaia:}
The Gaia satellite was launched at the end of 2013 \citep{Perryman:2001sp,Prusti:2016bjo}. This European Space Agency (ESA) mission Gaia provides data that allow us to determine with high accuracy positions, parallaxes and proper  motions  for more than $1$ billion sources. There have been two data releases GDR$1$ \citep{Brown:2016tnb} and GDR$2$ \citep{Brown:2018dum} of Gaia results. Using quasars \citet{Lindegren:2018cgr} found that the GDR$2$ suffer from the parallax zero point (ZP) error.  \citet{Riess:2020fzl} refer to this additional error as parallax offset because it is not a single value but depends on the color or/and magnitude of the source and its position on the sky.  \citet{Riess:2018byc} found that the parallax offset can be measured directly from the Cepheids, but with a reduced precision of the distance scale from GDR2. This reduction leads to a increased uncertainty of the determination of $H_0$ value by a factor of $2.5$. 

Recently the Gaia team presented the Gaia Early Data Release 3 (EDR3) (the full Gaia DR3 release is expected in 2022) \citep{gaiacollaboration2020gaia} with improved parallaxes since GDR$2$. Using the EDR$3$ parallaxes and Cepheid PL relation the latest analysis of the SH0ES Team \citep{Riess:2020fzl} achieved a precision of $1.0\%$ in the geometric calibration of Cepheid luminosities. The precision of the geometric calibration of Cepheids  will approach $0.5\%$ by Gaia DR4 \citep{Riess:2020fzl}. This higher precision will be sufficient to confirm the present $H_0$ tension.

\item

 {\bf James Webb Space Telescope:} 
The James Webb Space Telescope (JWST or 'Webb') \citep{2004SPIE.5487..550S,Gardner:2006ky} is a joint NASA-ESA-CSA (National Aeronautics and Space Administration -European Space Agency-Canadian Space Agency) large, cold (under $50\,K$), infrared optimized ($0.6< \lambda <28.3.0\,\mu m$), space telescope and its launch is currently planned for 31 October 2021. JWST is a scientific successor to HST and will extend its discoveries to higher redshifts. It is nearly twice as big as HST with $6.6\,m$ gold-plated primary mirror much larger than $2.4\,m$ of HST. 

The two main goals of this upcoming, next-generation telescope are to look much closer to the Big Bang and to investigate the light from the first stars and galaxies that formed in the Universe \citep[see][for other goals]{Gardner:2006ky}. The observational data of this mission will essentially enhance our understanding of the formation and evolution of galaxies, stars, and planetary systems. The JWST will detect galaxies out to a redshift of $z\geq15$. It will probably be able to detect Pop III stars (see Subsection \ref{Age of the Universe}) in the high-redshift galaxies \citep{Zackrisson:2011ct,Rydberg:2012ez} in a mass range $140-260\, M_{\odot}$ as pair-instability supernovae \citep{Whalen:2012yk,Hartwig:2017iev}. Various projects using the JWST observations will provide stronger nucleosynthesis constraints inside the first supernova \citep{Bromm:2013iya} and constraints on the nature of dark matter \citep{Schultz:2014eia,Maio:2014qwa}. These constraints and other unprecedented information from JWST could potentially help address the lithium problem, explain small-scale curiosities and improve constraints on the age of Universe. 
\item
{\bf Simons Observatory:} The  Simons Observatory (SO) \citep{Galitzki:2018wvp,Ade:2018sbj,Abitbol:2019nhf} is a next generation CMB ground-based experiment. SO consists of one  $6\,m$  Large  Aperture  Telescope (LAT) and three $0.42\,m$ Small Aperture Telescopes (SATs) at the Atacama Desert, Chile. It will provide more accurate measurements of the primary CMB temperature  and polarization signals. The main targets of SO as described by \citet{Ade:2018sbj} are: primordial perturbations, effective number of relativistic species, neutrino mass, deviations from $\Lambda$CDM, galaxy evolution (feedback efficiency and non-thermal pressure in massive halos) and reionization (measurement of duration). Also a goal of SO survey is to provide a catalog of $16.000$ galaxy clusters and more than $20.000$ extragalactic sources. The sky region from SO survey overlaps with many surveys such as LSST, DES, DESI and Euclid at different wavelengths \citep{Ade:2018sbj}. This overlap is extremely beneficial as it will allow data cross correlation tests \citep[for a detailed discussion, see][]{Dodelson:2016wal}. Like CMB-S4, SO will provide improved constraints on the effective number of relativistic species $N_{eff}$, the sum of the neutrino masses $\sum m_{\nu}$ and the dark energy equation of state $w_{DE}$ \citep[ see][for the forecast constraints on cosmological parameters]{Ade:2018sbj}. Also the SO and CMB-S4 experiments will measure the primordial tensor-to-scalar ratio $r$ to a target sensitivity of $\sigma_r\sim 0.002$ (for an $r= 0$ model). This will be an improvement by a factor of approximately $5$ compared to Planck sensitivity. In addition the uncertainty of the determinations of $H_0$ from SO will be two and five times better than that inferred from Planck and local direct  measurement respectively. Therefore, the SO data will enable us to improve constraints on extensions of $\Lambda$CDM which alleviate its tensions and curiosities. In addition the improved quality of lensing data from SO as well as CMB-S4 will improve our understanding of the CMB anisotropy anomalies.
\item
{\bf SPT-3G:} This is a third generation CMB experiment \citep{SPT-3G:2014dbx,SPT-3G:2021eoc}. It uses the third survey camera SPT-3G  which was installed on the South Pole Telescope (SPT) in 2017. The  SPT-3G  with the 10-meter diameter telescope targets at least $1,500\,deg^2$ region of low-foreground sky in three spectral bands centered at 95, 150, and $220 \,GHz$ with $\sim16,000$ detectors (10 times more than its predecessor SPTpol \citealt{2012SPIE.8452E..1EA,SPT:2015htm}).

Its scientific goals aim to constrain the physics of the cosmic inflation, to explore the neutrino sector, and to constrain the relativistic energy density of the Universe \citep{SPT-3G:2014dbx,SPT-3G:2021eoc}. The SPT-3G survey in combination with the deep and wide optical survey DES, will provide detailed data on $\sim 200\,Mpc$ scales which may be used to test General Relativity. The SPT-3G  will also provide stringent and improved constraints on the effective number of relativistic species, $N_{eff}$ and on the sum of the neutrino masses $\sum m_{\nu}$ by  synergy with Planck.

\item
{\bf GAUSS:} This space mission concept combines the WL and galaxy clustering  probes using three two-point correlation functions ($3\times2$pt analysis) of gravitational lensing and galaxy positions: the cosmic shear, the galaxy clustering and galaxy-galaxy lensing.  GAUSS aims to fully map the cosmic web up to redshift $z\sim 5$ and to provide a catalog with the spectroscopic redshifts and the shapes of $10$ billions of galaxies \citep{Blanchard:2021ffq} increased by a factor of approximately $10^3$ compared to DESI which measures the spectra of $35$ million  galaxies and quasars \citep{Aghamousa:2016zmz}. The very large sky coverage and the high galaxy density provided by the GAUSS will facilitate the construction of the 3D matter power spectrum of all scales (large and small) in detail. The $3\times2$pt correlation functions in combination with 3D matter power spectrum will provide stronger constraints and break parameter degeneracies \citep{Blanchard:2021ffq}. The constraining power of the GAUSS will be an order of magnitude larger than that of  any  currently planned projects such as Euclid and Vera C. Rubin Observatory.  

\item
{\bf Laser Interferometer Gravitational-Wave Observatory:} The Laser Interferometer Gravitational-Wave Observatory (LIGO) proposed by  \citet{Abramovici:1992ah} is a large-scale experiment that uses ground-based laser interferometers with $L=L_x=L_y=4\, km$ long orthogonal arms to detect GWs\footnote{Another class of GW detectors are the resonant mass antennas \citep{Pizzella:1997hx,Maggiore:1999vm,Aguiar:2010kn,Astone:2010mr} in the frequency range from $15\,Hz$ to few $kHz$. The principle of operation of mass resonance detectors is related to the periodic dimensional changes caused by the ripple effect of GWs on solid bodies. The Weber bar \citep{Weber:1960zz} is a first generation resonant mass detector, the ALLEGRO \citep{Mauceli:1995rm}, NAUTILUS \citep{Amaldi:1990kh,Astone:1996hf,Astone:1997gi}, EXPLORER \citep{ciufolini1997proceedings}, AURIGA \citep{Cerdonio:1997hz}, NIOBE \citep{Blair:1995wx,Heng:1996np} are the second generation and Mario Schenberg \citep{Aguiar:2002eq}, MiniGRAIL \citep{deWaard:2003ug}  are the third generation.}. 

There are two identical LIGO instruments, one in Hanford (LHO) and one in Livingston (LLO) separated by roughly $3000\, km$. The principle of operation of laser interferometers \citep{Gertsenshtein:1962kfm,Moss:1971ocz} is similar to that of a simple interferometer, such as that used by Michelson and Morley. Detection of GWs with strain amplitude $h\sim 10^{-21}$ by a ground detector with arms of length $L=4\,km$ requires length change measurement \citep{Gertsenshtein:1962kfm}
\be
\Delta L=\delta L_x-\delta L_y\sim h L\sim 4\cdot 10^{-18}\,m
\ee

For the period between 2002 and 2010, the two LIGO observatories were unable to detect GWs.  The detectors were later replaced by much improved Advanced LIGO versions \citep{TheLIGOScientific:2014jea,Harry:2010zz}. The improved detectors that officially went into operation in 2015 have about ten times the sensitivity to detect GWs in the frequency range around $\sim 100\, Hz$ compared to the initial LIGO interferometers  \citep{TheLIGOScientific:2014jea}. In addition Advanced  LIGO  extends  the low frequency end from $40\,Hz$ down to $10\,Hz$. Much of the research and development work for LIGO/Advanced LIGO projects  was based on the groundbreaking work of the GEO 600 detector \citep{Willke:2002bs,Affeldt:2014rza} which is a $600\,m$ interferometer in Hanover, Germany. 

On February 11, 2016 the LIGO Scientific Collaboration \citep{Abbott:2007kv} and the Virgo Collaboration \citep{Giazotto:1988gw,Acernese:2008zza} announced  the first directly observed  GWs from a signal detected on September 14, 2015 by the Advanced LIGO devices (the Virgo was not working at the time due to an upgrade). The detected signal was named GW150914 and its source was the merger of two stellar-mass BHs \citep{Abbott:2016blz}. 

The Advanced Virgo \citep{TheVirgo:2014hva} with $3\,km$ arm length interferometer contributes to the reliability of Advanced LIGO experimental device detections allowing for greater accuracy in locating the source in the sky (triangulation i.e 3-detector localization) \citep{Singer:2015ema} and more accurate reconstruction of the signal waveform  \citep[for the relevant method, see][]{Cornish:2014kda,Littenberg:2014oda,Cornish:2020dwh,Dalya:2020gra,Ghonge:2020suv}. For example, in the case of the event GW170814 the three detectors improved the sky localization of the source, reducing the area of the $90\%$ credible region from $1160\, deg^2$ using only the two LIGO detectors to $60\,deg^2$ using all three LIGO/Virgo detectors and reduced the luminosity distance uncertainty from $570_{-230}^{+300}\,Mpc$ to $540_{-210}^{+130}\,Mpc$ \citep{Abbott:2017oio}.

In 2019 the  Advanced  LIGO \citep{TheLIGOScientific:2014jea}, the Advanced Virgo \citep{TheVirgo:2014hva}  and the Japanese successor of the Tama300 \citep{Ando:2001ej,Arai:2008zzc}, Kamioka Gravitational (KAGRA) wave detector  \citep{Aso:2013eba,Somiya:2011np,Akutsu:2020zlw} (previously called LCGT \citealt{Kuroda:2010zzb}) signed collaboration agreement to begin joint observation. The LIGO, Virgo and KAGRA collaboration will be probably complemented by other interferometers like the planned Indian LIGO by the Indian Initiative in Gravitational Wave Observations (IndIGO) consortium \citep{Unnikrishnan:2013qwa}. In addition a future third-generation ground-based detector the Cosmic Explorer (CE) \citep{Evans:2016mbw,Reitze:2019dyk,Reitze:2019iox} is envisioned to begin operation in the 2030s in the USA. It will contribute to the GW Astronomy beyond LIGO. CE with ten times longer arms ($40\,km$) than Advanced LIGO’s will amplify the amplitude of the observed signals \citep{Essick:2017wyl,Chamberlain:2017fjl} and will significantly increase the sensitivity of the observations \citep{Reitze:2019iox,Reitze:2019dyk}.

Many events ($\sim 50$ compact binary coalescences) were observed by the  Advanced LIGO/Virgo  interferometers during three  observing run  periods (O1, O2 and O3)\footnote{For the first Gravitational Wave Transient Catalog (GWTC-1) during O1 and O2, see \citet{LIGOScientific:2018mvr}, for second Gravitational Wave Transient Catalog (GWTC-2) from the first part of the third observing run (O3a), see \citet{Abbott:2020niy} and for third Gravitational Wave Transient Catalog (GWTC-3) from the second part of the third observing run (O3a), see \citet{LIGOScientific:2021djp}}. The full three-detector network provided data which enabled the standard siren measurement of the Hubble constant $H_0$ (see Subsections \ref{Standard sirens: gravitational waves}). These data are not yet sufficiently constraining the Hubble constant but in the future they are expected to improve significantly.

\item
{\bf Laser Interferometer Space Antenna:}
The Laser Interferometer Space Antenna (LISA) \citep{Caprini:2015zlo,Audley:2017drz} is a large-scale space mission proposed by ESA, planned for launch in 2034. It will consist of three spacecrafts placed in an equilateral triangle with arms $2.5$ million kilometers long which will be placed near the Earth in a heliocentric orbit. In order to pave the way for the LISA mission ESA launched LISA Pathfinder in 2015 and it was operational from 2016 to 2017 \citep{Armano:2016bkm,Armano:2017bsn}. The results from scientific research show that LISA Pathfinder works exactly five times better than required, with a successful demonstration of the basic technologies for a large gravitational wave observatory.

LISA is designed to detect GWs in the frequency range from $0.1\, mHz$ to $10^{-1}\, Hz$ \citep{AmaroSeoane:2012km,AmaroSeoane:2012je} targeting very different source populations from ground-based detectors such as LIGO, Virgo and KAGRA which operate in the frequency range\footnote{The frequency range from $0.1\, mHz$ to $10^{-1}\, Hz$ is unobservable by any proposed ground based detectors, due to seismic noise.} from $10\, Hz$ to  $10^3\, Hz$ \citep{Aasi:2013wya}.

There are many different sources of  GWs  \citep[see][for a review of GW physics]{Sathyaprakash:2009xs,Guzzetti:2016mkm,Cai:2017cbj}. LIGO and Virgo can detect the merger events of binaries with  masses $\lesssim 100 M_{\odot}$ while LISA will be able to detect the merger of massive BHs ($10^5-10^7 M_{\odot}$) with higher signal-to-noise ratio (SNR) and thus to perform precision tests in the strong gravity regime of $\Lambda$CDM model. LISA will detect events lasting weeks, months or years allowing us to observe a much larger volume of the Universe. It  may improve our understanding of the early Universe. In addition the LISA mission will be able to detect sources like primordial BHs ($\sim 10^{-12} M_{\odot}$) which correspond to the mHz frequency \citep{Bartolo:2018evs,Bartolo:2018rku,Cai:2018dig}. This possibility can help to test primordial BH dark matter scenario. Finally, LISA and the Big Bang Observer (BBO) \citep{Crowder:2005nr,Harry:2006fi}, which is a proposed LISA's successor will detect many other known or currently unknown exotic sources. Thus it will enable us to explore alternative gravity theories and to address the problems of the $\Lambda$CDM cosmology.

\item
{\bf Taiji:}
Taiji \citep{Guo:2018npi,Hu:2017mde} meaning ‘supreme ultimate’ is a Chinese large-scale space mission, planned for launch in 2033. Like LISA, Taiji is a laser interferometric GW detector which will consist of three spacecraft placed in an equilateral triangle with arms $2$ million kilometers long in orbit around the Sun. Taiji will detect GWs in the frequency range from $0.1\, mHz$ to $10^{-1}\, Hz$.  Like LISA, Taiji can detect many possible GW sources such as a stochastic GW background generated in the early Universe and the merger of two super massive BHs.

A potential LISA-Taiji network  was explored by  \citet{Ruan:2019tje,Ruan:2020smc}. This network with a separation distance of about $0.7\, AU$ can accurately localize  the sky position of a GW source and may completely identify the host galaxy.

\item
{\bf TianQin :}
TianQin \citep{Luo:2015ght,Mei:2020lrl} is a Chinese large-scale space mission. It aims to launch a laser interferometric GW detector around 2035. Like other space-base observatories, TianQin observatory consist of three spacecrafts placed in an equilateral triangle with arms $\sim0.1 \,Gm$ long but in geocentric orbit with an  orbital radius of about $10^5\,km$ \citep{Luo:2015ght,Mei:2020lrl}. TianQin aims to detect GWs in the frequency range from $10^{-4}\,Hz$ to $1\, Hz$ (overlapping with that of LISA near $10^{-4}\,Hz$ and with that of DECIGO near $1\,Hz$). It will search for GW signals from various cosmological sources such as the inspiral of supermassive BBH \citep{Feng:2019wgq}, stellar-mass BBH \citep{Liu:2020eko}, the merger of massive BBHs \citep{Wang:2019ryf}  and stochastic GW background originating from primordial BHs \citep{Gong:2017qlj} and/or cosmic strings \citep{Olmez:2010bi}. As a precursor mission of TianQin, TianQin-1 experimental satellite has been launched on 20 December 2019.  The results from scientific research shows that TianQin-1 satellite has exceeded all of its mission requirements.

\item
{\bf Deci-hertz Interferometer Gravitational wave Observatory:}
The DECi-hertz Interferometer Gravitational wave Observatory (DECIGO) is a Japanese large-scale space mission \citep{Kawamura:2011zz,Kawamura:2008zza,Kawamura:2008zz,Kawamura:2006up} which was proposed by  \citet{Seto:2001qf} and is planned for launch in 2027. DECIGO consists of four clusters (with two of them at the same position) and each cluster consists of three spacecrafts placed in an equilateral triangle with $1000\, km$ arm lengths in heliocentric orbit \citep{Kawamura:2020pcg}. As a precursor  mission of DECIGO, B-DECIGO (smaller version of DECIGO) will be launched before 2030 with $100\, km$ arm lengths orbiting around the earth at $2000\,km$ altitude above the surface of the earth \citep{Sato:2017dkf,Kawamura:2018esd,Kawamura:2020pcg}.

DECIGO is designed to detect GWs in the frequency range from $0.1\,Hz$ to $10\, Hz$ which is located in a gap between the frequency band of the LISA/Taiji and ground-based detectors such as advanced LIGO, advanced Virgo, and KAGRA. It aims to observe the primordial gravitational waves i.e. the beginning of the universe ($10^{-36}-10^{-34}\,sec$ right after the birth of the Universe), the formation of giant black holes in the center of galaxies and the compact binaries, such as white dwarf binaries \citep{Farmer:2003pa}.

\item
{\bf Einstein Telescope:}
Einstein Telescope (ET) or Einstein Observatory is a European proposed underground laser interferometric GW detector \citep{Punturo:2010zz}. It will be located underground at a depth of about  $100-300\,m$ in order to reduce the seismic noises. ET will consist of three nested detectors placed in an equilateral triangle, each in turn composed of two interferometers with arms $10\, km$ long. Using two arms in each side of the triangle will enable the determination of the polarisation of GWs. As a third-generation observatory is targeting a sensitivity 10 times better than of current second-generation laser-interferometric detectors such as advanced LIGO, advanced Virgo, and KAGRA \citep{Hild:2009ns}. ET will reduce thermal noise compared to the first and second generations of GW detectors by operating the mirrors at cryogenic temperatures as low as $10\,K$ \citep{Hild:2010id}. It is planned to start observations in 2035 with two candidate sites: north of Lula in Sardinia (Italy) and in Meuse-Rhine Euroregion (the border area of Belgium, Germany, and the Netherlands) \citep{Amann:2020jgo}.                        
ET will detect GWs in the frequency range from $\sim1\,Hz$ to $\sim10\,kHz$. This will allow the detection of BNS up to a redshift of $z\sim2$, stellar-mass BBH at  $z\sim15$, and intermediate-mass BBH ($10^2 - 10^4 M_{\odot}$) at $z\sim5$. The observations of these standard sirens will be useful to calibrate the cosmic distance ladder and will improve the estimation of the Hubble constant.

\end{itemize}
Cosmology is entering an even more exciting era! The combination of the existing puzzling observational signals discussed in this review, along with the upcoming revolutionary improvement in the quality and quantity of data creates anticipation for exciting new effects and new physics discoveries in the coming two decades.
\\

\section*{Acknowledgements}
Special thanks are due to Subir Sarkar for extensive and detailed comments.  We also  thank George Alestas, Thomas Buchert, Eoin \'O Colg\'ain, Subinoy Das, Pablo Fosalba, Lavrentios  Kazantzidis, Pavel Kroupa, Robert Piccioni, Umesh Kumar Sharma, Shahin Sheikh-Jabbari, Sunny Vagnozzi and Wen Yin for useful comments. This research is co-financed by Greece and the European Union (European Social Fund-ESF) through the Operational Programme ”Human Resources Development, Education and Lifelong Learning 2014-2020” in the context of the project ”Scalar fields in Curved Spacetimes: Soliton Solutions, Observational Results and Gravitational Waves” (MIS 5047648). \\

\appendix
\section{Appendix}
\label{List of used acronyms}

In this appendix we present the list of used acronyms. \\

\begin{center}      

\centering       
\begin{longtable*}{p{2.1cm}p{0.0001cm}p{6.4cm}p{0.0001cm}p{1.8cm}p{0.0001cm}p{6.8cm} } 
\caption{List of  used acronyms.}
\label{acronym} \\
\hhline{=======}
\\
\multicolumn{1}{c}{\textbf{Acronym}} &\multicolumn{1}{c}{} &\multicolumn{1}{c}{\textbf{Meaning}} &\multicolumn{1}{c}{}& \multicolumn{1}{c}{\textbf{Acronym}}&\multicolumn{1}{c}{} &\multicolumn{1}{c}{\textbf{Meaning}} \\ 
\\
\hhline{=======} 
\endfirsthead

\multicolumn{7}{c}%

{{\tablename\ \thetable{} -- continued from previous page}} \\
\hhline{=======} 
\\
\multicolumn{1}{c}{\textbf{Acronym}} &\multicolumn{1}{c}{} &\multicolumn{1}{c}{\textbf{Meaning}} &\multicolumn{1}{c}{}& \multicolumn{1}{c}{\textbf{Acronym}}&\multicolumn{1}{c}{} &\multicolumn{1}{c}{\textbf{Meaning}} \\ 
\\
\hhline{=======}
\\
\endhead

\hline
\multicolumn{7}{r}{{Continued on next page}} \\ 
\hline
\endfoot 

\hhline{=======}
\endlastfoot
\\
        ACS      && Advanced Camera for Surveys     && KiDS    && Kilo Degree Survey   \\
        ACT    && Atacama Cosmology Telescope   &&   LAT   &&  Large  Aperture  Telescope  \\
      AEDGE       && Atomic  Experiments for Dark Matter and Gravity Exploration      && LDE    && Late Dark Energy    \\  
       AGB      &&  Asymptotic Giant Branch && LIGO     &&Laser Interferometer Gravitational-Wave Observatory    \\ 
        AGN     &&Active Galactic Nucleus   &&  LISA   &&  Laser Interferometer Space Antenna     \\
      AIC&& Akaike Information Criterion  &&   LMC    && Large Magellanic Cloud    \\ 
       AO && Adaptive Optics&&LOS  && Line-Of-Sight    \\
    AvERA &&  Average Expansion Rate Approximation  &&     LSS  &&  Large Scale Structure    \\ 
    BAO && Baryon Accoustic Oscillations   &&    LSST&& Large Synoptic Survey Telescope    \\
      BBH      && Binary Black Holes  &&  LwMPT       && Late $w-M$ Phantom Transition    \\  
     BBN      && Big Bang Nucleosynthesis   &&   MCP     && Megamaser Cosmology Project \\ 
     BBO     && Big Bang Observer      &&  MCT     && Multi-Cycle Treasury \\
    BH   &&   Black Hole    &&  MEDE && Modified Emergent Dark Energy  \\
    BIC     && Bayesian Information Criterion &&     MGS && Main Galaxy Sample  \\
     BNS     && Binary Neutron Stars &&  MM && Many Multiplet    \\ 
     BOSS &&  Baryon  Oscillation  Spectroscopic Survey && MST && Mass Sheet Transformation   \\
     BTFR      &&Baryonic Tully Fisher  Relation &&MW&& Milky Way\\
      CC   &&    Cluster Counts  &&   NANOGrav&& North American Nanohertz Observatory for Gravitational-waves  \\
      CCH &&    Cosmic CHronometric   &&      NASA&& National Aeronautics and Space Administration    \\
     CCHP  &&  Carnegie–Chicago Hubble Program &&   NEDE     && New Early Dark Energy   \\
     CDI     && Cold Dark matter  Isocurvature      &&  NIR     && Near InfraRed   \\ 
        CDM     && Cold Dark Matter       &&    NRAO&&  National Radio Astronomy Observatory \\  
        CE       &&  Cosmic Explorer     && NS    &&   Neutron Star \\  
     CFHTLenS      && Canada-France-Hawaii Telescope Lensing&&NVSS && NRAO VLA Sky Survey  \\
     CHP&& Carnegie Hubble Program &&    &&  \\
     CL  && Confidence Level      && PEDE       && Phenomenologically Emergent Dark Energy   \\ 
    CMB && Cosmic Microwave Backgroun      &&   PL     && Period–Luminosity  \\  
     COBE &&  Cosmic Background Explorer  &&   PTAs    &&  Pulsar Timing Arrays   \\
      COSMOGRAIL     && COSmological MOnitoring of GRAvItational Lenses  &&  QFT   && Quantum Field Theory \\
       CP     &&   Cosmological Principle&&  QSO      && Quasi-Stellar Object (quasar) \\ 
       CPL     && Chevallier - Polarski - Linder  &&  ROSAT  &&  ROentgen SATellite  \\ 
      CSA&& Canadian Space Agency&&  RSD && Redshift Space Distortions \\ 
     CSP&& Carnegie Supernova Project&& RVM   && Running Vacuum Model  \\
       DE  &&  Dark Energy                   &&  SATs       && Small Aperture Telescopes  \\ 
     DEBs&& Detached Eclipsing Binary stars &&  &&  \\  
     DECIGO     && DECi-hertz Interferometer Gravitational wave Observatory     &&  SBF     && Surface Brightness Fluctuations  \\
         DES     &&  Dark Energy Survey  && SDSS      && Sloan Digital Sky Survey   \\
    DESI &&  Dark  Energy  Spectroscopic  Instrument&&  SH0ES &&Supernovae $H_0$ for the Equation of State  \\
    DIC&& Deviance Information Criterion   &&     &&  \\
     DM  && Dark Matter                 &&    SKA   && Square Kilometre Array  \\   
     EBL      &&  Extragalactic Background Light  &&     SLACS  && Sloan  Lens  ACS  Survey  \\ 
     eBOSS      && Extended Baryon Oscillation Spectroscopic Survey  && SM      &&  Standard Model   \\
      EDE  && Early Dark Energy && SMBH    && SuperMassive  Black Hole \\
   EDGES && Experiment  to  Detect  the  Global  Epoch of Reionization Signature &&  SnIa && Supernova Type Ia   \\
      EDR     && Early Data Release &&  SneII &&  Supernovae Type II  \\
   EDS&&  Early Dark Sector&&  &&  \\
     EFTofLSS   && Effective Field Theory of Large-Scale Structure  &&   SNR    && Signal-to-Noise Ratio\\
     EM     &&  ElectroMagnetic    &&       SO &&Simons Observatory     \\
         EPTA &&  European Pulsar Timing Array   &&  SOLME && Stanford Optically Levitated Microsphere Experiment         \\
        eROSITA && extended ROentgen Survey with an  Imaging Telescope Array    &&  SPH&&  Smooth Particle Hydrodynamics   \\  
         ESA    &&  European Space Agency    &&    SPT  &&  South Pole Telescope    \\  
      ET   && Einstein Telescope  &&    STRIDES && STRong-lensing Insights into Dark Energy Survey  \\
       ETHOS    &&  Effective THeory Of Structure formation&& TBTF && Too Big To Fail  \\  
        FJ&&  Faber–Jackson  &&  TD     &&  Time-Delay    \\
     FLRW     &&Friedmann-Lema$\hat{ı}$tre-Roberson-Walker  &&   TDCOSMO    && Time-Delay COSMOgraphy    \\
 FMOS     &&  Fiber Multi-Object Spectrograph  &&   TDE    && Transitional Dark Energy  \\    
        FP     && Fundamental Plane   &&    TFR    &&  Tully-Fisher Relation  \\
       GAMA     && Galaxy  and Mass Assembly  &&  TGSS && TIFR GMRT Sky Survey  \\
         GAUSS&&    Gravitation And the Universe from large Scale-Structures  &&   TIFR  &&   Tata Institute of  Fundamental  Research   \\  
        GDR   && Gaia Data Release  &&   TPCF   &&  Two-Point Correlation Functions   \\
         GEDE  &&  Generalised Emergent Dark Energy  &&  TRGB     &&Tip of the Red Giant Branch   \\
         GEHR     && Giant Extragalactic  HII Region     &&  tSZ     && thermal Sunyaev-Zel’dovich   \\
     GGL   &&  Galaxy-Galaxy Lensing     &&UV && Ultraviolet   \\
     GMRT    &&Giant Metrewave Radio Telescope &&    UVES   && Ultraviolet and Visual Echelle  Spectrograph   \\
      GP  &&  Gaussian Process &&  VCDM     &&  Vacuum Cold Dark Matter   \\  
     GR    &&  General Relativity   &&  VHE    && Very High Energy \\  
     GRB   && Gamma-Ray Burst   &&  VIKING   &&VISTA Kilo-Degree Infrared Galaxy   \\ 
       GW     && Gravitational Waves  &&   VIMOS       &&  VIsible MultiObject Spectrograph     \\
      GWTC    && Gravitational Wave Transient Catalog   &&    VIPERS  && VIMOS Public Extra-galactic  Redshift Survey   \\
         HMF     && Halo Mass Function   &&  VISTA      && Visible and Infrared Survey Telescope for Astronomy \\
       HSC       &&  Subaru Hyper Suprime-Cam lensing survey  &&    VLT       && Very Large Telescope  \\
      HST     && Hubble Space Telescope&& VM  &&  Vacuum Metamorphosis   \\ 
       H$0$LiCOW     && $H_0$ Lenses in COSMOGRAIL’s  Wellspring   &&VSF   && Violent Star Formation \\
       ICM     &&  IntraCluster Medium  &&   VVDS       &&  VIMOS-VLT Deep  Survey  \\
     IDE      && Interacting Dark Energy  &&  WISE    && Wide-field Infrared  Survey  Explorer      \\
      IGM    &&InterGalactic Medium   &&   WL  &&Weak Lensing    \\
    ISW        &&  Integrated Sachs–Wolfe && WMAP && Wilkinson Microwave Anisotropy Probe   \\
       JWST  &&   James Webb Space Telescope   && WtG     &&  Weighting the Giant \\
      IndIGO       &&Indian Initiative in Gravitational wave Observations consortium && ZTF  && Zwicky Transient Facility  \\
     IPTA&& International Pulsar Timing Array&& $2$dFGRS&& $2$-degree Field Galaxy Redshift Survey \\
     JLA&& Joint Light-curve
Analysis&&      $2$dFlenS     &&  $2$-degree Field Lensing Survey    \\
      KAGRA     &&  Kamioka Gravitational   &&     $6$dFGS     &&  $6$-degree Field Galaxy Survey       \\

\end{longtable*}

\end{center}

\bibliography{bibliography}

\end{document}